\documentclass{elsarticle}
\usepackage{graphicx}
\usepackage{rotating}
\usepackage{amsmath}
\usepackage{amssymb}
\usepackage{xspace,color}
\usepackage{natbib}
\usepackage[linktocpage=true,unicode=true]{hyperref}
\usepackage{hypcap}
\usepackage[english,francais]{babel}

\graphicspath{{./Figs/}}
\newtheorem{e-proposition}[theorem]{Proposition}

\newtheorem{e-definition}[theorem]{Definition\rm}

\def\og{\leavevmode\raise.3ex\hbox{$\scriptscriptstyle\langle\!\langle$~}}
\def\fg{\leavevmode\raise.3ex\hbox{~$\!\scriptscriptstyle\,\rangle\!\rangle$}}
\newcommand{\beq}{\begin{equation}}
\newcommand{\eeq}{\end{equation}}
\newcommand{\bea}{\begin{eqnarray}}
\newcommand{\ena}{\end{eqnarray}}
\newcommand{\ie}{{\it i.e.}}
\newcommand{\lsim}{\mathrel{\mathop{\kern 0pt \rlap
{\raise.2ex\hbox{$<$}}}
\lower.9ex\hbox{\kern-.190em $\sim$}}}
\newcommand{\gsim}{\mathrel{\mathop{\kern 0pt \rlap
{\raise.2ex\hbox{$>$}}}
\lower.9ex\hbox{\kern-.190em $\sim$}}}
\newcommand{\aaa}{\hspace{0.3cm}}

\newcommand{\pbar}{\mbox{$\bar{\rm p}$}}

\newcommand{\nbar}{\mbox{$\bar{\rm n}$}}

\newcommand{\kX}{\mbox{$\mathbf{k}_{\rm X}$}}

 \newcommand{\kpbar}{\mbox{${\mathbf k}_{\pbar}$}}
 \newcommand{\knbar}{\mbox{${\mathbf k}_{\nbar}$}}
 \newcommand{\kdbar}{\mbox{${\mathbf k}_{\bar{\rm D}}$}}
\newcommand{\kdif}{\mbox{$\mathbf{\Delta}$}}
\newcommand{\STOT}{\mbox{$\sigma_{\rm tot}$}}

 \newcommand{\md}{\mbox{$m_{\bar{\rm D}}$}}
\newcommand{\pc}{\mbox{$P_{\rm coal}$}}

\newcommand{\nC}{\mbox{$n_{\chi}$}}
\newcommand{\Ep}{\mbox{$E_{\rm p}$}}
\newcommand{\Epbar}{\mbox{$E_{\pbar}$}}
 \newcommand{\Edbar}{\mbox{$E_{\bar{\rm D}}$}}

\newcommand{\mC}{\mbox{$m_{\chi}$}}

\newcommand{\lD}{\mbox{$\lambda_{\rm D}$}}

\definecolor{cyan}{cmyk}{1.,0.,0.,0.5}
\definecolor{vert}{cmyk}{0.5,0.,0.5,0.5}
\definecolor{magenta}{cmyk}{0.,1.,0.,0.5}
\definecolor{verdatre}{cmyk}{0.5,0.,0.5,0.5}
\definecolor{yellow}{cmyk}{0.,0.,1.,0.0}
\definecolor{rouge}{cmyk}{0.,0.4,0.6,0.0}
\definecolor{orange}{cmyk}{0.,0.5,0.5,0.}
\definecolor{violet}{rgb}{0.5,0.,0.5}
\newlength{\boxedparwidth}
\newcounter{toto}

\newcounter{tata}

\newenvironment{boxedtext}{
\begin{center}
\begin{tabular}{|@{\hspace{.5cm}}c@{\hspace{.5cm}}|}
\hline \\
\begin{minipage}[t]{\boxedparwidth}
\setlength{\parindent}{1cm}}
{\end{minipage} \\
\hline
\end{tabular}
\end{center}}
\newcommand{\ben}{\begin{eqnarray}}
\newcommand{\een}{\end{eqnarray}}
\newcommand{\bi}{\begin{itemize}}
\newcommand{\ei}{\end{itemize}}
\newcommand{\nn}{\nonumber}

\newcommand{\etc}{\mbox{\it etc.}}

\newcommand{\eg}{\mbox{\it e.g.}}

\newcommand{\citeeq}[1]{Eq.~(\ref{#1})}

\newcommand{\citesec}[1]{Sect.~\ref{#1}}

\newcommand{\citefig}[1]{Fig.~\ref{#1}}

\newcommand{\mybra}[1]{\mbox{$\langle {#1} |$}}
\newcommand{\myket}[1]{\mbox{$| {#1} \rangle$}}

\newcommand{\mplanck}{m_{\rm Pl}}

\newcommand{\mchi}{\mbox{$m_{\chi}$}}
\newcommand{\mchisq}{\mbox{$m_{\chi}^{2}$}}
\newcommand{\ohh}{\mbox{$\Omega_{\chi}h^2$}}

\newcommand{\sigv}{\mbox{$\langle \sigma_{\rm ann} v \rangle$}}
\newcommand{\rhosun}{\mbox{$\rho_\odot$}}

\newcommand{\mymean}[1]{\mbox{$ \langle {#1} \rangle $}}

\begin{document}
% You can place here the title of the dossier, if you know it,
% firstly in English, then in French
%\centerline{Dark Matter and Dark Energy/Mati\`ere et \'energie noires}
%
%%%%%%%%%%%%%%%%%%%%%%%%%%%%%%%%%%%%%%%%%%%%%%%%%%%%%%%%%%%%%%%%%%%%%%%%%%%%%%%%%%%%%%%%%%%%%%%%%%%
%%%%%%%%%%%%%%%%%%%%%%%%%%%%%%%%%%%%%%%%%%%%%%%%%%%%%%%%%%%%%%%%%%%%%%%%%%%%%%%%%%%%%%%%%%%%%%%%%%%
%
\begin{frontmatter}
% \author{Name\thanksref{label2}}
\selectlanguage{english}
\title{Dark Matter Indirect Signatures\\
--- \\
Signatures indirectes des Particules de Mati\`ere Noire}
% If all authors are at the same address, the [label1] can be suppressed
%
\selectlanguage{english}

\author[lupm]{Julien Lavalle}
\ead{lavalle@in2p3.fr}
\address[lupm]{Laboratoire Univers \& Particules de Montpellier (LUPM)\\
CNRS-IN2P3 \& Universit\'e Montpellier II (UMR-5299),
Place Eug\`ene Bataillon\\
F-34095 Montpellier Cedex 05 --- France}

\author[lapth]{Pierre Salati}
\ead{salati@lapp.in2p3.fr}
\address[lapth]{LAPTh, CNRS \& Universit\'e de Savoie\\ 
9, Chemin de Bellevue B.P.110, 
74941 Annecy-le-Vieux Cedex, France\\
Tel. +33 (0)4 50 09 16 90 and Fax +33 (0)4 50 09 89 13}

%
% \address[authorlabel2]{Address2}
%
%%%%%%%%%%%%%%%%%%%%%%%%%%%%%%%%%%%%%%%%%%%%%%%%%%%%%%%%%%%%%%%%%%%%%%%%%%%%%%%%%%%%%%%%%%%%%%%%%%%
%
\begin{abstract}
The astronomical dark matter could be made of weakly interacting and massive
particles. If so, these species would be abundant inside the Milky Way, where
they would continuously annihilate and produce cosmic rays. Those annihilation
products are potentially detectable at the Earth, and could provide indirect
clues for the presence of dark matter species within the Galaxy. We will
review here the various cosmic radiations which the dark matter can produce.
We will examine how they propagate throughout the Milky Way and compare the
dark matter yields with what pure astrophysical processes are expected to
generate. The presence of dark matter substructures might enhance the signals
and will be briefly discussed.
%
% {\it To cite this article: A. Name1, A. Name2, C. R. Physique 6 (2005).}
%
\vskip 0.5\baselineskip
\selectlanguage{francais}
\noindent{\bf R\'esum\'e}
\vskip 0.5\baselineskip
\noindent
%
%{\bf Signatures indirectes des particules de mati\`ere noire.}
%\keyword{Astronomical dark matter; Cosmic rays; Galactic antimatter; Gamma-rays; Neutrinos}
%\end{abstract}

\selectlanguage{francais}
%\begin{abstract}
%
La mati\`ere noire astronomique pourrait \^etre constitu\'ee de particules
massives aux interactions \'evanescentes. Si tel \'etait le cas, ces
particules se retrouveraient en abondance au sein de la Voie Lact\'ee
o\`u elles s'annihileraient en permanence, produisant de multiples
radiations cosmiques. Celles-ci sont \'eventuellement visibles de la
Terre et constituent d\`es lors des sortes d'empreintes spectrales,
v\'eritables signatures indirectes des candidats potentiels \`a la
mati\`ere noire galactique. Dans cet article, nous passons en revue
les diff\'erentes esp\`eces cosmiques susceptibles d'\^etre produites,
et comparons leur flux avec celui des radiations engendr\'ees par les
processus astrophysiques conventionnels. L'existence de condensations
de mati\`ere noire est bri\`evement discut\'ee. Le taux d'annihilation
pourrait \^etre amplifi\'e au sein de telles structures, conduisant \`a
des signatures indirectes plus intenses que dans le cas d'un halo
galactique lisse.
%
% {\it Pour citer cet article~: A. Name1, A. Name2, C. R. Physique 6 (2005).}
%
% Now keywords/mots-clés
\vskip 0.5\baselineskip \noindent
{\small{\it Mots-cl\'es~:} Mati\`ere noire astronomique~; Rayons cosmiques~;
Anti-mati\`ere galactique~; Rayons gamma~; Neutrinos}}
\end{abstract}
%
%%%%%%%%%%%%%%%%%%%%%%%%%%%%%%%%%%%%%%%%%%%%%%%%%%%%%%%%%%%%%%%%%%%%%%%%%%%%%%%%%%%%%%%%%%%%%%%%%%%
%
\end{frontmatter}
%
%%%%%%%%%%%%%%%%%%%%%%%%%%%%%%%%%%%%%%%%%%%%%%%%%%%%%%%%%%%%%%%%%%%%%%%%%%%%%%%%%%%%%%%%%%%%%%%%%%%
%%%%%%%%%%%%%%%%%%%%%%%%%%%%%%%%%%%%%%%%%%%%%%%%%%%%%%%%%%%%%%%%%%%%%%%%%%%%%%%%%%%%%%%%%%%%%%%%%%%
%
\selectlanguage{english}
%
%%%%%%%%%%%%%%%%%%%%%%%%%%%%%%%%%%%%%%%%%%%%%%%%%%%%%%%%%%%%%%%%%%%%%%%%%%%%%%%%%%%%%%%%%%%%%%%%%%%
%%%%%%%%%%%%%%%%%%%%%%%%%%%%%%%%%%%%%%%%%%%%%%%%%%%%%%%%%%%%%%%%%%%%%%%%%%%%%%%%%%%%%%%%%%%%%%%%%%%
%

{preprint: LAPTH-020/12 and LUPM:12-024}

\section{The messengers of dark matter annihilation}
\label{sec:introduction}

%
%General introduction.
%
The nature of the astronomical dark matter (DM) is still unknown.
This component, which contributes a quarter to the energy balance of
the universe, cannot be made of atoms and nuclei like ordinary matter.
An exciting, and quite plausible possibility, lies in the existence of
a population of weakly interacting and massive particles, dubbed the
acronym WIMP. Should DM species pervade the halo of the Milky Way,
their mutual annihilations would yield several indirect signatures.
These are potentially detectable at the Earth under the form of
spectral distortions appearing in various cosmic radiations
\beq
\chi + \chi \to q \bar{q} , W^{+} W^{-} , \ldots \to
\bar{p} , \bar{D} , e^{+} \, \gamma \, \& \, \nu's \;\; .
\eeq
Detection of the DM annihilation products has motivated the spectacular
development of several new experimental techniques.
Searches for antiprotons and positrons are performed by balloon and satellite-borne 
devices. Because the flux depends on the square of the WIMP density $\nC$, the 
limit which may be set on the annihilation cross section scales very roughly 
as $m_{\chi}^{2}$. That type of search is a priori mostly sensitive to small 
WIMP masses. Actually, the recent discovery of a positron excess above 
10~GeV by the PAMELA collaboration has triggered a lot of excitement 
in the field, and many new possibilities have been explored, which may 
lead to a strong signal even in the TeV window. Alas, the DM interpretation 
of the positron excess raises more problems than it solves, and the current
explanation of that anomaly lies in the existence of nearby pulsars.
High-energy photons are detected both by air Cherenkov telescopes (ACT)
and by satellite-borne instruments. The WIMP annihilation rate
and hence the gamma-ray signal both scale as $m_{\chi}^{-2}$. Because of
the background in which that signal is swamped, the experimental reach
on the annihilation cross section approximately scales as the mass $\mC$.
A smoking gun signature of the presence of DM particles would be the
detection of a hot spot in the gamma-ray sky.
The neutrino channel is mostly sensitive to large values of $\mC$.
The limit which may be set on the annihilation cross section does not
depend too much on the WIMP mass and this channel is complementary to
the other searches. Large areas under the ice cap of the South Pole
are equipped for the detection of high-energy up-going muons, and
the theoretical predictions start to be within reach and will soon
be checked.

%
%Annihilation of WIMPS and source term.
%
Various species can be produced by WIMP annihilations among which are
antimatter cosmic rays, high-energy photons and neutrinos. The
corresponding rate $q_{\rm DM}({\mathbf x} , E)$ for the production
of these particles depends on their energy $E$ and is related to the
WIMP annihilation cross section $\sigma_\mathrm{ann}$ through
\beq
q_{\rm DM}({\mathbf x},E) \, = \, \eta \;
{\left\langle \sigma_\mathrm{ann} v \right\rangle} \,
\left\{ {\displaystyle \frac{\rho({\mathbf x})}{m_{\chi}}} \right\}^{2} \,
f(E) \;\; .
\label{DM_source}
\eeq
The coefficient $\eta$ is a quantum factor equal to $1/2$ for a
self-conjugate particle like a Majorana fermion or to $1/4$ otherwise.
The annihilation cross section is averaged over the momenta of the
incoming DM particles to yield
${\left\langle \sigma_\mathrm{ann} v \right\rangle}$, whose value
depends on the specific microscopic interactions involved in the
annihilation process.
The DM density at location ${\mathbf x}$ is denoted by $\rho({\mathbf x})$,
while $f(E)$ stands generically for the energy distribution $dN/dE$ of the
species generated in a single annihilation event.

%
%Outline of the various sections.
%
Once produced, the cosmic radiations propagate within the Milky Way
and eventually reach the Earth. A key issue is the transport of charged
cosmic rays throughout the magnetic fields of the Galaxy. An overview
of the subject is given in section~\ref{sec:CR_model}.
This allows us to focus on antimatter particles. Although produced by
conventional astrophysical processes, these species are not abundant.
They are promising targets insofar as their spectra could be distorted
should additional (and exotic) sources, of DM origin in our case, operate.
Sections~\ref{sec:pbar}, \ref{sec:positrons} and \ref{sec:Dbar} are
respectively devoted to antiprotons, positrons and antideuterons.
Attention is paid to the astrophysical backgrounds inside
which the DM signals are buried. The PAMELA positron excess is also
discussed, and its interpretation in terms of WIMPs is shown to have
difficulties.
The gamma-ray and neutrino skies are the subjects of 
sections~\ref{sec:gammas} and \ref{sec:neut}.
Finally, the possibility that DM is clumpy is rapidly examined in
section~\ref{sec:pamela_vs_wimps}. The existence of substructures has
been extensively used in the literature as a pretext for arbitrarily
enhancing the fluxes from DM origin. These are generally depressingly
weaker than the backgrounds inside which they are hidden. However, a
recent analysis has established that the boost factor from DM clumpiness
cannot exceed on average a generous factor of $\sim 20$. Alternatively,
nothing precludes a DM clump to lie close to the Earth and produce
locally an intense flux of cosmic radiations. But the odds for this
to happen are not large.

For a more extensive view of the dark matter enigma, we refer the reader to
the complementary contributions of Silk \cite{silk}, Hooper \& Tait \cite{hoopertait}, 
and Armengaud \cite{armengaud} in the present issue.

%%%%%%%%%%%%%%%%%%%%%%%%%%%%%%%%%%%%%%%%%%%%%%%%%%%%%%%%%%%%%%%%%%%%%%%%%%%%%%%%%%%%%%%%%%%%%%%%%%%
%%%%%%%%%%%%%%%%%%%%%%%%%%%%%%%%%%%%%%%%%%%%%%%%%%%%%%%%%%%%%%%%%%%%%%%%%%%%%%%%%%%%%%%%%%%%%%%%%%%
%
\section{An overview of cosmic ray transport}
\label{sec:CR_model}
%
%
%We disregard photons and neutrinos in this discussion.
%
We will focus this section on the transport of charged cosmic rays.
Photons and neutrinos propagate along straight lines, and their fluxes
at a distance $r$ from a point source decrease as $r^{-2}$. Some
extinction can take place along the line of sight though, but this
is negligible for gamma-rays in the GeV to TeV energy range. The situation
for neutrinos is more complicated as oscillations among the various
families is known to occur in empty space, and is even resonantly
enhanced inside the Sun and the Earth. Neutrino transport is broached
in section~\ref{subsec:neut_osc}.

%
%Processes involved in cosmic ray transport.
%
Once charged cosmic rays have been accelerated, or created by spallations
of other nuclear species or through DM annihilation, they propagate through
the Galactic magnetic field and are deflected by its irregularities~: the
Alfv\'en waves. In the regime where the magnetic turbulence is strong, which
is actually the case for the Milky Way, cosmic ray (CR) transport needs to
be investigated numerically. Monte Carlo simulations \cite{2002PhRvD..65b3002C}
indicate that it is similar to space diffusion with a coefficient
\beq
K(E) \, = \, K_{0} \; \beta \;
\left( {\mathcal R}/{\rm 1 \; GV} \right)^{\delta} \;\; ,
\label{space_diffusion_coefficient}
\eeq
which increases as a power law with the rigidity ${\mathcal R} = {p}/{q}$
of the particle.
In addition, because the scattering centers drift inside the Milky Way with
a velocity $V_{a} \sim$ 20 to 100 km s$^{-1}$, a second order Fermi mechanism
is  responsible for some mild diffusive reacceleration. Its coefficient $K_{EE}$
depends on the particle velocity $\beta$ and total energy $E$, and is
related to the space diffusion coefficient $K(E)$ through
\beq
K_{EE} \, = \,
{\displaystyle \frac{2}{9}} \; V_{a}^{2} \;
{\displaystyle \frac{E^{2} \beta^{4}}{K(E)}} \;\; .
\eeq
In the case of positrons, diffusive reacceleration is completely dominated
by energy losses.
Finally, Galactic convection wipes cosmic rays away from the disk with a velocity
$V_{C} \sim$ 5 to 15 km s$^{-1}$.
%
%FFFFFFFFFFFFFFFFFFFFFFFFFFFFFFFFFFFFFFFFFFFFFFFFFFFFFFFFFFFFFFFFFFFFFFFFFFFFFFFFFFFFFFFFFFFFFFFFFF
\begin{figure}[h!]
\begin{center}
%\vskip 0.5cm
\noindent
\includegraphics[width=0.9\textwidth]{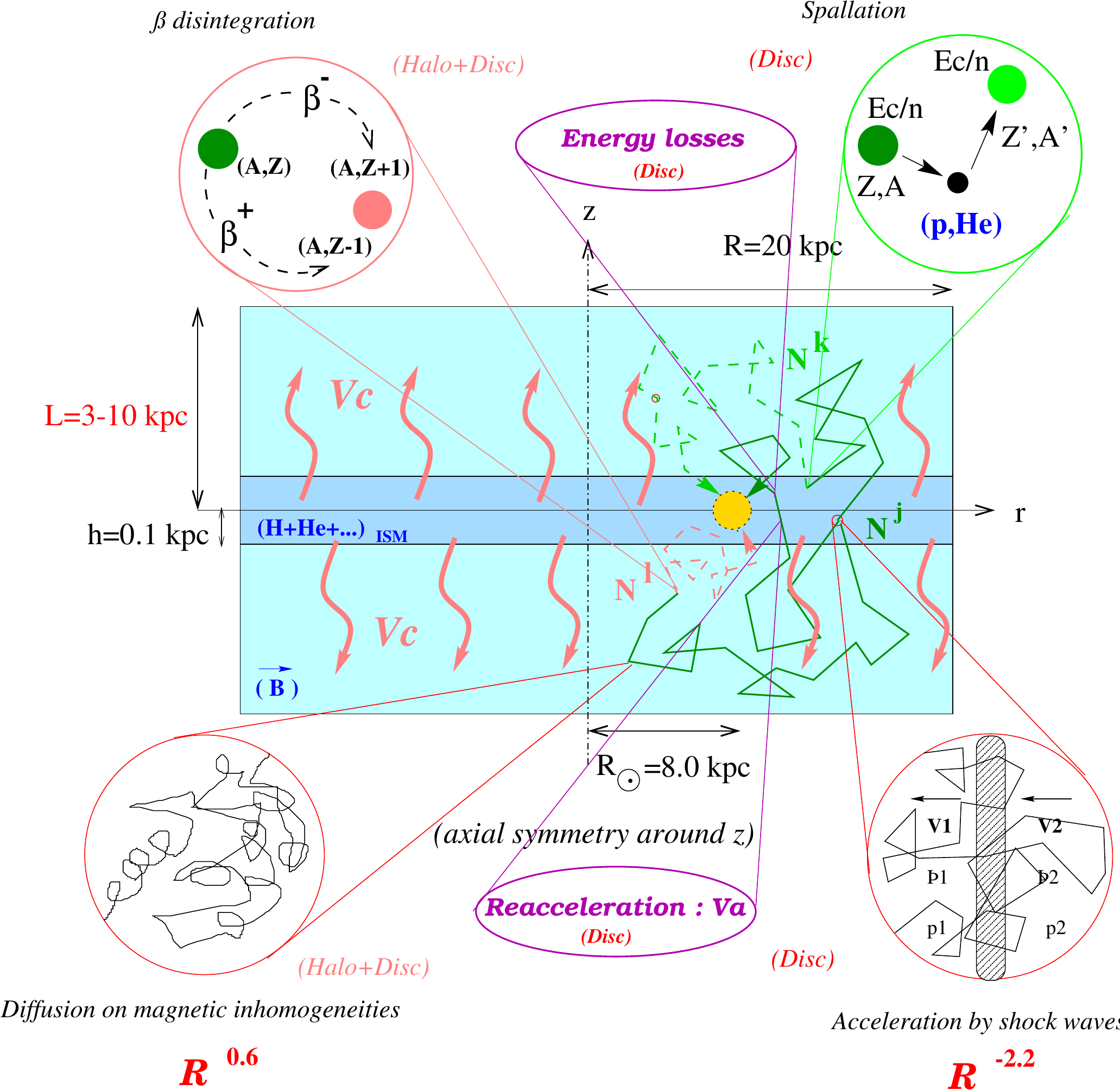}
\end{center}
%\vskip -0.5cm
\caption{
Schematic edge-on view of the Milky Way diffusive halo (DH) as seen by a cosmic ray
physicist. The stellar and gaseous disk is sandwiched between two thick layers which
contain turbulent magnetic fields.
After having been accelerated by supernova driven shock waves or produced by DM
species annihilating in the Galactic halo, cosmic rays diffuse on magnetic
inhomogeneities and are wiped away by a Galactic wind with velocity $V_{C}$.
They can lose energy and are also mildly subject to diffusive reacceleration.
The former process is by far the dominant one in the case of electrons and positrons.
This diagram has been borrowed from the review~\cite{2002astro.ph.12111M}.
}
\label{CR_camembert}
\end{figure}
%FFFFFFFFFFFFFFFFFFFFFFFFFFFFFFFFFFFFFFFFFFFFFFFFFFFFFFFFFFFFFFFFFFFFFFFFFFFFFFFFFFFFFFFFFFFFFFFFFF
%

%
%The diffusive halo or DH.
%
Radio observations indicate that the magnetic fields of galaxies are not
confined to the visible matter which they host, but extend far away in
space. An illustration is provided by the maps of the radio continuum
halo of NGC 4631, taken at 610 and 1412 MHz \cite{1977A&A....54..973E}. The synchrotron
emission from electrons spiraling inside the magnetic field of NGC 4631
probes the structure of its magnetic halo. This halo, inside which charged
particles are trapped and diffuse, extends well above and beneath the
luminous disk.
Inspired by this example, we can model the magnetic halo of our Galaxy
(the so-called diffusive halo or DH) as a thick disk which matches the
circular structure of the Milk Way, as shown in Fig.~\ref{CR_camembert}.
The Galactic disk of stars and gas, where primary cosmic rays are accelerated,
lies in the middle. Primary species, such as CR protons, helium nuclei
and electrons, are believed to be accelerated by the shock waves driven
by supernova explosions. These take place mostly in the Galactic disk
which extends radially 20 kpc from its center, and has a half-thickness $h$
of 100 pc. Confinement layers, where cosmic rays are trapped by diffusion,
lie above and beneath this thin disk of gas. The intergalactic medium starts
at the vertical boundaries $z = \pm L$, as well as beyond a radius of
$r = R \equiv 20$ kpc. Notice that the half-thickness $L$ of the diffusive halo
is not known and reasonable values range from 1 to 15 kpc.
The diffusion coefficient $K$ is assumed to be the same everywhere inside
the DH, whereas the convective velocity is exclusively vertical with component
$V_{C}(z) = V_{C} \; {\rm sign}(z)$. This Galactic wind, which is produced
by the bulk of the disk stars like the Sun, drifts away from its progenitors
along the vertical directions, hence the particular form assumed here for
$V_{C}$.

%
%The master equation.
%
The master equation for CR transport may be expressed as
\beq
{\displaystyle \frac{\partial \psi}{\partial t}} \, + \,
\partial_{z}(V_{C} \, \psi) \, - \, K  \, \Delta \psi \, + \,
\partial_{E} \{
b^{\rm loss}(E) \, \psi  \, - \, K_{EE}(E) \, \partial_{E} \psi \}
\, = \, q({\mathbf x},E) \;\; ,
\label{master_equation}
\eeq
where $\psi = dn / dE$ denotes the CR space and energy distribution function.
This relation applies to any species (protons, antiprotons or positrons)
as long as the rates for production $q$ and energy loss $b^{\rm loss}(E)$
are properly accounted for.
Most of the analyses devoted to Galactic cosmic rays are based on
the assumption that the acceleration and propagation of charged
particles have reached a steady state. The diffusion time of a 10~GeV
proton throughout the magnetic inhomogeneities of the Milky Way disk
is of order 7 million years whereas the rate of supernova explosions
is 1 to 3 every century.
%
% Diffusion timescale in MW disk is 1.80658e+01 [Myears] for a 3.00e+00 [GeV] proton
% Diffusion timescale in MW disk is 7.17970e+00 [Myears] for a 1.00e+01 [GeV] proton
% Diffusion timescale in MW disk is 1.42192e+00 [Myears] for a 1.00e+02 [GeV] proton
%
The case for steady state is made even stronger when the escape time
$\sim {L^{2}}/{K}$ from the DH as a whole is considered instead of the
Galactic disk residence time. A 10~GeV proton spends on average
$2.9 \times 10^{8}$ years before it escapes into intergalactic
space.
%
% Diffusion timescale in MW difusive halo is 7.22631e+02 [Myears] for a 3.00e+00 [GeV] proton
% Diffusion timescale in MW difusive halo is 2.87188e+02 [Myears] for a 1.00e+01 [GeV] proton
% Diffusion timescale in MW difusive halo is 5.68767e+01 [Myears] for a 1.00e+02 [GeV] proton
%
During that period, approximately 3 to 9 million supernova explosions have
taken place. It is then reasonable to describe these sources as if they
were continuously spread along the Galactic disk and were steadily
accelerating charged particles.
Different methods have been proposed to solve the CR diffusion
equation~(\ref{master_equation}). The purely numerical approach followed in
the GALPROP public code~\cite{1998ApJ...509..212S}\footnote{See the web site 
\url{http://galprop.stanford.edu/} and references therein.} is based on the well-known
Crank-Nicholson semi-implicit scheme, where the time-dependent equation
is evolved on a space and energy grid until convergence is reached. This
method allows to treat inhomogeneous cases where, for instance, $K$ or
$V_{C}$ depend on the location ${\mathbf x}$ within the DH. It is time
consuming though, and particular attention needs to be paid to the boundary
conditions. More recently, the DRAGON package \cite{2008JCAP...10..018E}\footnote{See the web site 
\url{http://www.desy.de/~maccione/DRAGON/} and references therein.} has been made
publicly available. Most of the building blocks of DRAGON come from GALPROP,
but this code relies on a faster solver.

%
%The semi-analytic methods used in USINE.
%
Another route has been taken by the USINE collaboration, based on two
semi-analytic methods.

%
%The Bessel expansion technique.
%
\noindent
{\bf (i)} The Bessel expansion technique takes advantage of the axial
symmetry of the DH and enforces a vanishing cosmic ray flux at
a distance $R = 20$ kpc from the rotation axis of the Galaxy. This
condition is actually implemented naturally by the following series
expansion for $\psi$
\beq
\psi({\mathbf x},E) \equiv \psi(r,z,E) \, = \,
{\displaystyle \sum_{i=1}^{+ \infty}} \; P_{i}(z,E) \,
J_{0}(\alpha_{i} \, r / R) \;\; .
\label{bessel_psi}
\eeq
The Bessel function of zeroth order $J_{0}$ vanishes at the points $\alpha_{i}$.
The radial dependence of $\psi$ is now taken into account by the set of its Bessel
transforms $P_{i}(z,E)$. The source term $q({\mathbf x},E) \equiv q(r,z,E)$ may
also be Bessel expanded into the corresponding functions $Q_{i}(z,E)$ so that
the master equation~(\ref{master_equation}) becomes
\beq
\partial_{z}(V_{C} \, P_{i}) \, - \,
K \, \partial_{z}^{2} P_{i} \, + \,
K    \left\{ {\displaystyle \frac{\alpha_{i}}{R}} \right\}^{2} \! P_{i} \, + \,
\partial_{E} \{ b^{\rm loss}(E) \, P_{i} \, - \,
K_{EE}(E) \, \partial_{E} P_{i} \}
\, = \, Q_{i}(z,E) \;\; .
\label{master_2}
\eeq
In the case of CR nuclei and antinuclei, energy losses and diffusive reacceleration
are confined inside the Galactic disk, considered here as infinitely thin, and an
effective term $2 \, h \, \delta(z)$ should be in factor of the energy derivative.

%
%The Green function technique.
%
\noindent
{\bf (ii)} The solution of the master equation~(\ref{master_equation}) may
also be generically expressed as the integral
\begin{equation}
\psi({\mathbf x}_{\odot},E) \, = \,
{\displaystyle \int} dE_{S}
{\displaystyle \int}_{\rm \!\! DH} \!\! d^{3}{\mathbf x}_{S} \;
G({\mathbf x}_{\odot} , E \leftarrow {\mathbf x}_{S} , E_{S}) \,
q({\mathbf x}_{S},E_{S}) \;\; .
\label{psi_convolution}
\end{equation}
The energy $E_{S}$ at the source runs over a range which depends on the
CR species and on the production mechanism. The space integral is performed
over the DH. The convolution~(\ref{psi_convolution}) involves
the Green function $G$ which describes the probability for a cosmic ray
produced at location ${\mathbf x}_{S}$ with energy $E_{S}$ to reach, for
instance, the Earth located at ${\mathbf x}_{\odot}$, where it is detected
with energy $E$.
The cosmic ray space and energy density $\psi$ can be translated into the
differential flux
$\Phi \equiv ({\beta}/{4 \pi}) \, \psi$
where $\beta$ stands for the particle velocity. This flux is expressed in
units of m$^{-2}$ s$^{-1}$ sr$^{-1}$ GeV$^{-1}$.
The flux of antinuclei or positrons produced by WIMP annihilations
may be written as the product
\beq
\Phi(\odot,E) \, = \, {\mathcal F} \,
{\displaystyle \int} dE_{S} \; f(E_{S}) \;
{I}(E , E_{S}) \;\; ,
\label{phi_CR_DM}
\eeq
where the information related to particle physics has been factored out in
\beq
{\mathcal F} \, = \, \eta \;
{\displaystyle \frac{\beta}{4 \pi}} \;
{\left\langle \sigma_\mathrm{ann} v \right\rangle} \,
\left\{ {\displaystyle \frac{\rho_{\odot}}{m_{\chi}}} \right\}^{2} \;\; .
\label{factor_F}
\eeq
The energy distribution $f(E_{S})$ describes the spectrum at the source and
depends on the details of the WIMP annihilation mechanism. The information
on the Galactic DM density profile $\rho$, as well as on the propagation of
cosmic rays within the Milky Way DH, is summarized in the halo integral
\beq
{I}(E , E_{S}) \, = \,
{\displaystyle \int}_{\rm \!\! DH} \!\! d^{3}{\mathbf x}_{S} \;
%G \left( \odot , E \leftarrow {\mathbf x}_{S} , E_{S} \right) \,
G({\mathbf x}_{\odot} , E \leftarrow {\mathbf x}_{S} , E_{S}) \,
\left\{ \! {\displaystyle \frac{\rho({\mathbf x}_{S})}{\rho_{\odot}}} \! \right\}^{2}
\;\; ,
\label{I_DH_a}
\eeq
where the solar neighborhood DM density is denoted by $\rho_{\odot}$.
The halo integral ${I}(E , E_{S})$ is a key ingredient for the derivation
of the flux at the Earth of an antimatter species produced inside
the Galactic DH by WIMP annihilations.
%
%The Green function and the horizon.
%
The spatial reach of the Green function $G$ depends on the nature of the
CR particle (antinuclei or positrons) and on the energies $E$ and $E_{S}$.
This range delineates the region of the Milky Way from which most of the
signal detected at the Earth originates. It corresponds to the extension
of the so-called horizon beyond which the Green function vanishes. The
antiproton horizon reaches the Galactic center at high energy, and the
antiproton flux at the Earth starts to be sensitive to the DM distribution
there. On the contrary, the positron horizon shrinks at high energy.
Above tens of GeV, positrons detected at the Earth originate from
its vicinity.

%
%TTTTTTTTTTTTTTTTTTTTTTTTTTTTTTTTTTTTTTTTTTTTTTTTTTTTTTTTTTTTTTTTTTTTTTTTTTTTTTTTTTTTTTTTTT
\begin{table}[h!]
\vskip 0.5cm
\begin{center}
\begin{tabular}{|c||c|c|c|c|c|c|}
\hline
Case  & $\delta$ & $K_0$ [kpc$^2$/Myr] & $L$ [kpc] & $V_{C}$ [km/s] & $V_{a}$ [km/s] \\
\hline \hline
MIN  & 0.85 &  0.0016 & 1  & 13.5 &  22.4 \\
MED  & 0.70 &  0.0112 & 4  & 12   &  52.9 \\
MAX  & 0.46 &  0.0765 & 15 &  5   & 117.6 \\
\hline
\end{tabular}
\end{center}
%\vskip -0.25cm
\caption{
Typical combinations of diffusion parameters that are compatible with the B/C
analysis \cite{2001ApJ...555..585M}. As shown in \cite{2004PhRvD..69f3501D}, these propagation
models correspond respectively to minimal, medium and maximal primary antiproton
fluxes.}
\label{tab_prop}
\vskip 0.5cm
\end{table}
%TTTTTTTTTTTTTTTTTTTTTTTTTTTTTTTTTTTTTTTTTTTTTTTTTTTTTTTTTTTTTTTTTTTTTTTTTTTTTTTTTTTTTTTTTT
%

%
%Constraining the propagation model from the B/C ratio.
%
The normalization coefficient $K_{0}$, the spectral index $\delta$, the
Galactic drift velocity $V_{C}$ and the Alfv\'en velocity $V_{a}$ are all
unknown. We have already mentioned that such is the case for the DH
half-thickness $L$. This situation can be remedied with the help of the
boron to carbon ratio B/C which is quite sensitive to CR transport
and which may be used as a constraint.
The three propagation models featured in table~\ref{tab_prop} have been drawn
from \cite{2004PhRvD..69f3501D}. The MED configuration provides the best fit to the B/C
measurements whereas the MIN and MAX models lead respectively to the minimal and
maximal allowed antiproton fluxes which can be produced by WIMP annihilations.
The three sets of parameters of table~\ref{tab_prop} belong to an ensemble
of more than 1,600 other models which have been shown in~\cite{2001ApJ...555..585M}
to be also compatible with the B/C ratio. That analysis has been recently improved
by several groups. In particular, the USINE collaboration has constrained
the CR propagation parameters with the help of a Markov chain Monte Carlo
technique~\cite{2010A&A...516A..66P} which allows to explore rapidly the parameter
space. The resulting best fit is somewhat different from what has been derived
ten years ago, but one of the important messages is the difficulty to get
a precise value for $L$ which still lies between 1 and 16~kpc.

%%%%%%%%%%%%%%%%%%%%%%%%%%%%%%%%%%%%%%%%%%%%%%%%%%%%%%%%%%%%%%%%%%%%%%%%%%%%%%%%%%%%%%%%%%%%%%%%%%%
%%%%%%%%%%%%%%%%%%%%%%%%%%%%%%%%%%%%%%%%%%%%%%%%%%%%%%%%%%%%%%%%%%%%%%%%%%%%%%%%%%%%%%%%%%%%%%%%%%%
%
%
\section{Antiprotons as a robust probe for DM species}
\label{sec:pbar}
%
%%%%%%%%%%%%%%%%%%%%%%%%%%%%%%%%%%%%%%%%%%%%%%%%%%%%%%%%%%%%%%%%%%%%%%%%%%%%%%%%%%%%%%%%%%%%%%%%%%%
%
\subsection{Calculation of the antiproton flux at the Earth}
\label{sec:pbar_calculation}
%
%The antiproton sources.
%
Antiprotons are produced during the collisions undergone by primary CR
nuclei on the interstellar gas, {\ie}, within the Galactic disk. Because they
are not directly injected in the interstellar medium (ISM) but are sourced by
primary species, these astrophysical antiprotons are dubbed secondaries. The
rate with which they are produced may be expressed as
\beq
q_{\bar{\rm p}}^{\rm sec}(r , E_{\bar{\rm p}}) \, = \,
{\displaystyle \int_{E^{0}_{\rm p}}^{+ \infty}} \,\, n_{\rm H} \times
\beta_{\rm p} \; \psi_{\rm p}(r , E_{\rm p}) \times dE_{\rm p} \times
%{\displaystyle \frac{d \sigma_{\rm p \, H \to \bar{p}}}{dE_{\bar{\rm p}}}}
%\left\{ E_{\rm p} \to E_{\bar{\rm p}} \right\}
{\displaystyle \frac{d \sigma}{dE_{\bar{\rm p}}}}(E_{\rm p} \to E_{\bar{\rm p}})
\;\; ,
\label{source_sec_pbar}
\eeq
in the case of interactions between CR protons and hydrogen atoms. The various
contributions from the spallations of interstellar H and He by CR protons and
alpha particles need to be taken into account. Details on how the cross sections
of these processes are parameterized are given in~\cite{1983JPhG....9.1289T} and
\cite{2007PhRvD..75h3006B}.
In addition to this conventional mechanism, antiprotons may be directly produced
as primary CR species by DM annihilation. The corresponding source term
$q_{\bar{\rm p}}^{\rm prim}(r,z,E)$ has already been discussed
and is generically given by expression~(\ref{DM_source}), where $f(E)$ stands
here for the antiproton spectrum $dN_{\bar{\rm p}}/dE_{\bar{\rm p}}$. Notice
that WIMP annihilations take place all over the diffusive halo and are not
restricted to the $z=0$ region. We therefore anticipate a different sensitivity
of that component to the CR propagation parameters than in the case of secondary
antiprotons.

%
%What happens then once the antiprotons have been produced.
%
Once produced, antiprotons propagate inside the DH. They can collide
elastically on interstellar H and He atoms. However, they are preferentially
scattered forward, so that these interactions are innocuous and will be
disregarded.
Antiprotons can also annihilate on interstellar H and He. This leads
to a negative source term
$- \, \Gamma_{\bar{\rm p}}^{\rm ann} \, \psi_{\bar{\rm p}}$, where the
annihilation rate $\Gamma_{\bar{\rm p}}^{\rm ann}$ is defined as
\beq
\Gamma_{\bar{\rm p}}^{\rm ann} \, = \,
\sigma_{\bar{\rm p} \, {\rm H}}^{\rm ann}  \, \beta_{\bar{\rm p}} \, n_{\rm H} \, + \,
\sigma_{\bar{\rm p} \, {\rm He}}^{\rm ann} \, \beta_{\bar{\rm p}} \, n_{\rm He} \;\; .
\eeq
The annihilation cross section $\sigma_{\bar{\rm p} \, {\rm H}}^{\rm ann}$
can be borrowed from \cite{1982PhRvD..26.1179T,1983JPhG....9..227T} and multiplied by a
factor of $4^{2/3} \sim 2.5$, taking into account the higher geometric
cross section, to get $\sigma_{\bar{\rm p} \, {\rm He}}^{\rm ann}$.
The average hydrogen $n_{\rm H}$ and helium $n_{\rm He}$ densities in
the Galactic disk can be respectively averaged to $0.9$ and $0.1$ cm$^{-3}$.
Last but not least, a tertiary component arises from the inelastic
and non-annihilating interactions which antiprotons undergo with the ISM.
Antiprotons can actually collide on a nucleon at rest and transfer enough
energy to excite it as a $\Delta$ resonance. This mechanism redistributes
antiprotons toward lower energies and flattens their spectrum as shown in
\cite{1999ApJ...526..215B}. It yields the source term
\begin{eqnarray}
q_{\bar{\rm p}}^{\rm ter}(r , E_{\bar{\rm p}}) & = &
{\displaystyle \int_{E_{\bar{\rm p}}}^{+ \infty}} \,\,
{\displaystyle
\frac{d \sigma_{\rm \bar{p} \, H \to \bar{p} \, X}}{dE_{\bar{\rm p}}}}
(E'_{\bar{\rm p}} \to E_{\bar{\rm p}}) \; n_{\rm H} \; \beta'_{\bar{\rm p}} \;
\psi_{\bar{\rm p}}(r , E'_{\bar{\rm p}}) \; dE'_{\bar{\rm p}} \nonumber \\
& - & \;\;
\sigma_{\rm \bar{p} \, H \to \bar{p} \, X}(E_{\bar{\rm p}}) \; n_{\rm H} \;
\beta_{\bar{\rm p}} \; \psi_{\bar{\rm p}}(r , E_{\bar{\rm p}}) \;\; ,
\label{tertiary}
\end{eqnarray}
where the inelastic and non-annihilating differential cross section
in this expression can be approximated by
${d \sigma_{\rm \bar{p} \, H \to \bar{p} \, X}}/{dE_{\bar{\rm p}}} \simeq
{\sigma_{\rm \bar{p} \, H \to \bar{p} \, X}}/{T'_{\bar{\rm p}}}$.
%
% \begin{equation}
% {\displaystyle
% \frac{d \sigma_{\rm \bar{p} \, H \to \bar{p} \, X}}{dE_{\bar{\rm p}}}}
% \, = \, {\displaystyle
% \frac{\sigma_{\rm \bar{p} \, H \to \bar{p} \, X}}{T'_{\bar{\rm p}}}} \;\; .
% \end{equation}
%
This parameterization can be improved by using the Anderson prescription
\cite{1967PhRvL..19..198A} as described in \cite{2005PhRvD..71h3013D,2008PhRvD..78d3506D}.
In relation~(\ref{tertiary}), the initial antiproton kinetic energy is
denoted by $T'_{\bar{\rm p}}$. In order to take into account elastic
scatterings on helium, one simply has to replace the hydrogen density by
$n_{\rm H} \, + \, 4^{2/3} \, n_{\rm He}$.

%
%How to reliably compute the background and the DM signal.
%
The antiproton background and DM signal are reliably computed with the
Bessel expansion approach. This method encodes directly the presence of
radial boundaries for the magnetic halo. The full expression for the
master equation describing the Bessel transformed antiproton distribution
function $\bar{P}_{i}(z,E)$ may be expressed as
\ben
& \partial_{z}( V_{C} \, \bar{P}_{i}) \; - \;
K \, \partial_{z}^{2} \bar{P}_{i}  + 
K \left\{
{\displaystyle \frac{\alpha_{i}^{2}}{R^{2}}} \right\} \bar{P}_{i}
\nn\\ 
& +  \, 2 \, h \, \delta(z) \,
\partial_{E} \left\{ b^{\rm loss}(E) \, \bar{P}_{i} \; - \;
K_{EE}(E) \, \partial_{E} \bar{P}_{i} \right\} \, = \nn \\
& -  \, 2 \, h \, \delta(z) \,
\Gamma_{\bar{\rm p}}^{\rm ann} \, \bar{P}_{i} \; + \;
Q_{\bar{\rm p} , i}^{\rm prim}(z,E) \; + \;
2 \, h \, \delta(z) \,
\left\{ Q_{\bar{\rm p} , i}^{\rm sec} + Q_{\bar{\rm p} , i}^{\rm ter} \right\} \;\; .
\label{master_susy}
\een
The various source terms have been specified above. Annihilations of
antiprotons take place inside the Galactic disk. This is also true for
the production of secondary and tertiary antiprotons. All those processes
involve the ISM. We have already mentioned, as a side remark on
equation~(\ref{master_2}), that energy losses and diffusive reacceleration are
also confined inside the Galactic disk. The approximation of infinitely
thinness for the disk leads to the effective term $2 \, h \, \delta(z)$
in the above expression.
Integrating it through the infinitely thin disk, along the vertical
axis $z$, leads finally to a diffusion equation
in energy which the Bessel transforms $\bar{P}_{i}(0,E)$
fulfill
\ben
& \bar{\cal A}_{i} \, \bar{P}_{i}(0,E) \; + \;
2 \, h \, \partial_{E} \left\{ b^{\rm loss}(E)
\! \right. \left. \!\! \bar{P}_{i}(0,E) \, - \,
K_{EE}(E) \, \partial_{E} \bar{P}_{i}(0,E) \right\}
\, = \nn \\
& 2 \, h \,
\left\{ Q_{\bar{\rm p} , i}^{\rm sec} + Q_{\bar{\rm p} , i}^{\rm ter} \right\}
\; + \;
2 \,
{\displaystyle \int_{0}^{L}} \, dz \;
Q_{\bar{\rm p} , i}^{\rm prim} \left( z , E \right) \;
e^{- \, {\displaystyle \frac{V_{C} z}{2 K}}} \;
{\mathcal F}_{i}(z) \;\; .
\label{master_final}
\een
The coefficients $\bar{\cal A}_{i}$ which appear in the above expression
are given by
\ben
& \bar{\cal A}_{i}(E) \, = \, V_{C} \, + \,
2 \, h \, \Gamma_{\bar{\rm p}}^{\rm ann}(E) \, + \,
K(E) \, S_{i} \coth \left\{ {\displaystyle \frac{S_{i} L}{2}} \right\}
\nn\\
&{\rm where}\;\;\;
S_{i}^{2} \, = \, ({V_{C}}/{K})^{2} + ({2 \, \alpha_{i}}/{R})^{2} \;\; ,
\een
while the vertical functions ${\mathcal F}_{i}(z)$ are defined as
\beq
{\mathcal F}_{i}(z) \, = \,
{\sinh \left\{ {\displaystyle \frac{S_{i}}{2}} (L-z) \right\}}
\, / \,
{\sinh \left\{ {\displaystyle \frac{S_{i}}{2}}  L \right\}} \;\; .
\label{F_de_i}
\eeq
For each Bessel order $i$, the integro-differential equation~(\ref{master_final})
is solved following the procedure explained in the appendix B of \cite{2001ApJ...563..172D}.
Evolving in time an initial burst with the help of a semi-implicit Crank-Nicholson
scheme allows to check the convergence of that procedure. The tertiary
component is computed by re-injecting several times the total antiproton
yield $\psi_{\bar{\rm p}}$ in the integral over energy of relation~(\ref{tertiary}).
Convergence is reached very rapidly.

%%%%%%%%%%%%%%%%%%%%%%%%%%%%%%%%%%%%%%%%%%%%%%%%%%%%%%%%%%%%%%%%%%%%%%%%%%%%%%%%%%%%%%%%%%%%%%%%%%%
%
\subsection{Antiproton background and signal}
\label{sec:pbar_fluxes}
%
%
%Background of secondary antiprotons.
%
The spallations of interstellar H and He by cosmic ray primaries, essentially
protons and alpha particles, produce an irreducible background of secondary
antiprotons inside which the signal from DM species is hidden. The precise
determination of this background is crucial in order to disentangle a possible
WIMP signature.
The semi-analytic treatment of CR propagation which has been discussed in
section~\ref{sec:CR_model}, and which is based on the Bessel
expansion~(\ref{bessel_psi}), is a convenient framework to derive the theoretical
uncertainties associated to the various parameters at stake, namely $K_{0}$,
$\delta$, $V_{a}$, $V_{C}$ and the DH half-thickness $L$. The space of these
propagation parameters has been extensively scanned~\cite{2001ApJ...555..585M} in order
to select the allowed regions where the predictions on B/C, a typical CR secondary
to primary ratio, match the observations. Several hundreds of different propagation
models have survived that test. The propagation parameters are thus only loosely
constrained by the CR nuclei abundances so far observed.
%
%FFFFFFFFFFFFFFFFFFFFFFFFFFFFFFFFFFFFFFFFFFFFFFFFFFFFFFFFFFFFFFFFFFFFFFFFFFFFFFFFFFFFFFFFFFFFFFFFFF
\begin{figure}[t!]
\begin{center}
%\vskip 0.5cm
\noindent
 \includegraphics[width=0.9\textwidth]{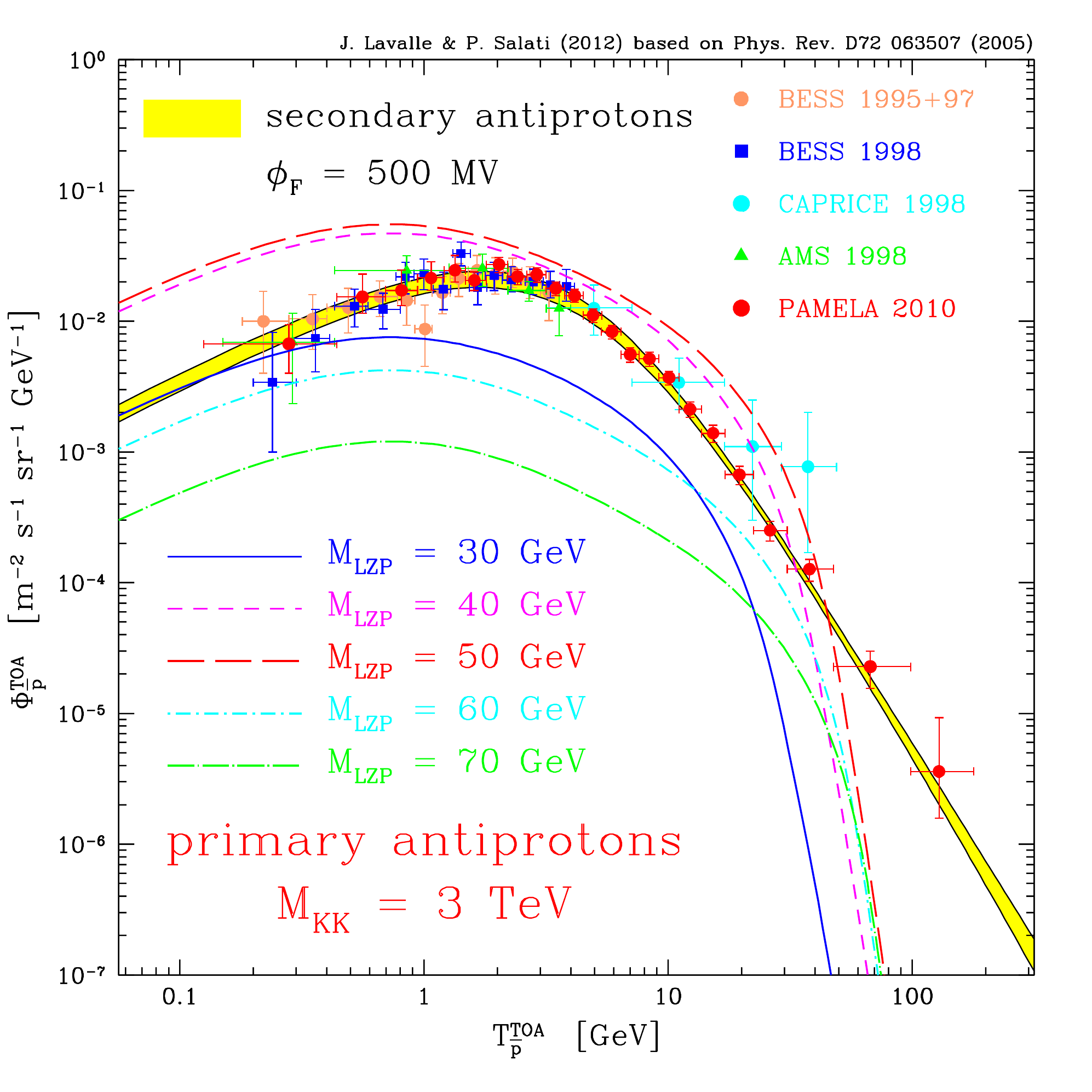}
\end{center}
%\vskip -0.5cm
\caption{
The astrophysical antiproton background lies within the yellow band whose
thickness indicates the theoretical uncertainty arising from CR propagation.
This band actually encompasses the secondary antiproton yields computed for
more than 1,600 different sets of CR propagation parameters, all compatible
with the B/C ratio.
The primary antiproton fluxes are produced by the annihilation of a Kaluza-Klein
WIMP that appears in the higher dimensional warped Grand Unified Theories of
\cite{2004PhRvL..93w1805A,2005JCAP...02..002A}. The mass of that DM species (dubbed
LZP) has been varied from 30 to 70 GeV with a Kaluza-Klein scale $M_{\rm KK}$ of
3 TeV. When the LZP mass is close to $M_{\rm Z^{0}} / 2$, the annihilation becomes
resonant and the primary signal may even exceed the background.
The MAX diffusion parameters of table~\ref{tab_prop} have been used for the LZP
antiproton spectra, with a canonical isothermal DM distribution.
This figure has been borrowed from~\cite{2005PhRvD..72f3507B}.
}
\label{fig:LZP}
\end{figure}
%FFFFFFFFFFFFFFFFFFFFFFFFFFFFFFFFFFFFFFFFFFFFFFFFFFFFFFFFFFFFFFFFFFFFFFFFFFFFFFFFFFFFFFFFFFFFFFFFFF
%
%
%Discussion of the yellow band in the antiproton figures.
%
The yellow band presented in each of the figures of this section is actually the envelope
of the secondary antiproton spectra computed with the set of those $\sim 1,600$ different
propagation models found in~\cite{2001ApJ...555..585M} to pass the B/C test. This band comprises
the theoretical uncertainty in the determination of the secondary antiproton flux. It is
confined by the MIN and MAX configurations of table~\ref{tab_prop}.
As a first observation, notice how narrow the uncertainty strip is between $\sim$ 10 and
100 GeV. In spite of this, the theoretical predictions are in remarkable agreement with
the recent antiproton measurements~\cite{2010PhRvL.105l1101A} of the PAMELA collaboration.
These span 4 orders of magnitude between 1 and 100~GeV. At higher energies, the yellow band
widens as a result of the energy dependence of the diffusion coefficient $K$. We expect
antiproton propagation to be dominated by pure diffusion in that energy range. A very
crude approximation for the antiproton Green function is obtained in the limit where
the DH is infinite and may be expressed as
\beq
G_{\bar{\rm p}}({\mathbf x}_{\odot} \leftarrow {\mathbf x}_{S}) \equiv
{\displaystyle \frac{1}{4 \, \pi \, K(E)}} \;
{\displaystyle \frac
{1}
{r_{\oplus}}} \;\; ,
\label{G_pbar_3D}
\eeq
where $r_{\oplus}$ denotes the distance between the Earth and the source. The antiproton
flux is expected to scale as ${\Phi_{\rm p}}/{K(E)} \propto E^{- \alpha - \delta}$, where
$\alpha$ is the spectral index of the proton flux at high energies. From the B/C analysis,
the spectral index $\delta$ of the diffusion coefficient $K$ may take any value between
$0.46$ and $0.85$. Its spread $\Delta \delta = 0.4$ translates for the antiproton flux
into a factor of 3 of uncertainty at 1 TeV. Notice that PAMELA and AMS-02 will considerably
improve the measurements of the CR nuclei abundances, with a determination of the B/C ratio
to a better accuracy and over a wider energy range than available so far. This will translate
into improved constraints on the propagation parameters and eventually into a thinner uncertainty
yellow strip for the antiproton astrophysical background.
%
%FFFFFFFFFFFFFFFFFFFFFFFFFFFFFFFFFFFFFFFFFFFFFFFFFFFFFFFFFFFFFFFFFFFFFFFFFFFFFFFFFFFFFFFFFFFFFFFFFF
\begin{sidewaysfigure}
\begin{center}
\centerline{
 \includegraphics[width=0.5\textwidth]{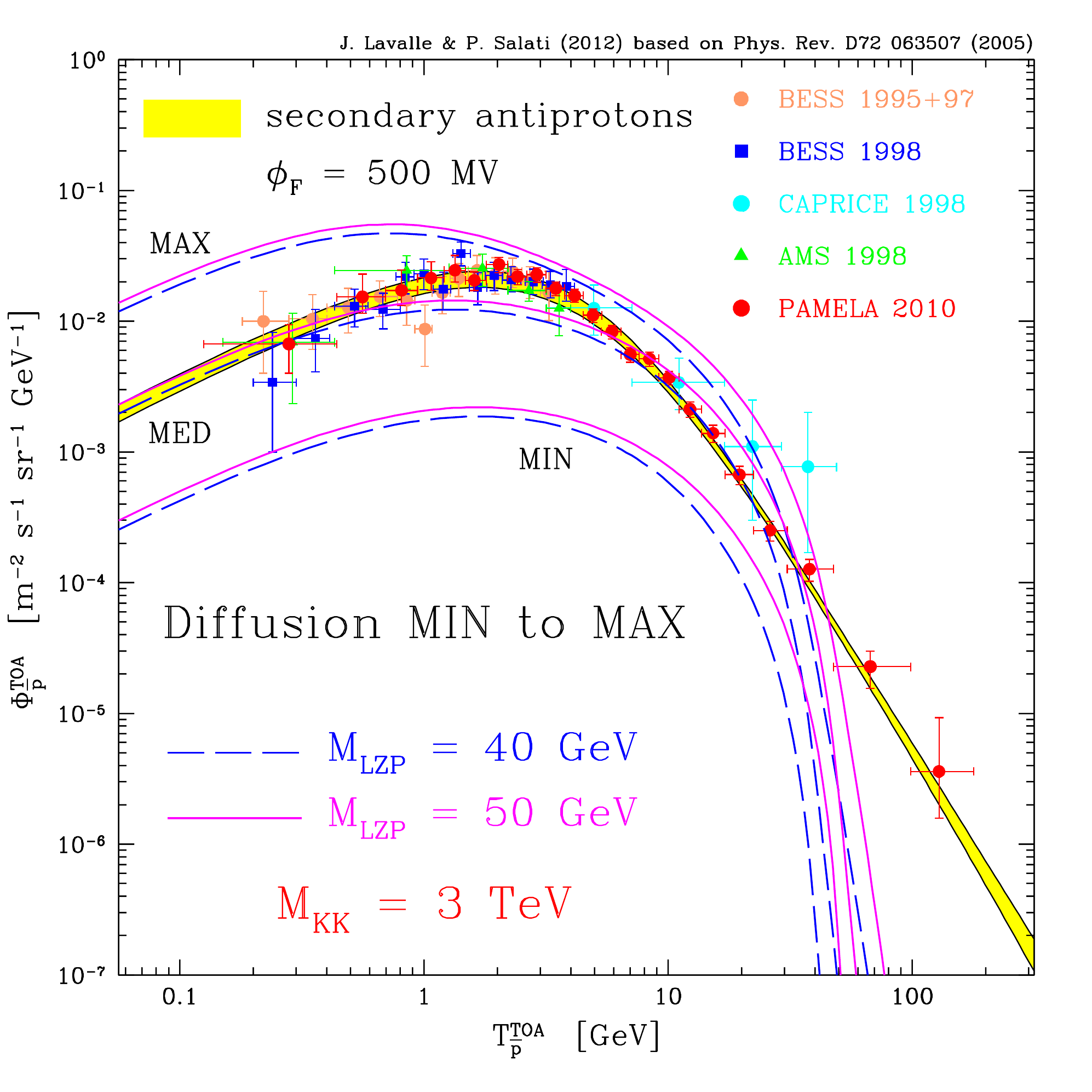}
 \includegraphics[width=0.5\textwidth]{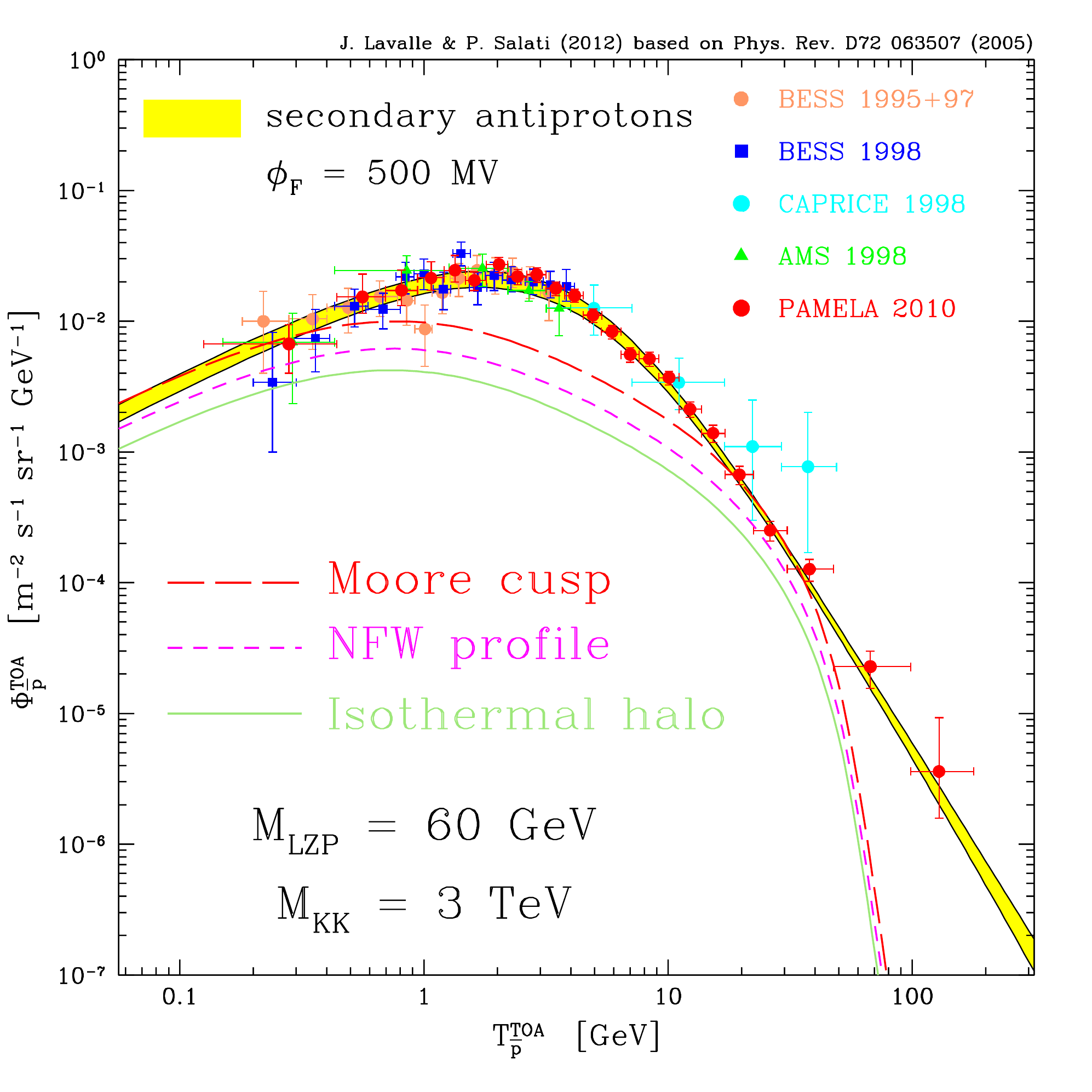}}
\end{center}
\caption{
{\bf Left panel} --
When the diffusion parameters are varied over the entire domain found to be compatible
with the B/C ratio, antiproton primary fluxes span two orders of magnitude while
the secondary component lies within the much narrower yellow band. The case of a
resonant LZP has been featured here with $M_{\rm LZP} = 40$ (blue dashed) and 50 GeV
(solid magenta). A canonical isothermal DM distribution has been assumed. The LZP signal
is well below the background in the case of the MIN model of
table~\ref{tab_prop}.
{\bf Right panel} --
The effect of the Galactic DM halo profile is investigated. The mass of the LZP has
been set equal to $M_{\rm LZP} = 60$ GeV with a Kaluza-Klein scale $M_{\rm KK}$ of 3 TeV.
The more divergent and concentrated the LZP distribution at the center of the Milky Way,
the larger the antiproton yield. That effect is particularly acute in this plot where the
MAX propagation parameters have been assumed.
These figures have been borrowed from~\cite{2005PhRvD..72f3507B}.
}
\label{fig:CR_DM_halo_effects}
\end{sidewaysfigure}
%FFFFFFFFFFFFFFFFFFFFFFFFFFFFFFFFFFFFFFFFFFFFFFFFFFFFFFFFFFFFFFFFFFFFFFFFFFFFFFFFFFFFFFFFFFFFFFFFFF
%

%
%Antiproton DM signal.
%
The antiproton signal from annihilating DM species leads to a primary component
directly produced throughout the DH. It depends on many unknown ingredients as
is clear from relation~(\ref{DM_source}).
The annihilation cross section ${\left\langle \sigma_\mathrm{ann} v \right\rangle}$
at freeze-out is related to the WIMP cosmological relic abundance through
\beq
\Omega_{\chi} h^{2} \, \simeq \,
{\displaystyle \frac{3 \times 10^{-27} \; {\rm cm^{3} \; s^{-1}}}
{\left\langle \sigma_\mathrm{ann} v \right\rangle}} \;\; .
\eeq
A value of $\Omega_{\rm \; DM} \sim 0.21$ translates into a typical WIMP annihilation
cross section of order $3~\times~10^{-26}$~cm$^{3}$~s$^{-1}$. Although this is strictly
true only at decoupling, we can keep in mind that value as a benchmark (see below though).
The antiproton spectrum $f(E)$ at the source is also model dependent. Antiprotons are
generated through quark or gauge boson jets. Since the WIMPs are at rest in the Galactic
frame, the antiproton spectrum tends to flatten at low energy. The production of tertiary
antiprotons leading to the same effect, we do expect a flat spectrum below a few GeV,
whatever the production mechanism. The primary and secondary fluxes displayed in
Fig.~\ref{fig:LZP} and \ref{fig:CR_DM_halo_effects} have actually fairly similar
spectra at low energy.
The WIMP mass is a key ingredient. The production rate $q_{\rm DM}$ scales as
${m_{\chi}^{-2}}$. The largest DM antiproton signals are generally expected
in the case of light DM candidates. For illustration, we have selected
among the realm of possible DM candidates a particular Kaluza-Klein WIMP which
appears in the context of higher dimensional warped Grand Unified Theories
\cite{2004PhRvL..93w1805A,2005JCAP...02..002A}. In these models, a stable KK fermion
can arise as a consequence of imposing proton stability in a way very reminiscent of
R-parity stabilizing the lightest supersymmetric particle in supersymmetric models.
The symmetry is called $Z_3$ and the Lightest $Z_3$ Particle (LZP) is stable since
it cannot decay into standard model particles. It is actually identified with a KK
Dirac right-handed (RH) neutrino with a mass in the 1 GeV to 1 TeV range. This RH
neutrino has gauge interactions in particular with additional KK $Z^{\prime}$ gauge
bosons. Nevertheless, its interactions with ordinary matter are feeble because they
involve heavy gauge bosons with a mass $M_{\rm KK} \gsim 3$ TeV.
The primary antiproton flux at the top of the atmosphere (TOA) is plotted in
Fig.~\ref{fig:LZP} as a function of antiproton kinetic energy for five different
values of the LZP mass. The most optimistic Galactic diffusion scheme MAX as well
as a canonical isothermal Galactic DM halo have been assumed for the primary signal.
The curves corresponding to $M_{\rm LZP} = 40$ (short dashed magenta) and 50 GeV
(long dashed red) exceed the background and should have already led to a detection
would our assumptions on Galactic diffusion and halo profile be correct. For
$M_{\rm LZP} = M_{\rm Z^{0}} / 2$, the LZP annihilation is actually driven by the
$Z$-resonance and is significantly enhanced.

%
%Discussion of figure~\ref{fig:CR_DM_halo_effects} -- CR effects.
%
The antiproton DM signal is more sensitive to Galactic CR propagation than the
secondary component. The latter is generated through the interactions of CR protons
and helium nuclei with the ISM, a process quite similar to the production of boron
through the spallation of carbon or nitrogen nuclei impinging on interstellar gas.
Although the subnuclear processes at stake are different, secondary antiprotons
share a few similarities with boron nuclei. In particular, the production site
is the Galactic disk in both cases. It is no surprise then if the yellow band of
uncertainty is so narrow once the B/C constraint has been taken into account.
Such is not the case for the antiproton DM signal which is produced all over
the magnetic halo, in regions far away from the disk. We do expect then a larger
variance for that signal than for the background as the CR propagation parameters
are varied over the range allowed by the B/C measurements. In the left panel of
Fig.~\ref{fig:CR_DM_halo_effects}, the primary antiproton flux produced by
a 40 or 50 GeV LZP species decreases by two orders of magnitude between the most
optimistic (MAX) and the most pessimistic (MIN) diffusion cases of
table~\ref{tab_prop}. In the last configuration, the antiproton signal is now well
below the background.

%
%Discussion of figure~\ref{fig:CR_DM_halo_effects} -- DM halo effects.
%
The distribution of DM inside galaxies is still an open question.
From one side, results from cosmological N-body simulations in $\Lambda$-CDM models
\cite{1997ApJ...490..493N,1999ApJ...524L..19M} indicate a universal and coreless DM density profile.
At small radii, the DM density diverges with the distance $r$ from the Galactic center
as $r^{- \gamma}$, with $\gamma \sim$ 1 to 1.5. Other results, also obtained from
simulations of halo formation, strongly disfavour a singularity as steep as 1.5 and
seem to point toward slopes logarithmically dependent on the distance from the Galactic
center and no steeper than $\sim$ 1.2~\cite{2004MNRAS.349.1039N,2004MNRAS.353..624D}.
On the other side, several analyses of rotational curves observed for galaxies of
different morphological types \cite{2004MNRAS.351..903G,2004MNRAS.353L..17D} put serious doubts
on the existence of DM cusps in the central regions of the considered objects. Instead
of a central singularity, these studies rather suggest a cored DM distribution, flattened
toward the central regions.
The DM profile inside the Milky Way can be parameterized by the generic distribution
\beq
%\rho_{\rm DM}(r) \, = \, \rho_{\odot} \,
\rho(r) \, = \, \rho_{\odot} \,
\left\{
{\displaystyle \frac{r_{\odot}}{r}} \right\}^{\gamma} \,
\left\{
{\displaystyle \frac{1 \, + \, \left( r_{\odot} / a \right)^{\alpha}}
{1 \, + \, \left( r / a \right)^{\alpha}}}
\right\}^{\left( \beta - \gamma \right) / \alpha} \;\; ,
\label{eq:profile}
\eeq
where $r_{\odot} = 8.5$ kpc is the distance of the solar system from the Galactic
center. The local DM density has been set equal to $\rho_{\odot} = 0.3$ GeV cm$^{-3}$ 
%% modif JL
--- note that 
this canonical value is slightly lower than more recent estimates made in 
\eg~Refs.~\cite{2010JCAP...08..004C,2010A&A...523A..83S,2011MNRAS.414.2446M,2012ApJ...756...89B}, 
which agree on a central value around 0.4 GeV/cm$^{-3}$.
%% end modif JL
In the case of the pseudo-isothermal profile~\cite{1980ApJS...44...73B}, the typical length scale 
$a$ is the radius of the central core. The profile indices $\alpha$, $\beta$ and $\gamma$ for
the DM distributions considered in Fig.~\ref{fig:LZP} and \ref{fig:CR_DM_halo_effects}
are indicated in table~\ref{tab:indices}.
%
%TTTTTTTTTTTTTTTTTTTTTTTTTTTTTTTTTTTTTTTTTTTTTTTTTTTTTTTTTTTTTTTTTTTTTTTTTTTTTTTTTTTTTTTTTT
\begin{table}[h!]
\vskip 0.5cm
\begin{center}
{\begin{tabular}{@{}|l||c|c|c|c|@{}}
\hline
Halo model & $\alpha$ & $\beta$ & $\gamma$ & $a$ [kpc] \\
\hline
\hline
Cored isothermal~\cite{1980ApJS...44...73B}
& {\aaa} 2 {\aaa} & {\aaa} 2 {\aaa} & {\aaa} 0 {\aaa} & {\aaa} 4 {\aaa} \\
Navarro, Frenk \& White~\cite{1997ApJ...490..493N}
&        1        &        3        &        1        &       25        \\
Moore~\cite{1999ApJ...524L..19M}
&        1.5      &        3        &        1.3      &       30        \\
\hline
\end{tabular}}
\end{center}
%\vskip -0.25cm
\caption{
Dark matter distribution profiles in the Milky Way.}
\label{tab:indices}
\vskip 0.5cm
\end{table}
%TTTTTTTTTTTTTTTTTTTTTTTTTTTTTTTTTTTTTTTTTTTTTTTTTTTTTTTTTTTTTTTTTTTTTTTTTTTTTTTTTTTTTTTTTT
%
As is clear in the right panel of Fig.~\ref{fig:CR_DM_halo_effects}, where the
case of a 60 GeV LZP is considered, the Galactic DM distribution $\rho({\mathbf x})$
is also a source of uncertainty in the calculation of the antiproton DM signal. The MAX
set of CR propagation parameters, which has been assumed in that example, makes it possible
for primary antiprotons produced at the central Galactic cusp to reach the solar circle.
The degeneracy among the various DM distributions is not lifted if the MIN propagation
model replaces the MAX configuration. In that case, the three colored curves are
one and the same.

%
%FFFFFFFFFFFFFFFFFFFFFFFFFFFFFFFFFFFFFFFFFFFFFFFFFFFFFFFFFFFFFFFFFFFFFFFFFFFFFFFFFFFFFFFFFFFFFFFFFF
\begin{figure}[t!]
\begin{center}
%\vskip 0.5cm
\noindent
 \includegraphics[width=0.9\textwidth]{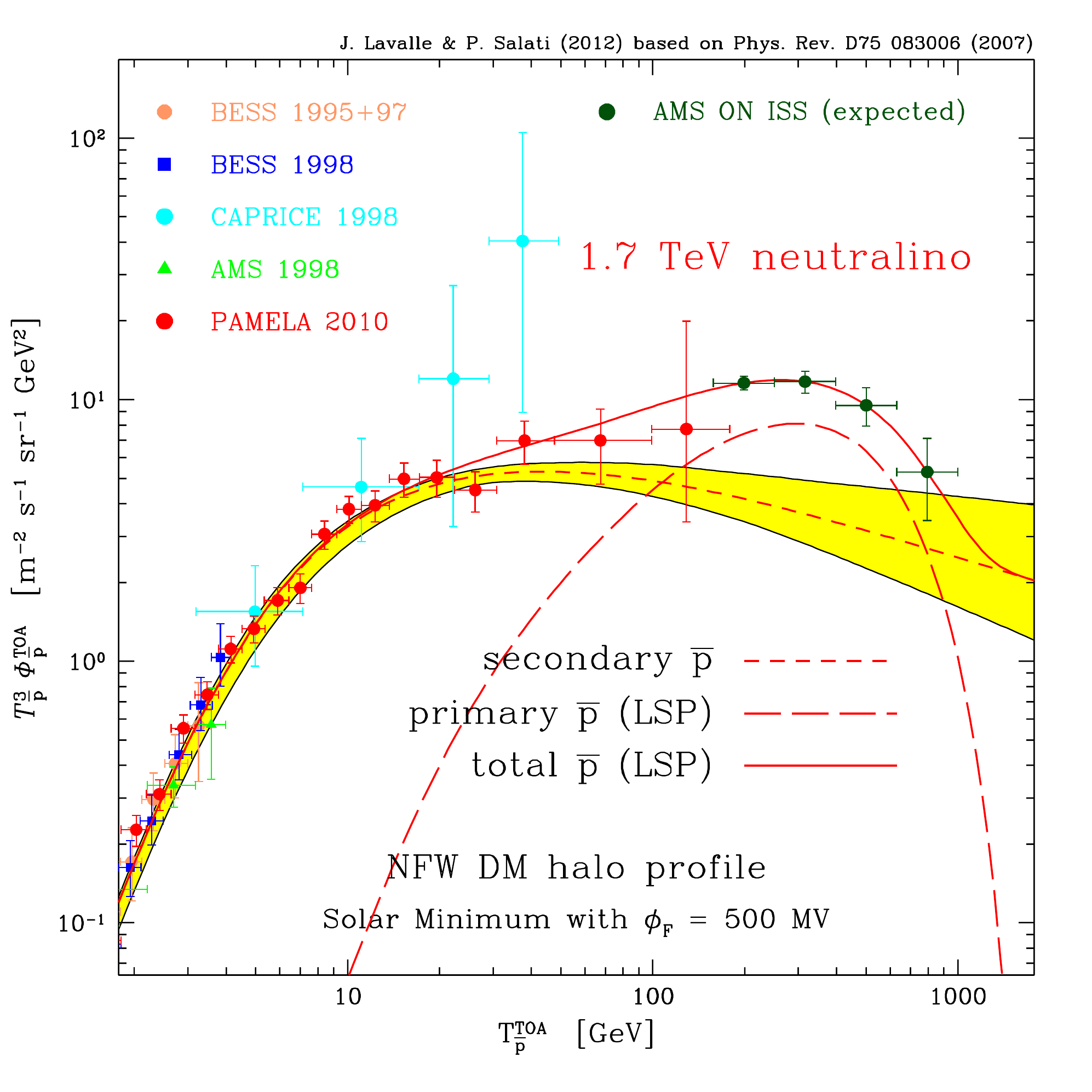}
\end{center}
%\vskip -0.5cm
\caption{
The yellow band features the expected antiproton background for the full range of
diffusion parameters allowed by the B/C ratio. A heavy WIMP is also considered.
This DM species is almost a pure Wino and its annihilation cross section is
significantly enhanced today by non-perturbative, binding energy effects.
The corresponding primary (long dashed) and total (solid) fluxes have been
derived for a NFW halo profile and for the MED set of diffusion parameters.
For illustration, a global boost factor of 2 has also been included in the
signal. The antiproton flux is compared to several measurements, whereas the
expected statistical error after 3 years of data sampling by AMS-02 is indicated.
This plot has been borrowed from~\cite{2007PhRvD..75h3006B}.
}
\label{fig:heavy_wino}
\end{figure}
%FFFFFFFFFFFFFFFFFFFFFFFFFFFFFFFFFFFFFFFFFFFFFFFFFFFFFFFFFFFFFFFFFFFFFFFFFFFFFFFFFFFFFFFFFFFFFFFFFF
%

%
%Importance des WIMPs de basse masse vis ˆ vis de COGENT.
%
Because the astrophysical background of secondary antiprotons is well under control
once the B/C constraint is taken into account, antiprotons offer a unique probe of
the presence of annihilating DM species within the Galactic halo. This is especially
true for light candidates. The DM signal is expected to be the strongest in that case. As a
matter of fact, recent observations collected by the direct detection experiments
CDMS-II~\cite{2010Sci...327.1619C} and CoGENT~\cite{2011PhRvL.106m1301A} are compatible with a
light WIMP with mass around 10~GeV. Should such a particle couple preferentially to
quarks, it would generate a sizable amount of antiprotons at low energy, and would
be excluded by observations as shown in~\cite{2010PhRvD..82h1302L}. A light neutralino
arises in supersymmetric extensions of the standard model with an extra gauge singlet
superfield. Generic models embedding an additional Majorana fermion associated to
two new scalar fields are constrained by antiprotons, although some regions of the
parameter space still pass the test \cite{2012NuPhB.854..738C}.

%
%Quid des WIMPs massifs ?
%
The annihilation rate of a heavy WIMP is suppressed by a factor of ${m_{\chi}^{-2}}$.
The antiproton DM signal is hidden in the background, unless the annihilation cross
section ${\left\langle \sigma_\mathrm{ann} v \right\rangle}$ is abnormally large.
This situation may happen in supersymmetric models where the neutralino is almost a pure
Wino, as expected for example in anomaly mediated supersymmetry breaking (AMSB) scenarios
\cite{2001JHEP...06..053U}. For Winos, the preferred mass from relic density requirements is 
peaked at about 1.7 TeV \cite{2004PhRvD..70i5004P}. Non-perturbative binding energy effects 
then result in greatly enhanced annihilation cross sections today, when the neutralinos have 
very small Galactic velocities \cite{2004PhRvL..92c1303H,2005PhRvD..71f3528H}. In this limit, heavy Winos annihilate 
almost exclusively into gauge bosons.
The case of a heavy Wino particle is featured in Fig.~\ref{fig:heavy_wino}. The
annihilation cross section today ${\left\langle \sigma_\mathrm{ann} v \right\rangle}$
is equal to $1.02 \times 10^{-24}$ cm$^3$ s$^{-1}$~\cite{2007PhRvD..75h3006B}, well above
the decoupling value. That species annihilate predominantly into $W^+W^-$ (79.9\%) and
$ZZ$ (20.1\%) gauge boson pairs. For illustration, a global boost factor of 2 has also
been included in the DM signal. The three last PAMELA data (red) points lie just between
the astrophysical background (yellow band) and the predicted signal (red solid).
The error bars are sufficiently large to still allow the presence of a 1.7 TeV Wino-like
species in the Galactic halo, although the pure astrophysical explanation is perfectly
compatible with the measurements. The AMS-02 experiment is now on board the ISS, and its
large acceptance and impressive statistics should allow to disentangle both possibilities,
as indicated by the theoretical dark-green data points.

%%%%%%%%%%%%%%%%%%%%%%%%%%%%%%%%%%%%%%%%%%%%%%%%%%%%%%%%%%%%%%%%%%%%%%%%%%%%%%%%%%%%%%%%%%%%%%%%%%%
%%%%%%%%%%%%%%%%%%%%%%%%%%%%%%%%%%%%%%%%%%%%%%%%%%%%%%%%%%%%%%%%%%%%%%%%%%%%%%%%%%%%%%%%%%%%%%%%%%%
%
% \newpage
%
\section{The PAMELA positron excess}
\label{sec:positrons}

The confirmation by the PAMELA collaboration~\cite{2009Natur.458..607A} of a positron
excess above 10~GeV has been triggering a lot of excitement in the field of particle
astrophysics since its announcement three years ago. This excess has been readily
considered as the first long waited hint of the presence of WIMPs in our Galaxy.
However, the DM candidates which can potentially lead to this positron anomaly must
have quite special properties. They are also severely constrained by radio and
gamma-ray observations, unless they are tightly packed inside improbable or bizarre
clumps. These species could also be unstable with abnormally long lifetimes.
In order to reach any conclusion, we need to investigate how positrons (and electrons)
propagate throughout the DH and we must calculate the astrophysical background to the
PAMELA signal.
Although this positron excess could be generated by annihilating or decaying DM particles,
William of Ockham would warn us that a more natural explanation is to be found in pulsars
for instance, and that {\sl entia non sunt multiplicanda praeter necessitatem}.

%%%%%%%%%%%%%%%%%%%%%%%%%%%%%%%%%%%%%%%%%%%%%%%%%%%%%%%%%%%%%%%%%%%%%%%%%%%%%%%%%%%%%%%%%%%%%%%%%%%
%
\subsection{Positron propagation and associated Green function}
\label{sec:pos_calculation}
%
%The master equation.
%
The propagation of CR positrons in the magnetic halo of the Galaxy differs from
that of nuclei in several respects.
Although space diffusion is still an essential ingredient common to all CR species,
positrons undergo mostly inverse Compton and synchrotron energy losses, as discussed
in~\cite{1998ApJ...493..694M} for instance, whereas nuclei and antinuclei are mostly
sensitive to the Galactic wind and to nuclear interactions as they cross the Milky
Way disk.
As a result, a positron line injected by a source leads to an extended positron
spectrum once propagated. This is at variance with most of nuclear species for which
energy losses can be neglected in a first approximation.
Consequently, the diffusion equation that relates the positron space and energy
density $\psi$ to the source term $q({\mathbf x},E)$ takes the form
\beq
- \, K_0 \! \left( \frac{E}{E_0} \right)^{\! \delta} \! \triangle \psi
\; + \;
{\displaystyle \frac{\partial}{\partial E}}
 \left\{  b^{\rm loss}(E)              \, \psi \right\} \, = \,
%\left\{ {\displaystyle \frac{dE}{dt}} \, \psi \right\} \, = \,
% q_{e^{+}}({\mathbf x},E) \;\; .
q({\mathbf x},E) \;\; .
\label{master_positron_2}
\eeq
%
%The diffusion term.
%
The first term is simply the diffusion coefficient written as
$K(E) \approx K_0 (E/E_0)^{\delta}$, where $E_0 \equiv 1$ GeV will be used
hereafter to keep track of the correct units.
%
%Energy loss term through synchrotron and inverse Compton processes.
%
The synchrotron and inverse Compton processes lead to energy losses with
the rate
\beq
 b^{\rm loss}(E) \equiv
{\displaystyle \frac{dE}{dt}} \, = \, - \,
{\displaystyle \frac{E_{0}}{\tau_{E}}} \, \epsilon^{2} \;\; ,
\eeq
where $\epsilon = {E}/{E_{0}}$ is the reduced positron energy. The energy
loss timescale $\tau_{E}$ is set equal to $0.6 \times 10^{16}$ s.
%
%The Baltz & Edsjo trick to solve the energy evolution.
%
The master equation~(\ref{master_positron_2}) can be solved with Baltz \&
Edsj\"{o}~\cite{1999PhRvD..59b3511B} clever trick which consists in translating
the energy $E$ into the pseudo-time
\begin{equation}
\tilde{t}(E) \, = \, \tau_{E} \times
\left\{
v(E) \, = \, {\displaystyle \frac{\epsilon^{\delta - 1}}{1 - \delta}}
\right\} \;\; .
\label{connection_E_pseudo_t}
\end{equation}
In this formalism, the energy losses which positrons experience boil down
to a mere evolution in the pseudo-time $\tilde{t}$ and the diffusion
equation~(\ref{master_positron_2}) simplifies into the well-known heat equation
\begin{equation}
{\displaystyle \frac{\partial \tilde{\psi}}{\partial \tilde{t}}} \ - \
K_{0} \, \Delta \tilde{\psi} \, = \,
\tilde{q}({\mathbf x},\tilde{t}) \;\; .
\label{master_positron_3}
\end{equation}
The positron density is now $\tilde{\psi} = \epsilon^{2} \, \psi$ whereas
the positron production rate has become
$\tilde{q}_{e^{+}} = \epsilon^{2 - \delta} \, q_{e^{+}}$. Both $\tilde{\psi}$
and $\tilde{q}$ have the same dimensions as before, because $\epsilon$ is
a dimensionless variable.
The Green function $\tilde{G}$ of the heat equation~(\ref{master_positron_3})
leads to the positron propagator through
\begin{equation}
G_{e^{+}} \!\! \left( {\mathbf x} , E \leftarrow {\mathbf x}_{S} , E_{S} \right) \, = \,
{\displaystyle \frac{\tau_{E}}{E_{0} \, \epsilon^{2}}} \;
\tilde{G} \left( {\mathbf x} , \tilde{t} \leftarrow {\mathbf x}_{S} , \tilde{t}_{S} \right)
\;\; ,
\label{positron_propagator}
\end{equation}
where the connection between the energy $E$ and pseudo-time $\tilde{t}$ is given by
relation~(\ref{connection_E_pseudo_t}). We are led to express the density of positrons
resulting from their transport within the Milky Way as the convolution
\begin{equation}
\psi_{e^{+}} \!\! \left( {\mathbf x} , E \right) \, = \,
{\displaystyle \int_{E_{S} = E}^{E_{S} = + \infty}} dE_{S} \;
{\displaystyle \int}_{\rm \!\! DH} \!\! d^{3}{\mathbf x}_{S} \;\;
G_{e^{+}} \!\! \left( {\mathbf x} , E \leftarrow {\mathbf x}_{S} , E_{S} \right) \;
q_{e^{+}} \!\! \left( {\mathbf x}_{S} , E_{S} \, \right) \;\; ,
\label{psi_convolution_positron}
\end{equation}
recovering thus the generic expression~(\ref{psi_convolution}). The positron propagator
$G_{e^{+}} \!\! \left( {\mathbf x} , E \leftarrow {\mathbf x}_{S} , E_{S} \right)$
measures the probability for a positron injected at $\mathbf{x}_{S}$ with energy
$E_{S}$ to reach the location $\mathbf{x}$ with the degraded energy $E \leq E_{S}$.
%
%Solution within an infinite magnetic halo.
%
In the 3D limiting case of an infinite magnetic halo, the heat Green function connecting
the source $\mathbf{x}_{S}$ to the Earth is the Gaussian distribution
\begin{equation}
\tilde{G}
\left( {\mathbf x}_{\odot} , \tilde{t} \leftarrow \mathbf{x}_{S} , \tilde{t}_{S} \right)
\, = \,
\Theta(\tilde{\tau}) \,
\left\{ \frac{1}{4 \, \pi \, K_{0} \, \tilde{\tau}} \right\}^{3/2}
\exp \left\{ - \,
{\displaystyle \frac{r_{\oplus}^{2}}{4 \, K_{0} \, \tilde{\tau}}} \right\} \;\; ,
\label{propagator_reduced_3D_lD_a}
\end{equation}
where $\tilde{\tau} = \tilde{t} - \tilde{t}_{S}$ is the typical duration, including
the diffusion process, over which the positron energy decreases from $E_{S}$ to $E$.
The distance between the Earth and the source is denoted by
\beq
r_{\oplus} = \left\{
(x_{\odot} - x_{S})^{2} + (y_{\odot} - y_{S})^{2} + (z_{\odot} - z_{S})^{2}
\right\}^{1/2} \;\; .
\eeq
The radial behavior of the heat propagator~(\ref{propagator_reduced_3D_lD_a}) suggests
to define the characteristic diffusion length $\lD = \sqrt{4 K_{0} \tilde{\tau}}$ which
sets the scale of the positron sphere, {\ie}, the region where most of the positrons
detected at the Earth are produced. It depends on the injected $E_{S}$ and detected
$E$ positron energies through the pseudo-time difference $\tilde{\tau}$. Above GeV
energies, $\lD$ is typically smaller than 5~kpc. Most of the positrons detected by
high-energy experiments like PAMELA have a local origin.

%
%Radial bondaries of the DH.
%
The magnetic halo is nevertheless finite, with radial and vertical boundaries.
Taking into account the former requires to use an expansion over Bessel functions
as explained in~\cite{2008PhRvD..77f3527D}. However, in most situations, it is safe to
ignore the radial boundaries and to picture the DH as an infinite slab with
half-thickness $L$. Sources located beyond $R = 20$~kpc should be disregarded
though, since the convolution~(\ref{psi_convolution_positron}) is only performed
over the DH.
%
%Vertical boundaries.
%
The infinite slab hypothesis allows the radial and vertical directions to be
disentangled and the reduced propagator $\tilde{G}$ may be expressed as
\begin{equation}
\tilde{G}
\left( {\mathbf x} , \tilde{t} \leftarrow {\mathbf x}_{S} , \tilde{t}_{S} \right)
\, = \,
{\displaystyle
 \frac{\Theta \! \left( \tilde{\tau} \right)}{4 \, \pi \, K_{0} \, \tilde{\tau}}}
%\frac{1}                                    {4 \, \pi \, K_{0} \, \tilde{\tau}}}
\; \exp \left\{ - \,
{\displaystyle \frac{r^{2}}{4 \, K_{0} \, \tilde{\tau}}}
\right\} \;
\tilde{V}
\left( z , \tilde{t} \leftarrow z_{S} , \tilde{t}_{S} \right) \;\; ,
\end{equation}
where $\tilde{\tau} =  \tilde{t} - \tilde{t}_{S}$ as before. The radial distance
between the source ${\mathbf x}_{S}$ and the point ${\mathbf x}$ of observation
is now defined as
$r = \{(x_{\odot} - x_{S})^{2} + (y_{\odot} - y_{S})^{2}\}^{1/2}$. The vertical
boundary conditions need to be implemented. Wherever the source inside the slab,
the positron density vanishes at $z = \pm L$.

\noindent {\bf (i)}
A first approach relies on the method of the so-called  electrical images and
has been discussed in~\cite{1999PhRvD..59b3511B}. Any point-like source inside the slab
is associated to the infinite series of its multiple images through the boundaries,
at $z = \pm L$, which act as mirrors. The n-th image is located at
$z_{n} = 2 \, n L \, + \, (-1)^{n} z_{S}$,
%
% \begin{equation}
% z_{n} \, = \,
% 2 \, L \, n \; + \; \left( -1 \right)^{n} \, z_{S} \;\; ,
% \end{equation}
%
and has a positive or negative contribution depending on whether $n$ is
even or odd. When the diffusion time $\tilde{\tau}$ is small, the vertical
function $\tilde{V}$ is well approximated by its infinite 1D limit
\begin{equation}
\tilde{V}
\left( z , \tilde{t} \leftarrow z_{S} , \tilde{t}_{S} \right)
\simeq
\mathcal{V}_{1D}
\left( z , \tilde{t} \leftarrow z_{S} , \tilde{t}_{S} \right) =
{\displaystyle \frac
 {\Theta \! \left( \tilde{\tau} \right)}
%{1}
{\sqrt{ 4 \, \pi \, K_{0} \, \tilde{\tau}}}} \;
\exp \left\{ - \, {\displaystyle
\frac{\left( z - z_{S} \right)^{2}}{4 \, K_{0} \, \tilde{\tau} \,}}
\right\} \;\; .
\label{propagator_reduced_1D}
\end{equation}
A relevant parameter is the ratio
$\zeta = {L^{2}}/({4 K_{0} \tilde{\tau}}) \equiv {L^{2}}/{\lambda_{\rm D}^{2}}$.
In the regime where it is much larger than 1, the propagation is insensitive to
the vertical boundaries. On the contrary, when $\zeta$ is much smaller than 1,
a large number of images needs to be taken into account in the sum
\begin{equation}
\tilde{V}
\left( z , \tilde{t} \leftarrow z_{S} , \tilde{t}_{S} \right) \, = \,
{\displaystyle \sum_{n \, = \, - \infty}^{+ \infty}} \,
\left( -1 \right)^{n} \;
\mathcal{V}_{1D}
\left( z , \tilde{t} \leftarrow z_{n} , \tilde{t}_{S} \right) \;\; ,
\label{V_image}
\end{equation}
and convergence may be a problem.

\noindent {\bf (ii)}
It is fortunate that a quite different approach is possible in that case.
The 1D version of equation~(\ref{master_positron_3}) actually looks like
the Schr\"{o}dinger equation (though in imaginary time) that accounts for
the behavior of a particle inside an infinitely deep 1D potential well
extending from $z = - L$ to $z = + L$. The eigenfunctions of the
associated Hamiltonian are given by
\begin{equation}
\varphi_{n}(z) \, = \, \sin
\left\{ k_{n} \left( L - \left| z \right| \right) \right\}
\;({\rm even})
\;\;\;{\rm and}\;\;\;
\varphi'_{n}(z) \, = \, \sin
\left\{ k'_{n} \left( L - z \right) \right\}
\;({\rm odd}) \;\; .
\end{equation}
They depend on the vertical coordinate $z$ through the wave-vectors
$k_{n} = {(n - 1/2) \pi}/{L}$ (even) and
$k'_{n} = {n     \, \pi}/{L}$ (odd). The vertical propagator may be expanded
as the series
\begin{equation}
\tilde{V}
\left( z , \tilde{t} \leftarrow z_{S} , \tilde{t}_{S} \right) =
{\displaystyle \sum_{n \, = \, 1}^{+ \infty}} \;\;
{\displaystyle \frac{1}{L}} \;
\left\{
e^{\displaystyle - \, \lambda_{n} \tilde{\tau}} \,
\varphi_{n} \left( z_{S} \right) \, \varphi_{n}(z)
\; + \;
e^{\displaystyle - \, \lambda'_{n} \tilde{\tau}} \,
\varphi'_{n} \left( z_{S} \right) \, \varphi'_{n}(z)
\right\} \;\; ,
\label{V_quantum}
\end{equation}
where the time constants $\lambda_{n}$ and $\lambda'_{n}$ are
respectively equal to $K_{0} \, {k_{n}^{2}}$ and $K_{0} \, {{k'}_{n}^{2}}$.
In the regime where $\zeta$ is much smaller than 1, {\ie}, for very large
values of the diffusion time $\tilde{\tau}$, just a few eigenfunctions
need to be considered in order for the sum~(\ref{V_quantum}) to converge.
Notice finally that the energies $E$ and $E_{S}$ always come into play in
the reduced propagator $\tilde{G}$ through the diffusion length $\lD$.

%%%%%%%%%%%%%%%%%%%%%%%%%%%%%%%%%%%%%%%%%%%%%%%%%%%%%%%%%%%%%%%%%%%%%%%%%%%%%%%%%%%%%%%%%%%%%%%%%%%
%
\subsection{The background of secondary positrons}
\label{sec:sec_pos_background}
%
%
%The various processes involved in the production p + H ===> e^{+} + X.
%
Like for antiprotons, an irreducible background of secondary positrons
is produced by primary CR nuclei colliding on the ISM.
The dominant mechanism is the collision of protons with hydrogen atoms at rest,
producing charged pions $\pi^{\pm}$ which decay into muons $\mu^{\pm}$. These are
also unstable and eventually lead to electrons and positrons through the chain
%
%TTTTTTTTTTTTTTTTTTTTTTTTTTTTTTTTTTTTTTTTTTTTTTTTTTTTTTTTTTTTTTTTTTTTTTTTTTTTTTTTTTTTTTTTTT
%
%%\refstepcounter{equation}(\theequation)\label{pp_pion_chain}
%%%
%%\begin{flushleft}
%%\begin{tabular}{c c c c c r}
%%{\hspace{0.6cm}}
%%${\rm p \; + \; H} \; \longrightarrow \; {\rm X} \; + \;$ &
%%$\pi^{\pm}$ & & & &
%%% {\hspace{1.0cm}
%%% \beq
%%% \label{pp_pion_chain}
%%% \eeq}
%%\\
%%& $\pi^{\pm}$ & $ \; \longrightarrow \; \nu_{\mu} \; + \;$
%%& $\mu^{\pm}$ & & \hspace{5.25cm} ({\normalsize {\ref{pp_pion_chain}}})
%%\\
%%& & & $\mu^{\pm}$ &
%%$\; \longrightarrow \; \nu_{\mu} \; + \; \nu_{e} \; + \; e^{\pm} \;\; .$ & \\
%%\end{tabular}
%%\vskip 0.2cm
%%\end{flushleft}
%TTTTTTTTTTTTTTTTTTTTTTTTTTTTTTTTTTTTTTTTTTTTTTTTTTTTTTTTTTTTTTTTTTTTTTTTTTTTTTTTTTTTTTTTTT
%
\begin{align}
{\rm p \; + \; H} \; \longrightarrow \; {\rm X} \; + \; & \pi^{\pm} & && &&\nn\\
& \pi^{\pm} & \; \longrightarrow \; \nu_{\mu} \; + \; & \mu^{\pm} & &&\nn\\  
&& & \mu^{\pm} & \; \longrightarrow \; \nu_{\mu} \; + \; \nu_{e} \; + \; e^{\pm} \;.
\label{pp_pion_chain}
\end{align}
In proton-proton collisions, pions can be produced in two different ways,
depending on the energy $E_{\rm p}$ of the incoming proton.
Below $\sim$ 3 GeV, one of the protons is predominantly excited to a $\Delta^{+}$
resonance which subsequently decays into a nucleon and a pion. As isospin is conserved,
the branching ratios into ${\rm p} + \pi^{0}$ and ${\rm n} + \pi^{+}$ are respectively
equal to ${2}/{3}$ and ${1}/{3}$.
Above $\sim$ 7 GeV, the pion production~(\ref{pp_pion_chain}) is well described in
the framework of the scaling model. Various parameterizations are given in the 
literature~\cite{1977PhRvD..15..820B,1983JPhG....9.1289T} for the Lorentz invariant pion
production cross section $E_{\pi} \, ({d^{3} \sigma}/{d^{3}p_{\pi}})$.
Positrons may also be produced through kaons generated in proton-proton collisions.
In a first chain of reactions, the kaon plays the same role as the pion in the set
of decays~(\ref{pp_pion_chain}). The branching ratio of the
$K^{\pm} \longrightarrow \nu_{\mu} + \mu^{\pm}$ decay channel is 63.5\%.
The kaon may also decay into a pair of pions, through the reaction
$K^{\pm} \longrightarrow \pi^{0} + \pi^{\pm}$, with a branching ratio of 21\%.
The charged pion $\pi^{+}$ subsequently follows the chain~(\ref{pp_pion_chain})
and yields a positron.
These two main kaon decay modes contribute together a few percent to the total
positron production differential cross section. Parameterizations of kaon production
in proton--proton interactions can also be found in~\cite{1983JPhG....9.1289T,1977PhRvD..15..820B}
for the scaling regime.
Useful parametric expressions for the yield and spectra of stable secondary species
produced in proton-proton collisions have been derived from experimental data and
summarized in \cite{2006ApJ...647..692K}.
Primary CR protons induce a production of positrons whose rate, per unit of volume
and energy, is given by the integral over proton energy
\beq
q_{e^{+}}^{\rm sec} \!\! \left( {\mathbf x} , E_{e} \, \right) \, = \,
4 \; \pi \; n_{\rm H}({\mathbf x}) \,
{\displaystyle \int} \, \Phi_{\rm p} \! \left( {\mathbf x} , E_{\rm p} \, \right)
\times dE_{\rm p} \times
{\displaystyle \frac{d\sigma}{dE_{e}}}(E_{\rm p} \to E_{e}) \;\; .
\label{source_sec_pos}
\eeq
As in the case of antiprotons, this relation can be generalized in order to
incorporate CR alpha particles as well as interstellar helium. The positron
production rate $q_{e^{+}}^{\rm sec}({\mathbf x} , E_{e})$ can be approximated
by its solar value since most of the positrons detected at the Earth originate
from the solar neighborhood. Folding that rate in
relation~(\ref{psi_convolution_positron}) yields the flux
\beq
\Phi_{e^{+}}^{\rm sec}(\odot , \epsilon \equiv {E_{e}}/{E_{0}}) \, = \,
{\displaystyle \frac{\beta_{e^{+}}}{4 \, \pi}} \times
{\displaystyle \frac{\tau_{E}}{\epsilon^{2}}} \times
{\displaystyle \int_{\epsilon}^{+ \infty}} d\epsilon_{S} \times
\tilde{I} \left( \lD \right) \times
q_{e^{+}}^{\rm sec} \!\! \left( \odot , \epsilon_{S} \, \right) \;\; .
\label{pos_flux_simple}
\eeq
The integral $\tilde{I}$ is the convolution of the reduced positron Green function
over the Galactic disk alone
\beq
\tilde{I} \left( \lD \right) \, = \,
{\displaystyle \int}_{\rm \!\!\! disk} \!\! d^{3}{\mathbf x}_{S} \;\;
\tilde{G} \left( {\mathbf x}_{\odot} \leftarrow {\mathbf x}_{S} ; \lD \right) \;\; .
\label{I_DH_b}
\eeq

%
%FFFFFFFFFFFFFFFFFFFFFFFFFFFFFFFFFFFFFFFFFFFFFFFFFFFFFFFFFFFFFFFFFFFFFFFFFFFFFFFFFFFFFFFFFFFFFFFFFF
\begin{figure}[t!]
\begin{center}
%\vskip 0.5cm
\noindent
 \includegraphics[width=0.9\textwidth]{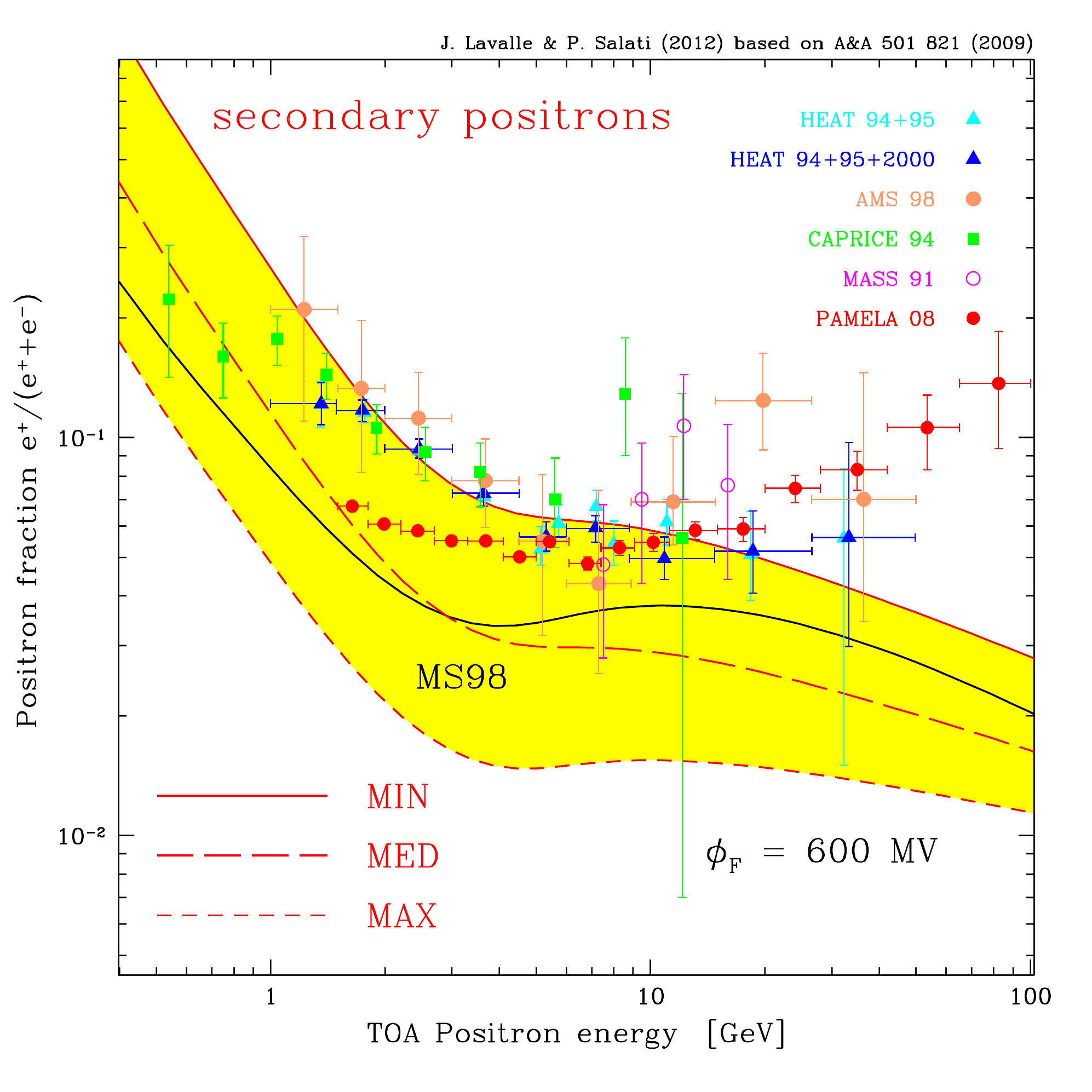}
\end{center}
%\vskip -0.5cm
\caption{
Positron fraction as a function of positron energy.
The yellow banana shape region encompasses the astrophysical background of secondary
positrons yielded by the 1,600 different configurations of CR propagation parameters,
shown in~\cite{2001ApJ...555..585M} to be compatible with the B/C data. This band is bounded
by the MAX (short dashed) and MIN (solid) curves, while the central long-dashed curve
stands for the MED model (see table~\ref{tab_prop}).
The nuclear cross sections have been parameterized according to~\cite{2006ApJ...647..692K}
%
%/Users/p_salati/__new_projects/secondary_positrons_08/production_cross_section/Kamae_matrix_080214"
%
whereas the recent PAMELA measurements of the CR proton and helium
spectra~\cite{2011Sci...332...69A} have been fitted by~\cite{2011arXiv1109.0834D}.
In the same figure, the positron fraction obtained with the positron flux calculated
by~\cite{1998ApJ...493..694M} and fitted by~\cite{1999PhRvD..59b3511B} is also indicated by the
solid black line labeled MS98.
The electron and positron flux enters in the denominator of the positron fraction.
Measurements from AMS-01~\cite{2002PhR...366..331A} and Fermi-LAT~\cite{2009PhRvL.102r1101A} have been
parameterized according to~\cite{2011MNRAS.414..985L}.
A solar modulation with a Fisk potential of 600~MV has been applied to the positron
flux. This corresponds to the level of solar activity during the data taking of AMS-01.
This figure has been borrowed from~\cite{2009A&A...501..821D}.
}
\label{fig:pos_banane}
\end{figure}
%FFFFFFFFFFFFFFFFFFFFFFFFFFFFFFFFFFFFFFFFFFFFFFFFFFFFFFFFFFFFFFFFFFFFFFFFFFFFFFFFFFFFFFFFFFFFFFFFFF
%
%Observations of the cosmic ray electron and positron flux at high energy.
%
As regards observations,
the ATIC~\cite{2008Natur.456..362C}, HESS~\cite{2008PhRvL.101z1104A,2009A&A...508..561A} and
Fermi~\cite{2009PhRvL.102r1101A} collaborations have reported measurements of the total
CR electron and positron flux at high energy. The Fermi-LAT instrument, in particular,
has collected high precision data between 20~GeV and 1~TeV, which can be fitted by the
simple power-law spectrum~\cite{2009APh....32..140G}
\beq
\Phi_{e^{\pm}} \, = \,
\left\{ 175.40 \pm 6.09 \;\; {\rm GeV^{-1} \, m^{-2} \, s^{-1} \, sr^{-1}} \right\} \;
\epsilon^{-(3.045 \pm 0.008)} \;\; ,
\label{fermi_LAT_phi_e}
\eeq
where $\epsilon = {E}/{E_{0}}$ is the reduced lepton energy.
Primary CR electrons are believed to originate from the interstellar medium, like
most of the primary CR nuclei, and are accelerated by supernova driven shock waves
which inject them with a spectral index $\alpha \sim 2.2 \pm 0.1$ inside the Milky Way
disk. They subsequently propagate within the magnetic halo and lose energy. A detailed
investigation~\cite{2010A&A...524A..51D} leads to a CR electron flux at the Earth of the form
\beq
\Phi_{e^{-}} \propto \epsilon^{\displaystyle - \alpha - 0.5 - {\delta}/{2}} \;\; ,
\eeq
with a spectral index of order $3.05 \pm 0.15$, in reasonable agreement with the
Fermi-LAT measurements should $\delta = 0.7$, as indicated by the MED model of
table~\ref{tab_prop}. Some authors~\cite{2004ApJ...613..962S} find that value too high.
It suffices then to increase $\alpha$ in order to match the Fermi-LAT observations.
The lepton spectrum contains also a small admixture of secondary electrons and positrons,
produced by CR primary nuclei impinging on interstellar gas. These spallation
reactions generate many charged pions which eventually decay into positrons and,
in a lesser extent, into electrons. The source spectral index is set by the CR
proton and helium fluxes. With a gross value of $\alpha = 2.7$, we expect
a spectral index of $3.55 \pm 0.05$ for secondary positrons.

%PAMELA positron excess.
%
The PAMELA collaboration has measured~\cite{2009Natur.458..607A} the positron fraction
${e^{+}}/({e^{-} + e^{+}})$ over a large energy range. From the above mentioned arguments,
we would naively expect that fraction to decrease with energy as $\epsilon^{-1/2}$.
A more refined analysis~\cite{2009A&A...501..821D} confirms that trend, as featured by the
banana shape region of Fig.~\ref{fig:pos_banane}. The thickness of the yellow band
gauges the uncertainties arising from CR transport. In spite of these, the positron
fraction is expected to decrease above 10~GeV. The PAMELA measurements establish on
the contrary that the positron fraction increases at high energy. This very important
observation is in contradiction with a pure secondary origin for the positrons. Primary
sources, which directly inject these particles in the interstellar medium, are necessary.

%%%%%%%%%%%%%%%%%%%%%%%%%%%%%%%%%%%%%%%%%%%%%%%%%%%%%%%%%%%%%%%%%%%%%%%%%%%%%%%%%%%%%%%%%%%%%%%%%%%
%
\subsection{The PAMELA excess as a signature for Galactic WIMPs}
\label{sec:pamela_vs_wimps}
%
%The PAMELA excess as a signature for WIMPs.
%
%FFFFFFFFFFFFFFFFFFFFFFFFFFFFFFFFFFFFFFFFFFFFFFFFFFFFFFFFFFFFFFFFFFFFFFFFFFFFFFFFFFFFFFFFFFFFFFFFFF
\begin{sidewaysfigure}
\begin{center}
\centerline{
 \includegraphics[width=0.51\textwidth]{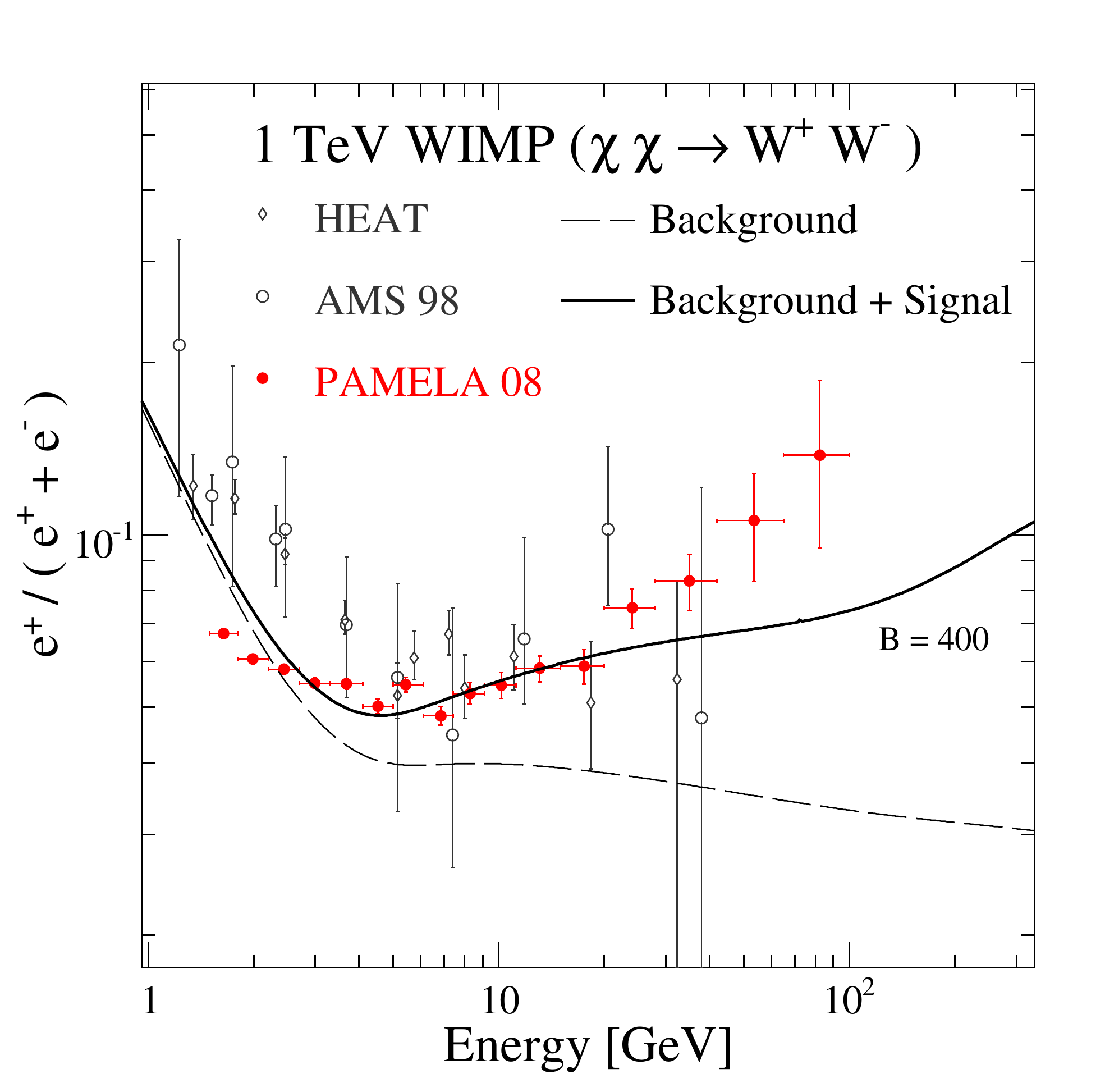}
 \includegraphics[width=0.49\textwidth]{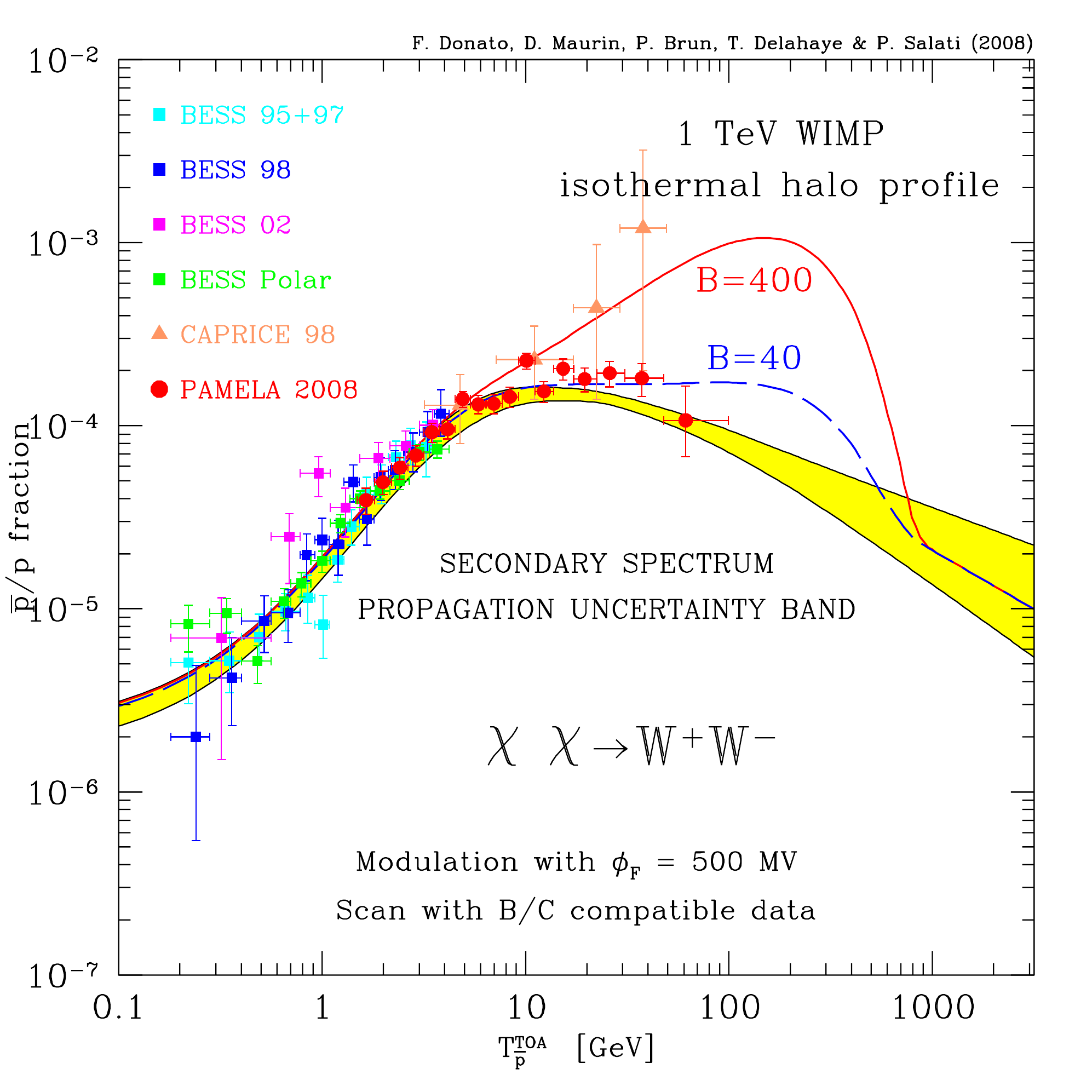}}
\end{center}
\caption{
The pedagogical example of a 1~TeV neutralino annihilating into a $W^{+} W^{-}$ gauge
boson pair is presented. In the left panel, the positron signal which the DM species
yield has been increased by a factor of 400, hence the solid curve and a marginal
agreement with the PAMELA data.
%
% Positron fraction data are from HEAT~\cite{1997ApJ...482L.191B},
% AMS-01~\cite{Aguilar:2007yf,Alcaraz:2000bf} and PAMELA~\cite{2009Natur.458..607A}.
%
If the so-called Sommerfeld effect~\cite{2004PhRvL..92c1303H,2005PhRvD..71f3528H} is invoked to explain such
a large enhancement of the annihilation cross section, the same boost applies to
antiprotons, and leads to an unacceptable distortion of their spectrum as indicated
by the red solid line in the right panel. The yellow band encompasses the models
shown in~\cite{2001ApJ...555..585M} to be compatible with the B/C data. The antiproton
red solid line (B=400) lies well above that astrophysical background.
Figures and caption are borrowed from~\cite{2009PhRvL.102g1301D}.
}
\label{fig:1_TEV_WIMP}
\end{sidewaysfigure}
%FFFFFFFFFFFFFFFFFFFFFFFFFFFFFFFFFFFFFFFFFFFFFFFFFFFFFFFFFFFFFFFFFFFFFFFFFFFFFFFFFFFFFFFFFFFFFFFFFF
%
The announcement of a positron excess by the PAMELA collaboration has triggered
a lot of excitement. This excess was actually considered as the first hint of the
presence of DM species in the Milky Way halo. Should WIMPs exist, their annihilations
would be a source of primary positrons. The production rate $q_{e^{+}}^{\rm DM}$, given
by relation~(\ref{DM_source}), yields the positron flux at the Earth
\beq
\Phi_{e^{+}}^{\rm DM}(\odot , \epsilon) \, = \, {\mathcal F} \times
{\displaystyle \frac{\tau_{E}}{\epsilon^{2}}} \times
{\displaystyle \int_{\epsilon}^{m_{\chi}/E_{0}}} d\epsilon_{S} \;\; f(\epsilon_{S})
\;\; \tilde{I}_{\rm DM} \left( \lD \right) \;\; .
\label{flux_positron_DM}
\eeq
The halo integral $\tilde{I}_{\rm DM}$~\cite{2008PhRvD..77f3527D} is the convolution of the
reduced positron propagator $\tilde{G}$ with the square of the Galactic DM density
\beq
\tilde{I}_{\rm DM} \left( \lD \right) \, = \,
{\displaystyle \int}_{\rm \!\!\! DH} \!\! d^{3}{\mathbf x}_{S} \;
\tilde{G} \left( {\mathbf x}_{\odot} \leftarrow {\mathbf x}_{S} ; \lD \right) \;
\left\{ \! {\displaystyle \frac{\rho({\mathbf x}_{S})}{\rho_{\odot}}} \! \right\}^{2}
\;\; .
\label{I_DH_c}
\eeq
The positron excess lies above 10~GeV. It can be generated by a particle with a mass
in the range 100~GeV to a few TeV, in good agreement with theoretical expectations for
WIMP masses.
If the DM species are thermally produced during the big bang, their relic abundance
matches the WMAP value~\cite{2011ApJS..192...18K} of $\Omega_{\chi} h^{2} = 0.1126 \pm 0.0036$
provided that their annihilation cross section, at the time of decoupling, is equal to
${\left\langle \sigma_\mathrm{ann} v \right\rangle} \sim 3 \times 10^{-26}$ cm$^{3}$ s$^{-1}$.
The energy distribution $f(\epsilon)$ of the positrons produced in a single annihilation
event is set by the WIMP type. High energy positrons (and electrons) cannot diffuse on
long distances, and those detected at the Earth must have been produced locally, hence a
DM density of $\rho_{\odot} = 0.3$ GeV cm$^{-3}$. Baring in mind these benchmark values,
one can estimate $q_{e^{+}}^{\rm DM}$ as well as the positron flux $\Phi_{e^{+}}^{\rm DM}$.
Alas, the signal derived for a large choice of DM candidates, and for various DM Galactic
distributions, is way too small to account for the observed excess. For a WIMP mass
$m_{\chi}$ of 1~TeV, $q_{e^{+}}^{\rm DM}$ needs to be enhanced
by a factor of $\sim 10^{3}$ in order to match the PAMELA measurements, as illustrated
in the left panel of Fig.~\ref{fig:1_TEV_WIMP}.

%
%Rate is too low and needs to be enhanced.
%
To save the WIMP explanation of the PAMELA positron anomaly, a first remedy is the
possibility that ${\left\langle \sigma_\mathrm{ann} v \right\rangle}$ is much larger
than what is currently assumed. Three directions at least have been explored so far.
To commence, WIMP decoupling in the early universe takes place conventionally during
a period of radiation domination. If the universe at that time is dominated by another
component, like a scalar field rolling down its 
potential~\cite{2003PhLB..571..121S,2003JCAP...11..006P},
the expansion rate is significantly increased and the same WIMP relic abundance $\Omega_{\chi}$
requires now a much larger annihilation cross section.
Another possibility is that WIMPs do not decouple from the primordial plasma. In this class
of less natural scenarios, a non-thermal production takes place through the decay of heavier
species, like gravitinos or moduli fields~\cite{2009PhLB..681..151K}. The WIMP relic abundance 
is no longer determined solely by ${\left\langle \sigma_\mathrm{ann} v \right\rangle}$. It also
depends on the production rate and lifetime of the unstable heavier states.
A final solution relies on the so-called Sommerfeld 
effect~\cite{2004PhRvL..92c1303H,2005PhRvD..71f3528H}. If the
DM species are massive enough and interact through the exchange of a light particle $\phi$,
they are attracted toward each other at sufficiently low velocities. Their wave functions
get focalized at the interaction point so that the annihilation cross section can be
significantly enhanced. Scenarios based on that effect have been
proposed~\cite{2008NuPhB.800..204C,2009PhRvD..79a5014A,2009PhLB..671..391P}, reviving the idea of
secluded dark matter~\cite{2008PhLB..662...53P}. WIMPs are still thermally produced in the
early universe. Their annihilation cross section at that time is equal to the canonical
value of $3 \times 10^{-26}$ cm$^{3}$ s$^{-1}$. As the universe expands, the average WIMP
velocity $\beta$ decreases. The Sommerfeld enhancement sets in as soon as the WIMP kinetic energy
is dominated by the interaction energy, {\ie}, when $\beta$ no longer exceeds the effective
WIMP coupling constant ${\alpha'}$. The cross section increases and may be today large enough
to account for the PAMELA positron excess. The enhancement saturates when the WIMP de Broglie
wavelength $\sim {1}/({\beta m_{\chi}})$ exceeds the range $\sim {1}/{m_{\phi}}$ of the
interaction. This occurs for $\beta \leq {m_{\phi}}/{m_{\chi}}$. The annihilation cross
section is also resonantly increased whenever
$\alpha' \, m_{\phi} \sim \alpha'^{2} \, {m_{\chi}} / {n^{2}}$, where $n$ is an
integer~\cite{2009PhRvD..79h3523L}.

%
%The antiproton constraint.
%
Irrespective of the mechanism enhancing
${\left\langle \sigma_\mathrm{ann} v \right\rangle}$, WIMPs should not overproduce
antiprotons~\cite{2009NuPhB.813....1C,2009PhRvL.102g1301D}. The annihilation channels leading to
antiprotons must be significantly suppressed. The antiproton to proton ratio
measured by the PAMELA collaboration~\cite{2009PhRvL.102e1101A} does not show actually any
excess and is consistent with a pure secondary origin, as featured in the right panel of
Fig.~\ref{fig:1_TEV_WIMP}. Therefore, besides an abnormally large annihilation cross
section, DM species preferentially annihilate into charged leptons, a feature which is
unusual in supersymmetry, but more typical of Kaluza-Klein theories. From a phenomenological
point of view, WIMP annihilations must proceed either directly to lepton pairs or, in some
models, through the production of the above mentioned light mediator bosons $\phi$ which
subsequently decay into leptons via
\beq
\chi \, + \, \chi \, \longrightarrow \, \phi \, + \, \phi \, \longrightarrow \,
l^{+} \; l^{-} \; l^{+} \; l^{-} \;\; .
\eeq

%
%Constraints from other messengers.
%
The possibility to explain the PAMELA positron excess by leptophilic DM particles with
enhanced annihilation cross section has been extensively investigated.
Depending on its magnitude, the injection of high-energy positrons and electrons is
associated to clear astrophysical signatures~\cite{2009ApJ...699L..59B}. As they spiral
along the Galactic magnetic field lines, these particles produce a radio emission through
synchrotron radiation. They also inverse Compton scatter (ICS) photons from the CMB and
stellar light. Finally, the charged leptons released by the WIMP annihilations can produce
final state radiation photons.
After an extensive
scan~\cite{2009JCAP...03..009B,2009PhRvD..79h1303B,2010NuPhB.831..178M} of the various 
possibilities, the region of the
$(m_{\chi} , \left\langle \sigma_\mathrm{ann} v \right\rangle)$ plane compatible with the
PAMELA and Fermi-LAT measurements is found to be excluded for most of the leptophilic DM
candidates. The most stringent limit arises from the synchrotron radio emission produced
at the Galactic center, a region where the DM distribution may be cuspy, and where magnetic
fields should be strong. That limit weakens considerably, though, as soon as a cored
isothermal DM distribution is preferred to the usual Einasto or NFW profile. In that case,
leptophilic WIMPs are not excluded by radio observations.
A more reliable tool is provided by the ICS of the positrons and electrons produced by WIMP
annihilation at high Galactic latitude, a region which is less subject to uncertainties than
the center of the Milky Way. A careful investigation~\cite{2010NuPhB.840..284C}, performed only
in the case where WIMPs annihilate directly into lepton pairs, excludes values of $m_{\chi}$
larger than 1~TeV.
Finally, another model independent analysis implies the injection of WIMP annihilation
products in the intergalactic medium at redshift $\sim 1000$. The recombination process is
affected and an imprint is left on the CMB, which may be searched for in the high precision
WMAP data~\cite{2011PhRvD..84b7302G}. Most of the PAMELA compatible leptophilic DM candidates still
survive that test~\cite{2009PhRvD..80d3526S}, but will be probed by the Planck mission.

%
%DM clumps in the $\Lambda$CDM universe.
%Part to be left if we do NOT have the section on DM substructures and boost factors.
%
Another way to increase the production rate of DM positrons relies on the existence of
substructures inside the smooth DM Galactic distribution. Because
$\left\langle \rho^{2}({\mathbf x}) \right\rangle$ is always larger than
$\left\langle \rho({\mathbf x})     \right\rangle^{2}$,
%
% $\bar{\rho_{\chi}^{2}}$ is larger than ${\bar{\rho_{\chi}}}^{2}$,
% $< \! \rho_{\chi}^{2} \! >$ is larger than $< \! \rho_{\chi} \! >^{2}$,
% $\left< \rho_{\chi}^{2} \right> \leq \left< \rho_{\chi} \right>^{2}$,
%
inhomogeneities tend to effectively enhance $q_{e^{+}}^{\rm DM}$, a quantity
proportional to the square of the DM density $\rho({\mathbf x})$. How significant is
that enhancement depends on the position of the clumps with respect to the smooth
component of the DM halo. The inner structure of clumps, as well as their mass
distribution, are also key factors. Finally, the lower mass cut-off and the
concentration-to-mass relation of the substructures need to be considered.
A statistical analysis is mandatory since we live inside one particular realization
of the clump distribution, to be taken out of an infinite set of similar realizations.
Using the tools specifically forged for computing the odds of that Galactic
lottery~\cite{2007A&A...462..827L}, a comprehensive analysis~\cite{2008A&A...479..427L}
indicates that the boost factor to be applied to $q_{e^{+}}^{\rm DM}$ in the case of
a ${\Lambda}$CDM universe cannot exceed at most a factor of $20$ at high energy. The
correct value is presumably much smaller, {\ie}, of order unity.

%
%The Galactic lottery.
%
Although a clumpy DM halo does not lead, on average, to an increase of the positron
flux at the Earth, the statistical variance of the boost factor is large at high energy,
hence the existence of configurations where $q_{e^{+}}^{\rm DM}$ could be significantly
enhanced. The possibility that a DM clump lies, for instance, in the vicinity of the solar
system has been suggested~\cite{2007MNRAS.374..455C,2009PhRvD..79j3513H} as an explanation for
the PAMELA positron anomaly. Such a local substructure would actually outshine the rest
of the Galactic DM distribution.
Alas, the probability that such a situation occurs in the DM halo of the Milky Way is
vanishingly small as demonstrated in an analysis~\cite{2009PhRvD..80c5023B} based on the results
of the cosmological N-body simulation Via Lactea~II~\cite{2008Natur.454..735D}. In the favorable
case of a 100~GeV DM species annihilating into $e^{\pm}$ pairs, the PAMELA positron excess
is best fitted by a subhalo with annihilation volume
\beq
\xi \equiv {\displaystyle {{\int}_{\rm \!\! clump}}} \,
\left\{ \! {\displaystyle \frac{\rho({\mathbf x})}{\rho_{\odot}}} \! \right\}^{2} \,
d^{3}{\mathbf{x}} \, \sim \, 114 \;\; {\rm kpc^{3}} \;\; ,
\eeq
located at 1.22~kpc from the Earth. That configuration has a probability of only 0.37\%
to occur. Other arrangements are even less probable.

%
%Unstable DM particles with VERY large lifetimes.
%
Finally, the possibility of unstable DM candidates with very large lifetimes has also
been explored (see for instance~\cite{2009JCAP...02..021I,2009JCAP...08..017I,2011JCAP...01..032G}).
Irrespective of the mechanism which produces them in the early universe, unstable species
could in principle generate the PAMELA signal. The production rate~(\ref{DM_source}) can
now be written as
\beq
q_{e^{+}}^{\rm DM}({\mathbf x}) \, = \, \tau_{\rm dec}^{-1} \,
\left\{ \! {\displaystyle \frac{\rho({\mathbf x})}{m_{\chi}}} \right\} \,
f(\epsilon) \;\; ,
\eeq
where the quantity
$\eta \,\left\langle \sigma_\mathrm{ann} v \right\rangle \, ({\rho}/{m_{\chi}})$ has been
replaced by the decay rate $\tau_{\rm dec}^{-1}$. For a 1~TeV WIMP, a lifetime of order
$2 \times 10^{26}$ sec is required to reproduce the positron excess.
The synchrotron radio emission from the Galactic center is no longer a problem, even in
the case of a NFW or Einasto profile, since the positron production rate from decaying DM
is proportional to the DM density $\rho$, and not to its square.
A detailed analysis~\cite{2010NuPhB.840..284C} excludes however unstable DM as it would
unacceptably contribute to the isotropic gamma-ray emission inferred from the Fermi data.
This limit does not apply to WIMPs predominantly  decaying into muon pairs.
The lifetime needs to be fine tuned though. Higher dimension operators must be invoked to
explain the large value of $\tau_{\rm dec}$ required by PAMELA. If a four-fermion point-like
coupling is responsible for the WIMP instability, we anticipate~\cite{2009PThPh.122..553C} a decay
rate $\propto {m_{\chi}^{5}}/{M^{4}}$, where $M$ denotes the typical scale of the underlying
high-energy theory. If factors of $2$ and $\pi$ are neglected, a GUT mass
$M \sim 2.3 \times 10^{16}$ GeV could lead to the required lifetime.
Notice that the DM species must also be leptophilic to prevent antiprotons from being
overproduced.

%
%Conclusion of the PAMELA hot excitement.
%
The CR lepton anomalies reported three years ago, and the PAMELA positron excess in
particular, have triggered a boiling activity in the field of particle astrophysics.
The possibility that these signals are produced by WIMPs annihilating or decaying in
the Milky Way halo is fading away. The DM explanation is fairly contrived and far from
the main stream. The positrons which the DM species are supposed to produce generate
in turn radiations which are not seen. As William of Ockham would have pointed out,
a more plausible explanation has to be found elsewhere. Actually, a simple one lies in
pulsars. These objects are known to exist, and they do inject (primary) positrons in the
Galactic disk. The lepton anomalies can be nicely explained by local pulsars and supernova
remnants, as pointed out by~\cite{2012CEJPh..10....1P} as well as by a recent and very
comprehensive~\cite{2010A&A...524A..51D} study.
%
%
%%%%%%%%%%%%%%%%%%%%%%%%%%%%%%%%%%%%%%%%%%%%%%%%%%%%%%%%%%%%%%%%%%%%%%%%%%%%%%%%%%%%%%%%%%%%%%%%%%%
%%%%%%%%%%%%%%%%%%%%%%%%%%%%%%%%%%%%%%%%%%%%%%%%%%%%%%%%%%%%%%%%%%%%%%%%%%%%%%%%%%%%%%%%%%%%%%%%%%%
%
\section{Antideuterons or the new challenge}
\label{sec:Dbar}
Antideuterons are the nuclei of antideuterium, the element that mirrors deuterium.
Although they have been seen in nuclear or lepton interactions at terrestrial
accelerators, antideuterons have not yet been detected in the cosmic radiation.
A small admixture is nevertheless expected to be astrophysically produced by the
spallations of CR protons and helium nuclei on the ISM. Any process that generates
at the same time an antiproton and an antineutron could lead to the production of
antideuterons. The requisite is the fusion of the antiproton and antineutron pair
into an antideuteron. This happens whenever the momenta of the antinucleons are
so close to each other that the kinetic energy of the pair, as seen in its rest
frame, is of the same order of magnitude as the antideuteron binding energy.

%
%Generality.
%
To quantify this effect, let us consider the collisions of CR high-energy protons
on hydrogen atoms at rest. The number $d{\cal N}_{\rm X}$ of antiprotons,
antineutrons or antideuterons generated in a single reaction, with momenta $\kX$,
is related to the differential production cross section through
\beq
d{\cal N}_{\rm X} \, = \, {\displaystyle \frac{1}{\STOT}} \,
{d^{3} \sigma_{\rm X}}(\sqrt{s} , \kX)
\;\; ,
\eeq
where $\STOT$ denotes the total proton-proton cross section. The total available
energy is $\sqrt{s}$. The corresponding differential probability for the production
of the species ${\rm X}$ is defined as
\beq
d{\cal N}_{\rm X} \, = \, {\cal F}_{\rm X} (\sqrt{s} , \kX) \, d^{3} \kX
\;\; .
\eeq
For antiprotons, for instance, it can be expressed in terms of the Lorentz invariant
cross section through
\beq
{\STOT} \; E_{\pbar} \; {\cal F}_{\pbar} (\sqrt{s} , \kpbar) \, = \,
\left. E_{\pbar} \; \frac{d^{3} \sigma_{\pbar}}{d^{3} \kpbar} \right|_{\rm LI}
\;\; .
\eeq
That cross section is experimentally well-known and a parameterization has been
given, for instance, in~\cite{1982PhRvD..26.1179T}. Isospin symmetry can be assumed,
furthermore, to hold. If so, the production of antineutrons is the same as for
antiprotons. The calculation of the probability for the formation of an antideuteron
can now proceed in two steps. We first need to estimate the probability for the
creation of an antiproton-antineutron pair. Then, the antinucleons merge together
to yield an antinucleus of deuterium.
%
%Factorization.
%
As explained in~\cite{1997PhLB..409..313C}, the production of two antinucleons
is assumed to be proportional to the square of the production of one of them.
The hypothesis that factorization of the probabilities holds is fairly well
established at high energies. For spallation reactions, however, the bulk
of the antiproton production takes place for an energy $\sqrt{s} \sim 10$ GeV
which turns out to be of the same order of magnitude as the antideuteron mass.
Pure factorization should break in that case as a result of energy conservation.
It needs to be slightly adjusted.
Following~\cite{1997PhLB..409..313C,2000PhRvD..62d3003D}, we may assume that the center
of mass energy available for the production of the second antinucleon is reduced
by twice the energy carried away by the first antinucleon
\beq
{\cal F}_{\pbar , \nbar} (\sqrt{s} , \kpbar , \knbar) \, = \,
{\displaystyle \frac{1}{2}} \,
{\cal F}_{\pbar} (\sqrt{s} , \kpbar) \,
{\cal F}_{\nbar} (\sqrt{s} - 2 E_{\pbar} , \knbar) \; + \;
\left( \pbar \leftrightarrow \nbar \right)
\;\; .
\eeq
%
%Coalescence.
%
Once the antiproton and the antineutron are formed, they combine
together to give an antideuteron with probability
\beq
{\cal F}_{\bar{\rm D}} (\sqrt{s} , \kdbar) \; d^{3} \kdbar \, = \,
{\displaystyle \int} \, d^{3} \kpbar \, d^{3} \knbar \;
{\cal C}(\kpbar , \knbar) \;
{\cal F}_{\pbar , \nbar} (\sqrt{s} , \kpbar , \knbar) \;\; .
\label{coalescence_1}
\eeq
The summation is performed on the antinucleon configurations for which
$\kpbar + \knbar = \kdbar$.
% \beq
% \kpbar + \knbar \; = \; \kdbar \;\; .
% \eeq
The coalescence function ${\cal C}(\kpbar , \knbar)$ describes the probability
for a ${\pbar}$-${\nbar}$ pair to yield by fusion an antideuteron. That function
depends actually on the difference
$\kpbar - \knbar = 2 \, \kdif$ between the antinucleon momenta so that
relation~(\ref{coalescence_1}) becomes
\beq
{\cal F}_{\bar{\rm D}} (\sqrt{s} , \kdbar) \, = \,
{\displaystyle \int} \, d^{3} \kdif \; {\cal C}(\kdif) \;
{\cal F}_{\pbar , \nbar} (\sqrt{s} ,
\kpbar = \frac{\kdbar}{2} + \kdif ,
\knbar = \frac{\kdbar}{2} - \kdif) \;\; .
\label{coalescence_2}
\eeq
In the center of mass frame of the proton-proton collision,
an energy of $\sim 3.7$ GeV is required to
form an antideuteron, to be compared to a binding energy of $B \sim 2.2$ MeV.
The coalescence function is therefore strongly peaked around
$\kdif = {\mathbf 0}$ and expression (\ref{coalescence_2}) simplifies into
\beq
{\cal F}_{\bar{\rm D}} (\sqrt{s} , \kdbar) \simeq \left\{
{\displaystyle \int} \, d^{3} \kdif \;\, {\cal C}(\kdif) \right\} \times
{\cal F}_{\pbar , \nbar} (\sqrt{s} ,
\kpbar = \frac{\kdbar}{2} , \knbar = \frac{\kdbar}{2}) \;\; ,
\eeq
where the probability for the formation of the ${\pbar}$-${\nbar}$ pair
has been factored out. The term in brackets may be estimated in the rest
frame of the antideuteron through the Lorentz invariant term
\beq
{\displaystyle \int} \, \frac{E_{\bar{\rm D}}}{E_{\pbar} \, E_{\nbar}} \,
d^{3} \kdif \;\, {\cal C}(\kdif) \simeq
\left( \frac{\md}{m_{\pbar} \, m_{\nbar}} \right) \,
\left(\frac{4}{3} \, \pi \, P_{\rm coal}^{\, 3} \right) \;\; .
\eeq 
In that frame, as already mentioned, the antinucleons merge together if
the momentum of the corresponding two-body reduced system is less than
some critical value $\pc$. That coalescence momentum is the only free
parameter of this simplistic factorization and coalescence scheme.
Theoretical values range from $\sqrt{m_{\rm p} B} \sim 46$ MeV, naively derived
from the antideuteron binding energy, up to 180 MeV as would follow from
a Hulthen parameterization of the deuterium wave function~\cite{Braun82}.
We therefore expect $\pc$ to lie somewhere in the range between 50 and
200~MeV. Direct comparison with accelerator data yields a phenomenological
value of 58~MeV as shown in~\cite{1997PhLB..409..313C,2000PhRvD..62d3003D}.
A recent upgrade~\cite{2005PhRvD..71h3013D,2008PhRvD..78d3506D} points toward
79~MeV.
The Lorentz invariant cross section for the production of antideuterons
can now be derived under the form
\bea
E_{\bar{\rm D}} \;
{\displaystyle \frac{d^3 \sigma_{\bar{\rm D}}}{d^{3} \kdbar}} & = &
\left( {\displaystyle \frac{\md}{m_{\pbar} \, m_{\nbar}}} \right) \,
\left({\displaystyle \frac{4}{3}} \, \pi \, P_{\rm coal}^{\, 3} \right) \times
{\displaystyle \frac{1}{2 \, \STOT}} \times \nonumber \\
& \times &
\left\{
E_{\pbar} \, {\displaystyle \frac{d^{3} \sigma_{\pbar}}{d^{3} \kpbar}}
(\sqrt{s} , \kpbar) \times
E_{\nbar} \, {\displaystyle \frac{d^{3} \sigma_{\nbar}}{d^{3} \knbar}}
(\sqrt{s} - 2 E_{\pbar} , \knbar)
\; + \;  \left( \pbar \leftrightarrow \nbar \right)
\right\}
\;\; . \nn \\
\label{LI_dbar_production}
\ena
A summation over the production angle $\theta$ in the Galactic frame leads
to the differential cross section
\beq
{\displaystyle \frac{d \sigma_{\bar{\rm D}}}{d \Edbar}}
\{ {\rm p}(\Ep) + {\rm H} \to \bar{\rm D}(\Edbar) \} \, = \, 2 \, \pi \, k_{\bar{\rm D}} \,
{\displaystyle \int_{0}^{\theta_{\rm M}}} \left. E_{\bar{\rm D}} \,
\frac{d^{3} \sigma_{\bar{\rm D}}}{d^{3} \kdbar} \right|_{\rm LI} \,
d(- \cos \theta) \;\; .
\label{integral_production_lab}
\eeq
In that frame, $\theta$ denotes the angle between the momenta of the incident
proton and produced antideuteron. It is integrated up to a maximal value
$\theta_{\rm M}$ set by the requirement that, in the center of mass
frame of the reaction, the antideuteron energy $E^{*}_{\bar{\rm D}}$ cannot
exceed the bound
\beq
E^{*}_{\bar{\rm D} , {\rm M}} \, = \,
{\displaystyle \frac{s \, - \, 16 \, m_{\rm p}^{2} \, + \, m_{\bar{\rm D}}^{2}}
{2 \sqrt{s}} } \;\; .
\eeq
The integral (\ref{integral_production_lab}) is performed at fixed antideuteron
energy $E_{\bar{\rm D}}^{2} = m_{\bar{\rm D}}^{2} + k_{\bar{\rm D}}^{2}$.

%
%FFFFFFFFFFFFFFFFFFFFFFFFFFFFFFFFFFFFFFFFFFFFFFFFFFFFFFFFFFFFFFFFFFFFFFFFFFFFFFFFFFFFFFFFFFFFFFFFFF
\begin{figure}[t!]
\begin{center}
%\vskip 0.5cm
\noindent
 \includegraphics[width=0.9\textwidth]{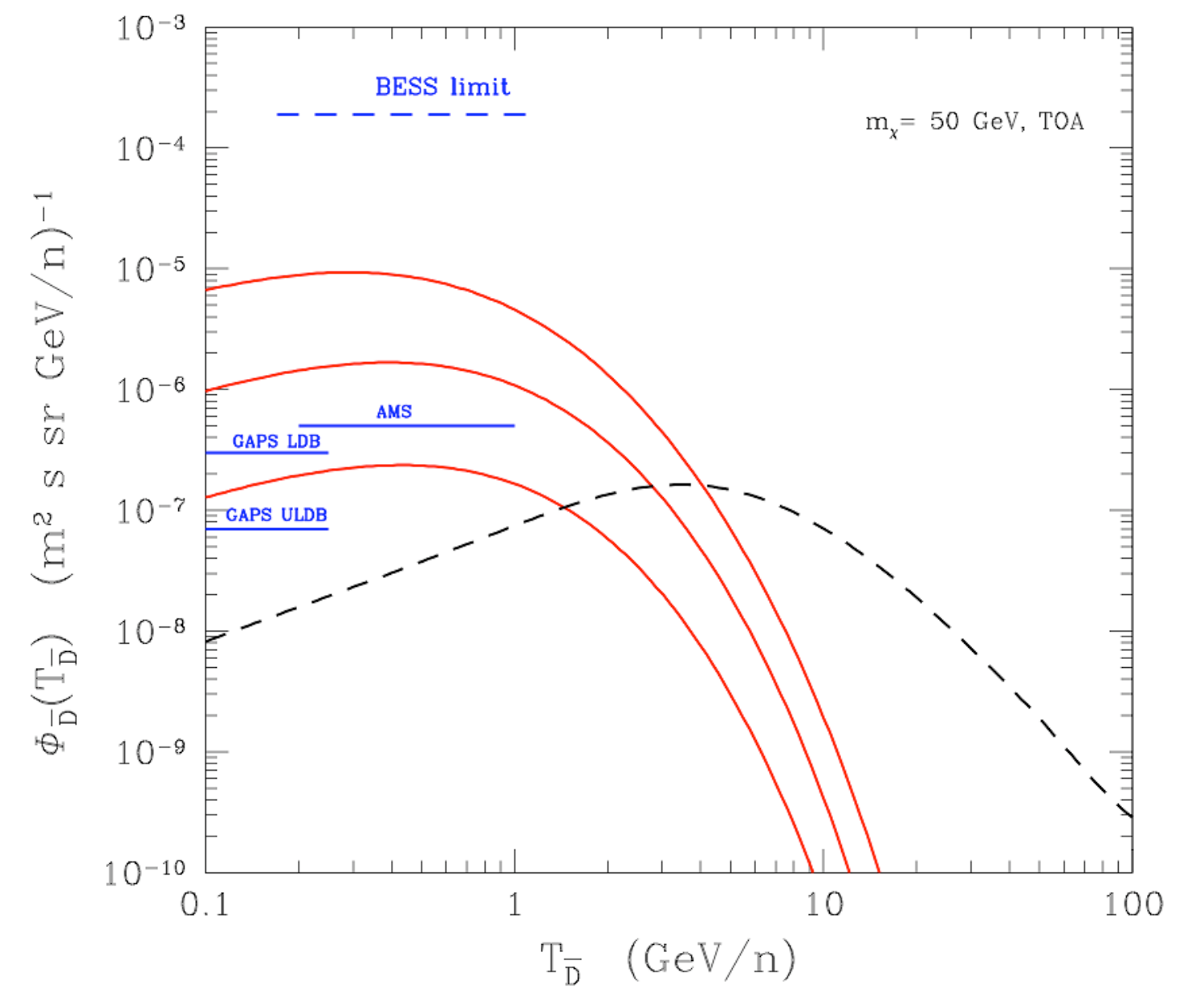}
\end{center}
%\vskip -0.5cm
\caption{
The various contributions to the antideuteron flux at the Earth,
modulated at solar minimum, are plotted as a function of kinetic
energy per nucleon $T_{\bar{\rm D}}$.
The astrophysical background (black dashed line) is generated by the
spallations of CR protons and helium nuclei on the ISM. This secondary
component is featured for the MED model of table~\ref{tab_prop}. Below
a few GeV/n, the interactions of CR antiprotons with the ISM need also
to be taken into account.
The DM signal (red solid curves) is derived for a $m_{\chi} = 50$~GeV
WIMP and for the three propagation models of table~\ref{tab_prop}.
The upper dashed horizontal line features the current BESS upper limit
on the search for CR antideuterons. The three horizontal solid (blue)
lines are the estimated sensitivities of forthcoming measurements which
will be performed by (from top to bottom) AMS-02~\cite{2002PhR...366..331A,2008ICRC....4..765C}
and GAPS on a long (LDB) and ultra-long (ULDB) duration balloon flights
\cite{2004NIMPB.214..122H,2006JCAP...01..007H,2008JPhCS.120d2011K,2010AdSpR..46.1349A}
This figure has been borrowed from~\cite{2008PhRvD..78d3506D}.
}
\label{fig:dbar_fut_exp}
\end{figure}
%FFFFFFFFFFFFFFFFFFFFFFFFFFFFFFFFFFFFFFFFFFFFFFFFFFFFFFFFFFFFFFFFFFFFFFFFFFFFFFFFFFFFFFFFFFFFFFFFFF
%

%
%DM species annihilations.
%
The above factorization and coalescence scheme can also be applied to WIMP
annihilation. In that case, the differential multiplicity for antiproton
production is expressed as
\beq
{\displaystyle \frac{dN_{\pbar}}{d \Epbar}} \, = \,
{\displaystyle \sum_{\rm F , h}} \; B_{\rm \chi h}^{\rm (F)} \;
{\displaystyle \frac{dN_{\pbar}^{\rm h}}{d \Epbar}} \;\; .
\eeq
The annihilation proceeds, through the various final states F, toward
the quark or the gluon h with the branching ratio $B_{\rm \chi h}^{\rm (F)}$.
Quarks or gluons may be directly produced when DM species self annihilate.
They may alternatively result from the intermediate production of a Higgs or
gauge boson as well as of a top quark. Each quark or gluon h generates in turn
a jet whose subsequent fragmentation and hadronization yields the antiproton
% energy spectrum ${dN_{\pbar}^{\rm h}} / {d \Epbar}$.
energy spectrum ${dN_{\bar{\rm p}}^{\rm h}} / {dE_{\bar{\rm p}}}$.
Because DM particles are at rest with respect to each other, the probability
to form, say, an antiproton with momentum $\kpbar$ is essentially  isotropic
\beq
{\displaystyle \frac{dN_{\pbar}}{d \Epbar}}(\chi + \chi \to \pbar + \ldots)
\, = \,
4 \pi \, k_{\pbar} \, \Epbar \; {\cal F}_{\pbar}(\sqrt{s} = 2 \mC , \Epbar)
\;\; .
\eeq
The factorization and coalescence scheme leads to the antideuteron differential
multiplicity
\beq
{\displaystyle \frac{dN_{\bar{\rm D}}}{d \Edbar}}\, = \,
\left( {\displaystyle \frac{4 \, P_{\rm coal}^{\, 3}}{3 \, k_{\bar{\rm D}}}} \right)
\,
\left( {\displaystyle \frac{\md}{m_{\pbar} \, m_{\nbar}}} \right)
\;
{\displaystyle \sum_{\rm F , h}} \; B_{\rm \chi h}^{\rm (F)} \,
\left\{
{\displaystyle \frac{dN_{\pbar}^{\rm h}}{d \Epbar}}
\left( \Epbar = \Edbar / 2 \right)
\right\}^{2} \;\; .
\label{dNdbar_on_dEdbar_susy}
\eeq
It may be expressed as a sum, extending over the various quarks and
gluons h, as well as over the different annihilation channels F, of the
square of the antiproton differential multiplicity. That sum is weighted
by the relevant branching ratios. The antineutron and antiproton differential
distributions have been assumed to be identical.

%
%Antideuteron signal and background.
%
The black dashed line of Fig.~\ref{fig:dbar_fut_exp} features the
astrophysical contribution to the antideuteron flux at the Earth.
It has been derived with the MED model of table~\ref{tab_prop}.
Varying the CR propagation parameters does not change much this
contribution, with a spread in possible values reaching at most
$\pm~40$-$50$~\% at energies below 1 GeV/n and decreasing down to
$\sim 15$~\% above 10~GeV/n, as shown in~\cite{2008PhRvD..78d3506D}.
Secondary antideuterons are not much affected by CR propagation
uncertainties. We already found a similar trend for secondary antiprotons.
Both species are produced by the interactions of primary CR nuclei on the
ISM, a process reminiscent of the fragmentation of CR carbon nuclei into
boron as they collide on interstellar gas. Constraining CR propagation
from the B/C ratio proves to be a powerful mean of predicting the flux
of background secondary antiprotons and antideuterons.
Such is not the case for the DM signal. The example of a 50~GeV neutralino
is presented in Fig.~\ref{fig:dbar_fut_exp} for the three CR propagation
configurations of table~\ref{tab_prop}. The spread of the red solid curves
reaches now two orders of magnitude. The larger the half-thickness $L$ of
the DH, the stronger the signal. The magnetic halo works like a fisher's
net inside which the products from DM annihilation are entrapped.
As noticed in~\cite{2000PhRvD..62d3003D}, the DM signal dominates at low energies,
below a few GeV/n, over the background. This may potentially help to
discriminate it from its astrophysical background, since very few secondary
antideuterons are expected in that energy range. Actually, these are produced
by energetic CR nuclei impinging upon atoms at rest of the ISM. The center of
mass frame of the reaction moves with respect to the Galaxy. Even though a
substantial amount of secondary antideuterons are produced with small
velocities in this frame, few remain at low energy after the proper boost
takes them back to the Galactic frame. On the contrary, DM species annihilate
at rest with respect to the Milky Way, hence a dominant contribution of primary
antideuterons below a few GeV/n.

%
%Flattening the secondary antideuteron spectrum at low energies.
%
A few processes, which were not considered in the exploratory work
of~\cite{2000PhRvD..62d3003D}, have been since then suspected to flatten the spectrum
of antideuterons by replenishing its low energy tail, erasing the above mentioned
discriminating difference between the primary and secondary components.
Inelastic but non-annihilating (INA) interactions of antideuterons with the
ISM may actually lead to such a flattening. In principle, these collisions are
expected to essentially break the antideuteron whose binding energy is very
small. That is why the INA cross section was disregarded in~\cite{2000PhRvD..62d3003D}.
But laboratory measurements of the INA cross section of antiprotons colliding
on deuterons yield~\cite{2005PhRvD..71h3013D} a non-vanishing, albeit small, value
of order 4~mb at most.
% for $\sigma_{\rm INA}^{\bar{\rm D}+{\rm p}} \leq 4$~mb.
%
Another source of flattening comes from CR {\bf antiprotons} impinging on
the ISM and producing antideuterons. This process cannot be ignored below
$\sim 1$~GeV/n, as featured in Fig.~1 of~\cite{2008PhRvD..78d3506D}.
Those two mechanisms (INA interactions and production from CR antiprotons)
generate tertiary antideuterons and a spectrum at low energy
somewhat flatter than previously anticipated~\cite{2000PhRvD..62d3003D}. The detailed
investigation carried out in~\cite{2008PhRvD..78d3506D} confirms nevertheless the
possibility of discriminating the DM signal from the astrophysical background,
as is clear in Fig.~\ref{fig:dbar_fut_exp}.
Low energy antideuterons provide then a unique tool to probe the presence
of DM particles within the Milky Way halo. The three horizontal solid blue lines
of Fig.~\ref{fig:dbar_fut_exp} indicate the potential reach of the forthcoming
measurements which will be performed by AMS-02~\cite{2002PhR...366..331A,2008ICRC....4..765C} and by
GAPS. The later is a detector with a very high rejection rate against antiprotons.
It will be flown on long (LDB) and ultra-long (ULDB) duration balloon flights
around Antartica~\cite{2004NIMPB.214..122H,2006JCAP...01..007H,2008JPhCS.120d2011K,2010AdSpR..46.1349A}.

%
%Hypothesis of isotropy revisited.
%
The hypothesis that factorization holds is certainly conservative. We have
naively assumed that both constituents of the antideuteron are independently
and isotropically distributed in the center of mass frame of the reaction. This
is certainly true for the first antinucleon and its associated jet. However,
once the axis of the pair of jets is determined, the second antinucleon tends
also to be aligned along that direction. Assuming that spherical symmetry holds
in that case leads to underestimate the probability of fusion. If both antinucleons
are back to back, they do not merge actually. But if they belong to the same jet,
their angular correlation is stronger than what has been assumed above, hence an
enhanced probability of fusion. The naive factorization and coalescence scheme
presented in this section has been recently improved by taking into account the
collimation of the antinucleons produced inside the same jet. The larger the jet
energy, the better the collimation and the easier the fusion.
The momenta of antinucleons appear to be more parallel in the Galactic
frame than in the jet frame, as a result of the Lorentz transformation that
connects both systems of coordinates.
Modeling hadronization inside jets with PYTHIA allows to take more accurately
into account the angular correlations between the antideuteron constituents
than in the naive isotropic scheme. The probability of fusion
increases~\cite{2010PhLB..683..248K} typically like the square of the jet energy.
The secondary component is mostly generated at low $\sqrt{s}$ and is not affected
by this effect. On the contrary, the collimation of the antinucleons produced
by WIMP annihilation leads to a stronger DM signal then previously estimated.
The primary component no longer decreases like ${1}/{m_{\chi}^{2}}$ and may even
overcome the background above tens of GeV, should the DM species be very heavy.
Notice however that a mass in excess of $\sim 10$~TeV does not seem very realistic.
Besides, such a DM species would still evade~\cite{2010JHEP...11..017C} the potential reach
of the above mentioned forthcoming experiments.

\section{Hot spots in the gamma-ray sky}
\label{sec:gammas}
%
%%%%%%%%%%%%%%%%%%%%%%%%%%%%%%%%%%%%%%%%%%%%%%%%%%%%%%%%%%%%%%%%%%%%%%%%%%%%%%%%%%%%%%%%%%%%%%%%%%%
%
\subsection{Production mechanisms and DM line of sight integral}
\label{sec:photon_introduction}
%
%Introduction
%
WIMP annihilations also generate high-energy photons whose energy distribution is
described by the function $f(E_{\gamma}) \equiv {dN_{\gamma}}/{dE_{\gamma}}$. The
corresponding flux at the Earth, from the direction toward which the unit vector
${\mathbf u}$ is pointing, is given by the product
\beq
\Phi_{\gamma}^{\rm DM}(E_{\gamma} , {\mathbf u}) \, = \,
\frac{\eta}{4 \pi} \,
\left\{ \! {\displaystyle
\frac{\left\langle \sigma_\mathrm{ann} v \right\rangle \, f(E_{\gamma})}{m_{\chi}^{2}}}
\! \right\} \times
{\displaystyle \int}_{\rm \!\! los} \, \rho^{2}({\mathbf x}) \,\, ds \;\; .
\label{flux_gamma_1}
\eeq
This formula is often seen as the emblem of particle astrophysics insofar as
it exhibits two distinct pieces. The first part is related to particle physics and
encodes information on the WIMP properties such as its mass and annihilation
cross section. The second term is clearly astrophysical in nature and deals with
the distribution of DM along the line of sight (los) toward which ${\mathbf u}$
is pointing.
%
%The particle physics part with the photon spectrum at the source
%
As regards the particle physics aspect, three different contributions to
$f(E_{\gamma})$ need to be considered.

\vskip 0.1cm
\noindent {\bf (i)}
The dominant source of high-energy photons is related to the production
of quarks and gauge bosons which subsequently fragment and decay into
secondary gamma-rays through essentially the two-photon decays of neutral
pions
\beq
\chi + \chi \to q \bar{q} , W^{+} W^{-} , \ldots \to \gamma + X \;\; .
\label{gamma_cont}
\eeq
For each annihilation channel, this leads to a continuum whose spectrum has
been parameterized in~\cite{2004PhRvD..70j3529F} with the generic form
\beq
{\displaystyle \frac{dN_{\gamma}^{\rm cont}}{dx}} \, = \,
x^{-1.5} \,
\exp \left( a + b x + c x^{2} + d x^{3} \right) \;\; ,
\eeq
where $x = E_{\gamma} / m_{\chi}$. This distribution exhibits a characteristic
$x^{-1.5}$ power law behavior for small values of $x$ and a smooth cut-off when
the photon energy is close to the WIMP mass.

\vskip 0.1cm
\noindent {\bf (ii)}
A particularly clear signal of the presence of DM species inside the Milky Way
halo is the production of monochromatic gamma-rays \cite{1988PhRvD..37.3737B} 
through the reaction
\beq
\chi + \chi \to \gamma + \gamma \;\; \& \;\; \gamma + Z^{0} \;\; .
\label{gamma_mono}
\eeq
This process gives rise to characteristic line signals which cannot be
mistaken for some conventional astrophysical source and which would
unequivocally signal the presence of an exotic component inside the
Galaxy, should a peak be detected in the high-energy spectrum. The energy of
the photons produced in reaction~(\ref{gamma_mono}) is respectively equal to
$E_{\gamma} = m_{\chi}$ and
$E_{\gamma} = m_{\chi} - ({m_{Z}^{2}}/{4 m_{\chi}})$.
Because WIMPs are electrically neutral, the production of monochromatic photons
is necessarily mediated by loop diagrams. It is generally suppressed and
the integrated photon yield amounts to $\sim 10^{-3}$ of the total. This leads
to a clear but faint signal which is beyond the reach of current detectors unless
the process is efficiently enhanced, as in the case of heavy Wino-like neutralinos
discussed in~\cite{2004PhRvL..92c1303H,2005PhRvD..71f3528H,2003PhRvD..67g5014H}.

\vskip 0.1cm
\noindent {\bf (iii)}
Finally, as already pointed out in~\cite{1989PhLB..225..372B}, a single photon may be
produced through internal bremsstrahlung as the WIMP pair annihilates. The gamma-ray is
radiated by the charged particles that are either exchanged (virtual internal bremsstrahlung
or VIB) or produced in the final state (final state radiation or FSR). This process
becomes particularly important for a sizable branching ratio into electron-positron
pairs, as in the case of MeV DM~\cite{2005PhRvL..94q1301B} or Kaluza-Klein inspired
models~\cite{2005PhRvL..94m1301B}. The FSR spectrum associated to the production of the
charged lepton pair $l^{+} l^{-}$ is, in leading logarithmic order, well approximated
by \cite{2005PhRvL..94m1301B,2005hep.ph....7194B}
\beq
{\displaystyle \frac{dN_{\gamma}^{\rm FSR}}{dx}} \, = \,
{\displaystyle \frac{d(\sigma_{l^{+} l^{-} \gamma} \, v)/dx}{\sigma_{l^{+} l^{-}} \, v}}
\simeq
{\displaystyle \frac{\alpha_{\rm \, em}}{\pi}} \;
{\displaystyle \frac{(x^{2} \, - \, 2x \, + \, 2)}{x}} \;
\ln \left\{ {\displaystyle \frac{m_{\chi}^{2}}{m_{l}^{2}}} \, (1 - x) \right\} \;\; .
\eeq
Final state radiation produces more photons than fragmentation does near the upper
edge $E_{\gamma} = m_{\chi}$ of the spectrum and is thus responsible for a
characteristic sharp cut-off there. For Wino-like heavy neutralinos, photons are
radiated by final $W^{+} W^{-}$ pairs as discussed in~\cite{2005PhRvL..95x1301B}.

\vskip 0.1cm
\noindent {\bf (iv)}
Photons can also be produced, albeit indirectly, by the inverse Compton (IC)
interactions of CR electrons and positrons on the Galactic radiation field.
The resulting gamma-ray flux is no longer given by relation~(\ref{flux_gamma_1})
but depends on the densities of CR electrons and positrons. These species will
be generically designated here as electrons. The IC process is very important
for DM species which preferentially annihilate into charged leptons, and most of
the models designed to explain the PAMELA positron anomaly are excluded by the
overproduction of IC photons to which they lead.
The scattering of an incoming electron with energy $E_{e}$ (and Lorentz factor
$\gamma_{e} \equiv {E_{e}}/{m_{e}}$) on a radiation field of incoming photons
with energies $E_{\rm in}$ generates the gamma-ray spectrum~\cite{1970RvMP...42..237B}
\ben
{\cal P}_{i}(E_{\gamma} , E_{e} , {\mathbf x}) \, = &\,
{\displaystyle \frac{3 \sigma_{\rm T}}{4 \gamma_{e}^{2}}} \,
{\displaystyle \int_{1/4 \gamma_{e}^{2}}^{1}} \! {\displaystyle \frac{dq}{q}} \,
{\displaystyle \frac{dn_{i}}{dE_{\rm in}}}\{E_{\rm in}(q) , {\mathbf x}\} \nn\\
& \times \left\{
2 q \ln q \, + \, q \, + \, 1 \, - \, 2 q^{2} \, + \,
{\displaystyle \frac{(1 - q) \Gamma_{e}^{2} q^{2}}{2 (1 + \Gamma_{e} q)}}
\right\} \;\; ,
\label{spectrum_IC}
\een
where $\sigma_{\rm T} = 0.665$~barn is the Thomson cross section. The parameter
$q$ is related to the energy $E_{\rm in}$ of the radiation field photon through
\beq
q \, = \,
{\displaystyle \frac{1}{\Gamma_{e}}} \,
{\displaystyle \frac{E_{\gamma}}{(E_{e} - E_{\gamma})}} \;\; .
\eeq
The dimensionless parameter
$\Gamma_{e} \equiv {4 E_{\rm in} \gamma_{e}}/{m_{e}}$ determines the regime
of the IC scattering, {\ie}, either ultra-relativistic ($\Gamma_{e} \gg 1$)
or non-relativistic ($\Gamma_{e} \ll 1$) in the Thomson limit.
Incoming photons belong to the CMB, the IR radiation field and the light
from stars. These three contributions are represented by the subscript $i$
in relation~(\ref{spectrum_IC}). The flux at the Earth of IC photons is
given by the convolution of the various contributions ${\cal P}_{i}$ with
the CR electron density $\psi_{e} \equiv {dn_{e}}/{dE_{e}}$ along the line
of sight
\beq
\Phi_{\gamma}^{\rm IC}(E_{\gamma} , {\mathbf u}) \, = \,
\frac{1}{4 \pi} \,
{\displaystyle \int}_{\rm \!\! los} \! ds \;
{\displaystyle \int_{m_{e}}^{m_{\chi}}} \, dE_{e} \times
\psi_{e}(E_{e} , {\mathbf x}) \times
{\displaystyle \sum_{i}} \;
{\cal P}_{i}(E_{\gamma} , E_{e} , {\mathbf x}) \;\; .
\label{flux_gamma_IC}
\eeq
The CR electron density $\psi_{e}$ is generated by the collisions of CR
protons and helium nuclei on the ISM (background) and also by WIMP
annihilations (DM signal). It needs to be calculated everywhere inside
the Galactic magnetic halo.

%
%Discussion of the integral along the line of sight J.
\vskip 0.1cm
In the case of the two-photon line, relation~(\ref{flux_gamma_1}) leads to the
estimate
\beq
\Phi_{\gamma}^{\rm DM} (E_{\gamma} = m_{\chi} , {\mathbf u}) \, = \,
1.88 \times 10^{-13} \;\; {\rm photons \;\; cm^{-2} \; s^{-1} \; sr^{-1}} \, \times \,
{\displaystyle \frac
{\left< \sigma_{\gamma \gamma} v \right>_{29}}{{m_{100}}^{2}}} \, \times \,
J({\mathbf u}) \;\; .
\label{flux_gamma_2}
\eeq
The annihilation cross section $\left< \sigma_{\gamma \gamma} v \right>$
and the WIMP mass $m_{\chi}$ are respectively expressed in units of
$10^{-29}$ cm$^{3}$ s$^{-1}$ and 100 GeV. A Majorana type DM species has been
assumed here with $\eta \equiv {1}/{2}$. The los integral $J({\mathbf u})$ has
been redefined as
\beq
J({\mathbf u}) \, = \,
\left\{ \rho_{\odot}^{2} \; r_{\odot} \right\}^{-1} \times
{\displaystyle \int}_{\rm \!\! los} \, \rho^{2}({\mathbf x}) \,\, ds \;\; ,
\eeq
with a solar neighborhood DM density of $\rho_{\odot} = 0.3$ GeV cm$^{-3}$.
The galactocentric distance $r_{\odot}$ of the solar system has been set
equal to $8.5$~kpc.
The los integral $J$ depends on the Galactic DM distribution. The various halo
models of table~\ref{tab:indices} yield similar values all over the sky, except
in the direction of the Galactic center where predictions can vary by several
orders of magnitude, depending on the assumed profile. As shown
in~\cite{2009PhLB..671...10R}, where it has been averaged over a solid angle of
$10^{-5}$~sr, $J$ is respectively equal to $30$ and $1.45 \times 10^{4}$ for
the cored isothermal and NFW distributions. In the case of a Moore profile which
has been adiabatically compressed by the collapse of the supermassive black hole
lying at the center of the Milky Way, $J$ may even reach the extreme value of
$3 \times 10^{8}$.
The astrophysical uncertainties are enormous. The possibility
of a strong signal has triggered a febrile activity around the Galactic
center and motivated observations focusing in that direction.
Let us concentrate, for instance, on the NFW profile of table~\ref{tab:indices}.
For small galactocentric distances, the DM distribution simplifies into
\beq
\rho(r) \, = \, \aleph \; \rho_{\odot} \;
\left\{
{\displaystyle \frac{r_{\odot}}{r}} \right\} \;\; ,
\label{eq:profile_NFW}
\eeq
where the normalization constant $\aleph \simeq 2.03$. The los integral $J$ is
found equal to ${\pi \, \aleph^{2}}/{\alpha}$ at an angular distance $\alpha$ from
the Galactic center. Once averaged over a disk with angular radius $\theta$, this
leads to $\langle J \rangle = {2  \pi \, \aleph^{2}}/{\alpha}$ and to a numerical
value of $\sim 1500 \times ({1^{\circ}}/{\theta})$.
%
%%%%%%%%%%%%%%%%%%%%%%%%%%%%%%%%%%%%%%%%%%%%%%%%%%%%%%%%%%%%%%%%%%%%%%%%%%%%%%%%%%%%%%%%%%%%%%%%%%%
%
\subsection{Atmospheric Cherenkov telescopes -- a new generation of instruments}
\label{sec:photon_GC}
%
%Discussion of the experimental part
%
%Instruments
%
High-energy photons can be detected by satellite borne instruments orbiting
the Earth and free, therefore, from the screening of the atmosphere. Because
of the reduced payload which can be carried up in orbit, the collecting area
of gamma-ray telescopes like Fermi-LAT~\footnote{See the web site \url{http://www-glast.stanford.edu/} and references therein.}
is quite reduced. Above a few tens of GeV, atmospheric Cherenkov telescopes (ACT)
come into play and offer a nice alternative which complements space observations.
The effective detecting area can be actually quite large insofar as any high-energy
gamma-ray impinging upon the upper atmosphere generates a shower, and is degraded
into many optical photons which ordinary telescopes spread on the ground can detect.
The tracks left in the focal planes of these telescopes give the direction of the
shower on the sky. Several instruments allow the stereoscopic reconstruction
of the direction of the initial particle. The Cherenkov light illuminates on the
ground a disk whose diameter is of order 250 m, hence a large effective area of
$\sim~0.05$~km$^{2}$ for each telescope, to be compared with the typical square
meter collecting surface of a space instrument. The intensity of the image depends
on the energy of the incident photon. The ACT technique is thus a powerful tool
allowing the observation of the high-energy gamma-ray sky.

%
%Backgrounds
%
However, as for any other observation, a gamma-ray signal is only detectable
as long as it emerges from the background. More specifically, the signal has
to be larger than the statistical fluctuations, also called the noise, of
the latter.
For atmospheric Cherenkov telescopes, the dominant background arises from
CR electrons that penetrate the atmosphere inside which they produce
electromagnetic showers of the same type as those induced by high-energy photons.
It is not possible to distinguish photons from electrons since both species
lead to the same light pattern on the ground. Fermi-LAT measurements of the CR
electron and positron flux can be fitted~\cite{2009APh....32..140G} by the
power-law~(\ref{fermi_LAT_phi_e}).
%
% The cosmic ray electron flux is given by~\cite{Nishimura:1980pz}
% \beq
% \Phi_{e^{-}}({E_{e}}) \, = \, 6.4 \times 10^{-2} \;\;
% {\rm GeV^{-1} \; cm^{-2} \; s^{-1} \; sr^{-1}} \;\;
% \left( {E_{e}} / {\rm 1 \; GeV} \right)^{-3.3 \pm 0.2} \;\; .
% \label{flux_electron}
% \eeq
%
Impinging CR hadrons also interact with the atmosphere. The showers
which they generate tend to develop at a lesser altitude and are more widely
spread on the ground than those of the electromagnetic type. Stereoscopy is
a powerful tool to discriminate hadrons from electrons and gamma-rays since
the pattern recognition of the light pool is then possible. Observations
performed between 50~GeV and 2~TeV yield a hadron flux~\cite{1972PhRvL..28..985R}
\beq
\Phi_{\rm had}(E) \, = \, 1.8 \;\;
{\rm GeV^{-1} \; cm^{-2} \; s^{-1} \; sr^{-1}} \;\;
\left( {E}/{\rm 1 \; GeV} \right)^{-2.75} \;\; .
\eeq
A small fraction of the hadron-induced showers are mistaken
for gamma-ray events though. This is the dominant source of background at
high energy as hadrons have a harder spectrum than electrons. The HESS
collaboration quotes a rejection factor of one misidentified event over
a sample of 300 showers generated by CR protons.
Satellite borne instruments do not suffer from the same problem. However,
point sources can still be buried inside a Galactic gamma-ray diffuse
emission produced, above 100~MeV, by the spallation of interstellar gas
by CR nuclei. The flux of this diffuse emission is given by the convolution
along the los of the hydrogen density $n_{\rm H}$ with the gamma-ray emissivity
$I_{\rm H}$ per hydrogen atom
\beq
\Phi_{\gamma}^{\rm sec}(E_{\gamma} , {\mathbf u}) \, = \,
{\displaystyle \int}_{\rm \!\! los} \, n_{\rm H}({\mathbf x}) \times
I_{\rm H}({\mathbf x} , E_{\gamma}) \times ds \;\; .
\label{sec_gamma_flux}
\eeq
The Galactic diffuse emission is made of secondary photons
resulting mostly from the interactions of high-energy CR protons and
helium nuclei on the ISM. This process has already been discussed for
antiprotons and positrons. It has been respectively described by the
source terms
$q_{\bar{\rm p}}^{\rm sec}({\mathbf x} , E_{\bar{\rm p}})$ in
relation~(\ref{source_sec_pbar}) of section~\ref{sec:pbar_calculation}
and
$q_{e^{+}}^{\rm sec}({\mathbf x} , E_{e})$ in relation~(\ref{source_sec_pos})
of section~\ref{sec:sec_pos_background}. The photon emissivity per hydrogen
atom is defined as
\beq
I_{\rm H}({\mathbf x} , E_{\gamma}) \, = \,
{\displaystyle \frac{q_{\gamma}^{\rm sec}({\mathbf x} , E_{\gamma})}
{4 \, \pi \, n_{\rm H}({\mathbf x})}} \;\; ,
\eeq
and may be written as the convolution over proton energy of the CR
proton flux with the differential photo-production cross section
of proton-proton interactions
\beq
I_{\rm H}({\mathbf x} , E_{\gamma}) \, = \,
{\displaystyle \int} \, \Phi_{\rm p} \! \left( {\mathbf x} , E_{\rm p} \, \right)
\times dE_{\rm p} \times
{\displaystyle \frac{d\sigma}{dE_{\gamma}}}(E_{\rm p} \to E_{\gamma}) \;\; .
\label{IH}
\eeq
The emissivity $I_{\rm H}$ is expressed in units of GeV$^{-1}$~s$^{-1}$~sr$^{-1}$
since it is essentially a production rate per unit of energy and solid angle.
In the solar neighborhood, the effective gamma-ray emissivity of the ISM, whose
atoms are illuminated by the local CR protons and helium nuclei, may be
approximated, between 3~GeV and 1~TeV, by the power law~\cite{2011A&A...531A..37D}
\beq
I_{\rm H}^{\rm \, eff}(\odot , E_{\gamma}) \, = \, 3.55 \times 10^{-27} \;\;
{\rm GeV^{-1} \; s^{-1} \; sr^{-1}} \;\;
\left( {E_{\gamma}}/{\rm 1 \; GeV} \right)^{-2.76} \;\; .
\label{emissivity_inner_M87}
\eeq
This effective emissivity is scaled back to each hydrogen atom and can be
readily used in expression~(\ref{sec_gamma_flux}) to infer the photon flux.
The Galactic diffuse emission dominates over an extragalactic component
which has been measured~\cite{1998ApJ...494..523S} with the EGRET instrument on board
the CGRO satellite
\beq
\Phi_{\gamma}^{\rm eg}(E_{\gamma}) \, = \,
7.32 \pm 0.34 \times 10^{-9} \;\;
{\rm MeV^{-1} \; cm^{-2} \; s^{-1} \; sr^{-1}} \;\;
\left( {E_{\gamma}}/{\rm 451 \; MeV} \right)^{-2.10 \pm 0.03} \;\; .
\eeq

%
%Sensitivity of a typical ACT
%
An atmospheric Cherenkov telescope of the HESS caliber has an effective detecting
area $\mathcal{S}$ of order 0.1~km$^{2}$ with four mirrors spread on a
300~m~$\times$~300~m square. One of the main targets of HESS is the Galactic
center where the putative WIMPs might have collapsed, producing a hot spot
in the gamma-ray sky. We will assume in what follows that the DM profile is
given by the NFW distribution of table~\ref{tab:indices}. The effective
time $\mathcal{T}$ during which the observation of the Galactic center is
performed, disregarding the periods of daylight as well as the nights during
which the Moon shines, will be taken to be a month. We infer an approximate
acceptance of
\beq
\mathcal{S} \times \mathcal{T} \approx 0.01 \;\;
{\rm km^{2} \; yr} \;\; .
\eeq
Let us also assume that this HESS type telescope monitors a circular field
of view with angular radius $\theta \sim 1^{\circ}$ surrounding the center
of the Milky Way. For illustration purposes, the self-annihilation of WIMPs
into photon pairs is the only process in which we will be interested here.
It produces monochromatic gamma-rays with energy $E_{\gamma} = m_{\chi}$.
The number of line photons collected during such a run is
\beq
N_{\gamma}^{\rm DM} \, = \,
850 \;\; {\rm photons} \, \times \,
{\displaystyle \frac
{\left< \sigma_{\gamma \gamma} v \right>_{29}}{{m_{100}}^{2}}} \, \times \,
\left\{ \! {\displaystyle \frac{\theta}{1^{\circ}}} \! \right\} \;\; .
\eeq
These monochromatic gamma-rays are detected within some energy bin whose
width is set by the resolution of the telescope. We may safely take an energy
resolution ${\sigma(E)}/{E}$ of order 10\%, to be compared to the value of 15\%
in the case of HESS. The energy bin that contains the line has thus a width
$\Delta E_{\gamma}$ of order $0.1 \times m_{\chi}$. The better the energy resolution,
the narrower the line bin and the more visible the peak in the photon spectrum.
As discussed above, all the bins are filled up predominantly by misidentified
hadron and electron events. The monochromatic signal from annihilating DM species
is detectable only if it exceeds the statistical fluctuations of that background.
If electron induced showers are assumed to be the only source of background,
although this may not be true at high energy where hadrons come also into play,
the number of background events collected inside the line bin, during the run, 
amounts to
\beq
N_{\gamma}^{\rm BK} \, = \, 4.3 \times 10^{5} \;\; {\rm photons} \, \times \,
(m_{100})^{-2.045} \, \times \,
\left\{ \! {\displaystyle \frac{\theta}{1^{\circ}}} \! \right\}^{2} \;\; .
\eeq
The DM line signal $N_{\gamma}^{\rm DM}$ is deeply swamped into the
background $N_{\gamma}^{\rm BK}$ and seems hopelessly out of reach.
However, electronic and hadronic events are homogeneously spread on the sky
since these cosmic radiations are isotropic at the Earth. Changing the direction
toward which the telescope is pointing does not affect the number of background
events. On the contrary, the line signal disappears as soon as the field of view
no longer encompasses the Galactic center. By alternatively pointing the telescope
on and off the source makes it possible to detect the line signal provided that
it exceeds the statistical background fluctuations. A good estimate for these
is given by the Poisson noise $(N_{\gamma}^{\rm BK})^{1/2}$. Detection of a signal
with a significance of n is therefore achieved whenever the signal to noise ratio
is equal to
\beq
{N_{\gamma}^{\rm DM}}/{\sqrt{N_{\gamma}^{\rm BK}}} \, = \, {\rm n} \;\; .
\eeq
The HESS like telescope of this example would detect at the 3~$\sigma$
level a NFW distribution of DM species at the Galactic center should their
mass and two-photon annihilation cross section fulfill the condition
\beq
\left< \sigma_{\gamma \gamma} v \right> \, \geq \,
2.3 \times 10^{-29} \;\; {\rm cm^{3} \; s^{-1}} \, \times \,
(m_{100})^{0.98} \;\; .
\eeq
%
%A short paragraph on what HESS has (not) discovered at the Galactic center.
%
The HESS collaboration has observed the Galactic center with unprecedented
accuracy~\cite{2004A&A...425L..13A,2006Natur.439..695A} above 160~GeV.
A strong point-like source called HESS J1745--290 is detected at the positions
of the supermassive black hole Sagittarius A$^{*}$ and supernova remnant
Sagittarius A East. Its spectral index is $\sim 2.25$.
An important gamma-ray diffuse emission is also seen in the same direction,
with a similar spectral index. It is clearly associated with the band of
molecular clouds lying in the central region and mapped from their CS line.
These clouds have been recently penetrated by CR protons and nuclei accelerated
by a nearby supernova event. The correlation between the intensity of the TeV
diffuse emission and the gas column density of the clouds is striking and
suggests a uniform density of cosmic rays.
No line is observed though. Moreover, the gamma-ray spectrum is too hard to be
compatible with a WIMP annihilation signal. As shown in~\cite{2005PhRvL..94m1301B},
a Kaluza-Klein DM species could still yield the same flat energy distribution
if the contributions from internal bremsstrahlung and $\tau$ lepton decays are
included. The price to pay, however, is an unacceptably large mass $m_{\chi}$
of 10~TeV.
A contribution from DM annihilation cannot be ruled out either, provided that
it contributes less than 10\% of the signal.

\subsection{Current observations, constraints and prospects}
\label{subsec:gamma_res}
The launch of the Fermi satellite in 2009 has considerably improved the picture in the field
of indirect DM searches. The LAT instrument onboard the Fermi satellite has allowed for 
a tremendous recording of the gamma-ray sky with unprecedented statistics and resolution, 
overtopping the amount and quality of the data collected by previous space experiments, 
like EGRET or AGILE. Data taking is expected to last until 2014. The effective detection area 
reaches $\sim$10,000 ${\rm cm^2}$ over an energy range spreading from $\sim$30 MeV to 
$\sim$300 GeV, with an angular resolution $\sim 0.1^\circ$ in a field of view of $\sim 1/5$ of the
sky. Such a low threshold together with the capability of observing the full sky in a short time 
make the Fermi-LAT instrument better than current ground-based gamma-ray telescopes for DM
searches. The latter are actually complementary since they provide a better sensitivity 
for energies larger than 100-200 GeV, but only for rather localized sources (ground-based 
telescopes are optimized for a pointing mode). We still note that much larger Cherenkov telescope 
arrays may reach the required sensitivity in the future to start probing DM signals in 
several classes of targets efficiently \cite{2011PhRvD..83d5024B}. Consequently, the following 
discussion will be strongly biased toward the Fermi-LAT data.

We will discuss the current results and prospects target by target, reviewing some 
predictions and constraints for DM signals from (i) the Galactic 
center, (ii) the diffuse Galactic and extragalactic emissions, and (iii) extragalactic 
sources (dwarf spheroidal galaxies, neighbor galaxies, and galaxy clusters). We emphasize
that the observational constraints associated with these targets only concern the s-wave part
of the annihilation cross section, while the p-wave part is also often relevant in setting
the cosmological DM abundance. Hence, the canonical value of $\sigv = 3\times 10^{-26}$ 
cm$^3$/s that will often appear in the following has to be taken with caution, keeping the 
previous remark in mind.
\subsubsection{The Galactic center}
\label{subsubsec:gamma_gc}
The Milky Way is obviously the biggest nearby DM reservoir where to seek DM
annihilation traces. Since DM is expected to concentrate at the centers of structures, the 
Galactic center (GC), which is located at $\sim 8$ kpc from the Sun, should be the most luminous
known DM source for the Earthian observer. This region was actually observed
with the EGRET telescope onboard the CGRO satellite, though with a poor angular resolution of 
$\sim 1^\circ$, which indeed led to the detection a gamma-ray 
signal in the 0.03-10 GeV energy range \cite{1998A&A...335..161M}. Interpretations in terms of DM
annihilation were performed for supersymmetric candidates (\eg~\cite{2004APh....21..267C}), 
but the main difficulty which arose (but was expected) was to distinguish between this putative
exotic origin and an astrophysical origin of the signal. This points toward the challenging 
character of looking for DM annihilation traces toward the GC: this region is
dominated by baryons, and high-energy astrophysical processes are known to take place (potential 
sites of CR acceleration, pion production through interactions of CRs with the ISM gas, inverse 
Compton from high-energy electrons, \etc). Anyway, the authors of Ref. \cite{2004APh....21..267C}
reached the conclusion that DM annihilation could provide a good fit to the EGRET data
when taking neutralinos in the 50-100 GeV mass range (50 GeV being the lower limit they fixed to 
their scan over parameters), but only for central densities larger than usually expected. Adopting 
an NFW DM profile normalized to $\rhosun = 0.3$ GeV/cm$^3$ at the Sun's position (taken at 
8.5 kpc from the GC), they fell short by 1 order of magnitude, which they argue could be compensated
by a more spiky profile resulting from baryon cooling effects known as adiabatic 
contraction \cite{1986ApJ...301...27B,1999PhRvL..83.1719G,2004ApJ...616...16G}. We note
that the current state-of-the-art numerical works in cosmological structure formation now seem
to favor a picture where feedback effects dominate over cooling, making contraction 
scenarios much less motivated (\eg~\cite{2010Natur.463..203G}).

The GC was further observed, though at higher energies, with the HESS ground-based telescope, and 
a signal was reported in the 0.25-10 TeV energy range ~\cite{2004A&A...425L..13A}. This signal is 
consistent with a point-like source coincident within 1' of Sgr A$^*$ (the data were taken with 
an angular resolution of $\sim 0.1^\circ$). The DM interpretation is made difficult because 
of the hard spectrum, well fitted with a power-law of spectral index $\sim -2.2$
up to 10 TeV, which implies very massive DM particles if one assumes their annihilation to 
be the main photon source. The exercise was still performed for supersymmetric as well as 
Kaluza-Klein candidates (\eg~\cite{2005PhRvL..94m1301B} and \cite{2005PhRvL..95x1301B}, 
respectively), where multi-TeV mass particles were found able to saturate the signal, though with a 
loose fit and to the price of an arbitrarily large boost factor of $\sim 1000$. Such an 
amplification is generally difficult to explain in terms of DM distribution, but 
non-perturbative effects were pointed out to significantly increase the annihilation cross section 
of such heavy particles \cite{2005PhRvD..71f3528H}, though by a factor limited to $\lesssim 100$ to 
simultaneously achieve the correct relic density \cite{2010PhRvL.104o1301F}. Nevertheless, as for 
the EGRET signal, astrophysical scenarios were also shown to be plausible 
(\eg~\cite{2005ApJ...619..306A,2011ApJ...726...60C}), making the DM scenario less 
attractive. Despite the absence of a robust smoking gun for DM annihilation, the HESS 
data have still been shown very powerful to constrain models proposed to fit the PAMELA data;
examples can be found in \cite{2009JCAP...03..009B,2010JCAP...11..041A}.

\begin{figure*}[t!]%[htp]
\centering
\begin{minipage}[t]{0.49\textwidth}
\vspace{0pt}
\includegraphics[width=\textwidth]{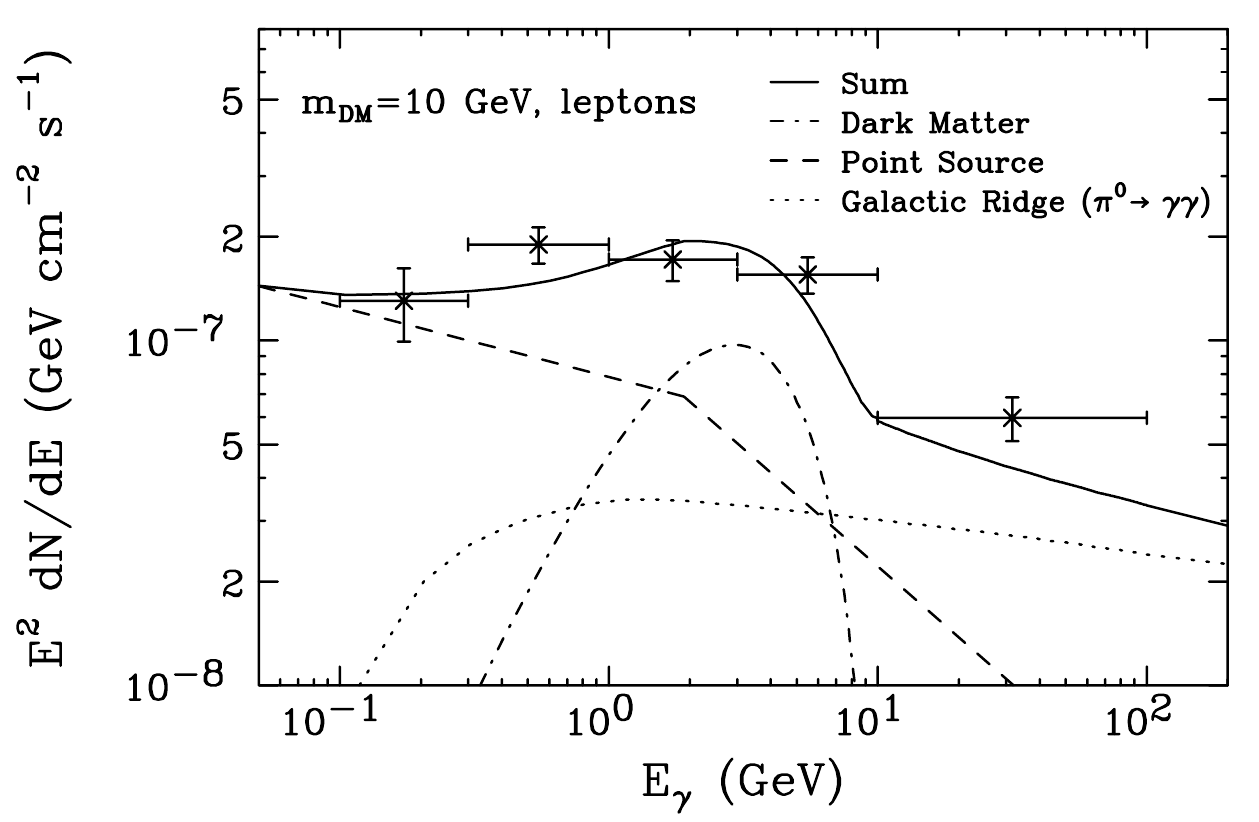}
\end{minipage}
\begin{minipage}[t]{0.49\textwidth}
\vspace{0pt}
\caption{Gamma-ray excess found toward the GC in 
\cite{2011PhLB..697..412H,2011PhRvD..84l3005H}, and tentative interpretation in terms of WIMPs
annihilating into pairs of tau leptons. The plot is taken from Ref. \cite{2011PhRvD..84l3005H}}
\label{fig:gc_hooper}
\end{minipage}
\end{figure*}

More recently, several studies were conducted relying on the publicly available Fermi data. 
The authors of Refs.~\cite{2011PhLB..697..412H,2011PhRvD..84l3005H} have found evidences for 
an extended source centered around the GC, the intensity of which is seemingly hard to explain in
terms of the point source located at the GC, probably also responsible for the TeV emission.
These authors have attempted a DM interpretation of the signal, suggesting that the spatial
and spectral fits to the data improved significantly by adding a $\sim 7-10$ GeV WIMP annihilating 
preferentially into tauons, and distributed according to a rather spiky profile scaling 
like $\rho \propto r^{-1.3}$ (normalized to 0.4 GeV/cm$^3$ locally). They need an annihilation
cross section of $\sigv \sim 7 \times 10^{-27}$ cm$^3$/s, rather consistent with cosmological
requirements. Moreover, they emphasized that the mass range is consistent with the recent hints
coming from direct detection experiments --- their result is shown in \citefig{fig:gc_hooper}. 
Nevertheless, a criticism to this analysis can be found in \cite{2011PhLB..705..165B}, where
no net excess is found with respect to canonical models of diffuse emission. More generally, we 
note that the GC is a region very complicated to model, though as fascinating as challenging, which 
makes DM searches very difficult toward this target unless very specific signatures are 
observed. Indeed, finding excesses with respect to {\em a fortiori} non-perfect descriptions of 
the region is not that a surprise. This statement is nicely reinforced by the recent finding, in 
the Fermi data, of large scale bubbles extending away from the GC \cite{2010ApJ...724.1044S}, to 
which astrophysical explanations have been
proposed \cite{2010ApJ...724.1044S,2011PhRvL.106j1102C,2011PhRvL.107i1101M} --- while the 
believer's eyes could see a DM signal, as apparent as in Fig. 7 of 
\cite{2011PhRvD..83b3518P}. Hence, strong improvements in the understanding of CR acceleration and 
transport, in the modeling of the
interstellar material, and in the census of the CR sources there, are clearly necessary to perform 
more reliable and detailed analyses of the GC and find DM signals in the near future --- 
complementary with other targets and messengers. The 
encouraging point is that GC observations at gamma-ray frequencies still do not exclude a DM
contribution. Regarding the experimental forecasts, a keypoint will be to improve the 
angular resolution and sensitivity in the 1-100 GeV energy range, which might be partly achieved 
with the new or coming generation of Cherenkov telescopes, like HESS-2 and CTA.

Finally, for completeness, we may also mention the possibility that DM annihilation
could also contribute to the 511 keV emission extended around the GC, which was measured to an 
unprecedented precision with the Integral satellite \cite{2005A&A...441..513K}. This would imply 
very light scalar DM candidates in the MeV mass range, which is formally not excluded 
\cite{2004PhRvL..92j1301B}. Again, many classical alternatives exist in astrophysics, which makes 
it difficult to unambiguously discuss this intense emission in terms of DM. A specific 
review on this topic can be found in Ref.~\cite{2011RvMP...83.1001P} for further details.

\subsubsection{Diffuse Galactic and extra-galactic emissions}
\label{subsubsec:gamma_diff}
Fermi-LAT observations of the full gamma-ray sky have allowed to scrutinize the properties of the
Galactic diffuse emission, as well as those of the isotropic extra-galactic component 
(see \eg~\cite{2009PhRvL.103y1101A,2012ApJ...750....3A}). DM annihilation in the Galactic 
halo is also expected to contribute to both these diffusion emissions, which makes the all-sky data
suitable for searches for departure from conventional astrophysical predictions. This again 
necessitates reliable models of astrophysical foregrounds, and significant efforts are currently 
made to improve them (see \eg~\cite{2012ApJ...750....3A,2011A&A...531A..37D} for the Galactic 
component, and \cite{2011PhRvD..84j3007A} for the extragalactic one).

To circumvent spurious effects coming from the potential use of an imperfect background model,
a first strategy consists in looking for signatures which are very specific to DM: 
gamma-ray lines \cite{1988PhRvD..37.3737B,1989PhRvD..40.3168B,1994APh.....2..261B}, or sharp 
spectral features stemming from internal bremsstrahlung 
diagrams~\cite{2005PhRvL..94q1301B,2005PhRvL..95x1301B,2008JHEP...01..049B}. The former signals are 
loop suppressed, while the latter come with significant amplitudes, inducing a sharp spectral bump 
close to the WIMP mass. 

Gamma-ray line searches in the Fermi data have been performed by several 
groups, \eg~\cite{2010PhRvL.104i1302A,2011JCAP...05..027V}. No signal was found, and limits have 
been derived for annihilating as well as decaying DM. The results obtained in 
Ref.~\cite{2011JCAP...05..027V}, valid in the 1-300 GeV energy range, are shown in 
\citefig{fig:gamma_lines}. The authors assumed an NFW halo, with a local density of 
$\rhosun = 0.4$ GeV/cm$^3$, and a scale radius of $r_s = 20$ kpc (variations of the halo shape 
appear as blue bands in the plots), and derived limits for
the annihilation into a pair of photons, and for the decay into a photon and a neutrino. It 
is an easy exercise to recast their results for other final states (\eg\ $\gamma\, X$, where
$X$ is any massive particle). The obtained limits are not of concern for most of supersymmetric
models leading to annihilating neutralino DM, while they become stringent for more
effective models \cite{2011NuPhB.844...55G}. However, they strongly constrain the
degree of R-parity breaking in scenarios of decaying gravitino DM, as well as
on the stau NLSP lifetime, whenever relevant. We refer the reader to the original paper
\cite{2011JCAP...05..027V} for more details.

\begin{figure*}[t]%[htp]
\centering
\includegraphics[width=0.49\textwidth]{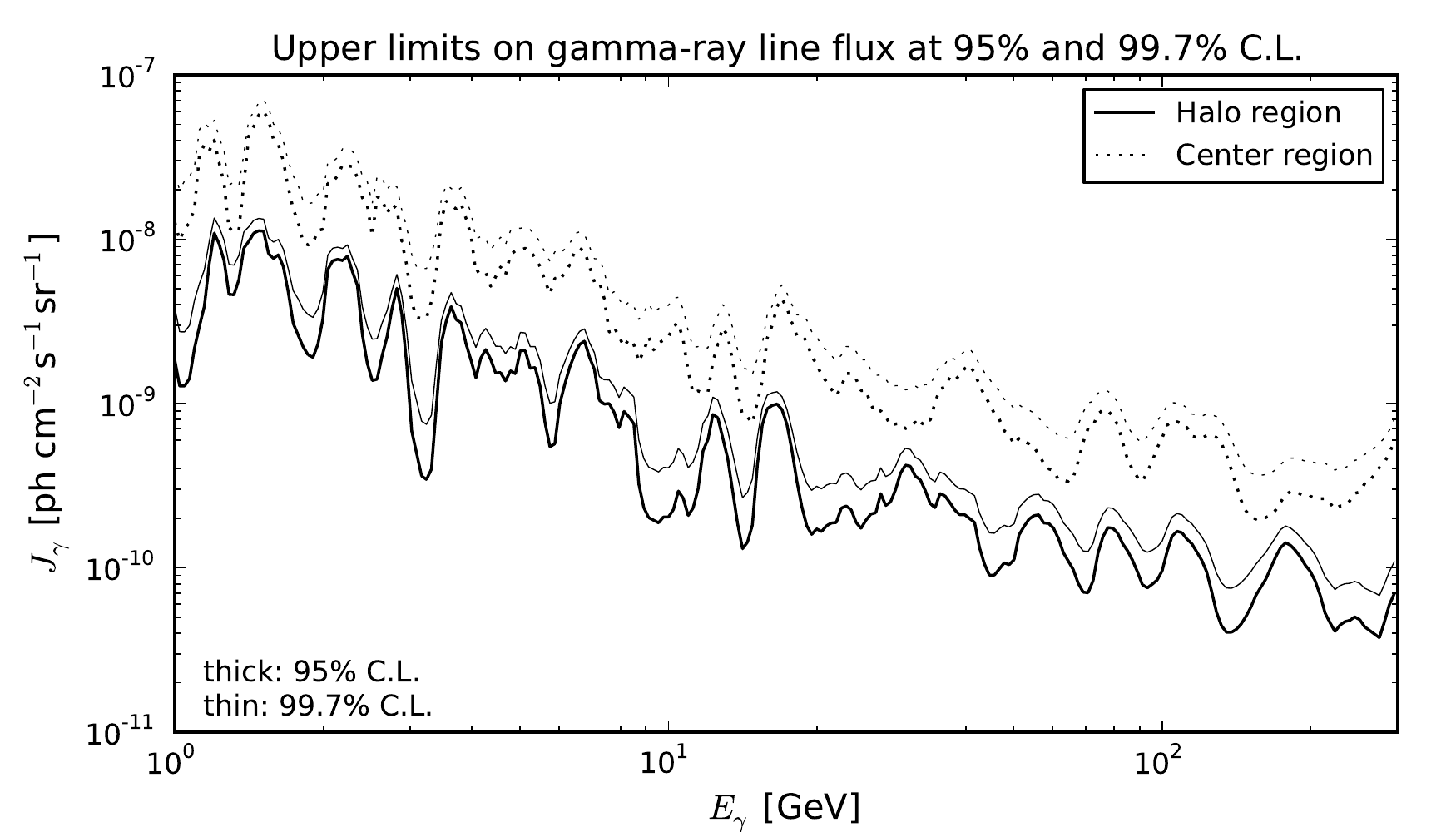}
\includegraphics[width=0.49\textwidth]{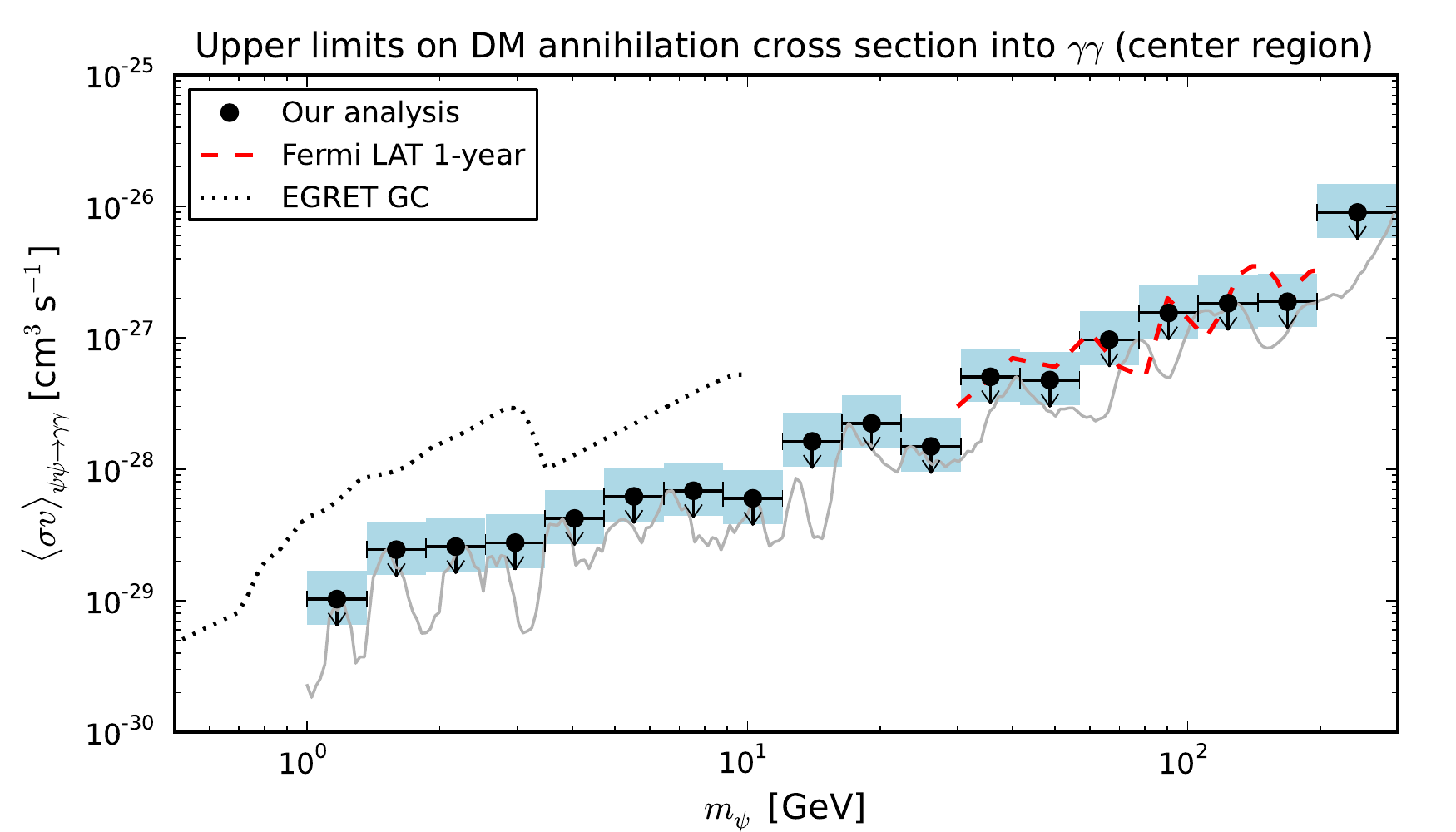}
\begin{minipage}[t]{0.49\textwidth}
\vspace{0pt}
\includegraphics[width=\textwidth]{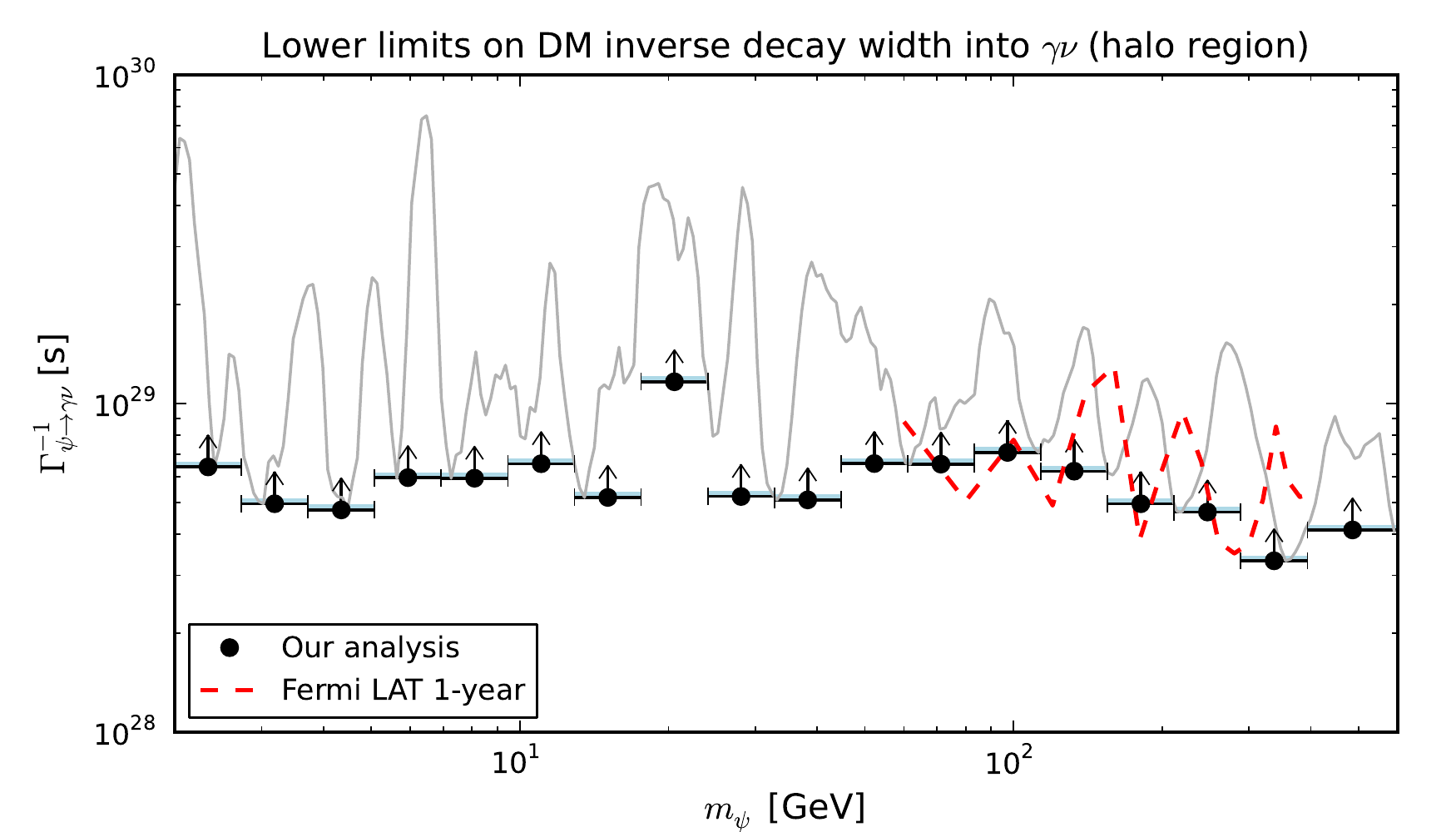}
\end{minipage}
\begin{minipage}[t]{0.49\textwidth}
\vspace{0pt} 
\caption{Results taken from Ref.~\cite{2011JCAP...05..027V} (the results obtained in 
Ref.~\cite{2010PhRvL.104i1302A} are shown as red dashed curves). Top left: model-independent limits 
on the gamma-ray line flux as a function of energy. Top right: limits in terms of DM
annihilation cross section into a pair of photons. Bottom left: limits in terms of DM decay
rate (decay into a photon and a neutrino).}
\label{fig:gamma_lines}
\end{minipage}
\end{figure*}
%%
%%\begin{sidewaysfigure}[t!]%[htp]
%%\centering
%%\includegraphics[width=0.32\textwidth]{fig_gamma_lines_flux.pdf}
%%\includegraphics[width=0.32\textwidth]{fig_gamma_lines_adm.pdf}
%%\includegraphics[width=0.32\textwidth]{fig_gamma_lines_ddm.pdf}
%%\caption{Results taken from Ref.~\cite{2011JCAP...05..027V} (the results obtained in 
%%Ref.~\cite{2010PhRvL.104i1302A} are shown as red dashed curves). Top left: model-independent limits 
%%on the gamma-ray line flux as a function of energy. Top right: limits in terms of DM
%%annihilation cross section into a pair of photons. Bottom left: limits in terms of DM decay
%%rate (decay into a photon and a neutrino).}
%%\label{fig:gamma_lines}
%%\end{sidewaysfigure}
%%
%%
\begin{figure*}[t]%[htp]
 \centering
\includegraphics[width=0.49\textwidth]{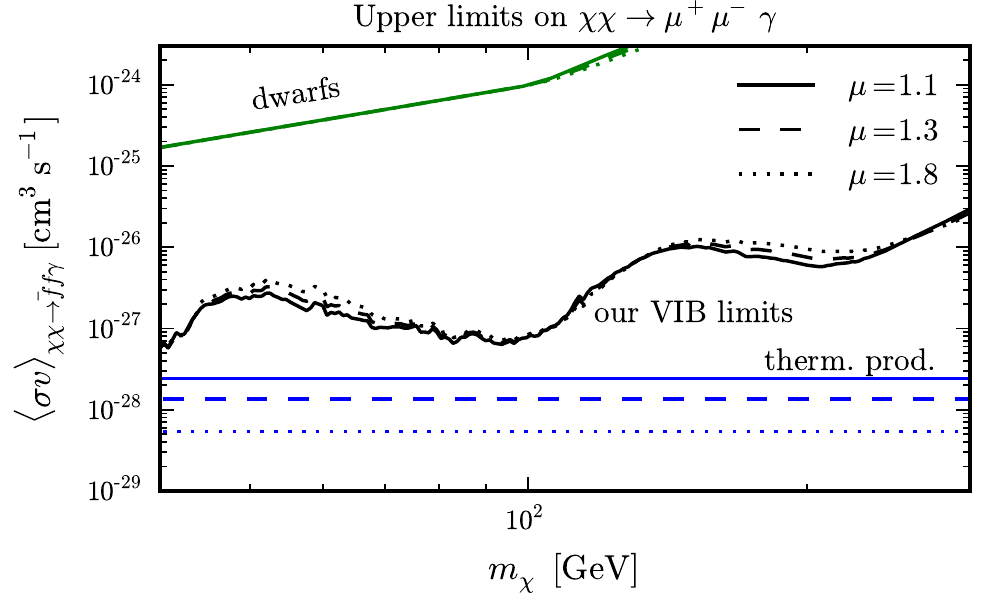}
\includegraphics[width=0.49\textwidth]{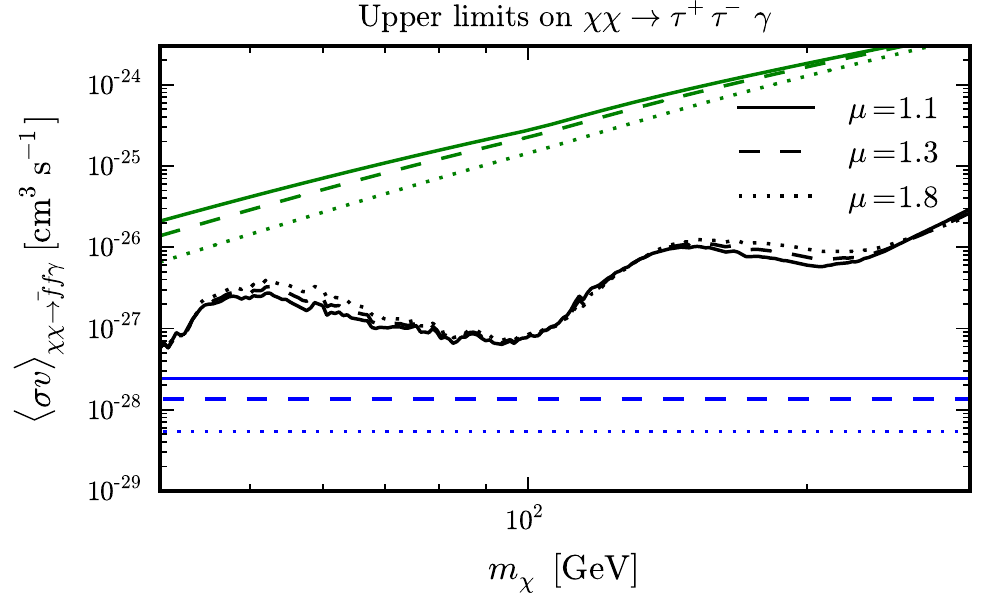}
\begin{minipage}[t]{0.49\textwidth}
\vspace{0pt}
\includegraphics[width=\textwidth]{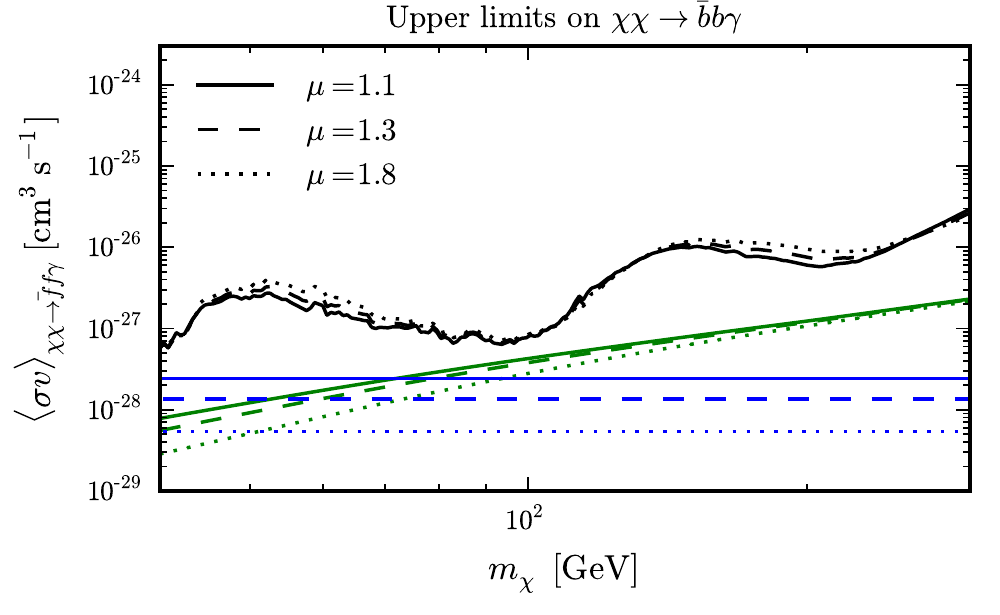}
\end{minipage}
\begin{minipage}[t]{0.49\textwidth}
\vspace{0pt} 
\caption{Limits obtained on the 3-body annihilation cross section for leptonic final states plus 
a VIB photon. These limits were derived from the Fermi data and the plots taken from 
Ref.~\cite{2012JCAP...07..054B}. They are compared to complementary limits obtained in 
Ref.~\cite{2011PhRvL.107x1303G} from a study of dwarf spheroidal galaxies.}
\label{fig:gamma_vib}
\end{minipage}
\end{figure*}

Further limits using three-body annihilation channels including virtual internal 
bremsstrahlung (VIB) photons\footnotemark were derived by the authors of 
Ref.~\cite{2012JCAP...07..054B}. They assumed an NFW halo profile with $\rhosun = 0.4$ GeV/cm$^3$ 
and $r_s=20$ kpc, and parameterized the VIB contribution in terms of the mass splitting between
the mediator and the DM particle $\mu \equiv (m_m/\mchi)^2$ (larger than 1 in the
R-parity conserved supersymmetric context), the larger the values of $\mu$ the smaller the
annihilation rates; though values close to 1 may correspond to co-annihilation regions.
They selected a large region around the GC to maximize the signal-to-noise ratio, and performed
a detailed analysis of the diffuse emission. While they found a 3.1-$\sigma$ hint for a signal 
corresponding to a WIMP mass of $\sim 150$ GeV, they adopted a conservative approach in deriving
limits on the 3-body annihilation cross section. Their results are shown in \citefig{fig:gamma_vib},
where the blue horizontal lines are upper bounds corresponding to the requirement that the 3-body 
annihilation cross section (an s-wave), when taken alone, is such that the relic abundance 
$\ohh \geq 0.1$; indeed, 2-body contributions are expected, which should reduce the relic density
significantly (VIB contributions, though s-waves, come with a $\alpha_{\rm em }/\pi$ suppression
factor). The green curves allow comparisons with the limits obtained from dwarf spheroidal galaxies
in Ref.~\cite{2011PhRvL.107x1303G}, and correspond to requiring $\sigv/(5\times10^{-30}{\rm cm^3/s})
\leq 8\,\pi\,(\mchi/{\rm GeV})^2/N_\gamma^{\rm tot}$, where $N_\gamma^{\rm tot}$ is the total number of 
gamma-rays per annihilation (including the 2-body contribution). We note that the limits 
obtained for leptonic channels are generally better than those derived from dwarf galaxies.

\footnotetext{As already mentioned in section~\ref{sec:photon_introduction},
internal bremsstrahlung photons are made of {\em final state radiation} (FSR) and
{\em virtual internal bremsstrahlung} (VIB) photons, the latter being associated with the radiation
of a $t$-channel charged mediator (\eg~any sfermion in the supersymmetric context). These
contributions can be separated in a gauge invariant way \cite{2008JHEP...01..049B}.}

Finally, the diffuse gamma-ray flux measured by Fermi
\cite{2009PhRvL.103y1101A,2009ApJ...703.1249A,2010PhRvL.104j1101A,2010JCAP...04..014A,2012ApJ...750....3A}
has also been used to put absolute 
upper limits on the continuum gamma-ray fluxes expected for most of DM candidates.
We may illustrate the main results, though not exhaustively, by showing the constraints
obtained in Ref.~\cite{2012PhRvD..85d3509A}, where the authors used the estimate of the isotropic
diffuse gamma-ray flux performed by the Fermi collaboration \cite{2010PhRvL.104j1101A} to constrain
DM contributions at the Galactic scale (one can also use the extragalactic contribution,
though less powerful and more affected by theoretical uncertainties 
\cite{2012PhRvD..85d3509A,2011JCAP...01..011A,2012MNRAS.421L..87S}). In addition to the DM
induced prompt continuum gamma-ray emissions, the authors included the inverse Compton contributions
for all leptonic annihilation channels. Their
results are reported in \citefig{fig:gamma_isotr}, where the $b \, \bar{b}$, $\tau^+ \tau^-$, and
$\mu^+ \mu^-$ annihilation channels are considered as representative of most of DM
models. The plots contain complementary limits
from studies of dwarf galaxies and of the GC 
\cite{2010ApJ...712..147A,2011PhRvL.107x1302A,2011PhRvL.106p1301A,2012JCAP...01..041A}, and 
also provide forecast limits for a 5-year Fermi-LAT sensitivity. We see that the Galactic diffuse
emission leads to constraints at the limit of the interesting supersymmetric parameter space,
assuming canonical cosmology and particle phenomenology, delineated as in 
Ref.~\cite{2011PhRvD..83d5024B}. On the other hand, as illustrated in the plots, such data make it 
possible to exclude more contrived configurations, for example those proposed to explain the 
so-called PAMELA excess \cite{2009JHEP...12..052M,2010JCAP...03..014P}. Many similar results
have been reached independently, that we cannot, unfortunately, exhaustively cite here. 
Nevertheless, we emphasize that a reliable and global picture 
of the diffuse Galactic gamma-ray sky will only come out after important refinements in the 
astrophysical background modeling are achieved. These are now necessary for seeking 
DM annihilation traces into more detail.

\begin{figure*}[t]%[htp]
 \centering
\begin{minipage}[t]{0.52\textwidth}
\vspace{0pt}
\includegraphics[width=\textwidth]{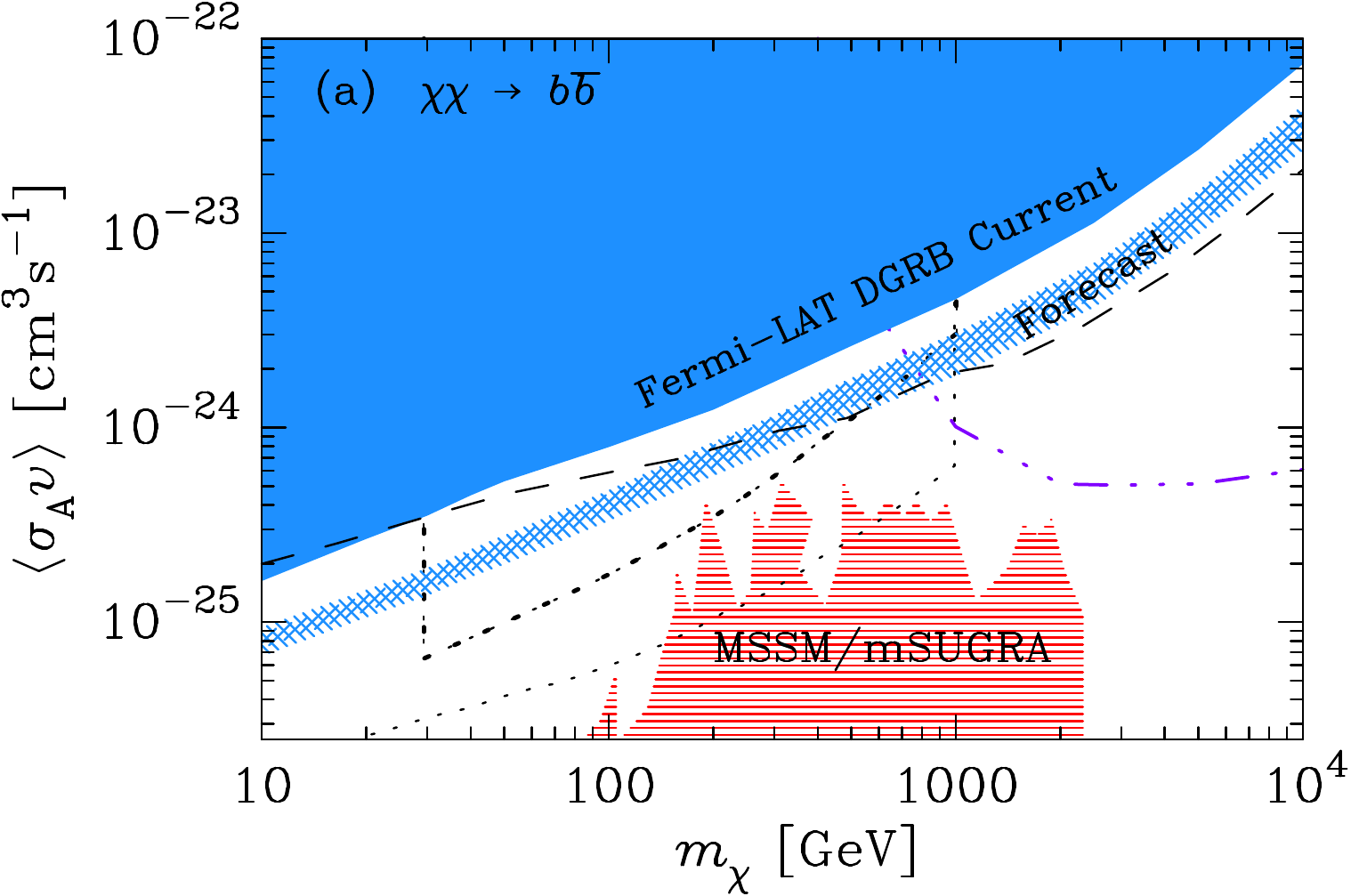}
\includegraphics[width=\textwidth]{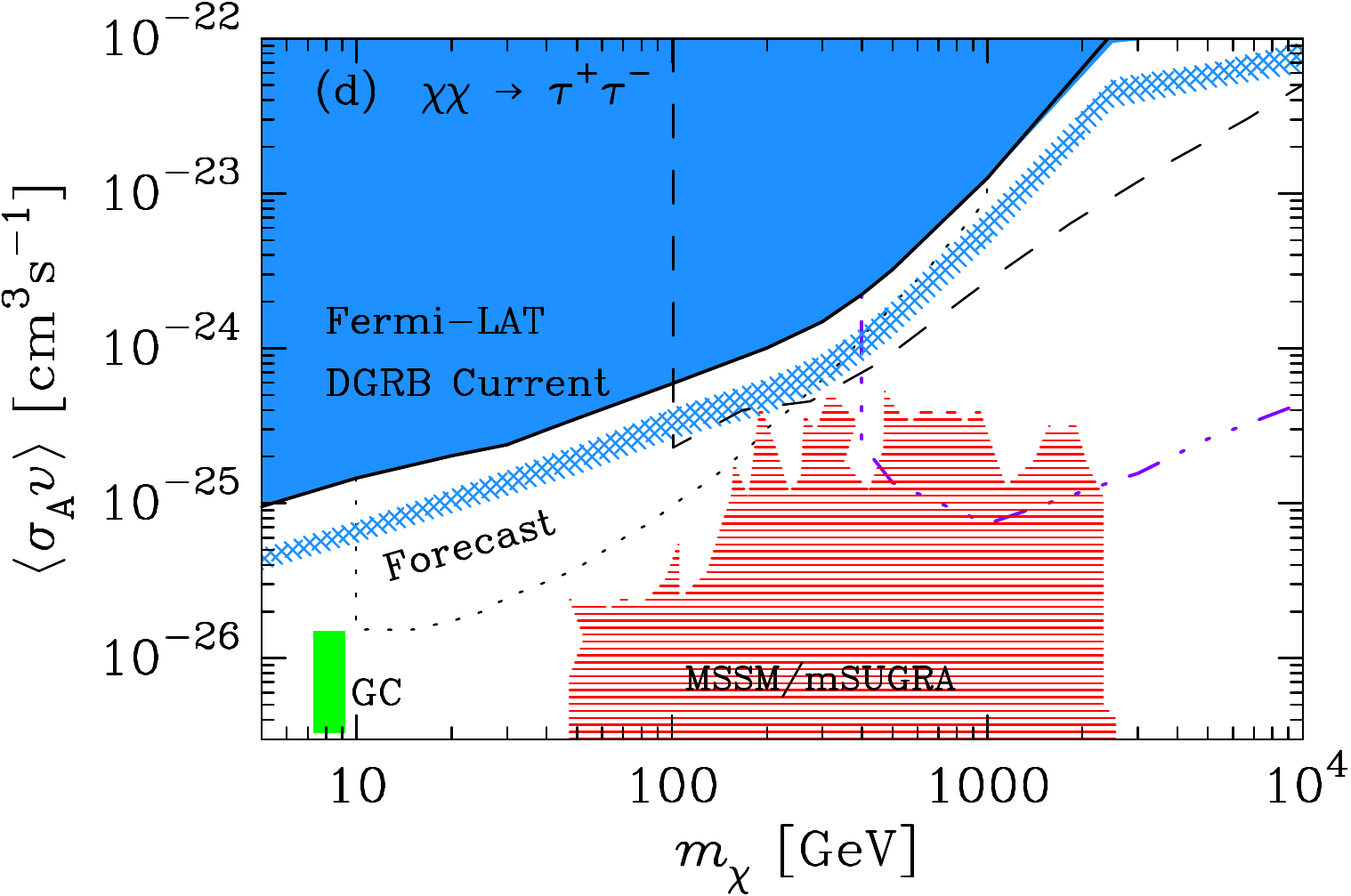}
\includegraphics[width=\textwidth]{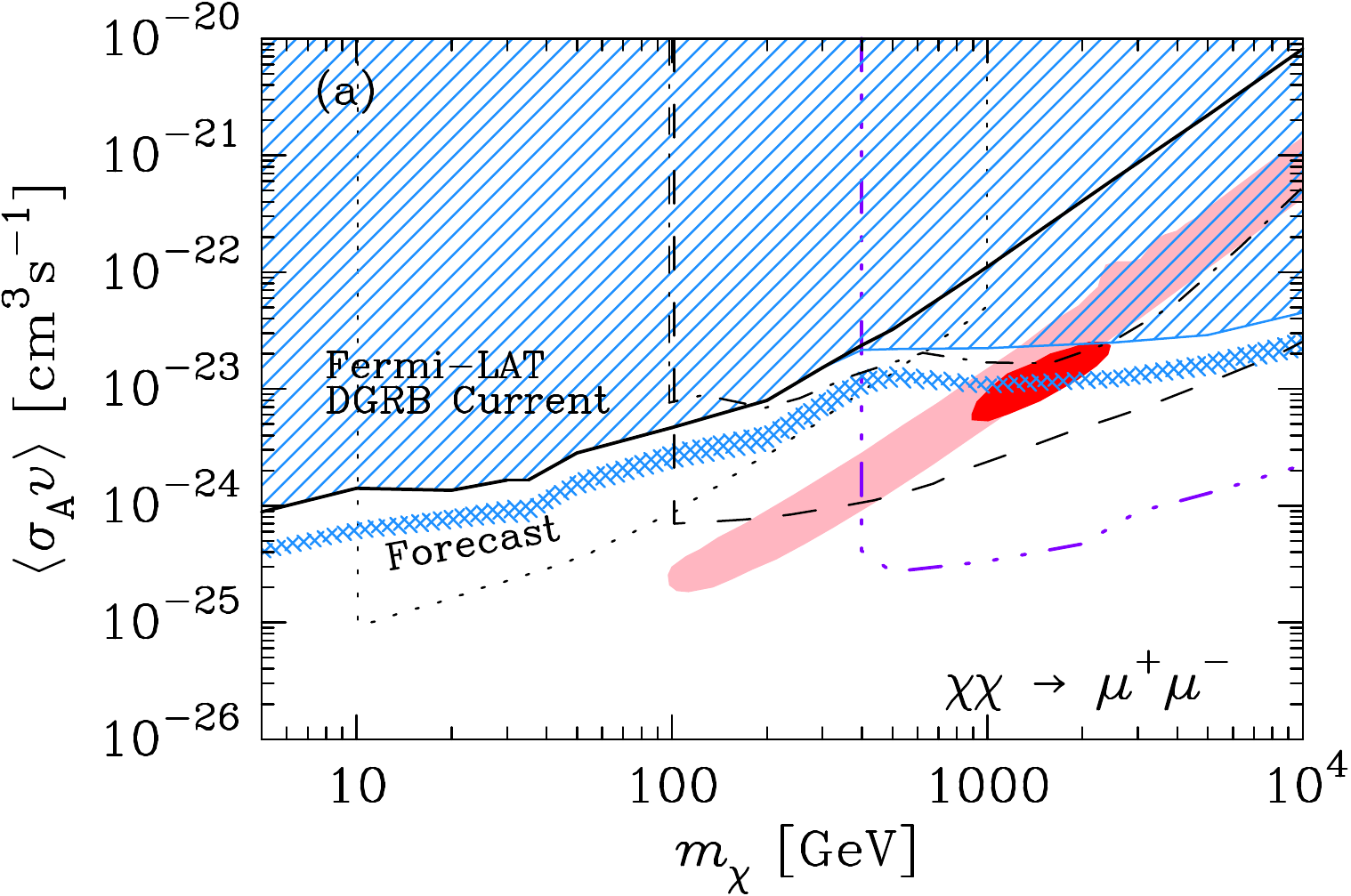}
\end{minipage}
\begin{minipage}[t]{0.47\textwidth}
\vspace{0pt} 
\caption{Plots taken from Ref.~\cite{2012PhRvD..85d3509A}. The dashed lines show the complementary 
limits derived from Fermi observations of the Draco dwarf galaxy in \cite{2010ApJ...712..147A}; 
the thin dotted lines show those obtained by the stacking of dwarf galaxies in 
\cite{2011PhRvL.107x1302A}, except for the $\mu^+ \mu^-$ case, where they correspond to the radio 
constraints found in \cite{2009JHEP...12..052M}. The triple-dot-dashed lines are limits 
corresponding to the HESS observations of the GC \cite{2011PhRvL.106p1301A,2012JCAP...01..041A}. 
The small green rectangle shows the parameter space consistent with the controversial gamma-ray 
excess found
toward the GC in \cite{2011PhLB..697..412H,2011PhRvD..84l3005H}. The solid line shows the Fermi 
limit without the inverse Compton contribution from leptonic annihilation channels.
The red MSSM/MSUGRA regions were derived in \cite{2011PhRvD..83d5024B}. The shade pink and red 
regions ($\mu^+ \mu^-$ case) are those which are consistent with a DM interpretation of the
positron fraction measured by PAMELA and of the electron+positron spectrum measured by Fermi 
\cite{2009JHEP...12..052M,2010JCAP...03..014P}.}
\label{fig:gamma_isotr}
\end{minipage}
\end{figure*}

\subsubsection{Extragalactic sources}
\label{subsubsec:gamma_extrag}

DM is also sought in extragalactic objects, like in nearby dwarf galaxies, 
galaxies, and clusters of galaxies.

The first class of objects, dwarf galaxies, is likely the most promising in terms of DM
detection because these targets are strongly DM dominated, with no expected high-energy 
astrophysical phenomena (\ie~background free, though this has to be weighted by their positions in 
the sky where the Galactic foreground may be strong). Kinematic data are also available which allow 
to constrain the DM distributions in several systems, and it is shown that when using
the current angular resolutions of gamma-ray telescopes at the ground or in space (say $\sim 5'$),
the theoretical errors expected after the angular average of the squared DM density in each
object, \eg~stemming from the large uncertainties in the inner slope, are actually rather small 
\cite{2011MNRAS.418.1526C,2011ApJ...733L..46W}. Several observations of the so-called
canonical dwarf galaxies (less than a ten of objects) were performed in the past by 
ground-based telescopes, but the limited observation times and high-energy threshold induced too 
weak sensitivities to get interesting limits 
(\eg~\cite{2008ApJ...679..428A,2010ApJ...720.1174A,2011APh....34..608H}).
As for other targets, the Fermi advent has considerably improved the detection potential. So
far, yet, no gamma-ray signal originating from a dwarf galaxy has ever been observed, leading
several groups to compete to get the best limits (\eg~\cite{2010ApJ...712..147A,2011PhRvL.107x1302A,2011PhRvL.107x1303G,2012arXiv1203.2954C}). In \citefig{fig:dsph}, we show the results obtained
for individual dwarf galaxies in \cite{2010ApJ...712..147A} (left panel), corresponding to 11 
months of survey mode operation with Fermi, and those obtained by stacking all individual 
contributions in \cite{2011PhRvL.107x1303G}, corresponding to about 2 years of Fermi data.
In the left panel, we see that Draco provides the best limit, though it lies at about one order
of magnitude away from the interesting parameter space. On the other hand, the stacking analysis
is shown a very powerful alternative, which currently yields the best limits on the s-wave
part of the WIMP annihilation cross section. We notice that despite the rather large systematic
errors coming from uncertainties in the DM distribution inside the objects (encoded as the
shaded areas in the plot), WIMP masses below $\sim 50$ GeV annihilating preferentially into 
tau leptons or $b$ quarks at the canonical rate start to be in serious tension with the data.
Beside exclusion prospects, this is also very encouraging in terms of detection prospects.

As regards expectations for the future, we refer the reader to Ref.~\cite{2011MNRAS.418.1526C},
where the authors show that future ground-based observatories, like CTA, will allow for
interesting complementary surveys with respect to Fermi, though further observations with Fermi
are likely the best strategy for discovery purposes --- 10 years of Fermi data would really
drive us in the ballpark, though such a duration currently appears as unrealistic in terms of 
funding (the official end of Fermi operations is June 2012, though 2 supplementary years seem 
granted).

\begin{figure*}[t!]%[htp]
\centering
%\begin{minipage}[t]{0.49\textwidth}
%\vspace{0pt}
%\includegraphics[width=\textwidth]{fig_gamma_stacked_dsph.pdf}
\includegraphics[width=0.49\textwidth]{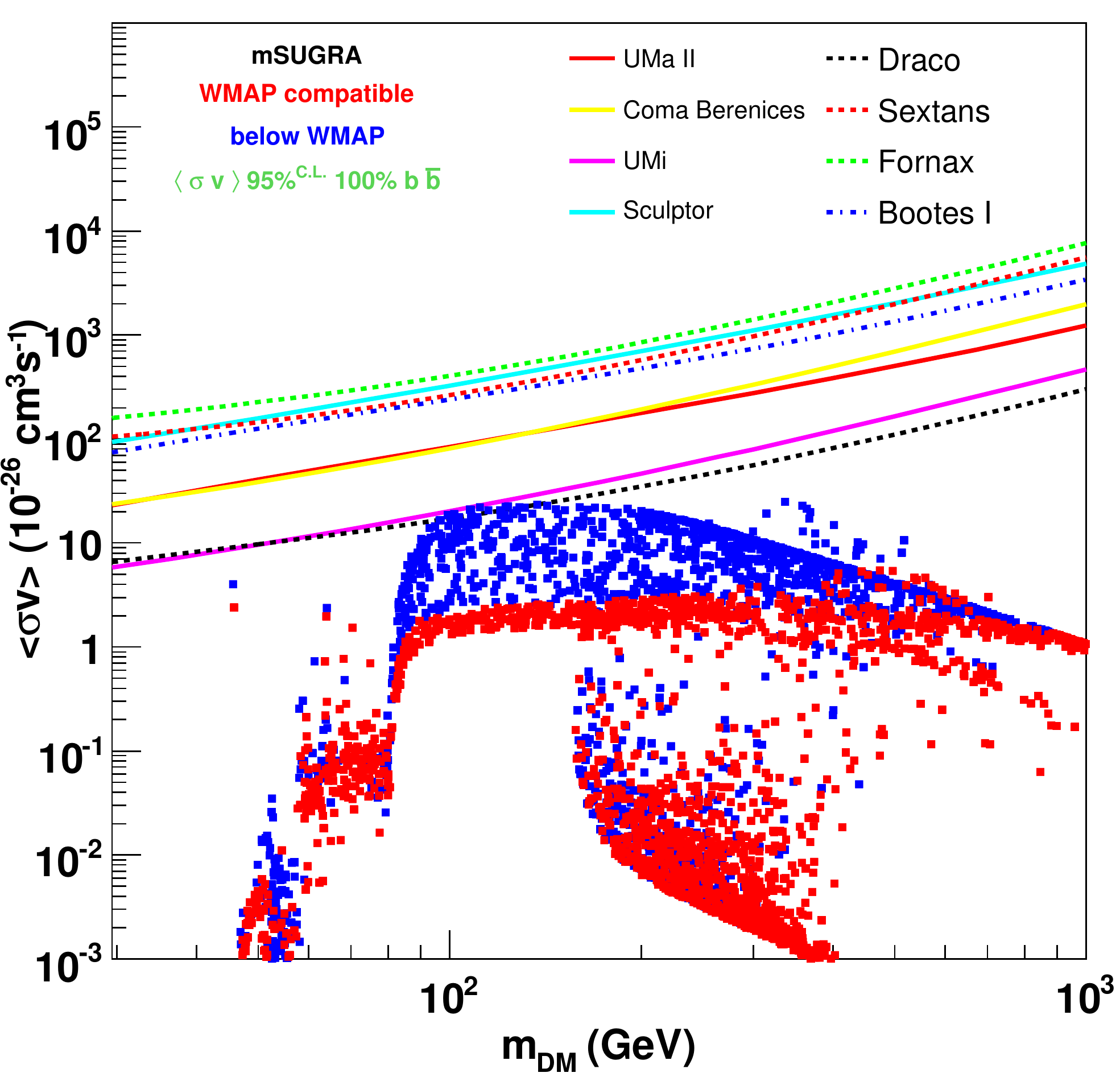}
\includegraphics[width=0.49\textwidth]{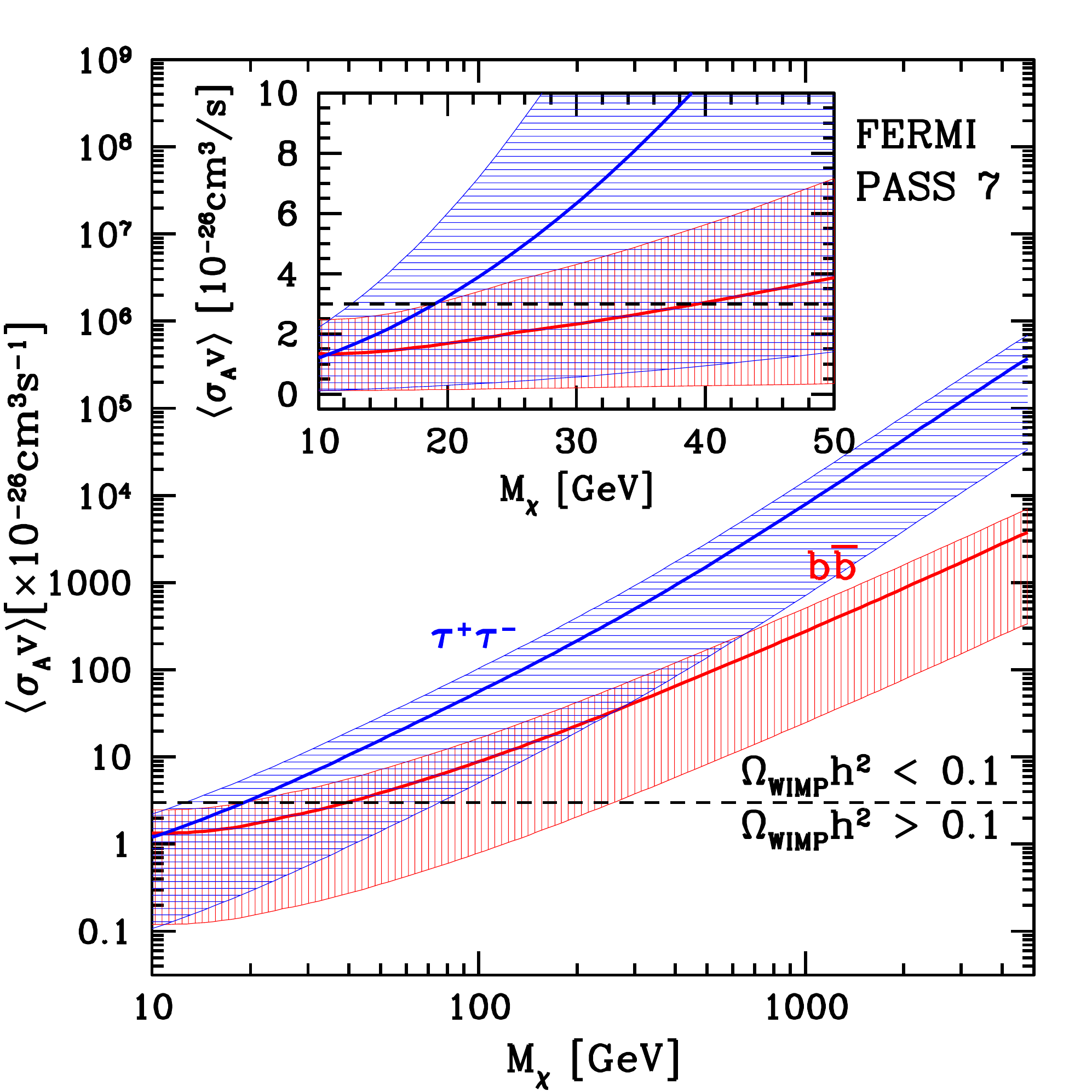}
%\end{minipage}
%\begin{minipage}[t]{0.49\textwidth}
%\vspace{0pt}
\caption{Left: indicative limits on the constrained MSSM parameter space from several dwarf 
galaxies, assuming a generic $b \, \bar{b}$ spectrum; taken from \cite{2010ApJ...712..147A}. 
Right: limits after stacking individual limits, with a self-consistent treatment of statistical 
and systematic errors (the shaded areas show the systematic errors coming from uncertainties in
the DM distribution inside the objects); taken from \cite{2011PhRvL.107x1303G}.}
\label{fig:dsph}
%\end{minipage}
\end{figure*}

For completeness, it is important to mention that other extragalactic objects might also 
be interesting sources for DM searches. Neighbor galaxies, like M31 (Andromeda), are 
potentially good targets \cite{2000PhRvD..61b3514B,2004APh....20..467F,2004PhRvD..70j3529F}, but 
as for the Galaxy, it is rather difficult to predict the astrophysical background in these 
objects, where, at variance with dwarf galaxies, high-energy phenomena are expected. Several 
unsuccessful observations have been performed in the past, \eg~\cite{2003A&A...400..153A,2006A&A...450....1L,2008ApJ...678..594W},
but the constraints obtained on DM models were 
insignificant. Nevertheless, some giant galaxies, like M87 and NGC 1275 were recently observed at 
TeV gamma-ray energies, but with strong variabilities 
(\eg~\cite{2006Sci...314.1424A,2012ApJ...746..141A,2012A&A...539L...2A}). The latter objects usually
belong to the broader class of radio emitters, which allow for rich multiwalength analyses. 
Nevertheless, the observed variability makes it very difficult to try to interpret any signal in 
terms of DM annihilation or decay in these sources. The Fermi satellite has more
recently allowed the first detections in gamma-ray frequencies of nearby galactic objects,
M31 \cite{2010A&A...523L...2A}, the SMC \cite{2010A&A...523A..46A} and LMC 
\cite{2010A&A...512A...7A}. As already emphasized above, DM interpretations are difficult 
because of the presence of CR accelerators and gas in these objects, and these studies 
actually barely addressed the DM issue.

Finally, many efforts have been recently done to check whether galaxy clusters could offer 
additional targets for DM searches (\eg~\cite{2006A&A...455...21C,2011PhRvD..84l3509P,2011JCAP...12..011S,2012JCAP...01..042H,2012PhRvD..85f3517C,2012arXiv1203.1165N,2012arXiv1203.1166M}).
Limits on DM candidates were again derived from the Fermi data 
\cite{2010JCAP...05..025A,2011arXiv1110.6863Z,2012JCAP...01..042H}, and we may also mention a 
recent claim for the detection of gamma-ray signals from a few clusters \cite{2012arXiv1201.1003H}.
Nevertheless, we know that high-energy processes are at work in galaxy clusters, 
and it is still difficult to infer and predict the related astrophysical background 
\cite{2010MNRAS.409..449P,2011PhRvD..84l3509P}. Moreover, we stress that galaxy clusters
host giant elliptical galaxies, like the M87 and NGC 1275 galaxies discussed above. Since
the latter have been demonstrated to be transient TeV emitters, the relevance of 
galaxy clusters for DM searches might fairly be questioned, though their observations
remain solidly motivated from many other astrophysical arguments. Deeper investigations
appear to be necessary, which could for instance rely on the state-of-the-art cosmological
simulations including baryon dynamics and star formation.

\section{Indirect dark matter searches with neutrinos}
\label{sec:neut}
%%
%% Intro
High-energy neutrinos are interesting messengers to seek for DM annihilation traces.
They can be produced through the fragmentation or decay of unstable primary annihilation
products, or directly in some more rare cases. Direct production is generically suppressed 
if DM is made of Majorana fermions because of helicity suppression. Nevertheless, 
it may naturally appear in other cases, like in extra-dimensional models. Beside being complementary
to other astrophysical messengers as tracers of DM annihilation in the Galactic halo,
neutrinos do provide an additional and very specific signature: they can also trace the 
gravitational capture of DM in celestial bodies, in particular in the Sun or the Earth. 
Detecting high-energy neutrinos from the Sun or the Earth would represent a very clean signature of 
DM capture and subsequent annihilation, hard to infer from other known phenomena. This 
idea was proposed in the mid 80's by several groups, notably by Press \& 
Spergel~\cite{1985ApJ...296..679P} and Silk, Olive \& Srednicki~\cite{1985PhRvL..55..257S} for the 
Sun, and by Freese~\cite{1986PhLB..167..295F}, Krauss, Srednicki \& 
Wilczek~\cite{1986PhRvD..33.2079K}, and Gaisser, Steigman \& Tilav~\cite{1986PhRvD..34.2206G} for 
the Earth.

%% Generalities
Active neutrinos are almost massless and thereby travel in the skies like photons: their sources can
be traced back, which is of particular interest for DM searches. Neutrinos are less 
subject to absorption than photons but in the meantime more difficult to collect on Earth. The 
detection of astrophysical neutrinos is indeed very challenging because of their weak cross section 
with ordinary matter, which entails gigantic detectors, and because of the copious diffuse 
atmospheric neutrino background coming from interactions of cosmic rays with the atmosphere. So 
far, the history of neutrino astronomy relies on confirmed observations of two objects only: the 
Sun, from which the electron-to-muon neutrino flavor oscillation was proven, and the supernova 
SN-1987A, the explosion of which allowed to derive a robust upper limit on the absolute neutrino 
mass scale. These sources were observed in the keV-MeV energy range. At higher energies, more global
(and indirect) astrophysical pieces of information have been delivered from \eg~measurements of the 
atmospheric neutrino anisotropy in the TeV energy range, related to the anisotropy in the 
CR flux. 

%% Detection principles and experiments
There are currently two main types of so-called neutrino telescopes in operation: 
MeV detectors historically optimized for solar/atmospheric electron/muon neutrinos --- 
\eg~Super-Kamiokande\footnote{\url{www-sk.icrr.u-tokyo.ac.jp/sk/index-e.html}} (Super-K hereafter)
 --- and GeV-PeV detectors mostly designed for muon 
neutrinos\footnote{Electron and tau neutrinos can
also be reconstructed in some cases.} originating in high-energy astrophysical phenomena 
(supernova remnants, active galactic nuclei, DM annihilation, \etc) --- 
\eg~AMANDA\footnote{\url{http://amanda.uci.edu/}}-IceCube\footnote{\url{http://icecube.wisc.edu/}}
and ANTARES\footnote{\url{http://antares.in2p3.fr/}}-KM3\footnote{\url{http://www.km3net.org/home.php}}.
These detectors are installed deep underground for protection against the CR
background, and when dedicated to studies of astrophysical sources, the Earth itself is 
used to convert muon neutrinos into upward-going muons, the reconstructed directions of which
allow a separation from the downward-going muons produced by interactions of cosmic
rays in the atmosphere. Indeed, muons, when crossing such a detector, may produce Cherenkov
light in the ice (AMANDA-IceCube are based at the South-Pole) or water (a tank of 50,000
tons of purified water for SuperK, 
the Mediterranean sea for ANTARES-KM3) that can be collected with arrays of photo-multipliers, 
allowing for direction reconstruction. The main background associated with this detection
strategy comes from the atmospheric neutrinos generated by CR interactions in the
atmosphere; although the amplitude of this background is large in the GeV-TeV range and makes 
neutrino astronomy challenging, predictions and measurements are available and under reasonable 
control (see~\eg~\cite{2004PhRvD..70b3006B,2011PhRvD..83l3001H}). Very approximately,
the atmospheric muon neutrino flux is of order $\phi_{\rm bg}(E)\approx 6 \times 10^{-3}
(E/100\,{\rm GeV})^{-3.7} \,{\rm m^{-2}s^{-1}sr^{-1}}$. Assuming a full time exposure, an
effective detection area of $\sim 10^{-2} \,{\rm m^{2}}$ (expected for a km-size neutrino telescope) 
and an optimistic flat angular resolution of 
$1^\circ$~\cite{2008JPhCS.136d2063C,2010RScI...81h1101H}, this corresponds 
to ${\cal O }(1-10)$ background event per year for a point-like source, among ${\cal O }(10^4)$ 
over half the sky.

In the following, we introduce the treatment of neutrino propagation (in vacuum or/and matter) 
necessary for proper predictions in the present context. Then, we shortly discuss DM
searches with high-energy neutrinos from the Galactic halo before delving into more details the 
case of solar neutrinos. We illustrate each paragraph with a selection of recent results and sketch 
the main prospects.
%
%
%
%##########
\subsection{Neutrino oscillations, interactions with matter, and muon flux at detectors}
\label{subsec:neut_osc}
Before discussing the Galactic or solar origin of the DM-induced neutrinos, it is 
interesting to review how they propagate to the Earth, and how one can predict the flux 
converted into muons at detectors. A fundamental feature of neutrino phenomenology is flavor 
oscillation, which is a consequence of their non-zero masses; interaction and mass eigenstates are 
different \cite{1968JETP...26..984P}. This of course must be accounted for in predictions, 
since high-energy neutrino telescopes are mostly designed for muon neutrinos. All cases are relevant
here: oscillation in vacuum (Galactic halo neutrinos), to which one may further have to combine 
oscillation in matter (neutrinos from the Sun or the Earth). In the latter case, additional effects
implying neutral and charge currents (absorption and re-emission) must also be plugged in. In the 
following, we ignore the potential effect of putative sterile neutrinos. We refer the reader to 
Refs. \cite{2006hep.ph....6054S,2008PhR...460....1G,2008PhR...458..173B} for extensive reviews.

The neutrino oscillation framework (in vacuum) is usually based on a few ingredients: the mixing 
matrix $U$ (including unknown but possible CP-violating phases), from which one can express a 
flavor (mass) eigenstate as a linear combination of the mass (flavor) eigenstates, and a 
classical quantum evolution equation assuming plane waves:
\ben
\myket{\nu_\alpha} = U_{\alpha i}^* \, \myket{\nu_i}\,,\;\;\; 
\myket{\nu_i(t)} = e^{- i \, E_i \, t} \myket{\nu_i(0)} \,,
\een
where Greek indices refer to the flavor basis ($e,\mu,\tau$) while Latin indices refer to the mass 
basis (1,2,3). The oscillation probability between flavors $\alpha$ and $\beta$ is then merely 
given by ${\cal P}_{\alpha\beta} = |\mybra{\nu_\beta} \myket{\nu_\alpha}|^2$. Since masses are small,
one can use the relativistic approximation such that $E_i \simeq p_i \simeq p \simeq E 
\simeq p + m_i^2/2E$, and the probability reads
\ben
{\cal P}_{\alpha\beta}(E,L=c\,t) = &\delta_{\alpha\beta} 
 - 4 \sum_{j>i} \,{\cal R}e (U_{\alpha i}^* \,U_{\beta i} \,U_{\alpha j} \,U_{\beta j}^*) 
\sin^2\left\{  \frac{\delta m_{ij}^2 \,L}{4 \, E_\nu}\right\} \\
& \pm 2 \sum_{j>i} \,{\cal I}m (U_{\alpha i}^* \,U_{\beta i} \,U_{\alpha j} \,U_{\beta j}^*) 
\sin\left\{  \frac{\delta m_{ij}^2 \,L}{2 \, E_\nu}\right\}\,.\nn
\een
When distances get tremendous, \eg~when neutrinos originate in DM annihilation in the 
Galactic halo, the oscillation effects in vacuum average out and the probability is merely related 
to the raw mixing angles:
\ben
\mymean{{\cal P}_{\alpha\beta}(E)}
\overset{L \gg 1 \,{\rm AU}}{\longrightarrow} & \delta_{\alpha\beta} - 
2 \sum_{j>i} \,{\cal R}e (U_{\alpha i}^*  \,U_{\beta i} \,U_{\alpha j} \,U_{\beta j}^* )\,.
\een
Assuming no CP-violating phases, $\theta_{23}\simeq \pi/4$ and 
$\theta_{13}\simeq 0$, the neutrino flux at the Earth may be expressed 
as~\cite{2003PhRvD..67g3024B}
\ben
\phi_{\nu_e} &\approx& \phi_{\nu_e}^0 - \frac{1}{4}\,\bar\phi_\nu \\
\phi_{\nu_\mu} &\approx& \phi_{\nu_\tau} \approx  \frac{1}{4}\,(\phi_{\nu_\mu}^0+\phi_{\nu_\tau}^0) + 
\frac{1}{8}\,\bar\phi_\nu \nn \\
\bar\phi_\nu &\equiv& \sin^2 \theta_{12}(2\,\phi_{\nu_e}^0 - \phi_{\nu_\mu}^0-\phi_{\nu_\tau}^0)\nn\,,
\een
where any $\phi^0$ denotes the flux without oscillation. 

When neutrinos travel a significant distance inside a dense medium, the 
Mikheyev-Smirnov-Wolfenstein (MSW) effect \cite{1978PhRvD..17.2369W,1986NCimC...9...17M} must also 
be included --- this is relevant to neutrinos crossing the Sun or the Earth. Then, the above 
formalism has to be enriched with a new interacting term in the Hamiltonian which modifies the 
dispersion relation \cite{2000PhLB..474..153O,2008JCAP...01..021B}, 
and the flavor vector $\mathbf{\nu}=(\nu_e,\nu_\mu,\nu_\tau)^t$ now evolves according to
\ben
\label{eq:neut_matter}
\mathbf{\nu}(t) &=& e^{- i \, H \, t} \, \mathbf{\nu}(0)\,,\\
H &=&\frac{1}{2\,E} \, U\,{\rm diag}(0,\Delta m_{21}^2,\Delta m_{31}^2)\,U^\dagger + 
{\rm diag}(\sqrt{2}\,G_F\,n_e,0,0)\nn\,,
\een
where $H$ is the Hamiltonian expressed in the flavor basis, $G_F$ is the Fermi constant 
and $n_e$ is the electron density. The flavor asymmetry comes from that in coherent charged current 
interactions, electron neutrinos see the electrons present in the medium while the other neutrinos 
do not, leading to a flavor-dependent refraction index. One must also include other charged and 
neutral current interactions with nucleons (inelastic scattering), which get very important at high 
energy and mostly induce an energy redistribution toward lower energies because of successive 
absorption and re-emission processes (the conversion of tau neutrinos into tauons also induces extra
flavor injection and a coupling in the evolution between neutrino and anti-neutrino species). This 
can be worked out with Monte Carlo methods \cite{2008JCAP...01..021B}. An elegant and 
self-consistent alternative to such a two-fold treatment is that of the density matrix formalism 
\cite{2006hep.ph....6054S,2005NuPhB.727...99C,2007PhRvD..76i5008B}, which combines oscillation
and absorption, and provides identical results. In that case, the flavor density matrix 
$\rho_\nu$ obeys the following evolution equation:
\ben
\label{eq:neut_dens_matrix_evol}
\frac{d\rho_\nu}{dl} = -i \left[H,\rho_\nu\right] 
+ \frac{d\rho_\nu}{dl}\Big|_{\rm NC}
+ \frac{d\rho_\nu}{dl}\Big|_{\rm CC}
+ \frac{d\rho_\nu}{dl}\Big|_{\rm in}\,,
\een
where, in the right-hand side, the terms respectively describe the neutral current effects,
the charged current effects, and the neutrino injection induced by DM annihilation.
We emphasize that the injection term in a dense medium is different from what is obtained for 
annihilation in the Galactic halo, since some of the fragmentation or decay products of the 
primary particles, or even the primary particles themselves, can be absorbed before cascading. 
This is the case, for instance, for muons. Irrespective of the method, the full calculation has to 
be performed numerically. Furthermore, detailed models of the Sun or the Earth, when concerned, 
have to be used to get accurate predictions.

An important outcome is that for neutrinos from the Sun, absorption is responsible for an 
exponential energy cut-off above 100 GeV in the differential flux. For neutrinos injected at the
center of the Earth, the cut-off is somewhat higher, around 10 TeV. Moreover, it turns out
that oscillation in the Sun mixes tau and muon neutrinos; oscillation only affects electron 
neutrinos in vacuum on their way to the Earth. This is is illustrated in \citefig{fig:neut_osc}, 
which was extracted from 
Ref.~\cite{2008JCAP...01..021B}. In the top left panel, one can see the injected neutrino spectra
at the Sun's core, where the annihilation of 250 GeV WIMPs into $\tau^+\tau^-$ was assumed. In the
top right panel, propagation effects (oscillation and weak currents effects) are shown by evolving
the injected spectra from the Sun's core to its surface; the mixing between tauon and muon neutrinos
is made clear, as well as energy redistribution toward low energies, due to weak scatterings; the
production of secondary muon and electron neutrinos is due to tauon decays, which are generated
from tau neutrinos through charged current interactions. Oscillation in vacuum between
the Sun and the Earth (the distance is fixed to 1 AU) further results in slightly mixing electron 
neutrinos with the other flavors, as sketched in the bottom left panel. The final neutrino 
spectra at the detector are shown in the bottom right panel, where the oscillation pattern
appears as completely smeared out. This does not come from additional mixing in the Earth, but from
a time average performed over a 1 year cycle, during which the variation in the distance between 
the Earth and the Sun is enough to erase the oscillating features. Therefore, the net propagation
effects for neutrinos coming from the Sun mostly translate into an amplitude rescaling, as in the 
case of Galactic neutrinos, and a high-energy damping due to efficient absorption in the Sun above 
100 GeV.
\begin{figure*}[t!]%[htp]
 \centering
\includegraphics[width=0.49\textwidth]{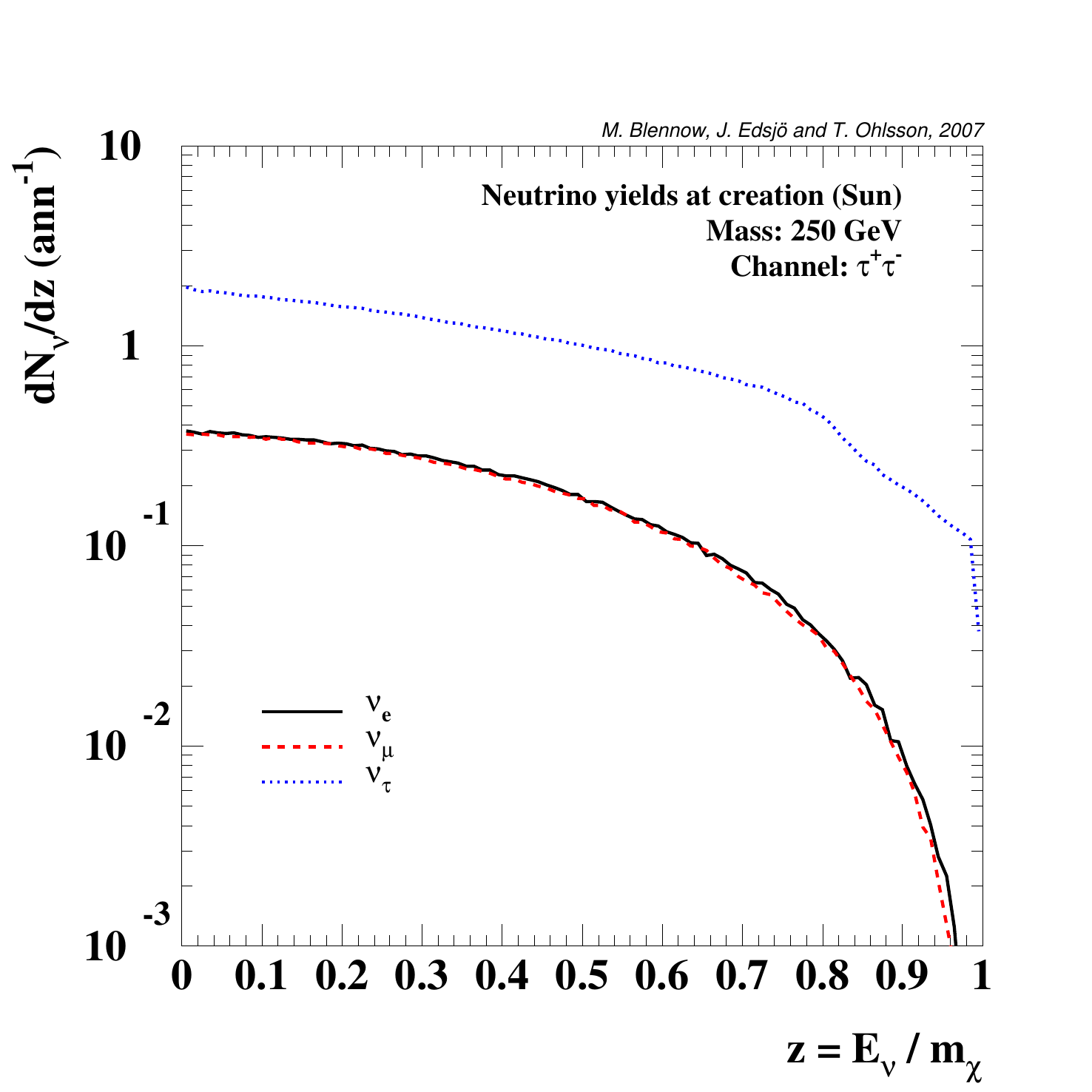}
\includegraphics[width=0.49\textwidth]{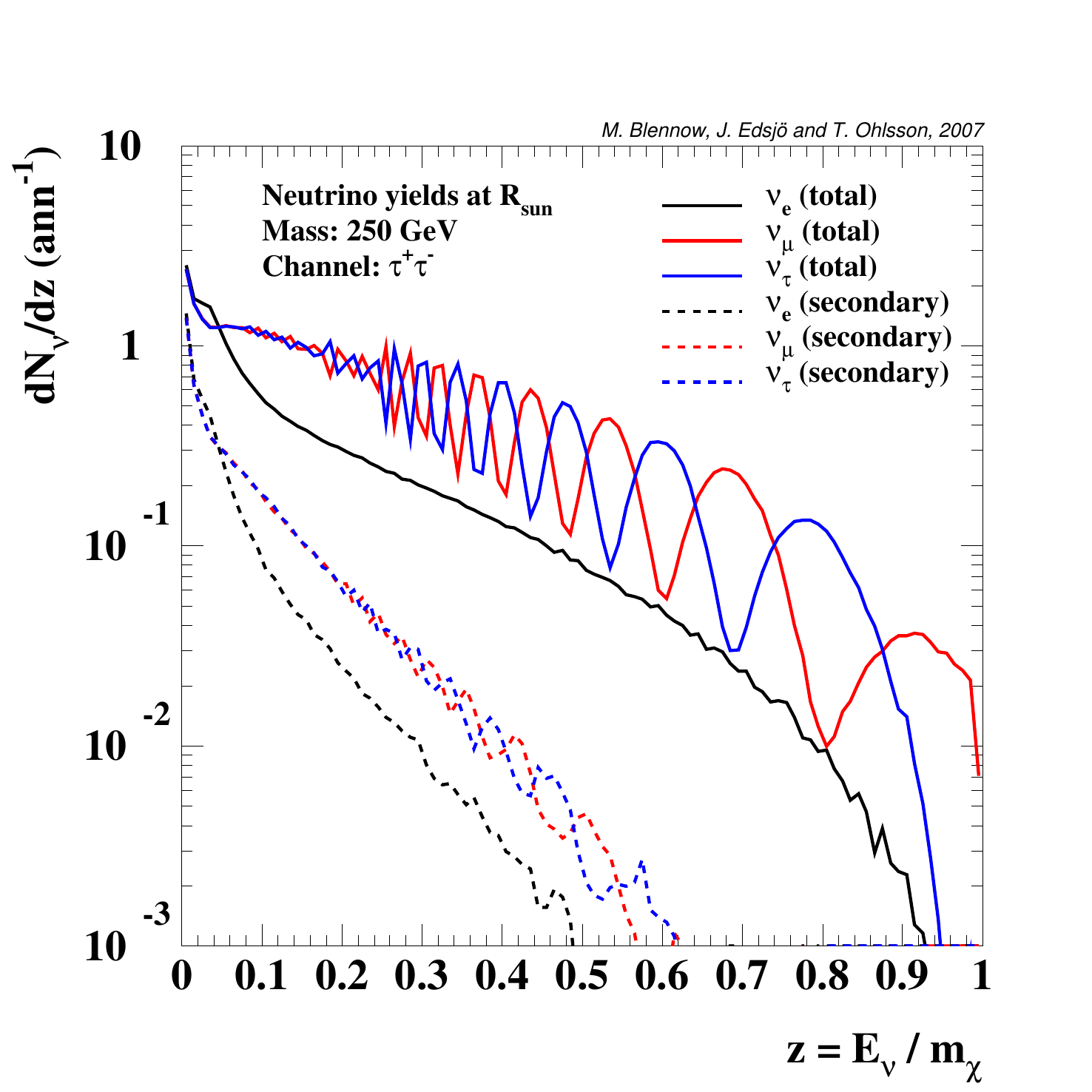}
\includegraphics[width=0.49\textwidth]{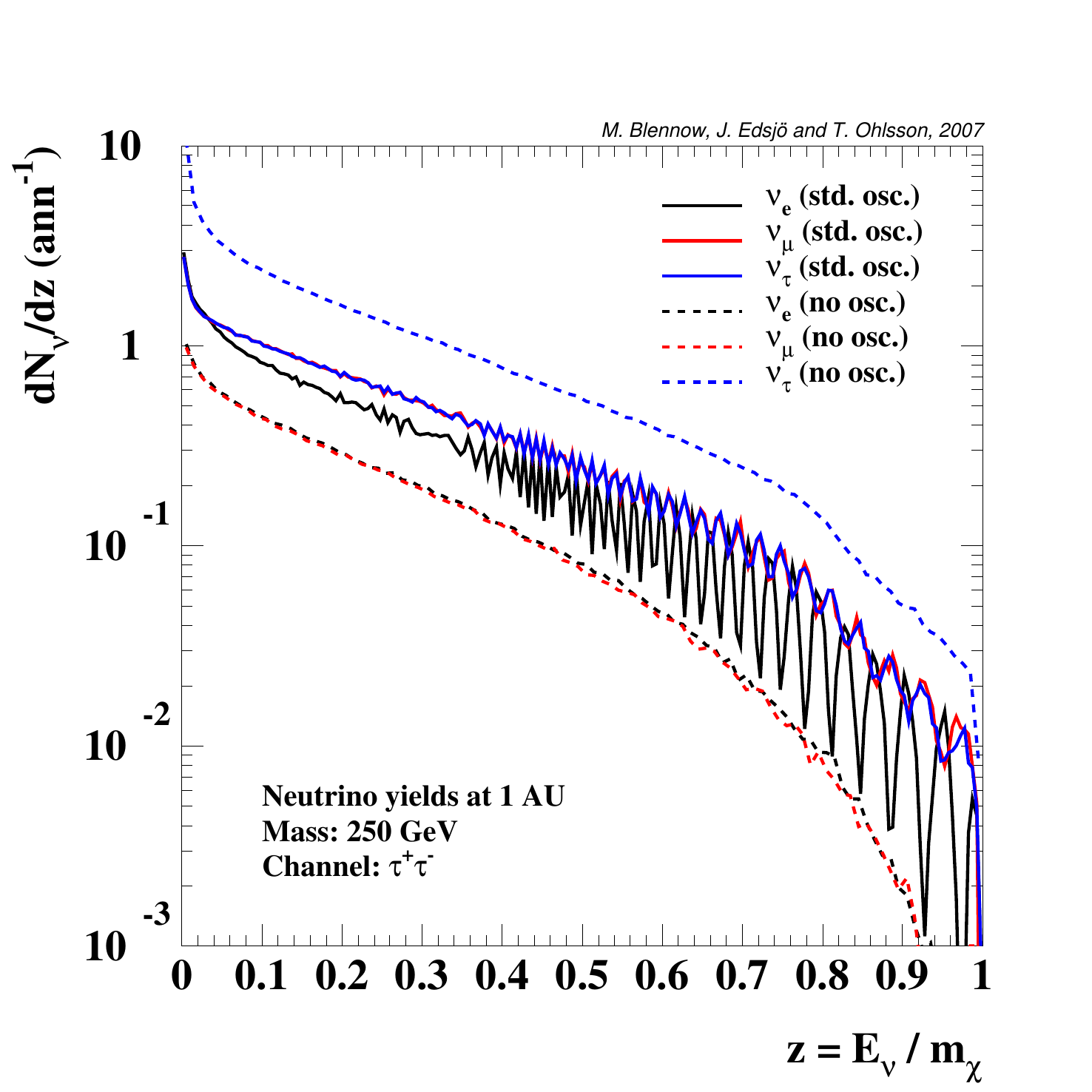}
\includegraphics[width=0.49\textwidth]{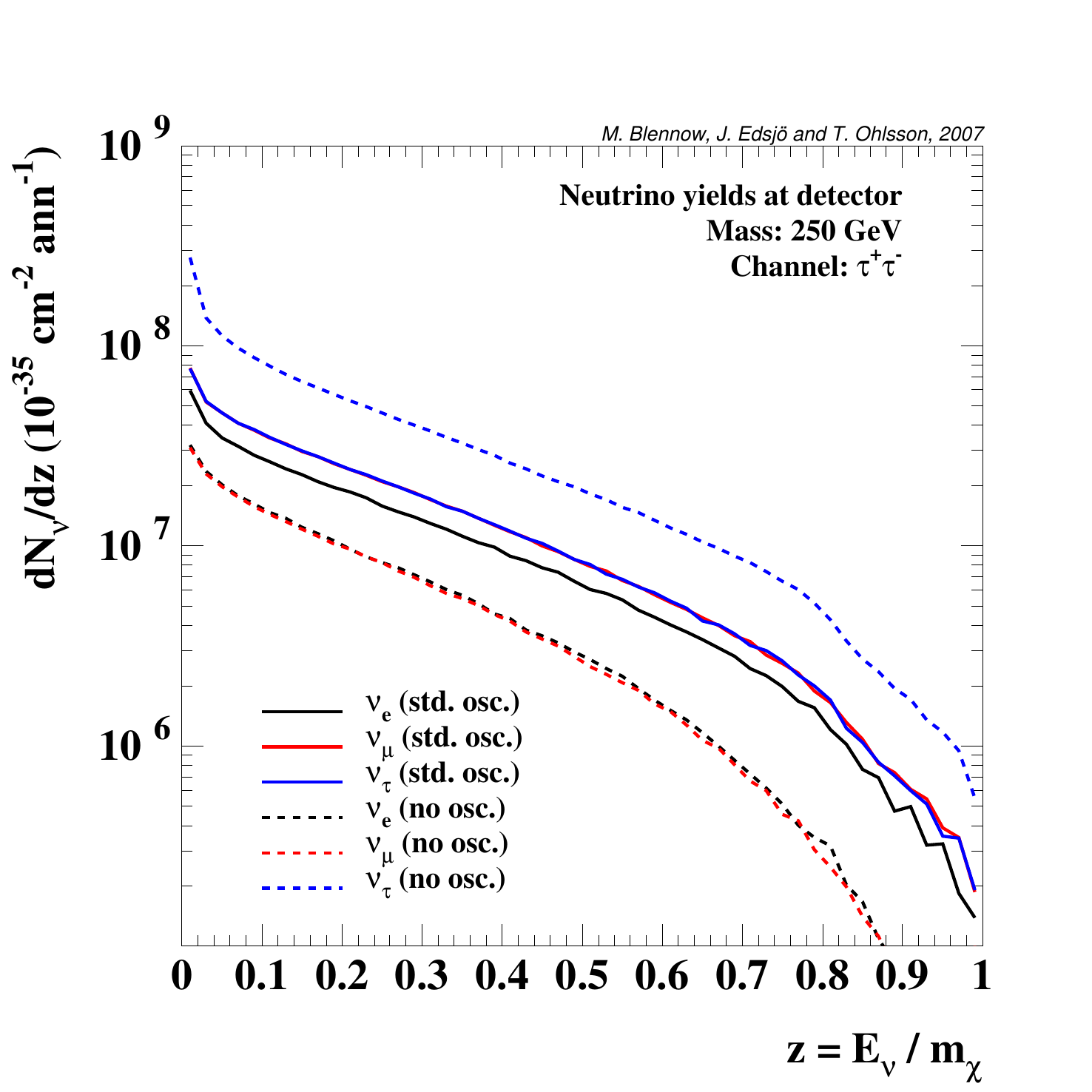}
\caption{Evolution of the neutrino spectra between the Sun's core (top left), the Sun's surface
(top right), the Earth's surface (bottom left), and the detector (bottom right). Legends
are explicit, and more details are given in the text. These plots have been borrowed from
Ref.~\cite{2008JCAP...01..021B}.}
\label{fig:neut_osc}
\end{figure*}

Once the neutrino flux at sea level is predicted, one still has to convert it in terms of muon
flux at detectors (or other observables related to neutrinos, \eg~hadronic showers). Again, 
the full flavor evolution given \eg~in \citeeq{eq:neut_dens_matrix_evol} must be adapted to the 
Earth medium to determine the neutrino flux at a certain distance from the detector. For a 
muon-optimized detector, the relevant distance is the mean free path of a muon in the
surrounding material, typically 1 km for a TeV muon travelling in the ice. Another important
point to take into account is the time and detector-location dependence of the neutrino arrival 
direction. Indeed, only upward-going muons can be used for astrophysical searches, and
detection efficiency is angular dependent. Therefore, the effective detection area 
${\cal A}_{\rm eff}$ is a quantity somewhat tricky to extract for neutrino telescopes. It 
basically depends on the neutrino flavor, energy and direction, and has to be determined
for each source. Anyway, for a given astrophysical source located at point ${\mathbf x}_\star$,
the detector efficiency might be time-averaged and the differential detected event rate then merely 
reads
\ben
% \frac{dN_\nu(\vec{x}_\star)}{dt} = \int_{E_{\nu,{\rm th}}} dE_\nu \frac{d\phi_\nu}{dE_\nu } 
% \mymean{{\cal A}_{\rm eff}(E_\nu,\vec{x}_\star)}\,.
\frac{dN_\nu({\mathbf x}_\star)}{dt} = \int_{E_{\nu,{\rm th}}} \! dE_\nu \,
\frac{d\phi_\nu}{dE_\nu } \, \mymean{{\cal A}_{\rm eff}(E_\nu,{\mathbf x}_\star)}\,.
\een
The effective detection area is typically of order 
$\mymean{{\cal A}_{\rm eff}} \approx 10^{-3}$ to $10^{-2}$ ${\rm m^2}$ $(E_\nu/300\,{\rm GeV})^2$ for
km-size instruments~\cite{2008JPhCS.136d2063C}, where one power of energy comes from the neutrino 
conversion into muon and the other from the muon energy loss converted into Cherenkov photons. To 
illustrate how challenging detecting astrophysical neutrinos is, this has to be compared with 
the usual effective areas encountered in gamma-ray astronomy. For instance, the Fermi-LAT 
instrument\footnote{\url{http://www-glast.stanford.edu/}} has an effective area of 
$\sim0.5-1  \,{\rm m^2}$ in the 0.1-100 GeV energy range.
%
%
%##########
\subsection{Neutrinos from DM annihilation in the Galactic halo}
\label{subsec:neut_gal}
Since neutrinos are almost massless, they travel along geodesics like photons, and may therefore 
trace the (squared) DM distribution in the Galaxy or extra-galactic objects. Thus, as for 
indirect DM searches with gamma-rays, the Galactic center on the one hand, and neighbor 
dwarf galaxies on the other hand, are very good targets. In these specific cases, the neutrino flux 
prediction can be written the same way as for gamma-rays, except for the possibility of flavor 
oscillations (see \citesec{subsec:neut_osc}). For an arbitrary flavor index $\alpha$, it reads:
\ben
\frac{d\phi_\alpha(E)}{dE} &=& \sum_{\beta=e,\mu,\tau} {\cal P}_{\alpha\beta}(E) \,
\frac{\eta}{4 \, \pi} \,
\frac{\sigv_\beta}{\mchisq} \,
\int_{\delta \Omega_{\rm obs}} \!\! d\Omega \, 
\int_{\rm los} \, dl \; \rho^2(l) \;\; ,
\een
where ${\cal P}$ is the flavor oscillation probability, and $\rho(l)$ is the DM mass 
density along the line-of-sight $l$ in the chosen direction. The integral is performed over a 
cone of angle $\delta \Omega_{\rm obs}$ corresponding to the angular resolution of the telescope 
(assumed flat for simplicity here). This is therefore similar to the flux prediction made for 
gamma-rays except for the relevant angular size. Indeed, neutrino telescopes have generically 
angular resolutions worse than gamma-ray telescopes, ranging from a few degrees around a few GeV 
down to half a degree above a few TeV. Still, the predicted fluxes do not suffer from high-energy
absorption like in the Sun, which is of interest for km-size neutrino telescopes, the effective 
detection areas of which typically increase like $E^2$.

Nevertheless, the predictions obtained for the Galactic center are generically small compared to the
current experimental sensitivities, typically of the order of 0.1-1 up-going muon at detectors per 
km$^2$ and per year, much smaller than the annihilation signal expected from the Sun as it will be 
shown below (see~\eg~\cite{2004PhRvD..70f3503B,2010PhRvD..81i6007E}). Such predictions are usually 
based upon assuming a generic NFW DM profile, with an optimistic angular average of 
1$^\circ$ around 
the Galactic center. The most optimistic estimate arises when requiring the potential complementary
gamma-ray flux (model-dependent approach) to saturate the gamma-ray signal already measured at the 
Galactic center by different experiments, \eg~EGRET, HESS, Fermi. Proceeding so implicitly accounts 
for the still large uncertainties in our knowledge of the central DM distribution, which 
might be more cuspy than expected (\eg~because of the adiabatic compression triggered by baryon 
cooling), inducing fluxes significantly larger, around 100 up-going muons per km$^2$ and per year 
\cite{2004PhRvD..70f3503B}.

Conservative limits have been obtained for generic WIMPs fully annihilating into 
specific final states by the IceCube Collaboration \cite{2011PhRvD..84b2004A}. They are reported in 
the $\sigv$-$\mchi$ plane in~\citefig{fig:neut_limits_gc} and shown to be rather far away from 
the relevant WIMP phase-space. This is essentially due to the poor exposure of IceCube to the 
Galactic center, since the experiment is located at the South Pole. Nevertheless, the analysis 
performed by the collaboration somehow compensates for this limitation by focusing on the 
high-latitude flux predictions, which are less subject to theoretical errors. Although this 
strongly damps the predictions, the DM density profile is in turn much less uncertain 
away than close to the Galactic center.
\begin{figure*}[t!]%[htp]
\centering
\begin{minipage}[t]{0.49\textwidth}
\vspace{0pt}
\includegraphics[width=\textwidth]{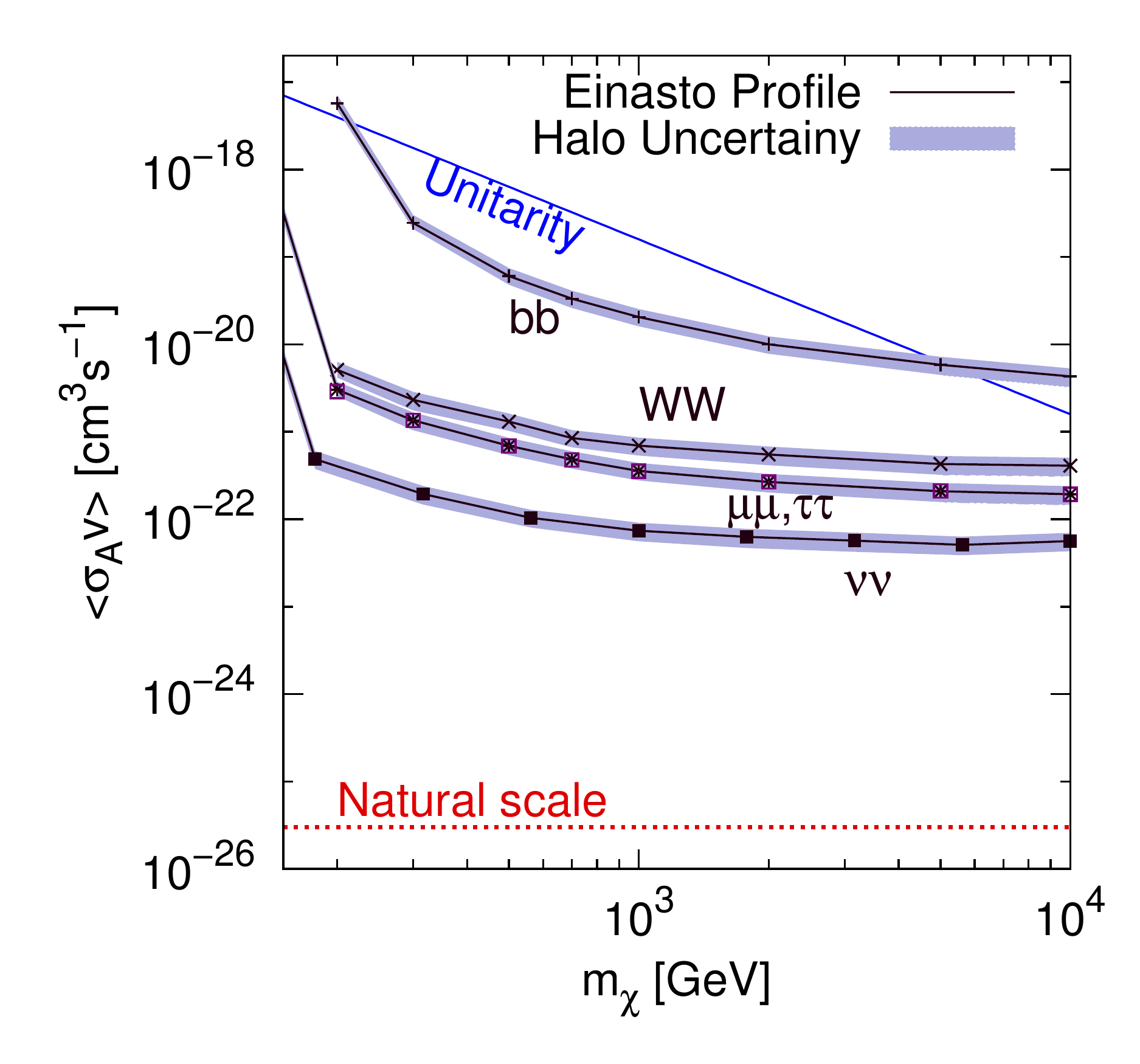}
\end{minipage}
\begin{minipage}[t]{0.49\textwidth}
\vspace{0pt}
\caption{Limits on the WIMP annihilation cross section obtained from observations of the 
Galactic halo at intermediate latitudes performed with the IceCube neutrino telescope. The unitarity
bound derived in Ref.~\cite{1990PhRvL..64..615G} is also shown. See the text for more details. 
This plot has been taken from Ref.~\cite{2011PhRvD..84b2004A}.}
\label{fig:neut_limits_gc}
\end{minipage}
\end{figure*}

Finally, we note that other interesting and similar DM sources are the neighbor
dwarf galaxies, which are DM dominated objects. Nevertheless, expectations are rather 
modest, and unless for TeV WIMP masses and decade timescale exposure, neutrino telescopes are poorly
sensitive to such objects \cite{2010PhRvD..81h3506S}.
%
%
%##########
\subsection{Neutrinos from DM annihilation in the Sun or the Earth}
\label{subsec:neut_sun}
The interaction properties of DM with ordinary matter open the possibility of its 
gravitational capture and further accumulation in celestial bodies, in particular in the Earth and 
in the Sun which are the closest potential reservoirs. Indeed, DM particles can lose 
kinetic energy through elastic scattering with ordinary matter and may thus remain gravitationally 
bound to the encountered object --- this happens when the final velocity is lower than the so-called
escape velocity. If sufficiently accumulated at the center, annihilation turns on and 
injects high-energy standard particles in the dense medium, among which neutrinos are the only 
species that can escape. As we will review below, this implies that the annihilation rate is not 
fixed by the (squared) average local DM density around the body, but is instead intimately 
linked to the capture rate. Once produced in the object, neutrinos experience the oscillation and 
absorption pattern discussed in the previous section before reaching the detector. We will focus 
on the Sun in 
the following, as a template example. For even more insights, we refer the reader to the seminal
works in Refs.~\cite{1985ApJ...296..679P,1985PhRvL..55..257S,1986PhLB..167..295F,1986PhRvD..33.2079K,1986PhRvD..34.2206G,1987NuPhB.283..681G,1987ApJ...321..560G}, and also to subsequent developments 
found in \eg~Refs.~\cite{1995PhLB..357..595E,1995APh.....3...65B,1996APh.....5..333B,1997PhDT.........5E,1998PhRvD..58j3519B,2004JCAP...07..008G,2006PhRvD..73l3507H,2007PhRvD..76i5008B,2008JCAP...01..021B,2009PhRvD..80i5019N,2009JCAP...01..032F,2011NuPhB.850..505K,2011JHEP...11..133F}. Dedicated 
and sophisticated numerical codes can be found in 
Refs.~\cite{2004JCAP...07..008G,2008JCAP...01..021B}.
%
%
%
%##########
\subsubsection{DM capture}
\label{subsubsec:dm_capt}
A pedagogical introduction to DM capture can be found in \cite{1996PhR...267..195J}. In the
Sun, the total number $N$ of DM particles evolves with time and depends on capture, 
annihilation, and escape, respectively as follows:
\ben
\label{eq:capt_sun}
\frac{dN}{dt} = C_c - C_a\,N^2 - C_e \,N \,,
\een
although the latter can safely be neglected for particle masses above a few GeV
\cite{1987NuPhB.283..681G,1987ApJ...321..560G}. The capture rate $\Gamma_{c}$ is readily identified 
to parameter $C_c$ and depends on the elastic scattering cross section between DM and 
ordinary matter, while the annihilation rate reads 
\ben
\Gamma_{a}(t) = \eta \,
\int_{\rm Sun} \! d^3{\mathbf x} \,
\sigv \,
% {\left\langle \sigma_\mathrm{ann} v \right\rangle} \,
n^{2}(t,{\mathbf x}) = \frac{C_a}{2}\,N^2\,, 
\een
where $n$ is the (time-dependent) number density of DM particles. We will further explain 
how to evaluate these parameters. Before, let's have a closer look to \citeeq{eq:capt_sun}. 
Interestingly, we note that when dynamical equilibrium is reached, 
\ie~$dN_\chi/dt = 0$, the annihilation rate is then fully fixed by the capture rate, with 
$\Gamma_{a} = C_{c}/2$. Nevertheless, this clearly depends on the annihilation parameter 
$C_a$ and the capture rate, and the general solution to \citeeq{eq:capt_sun} is given by:
\ben
\label{eq:capt_sol_N}
N(t) &=& \sqrt{\frac{C_c}{C_a}} \tanh\left\{ \frac{t}{\tau_c}\right\}
\overset{t\gg \tau_c}{\longrightarrow} \sqrt{\frac{C_c}{C_a}}\,,  \\
\tau_c &\equiv& (C_{c} \, C_a )^{-1/2}\,.\nn
\een
For neutrino detection today, this equation has to be evaluated at a time corresponding to the age 
of the Sun, $t=t_\odot = 4.6$ Gyr. Dynamical equilibrium is reached when $t_\odot/\tau_c\gg 1$, which
is usually the case for most DM candidates. Indeed, typical values are of order 
$\tau_{c,\odot}\approx 0.001 \, \tau_{c,\oplus}\approx 10^8$ yr --- equilibrium is thereby
barely reached in the Earth, while generically expected in the Sun. More precisely for the Sun, 
we have
$t_\odot/\tau_{c,\odot} \approx
\sqrt{(C_c/10^{25} \, {\rm s^{-1}}) (\sigv/10^{-30} \, {\rm cm^3 \, s^{-1}})}$,
which means that even with an annihilation cross section much lower today 
than the one relevant at thermal decoupling in the early universe (this is the case when 
annihilation is p-wave dominated), the annihilation rate in the Sun can still be maximal. 
Indeed, in terms of annihilation rate, \citeeq{eq:capt_sol_N} gives
\ben
\label{eq:capt_sol_Gamma}
\Gamma_a(t) &=& \frac{C_c}{2} \, \tanh^2\left\{ \frac{t}{\tau_c}\right\}
\overset{t\gg \tau_c}{\longrightarrow} \frac{C_c}{2}\,,
\een
which is independent of the annihilation cross section when equilibrium is reached. We therefore
anticipate that the solar neutrino flux can provide very strong constraints to p-wave annihilation
cross sections if the elastic scattering cross section is large enough, which is not the 
case with other indirect detection messengers. Given a point-like reservoir located at distance 
$d$ from the Earth, the neutrino flux reads
\ben
\phi_\nu(E) = \frac{\epsilon_\nu(E) \, \Gamma_a}{4 \,\pi \, d^2} \,\frac{dN_\nu}{dE}\,,
\een
where $dN_\nu/dE$ is the neutrino spectrum injected at the source and $\epsilon_\nu(E)$ 
accounts for the oscillation and absorption effects discussed in the previous section, which can 
significantly alter the original spectrum. The point-like approximation is usually very good for the
Sun given the poor angular resolution achieved in neutrino astronomy, typically 1$^\circ$ or more 
for GeV-TeV neutrinos; this ceases to be the case for the Earth. In the following, we discuss the 
calculations of the annihilation parameter $C_a$ and the capture rate $C_c$ which both set the 
annihilation rate $\Gamma_a$.

To derive $C_a$, one has first to guess the spatial distribution of DM in the Sun. 
Spherical symmetry applies, and if dynamical relaxation is assumed, then the spatial distribution
is fixed by the local gravitational potential $\Phi$ and the average kinetic energy (\ie~the average
temperature $\bar T$) \cite{1987NuPhB.283..681G} according to
\ben
n(t,{\mathbf x}) &=& n(t,r) = n_0(t) \, e^{-\mchi\,\Phi(r)/ \bar T}\,,\\
\Phi(r) &=& \int_0^r dr' \, \frac{G_N\,M(r')}{r'^2}\,,
\;\;\; %\nn\\
M(r) = 4\,\pi \int_0^r dr'\, r'^2 \, \rho_{\rm sun} (r')\,,\nn
\een
where $n_0$ is the average density at the core, $G_N$ is Newton's constant, $M(r)$ is the solar mass
within radius $r$, and $\rho_{\rm sun}$ is the solar mass density. Since $N = 4\,\pi \int_0^{R_{\rm sun}} 
dr\, r^2\, n(t,r)$, it comes
\ben
C_a = \frac{\int_0^{R_{\rm sun}} dr \, r^2\, \sigv \, n^2(t,r) } {N^2} 
=  \sigv \, \frac{\int_0^{R_{\rm sun}} dr \, r^2 \, e^{-2 \, \mchi\,\Phi(r) / \bar T}}
  {\left[ \int_0^{R_{\rm sun}} dr \, r^2 \, e^{- \mchi\,\Phi(r)/ \bar T} \right]^2} \,,
\een
and we remark that we can express $C_a$ in terms of effective volumes as
\ben
C_a &=& \sigv \, \frac{V_2}{V_1^2}\,,\\
V_n &\equiv& \left[ \frac{3\, \mplanck^2 \, T_{\rm sun} }{ 2\,n\,\mchi \,\rho_{\rm sun} }
\right]^{3/2}\simeq 6.6\times 10^{28}\left[ \frac{n\,\mchi}{10\,{\rm GeV}}\right]^{-3/2} 
\,{\rm cm^3}\,,\nn
\een
where the effective volume $V_n$ has an explicit dependence in the solar core temperature 
$T_{\rm sun}$ and density $\rho_{\rm sun}$ \cite{1996PhR...267..195J}. For the Earth, this volume
is $V_n \simeq 2.3 \times 10^{25} \, (n \, \mchi / 10 \, {\rm GeV})^{-3/2}$ cm$^3$.
Here, we have implicitly assumed that DM is concentrated at the Sun's core, and taken
the corresponding constant values for the solar mass density and temperature. Instead,
we may also evaluate these values at the mean orbital radius $\bar r$~\cite{2011NuPhB.850..505K}, so
that we now have to solve the equation for $C_a$ using $\rho_{\rm sun}(r) = \rho_{\rm sun}(\bar r)$, 
and $T(r) = T(\bar r)$. The mean orbital radius can be determined from the following implicit 
equation:
\ben
\bar r 
= %&=&
\frac{\int_0^{R_{\rm sun}} d^3{\mathbf x} \, r \, n(t,r)}
     {\int_0^{R_{\rm sun}} d^3{\mathbf x} \, n(t,r)}
= \frac{\int_0^{R_{\rm sun}} dr \, r^3 \, e^{- \, \mchi\,\Phi(r)/ \bar T(\bar r)}}
     {\int_0^{R_{\rm sun}} dr \, r^2 \, e^{- \, \mchi\,\Phi(r)/ \bar T(\bar r)}} 
%\,,\\
= % &=& 
\sqrt{\frac{6\,T(\bar r)}{\pi^2 \, G_N\, \rho_{\rm sun}(\bar r) \,\mchi}}\,.
%\nn
\een
This equation can be solved numerically. Then, the solution for $C_a$ is analytic and reads:
\ben
C_a = \left[ \frac{\sqrt{2}}{\pi \, \bar r}\right]^3 \, \sigv\,.
\een
The authors of Ref.~\cite{2011NuPhB.850..505K} have found that for masses larger than 1 GeV,
a good numerical approximation to the previous result is given by
\ben
C_a \simeq 4.5 \times 10^{-30} \, {\rm cm^{-3}}\,
\left[ \frac{\mchi \, - \, 0.6 \, {\rm GeV}}{10  \, {\rm GeV}} \right]^{3/2} \sigv\,.
\een
Here, the thermal average of the annihilation cross section $\sigv$ should be performed 
according to a Maxwellian with the core temperature $T(\bar r)\approx 10^7$ K $\approx 1$ keV as 
the characteristic temperature.

We now turn to the determination of the DM capture rate, which does depend on the 
scattering cross section and on the elemental distribution in the solar 
core~\cite{1987ApJ...321..560G}:
\ben
\Gamma_{c,\odot} = 4\, \pi \int_0^{R_{\rm sun}}dr \; r^2 \,
\left\{ \sum_i \, \frac{d\Gamma_{c,i}(r)}{dV} \right\}
\,.
\een
It involves the differential capture rate per shell volume unit induced by a solar element of
label $i$:
\ben
\label{eq:diff_capt}
\frac{d\Gamma_{c,i}(r)}{dV} = \frac{\rho_\chi\,\rho_i(r) \,\sigma_i}{2\, \mchi \, \mu_i^2}
\int_0^{w_{\rm max}} \! dw \; \frac{f(w)}{w} \int_{Q_{\rm min}}^{Q_{\rm max}} dQ \; F_i^2(Q)\,,
\een
where $\rho_\chi$ is the local DM mass density, $\rho_i(r)$ the element mass density in the
spherical shell at radius $r$, $\sigma_i$ the WIMP-element scattering cross section, 
$\mu_i$ the WIMP-element reduced mass, $f(w)$ the normalized velocity distribution in the solar
frame, $w$ the velocity as unaffected by the solar gravitational potential, $F(Q)$ the nuclear
form factor such that $F(0)=1$, and $Q$ the recoil energy\footnote{Form factors are different for 
scalar and spin-dependent interactions. Common
references are \cite{1987ApJ...321..571G,1996APh.....6...87L} for the former and 
\cite{1997PhRvC..56..535R} for the latter.}. The dependence on the 
elastic cross section and on the local WIMP density makes clear the correlation arising between
the neutrino signal expected from the Sun/Earth and the event rate predicted at direct 
detection experiments, especially when dynamical equilibrium is reached, in which case the 
annihilation rate is fully set by the capture rate. Additionally, we underline that given 
equivalent strengths for scalar (spin-independent) and axial-vector (spin-dependent) WIMP-nucleus 
interactions, capture in the Sun is usually dominated by the latter because of the prominence of 
light elements in the core (mostly hydrogen), while the former plays an important role for capture 
in the Earth.

In \citeeq{eq:diff_capt}, the first integral is performed over the Galactic velocity $w$, which can 
easily be related to the velocity in the Sun $v$. Indeed, energy conservation tells us that before 
the elastic scattering off a nucleus, the global WIMP kinetic energy $\propto \mchi \, w^2$ remains 
constant. Besides, when the WIMP falls in the Sun's gravitational potential, gravitational energy 
gets locally converted into kinetic energy. Therefore, the WIMP velocity in the Sun is given by 
$v^2 = w^2 + v_{\rm esc}^2(r)$, where the escape velocity $v_{\rm esc}$ is defined as the velocity at 
which the kinetic energy equals the gravitational potential.
% \ben
% v_{\rm esc}(r) = \sqrt{\frac{2\,G_N\,M(r)}{r}} 
% = 617.8 \, {\rm km/s} \, \sqrt{\frac{M(r)/M_\odot}{r/R_\odot}}\,,
% \een
% where $M_\odot$ and $R_\odot$ are the solar mass and radius, respectively.
The escape velocity at the Sun's surface is $617.8$ km/s, to be compared to
a value of $11.87$ km/s for the Earth. A Maxwellian velocity distribution is usually 
assumed, which (in the Sun's frame) is given by
\ben
\frac{f(w)}{w} = 
\sqrt{\frac{3}{2\,\pi}} \, \frac{1}{\sigma_v\,v_\odot} \,
\left[ \exp\left\{ -\frac{3(w-v_\odot)^2 }{2\sigma_v^2}\right\} - 
\exp\left\{ -\frac{3(w+v_\odot)^2 }{2\sigma_v^2} \right\}  \right]\,,
\een
where $v_\odot \simeq 220$ km/s is the velocity of the Sun in the Galactic frame, and 
$\sigma_v \simeq 270$ km/s is the velocity dispersion of DM species.

Kinematics must be inspected carefully, as detailed in 
\cite{1987ApJ...321..560G,1987ApJ...321..571G,1988ApJ...328..919G} and very clearly summarized 
in~\cite{1996PhR...267..195J}. A WIMP is captured as soon as its velocity is less than
$v_{\rm esc}$ after scattering. Defining
\ben
\beta_{\pm} \equiv \frac{4\,\mchi \, m_i}{(\mchi \pm m_i)^2}\,,
\een
the fraction of WIMP energy lost in the collision, $f_Q = Q/E$, must lie in the range
\ben
0 \leq & f_Q & \leq \beta_+\,,\\
\mchi \, w^2 \leq & 2\, Q & \leq  \beta_+ \, \mchi \, (w^2 + v_{\rm esc}^2)\,,\nn
\een
for the WIMP to be captured. At this stage, it is interesting to remark that for an isotropic
scattering (disregarding the form factor suppression for a moment), the probability ${\cal P}$ for 
a WIMP to scatter down to a velocity smaller than $v_{\rm esc}$ should be flat in the above range
such that~\cite{1996PhR...267..195J}
\ben
{\cal P}(w) = \frac{1}{(w^2 + v_{\rm esc}^2)}\,\left\{ v_{\rm esc}^2- \frac{w}{\beta_-} \right\}
\,\theta \left[ v_{\rm esc}^2- \frac{w}{\beta_-} \right]\,,
\een
where $\theta$ is the Heaviside function. Incidentally, this equation shows that the 
probability of capture is maximized when $\beta_-$ is maximized, \ie~when $\mchi \approx m_i$. This 
actually corresponds to resonance effects, which depend on both the elemental distribution in the 
celestial reservoir and the escape velocity. This has been intensely discussed in 
Ref.~\cite{1987ApJ...321..571G} where these resonance effects were predicted to be important for 
capture in the Earth, due to the larger relative density of oxygen and metals.

Kinematics further helps define the ranges over which the integrals of \citeeq{eq:diff_capt} 
are performed:
\ben
w_{\rm max}(r) &=& \sqrt{\frac{4\,\mchi\,m_i}{(\mchi-m_i)^2}} \; v_{\rm esc}(r)\,,\\
Q_{\rm min} &=&  \frac{1}{2} \, \mchi\, w^2\,,\nn\\
Q_{\rm max} &=&  \sqrt{\frac{4\,\mchi\,m_i}{(\mchi+m_i)^2}} \; \mchi \, v^2\,.\nn
\een
From these integral boundaries, we understand why a departure from the Maxwellian
law can have a significant impact on the predictions, as discussed into more details in
\eg~Ref.~\cite{2010PhRvD..82b3534L}.
%
%
%
%##########
\subsubsection{Current limits on DM models and prospects}
\label{subsubsec:sun_res}
We start by emphasizing that in contrast with other indirect detection messengers, the solar 
neutrino signature is one of the best to discover DM
annihilation because predictions do not suffer potentially big theoretical errors.
Indeed, they mostly rely (i) on the knowledge of the local DM density and dynamics,
which can be constrained reasonably well, (ii) on neutrino propagation in vacuum and matter,
which is under control, and (iii) on the DM particle mass, and its annihilation
and interaction properties. The latter point can be fixed for each DM model. The
most important uncertainty affects the evaluation of the WIMP-atom elastic scattering cross 
section, because of uncertainties in the estimates of the spin contents of nucleons
(spin-dependent interactions) and of the nucleonic matrix elements (scalar interactions) --- 
see \eg~\cite{2008PhRvD..77f5026E}. Nevertheless, there are hopes to reduce these uncertainties 
in the near future thanks to improvements in hadronic physics or QCD, for which experimental data 
will come from existing colliders. Finally, as another big advantage over other astrophysical 
messengers, the main neutrino background is well measured and also under control.

Many predictions of the neutrino fluxes from the Sun and the Earth have been derived in the context 
of supersymmetric DM~\cite{1996PhR...267..195J}. The reader can find extensive examples 
in~\eg~\cite{1997PhDT.........5E,1998PhRvD..58j3519B,2001PhRvD..63d5024F,2002EPJC...26..111B,2009JCAP...08..034T}. 
Predictions in the frame of extra-dimensional theories are quantitatively similar 
(see \eg~\cite{2003PhRvD..67e5003H,2010JCAP...01..018B}). 
These predictions mostly concern the muon neutrino flux\footnotemark, which can be reconstructed at 
detectors as upward-going muons after charged current conversion. The predicted fluxes, usually
higher if induced by DM annihilation in the Sun than in the Earth's interior, can
reach $10^{11}$-$10^{12}$ upgoing muon neutrinos per year and square kilometer above a 
detection threshold energy of $E_{\nu}^{\rm th}\approx 10$ GeV, which translates into an upgoing 
muon flux of $10^{2}$-$10^{3}\,{\rm km^{-2}yr^{-1}}$ at detectors. The full energy dependence has 
further to be taken into account properly to convolve these expectations with the detector 
efficiencies and to compare them with the background. For DM searches toward the Sun, we 
note that in addition to the atmospheric background, a contribution is also expected from 
interactions of cosmic rays in the Sun itself, although it has been shown
negligible~\cite{1991ApJ...382..652S,1996PhRvD..54.4385I}.
\footnotetext{See Ref.~\cite{2011JHEP...11..133F} for prospects with tau neutrinos.}
\begin{figure*}[t!]%[htp]
\centering
\begin{minipage}[t]{0.49\textwidth}
\vspace{0pt} 
\includegraphics[width=\textwidth]{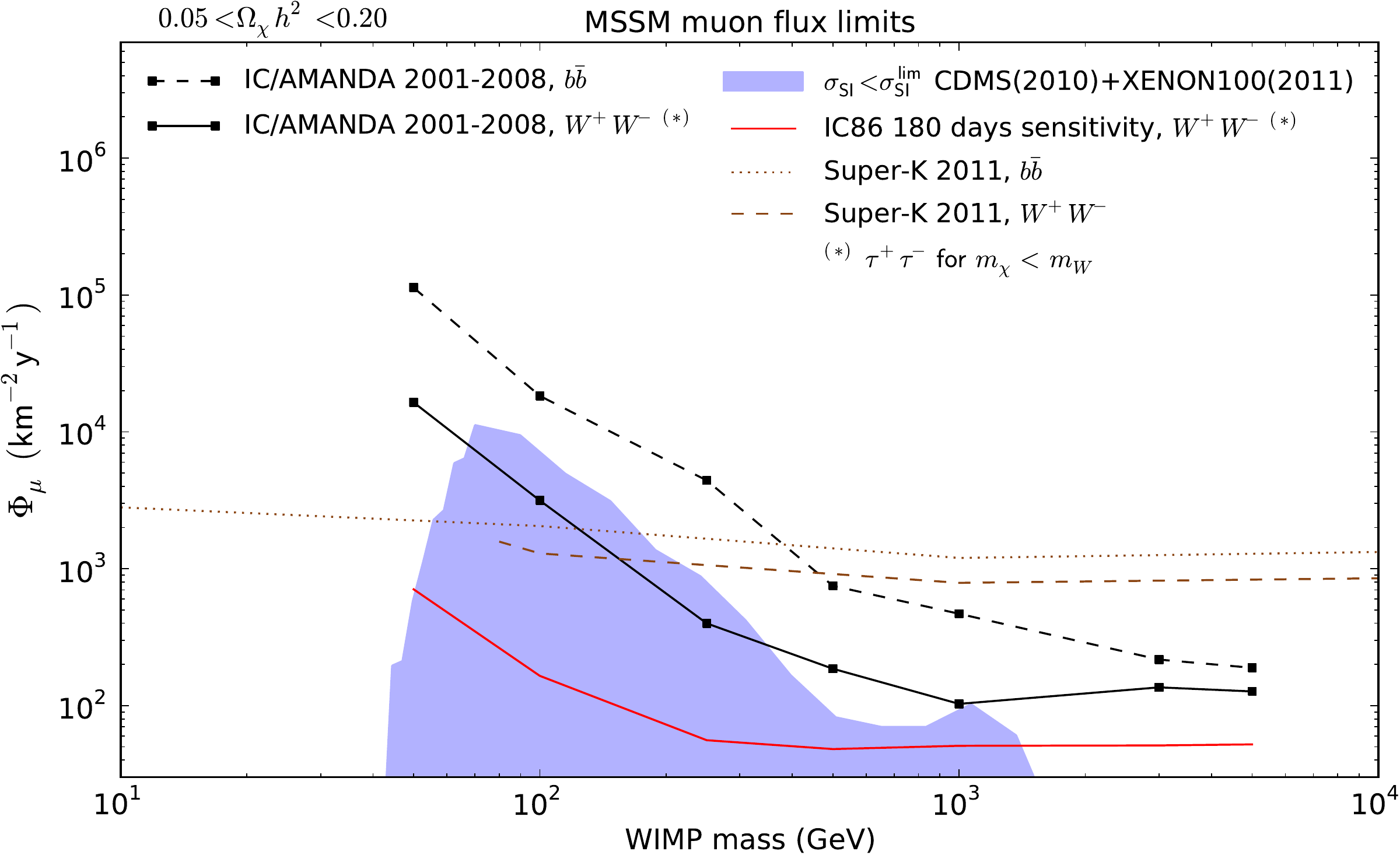}
\end{minipage}
\begin{minipage}[t]{0.49\textwidth}
 \vspace{0pt} 
\caption{Top left: Muon flux predictions for supersymmetric DM models and limits from
IceCube and Super-K. Bottom left: Corresponding limits in terms of spin-dependent
WIMP-nucleon cross section. Bottom right: Corresponding limits in terms of spin-independent
WIMP-nucleon cross section.}
\label{fig:neut_limits_sun}
\end{minipage}
\includegraphics[width=0.49\textwidth]{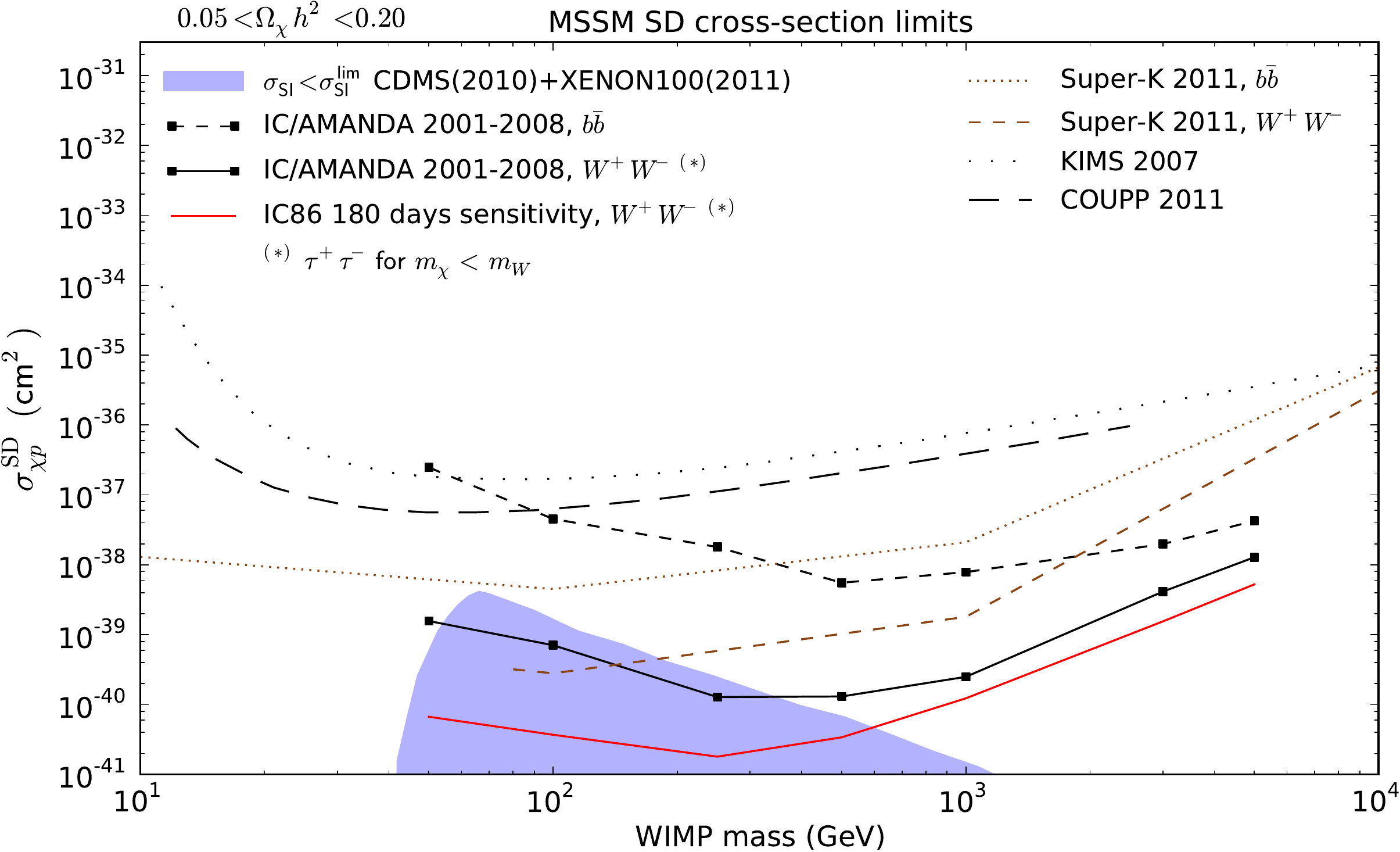}
\includegraphics[width=0.49\textwidth]{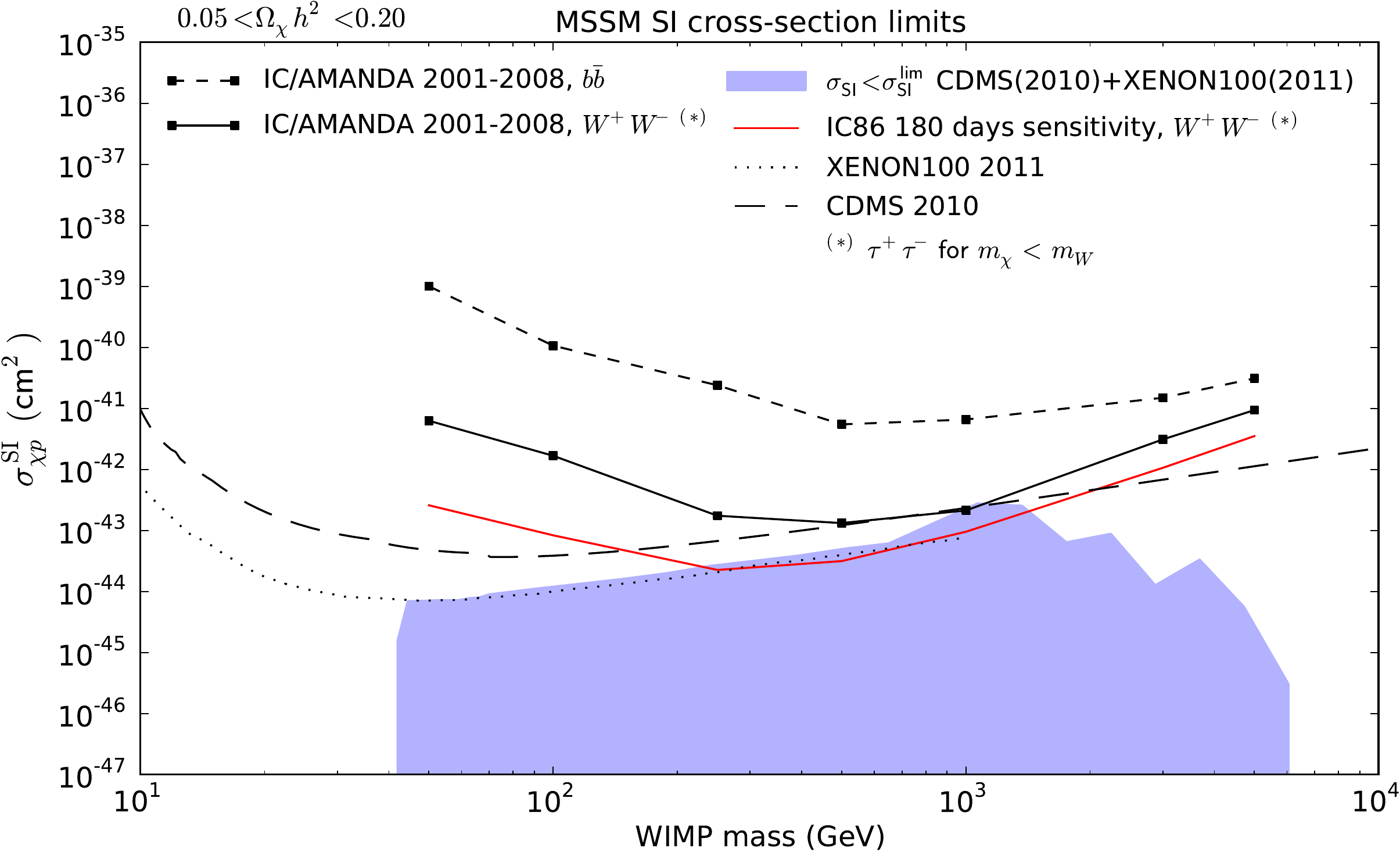}
\end{figure*}

Disregarding the details of predictions, we emphasize that such a muon flux level is 
within reach of some current experiments like Super-K~\cite{2011ApJ...742...78T} 
and IceCube~\cite{2012PhRvD..85d2002A}, which have already derived interesting limits.
These are shown in terms of muon flux in the top left panel of 
\citefig{fig:neut_limits_sun}, taken from Ref.~\cite{2012PhRvD..85d2002A}, on top of predictions 
associated with DM candidates arising in the minimal supersymmetric model (represented as 
the shaded area in the plot). Note that a broad range for the relic abundance was considered, down 
to a value as low as $\ohh = 0.05$. This implies that the reported shaded area contains models with 
optimistically large cross sections, most of which corresponding to the largest predicted muon 
fluxes. Some direct detection constraints have been included, coming from both a 
combined analysis of the CDMS and EDELWEISS data \cite{2011PhRvD..84a1102A} and from XENON 100 
\cite{2011PhRvL.107m1302A}. Two types of limits appear in the panel, one assuming a soft 
neutrino spectrum arising in annihilation into $b$ quarks, and another one assuming a harder 
neutrino spectrum originating in annihilation into $W$ bosons; the latter case is obviously more 
constraining. The supersymmetric models in the shaded zone, in contrast, usually come with more
complex spectra, in between these two cases. While the derived limits are shown to start tickling
the upper part of the relevant parameter space, the sensitivity of the complete version of IceCube
is drawn as the red solid curve for a duty cycle of 180 days. It clearly demonstrates the potential
of neutrino telescopes to survey a large part of the supersymmetric parameter space.
\begin{figure*}[t!]%[htp]
 \centering
\includegraphics[width=0.49\columnwidth]{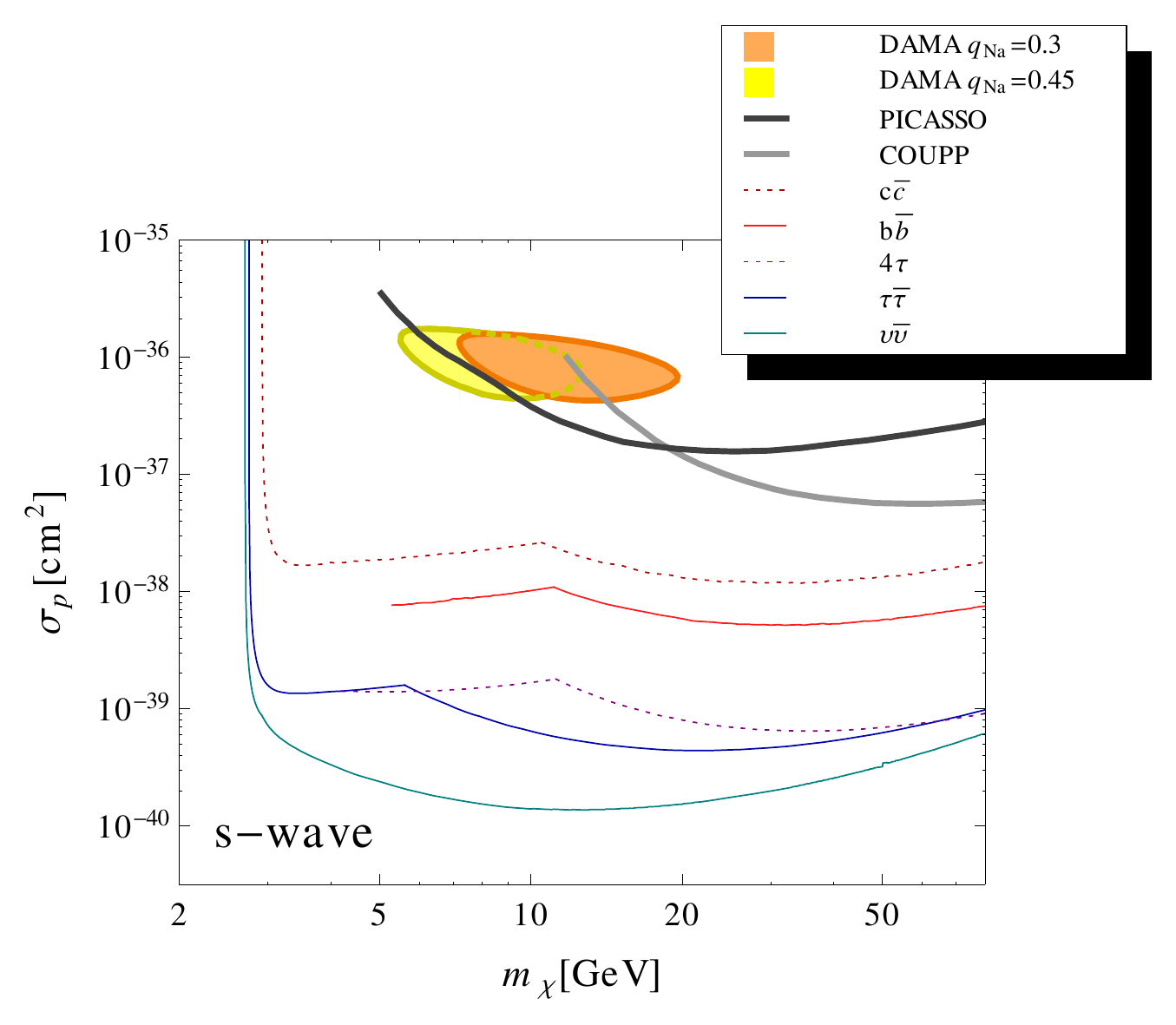}
\includegraphics[width=0.49\columnwidth]{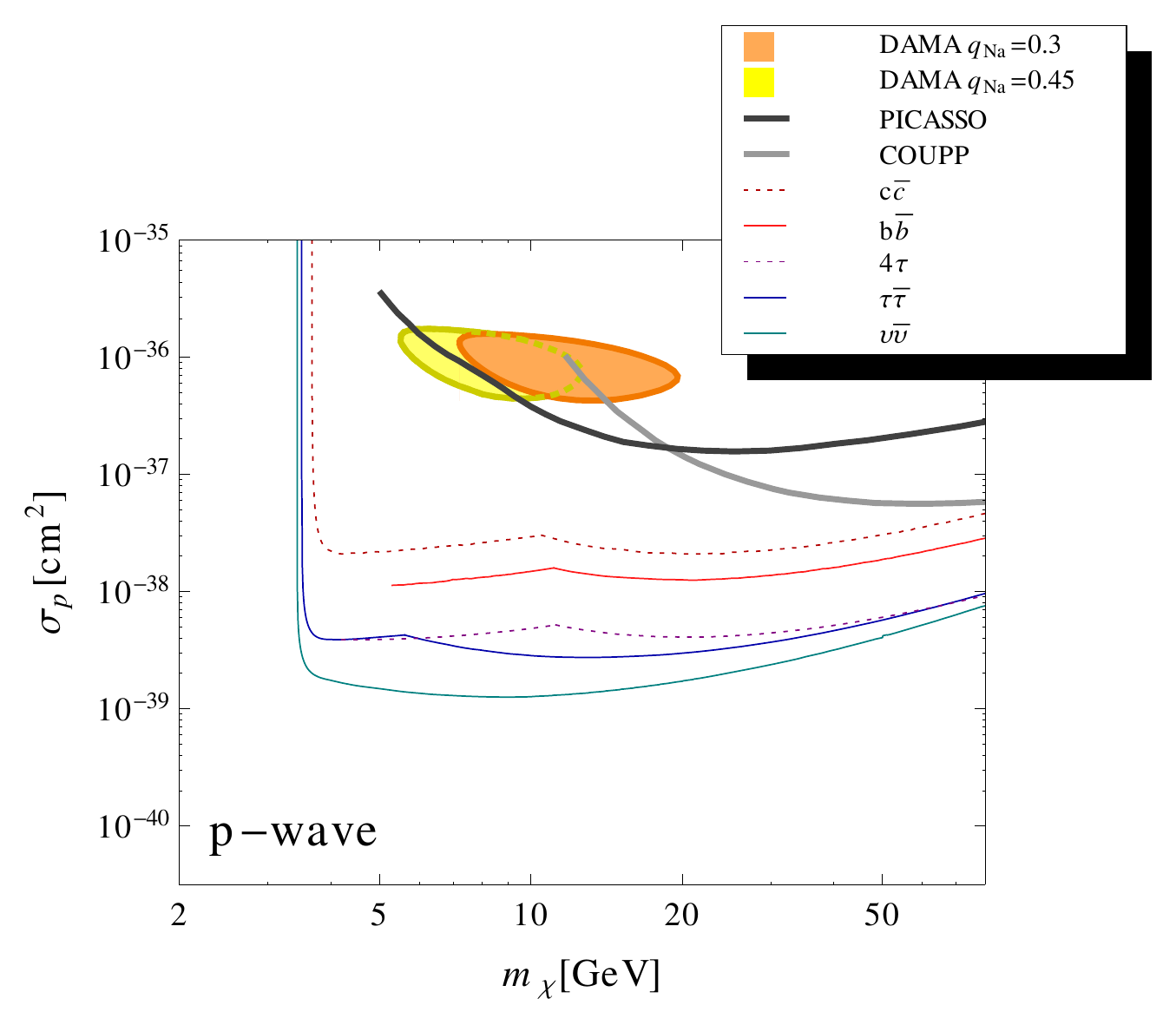}
\includegraphics[width=0.49\columnwidth]{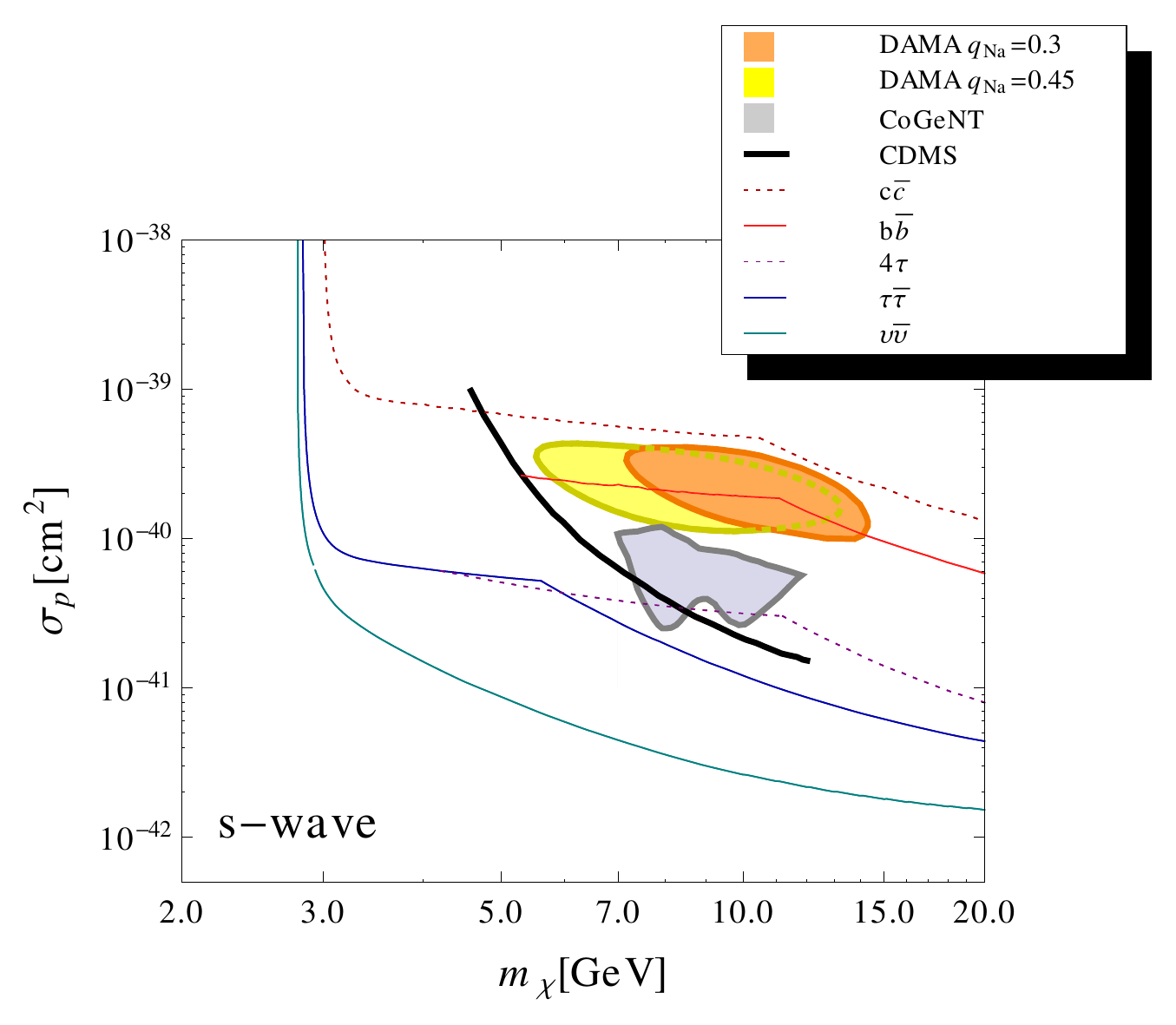}
\includegraphics[width=0.49\columnwidth]{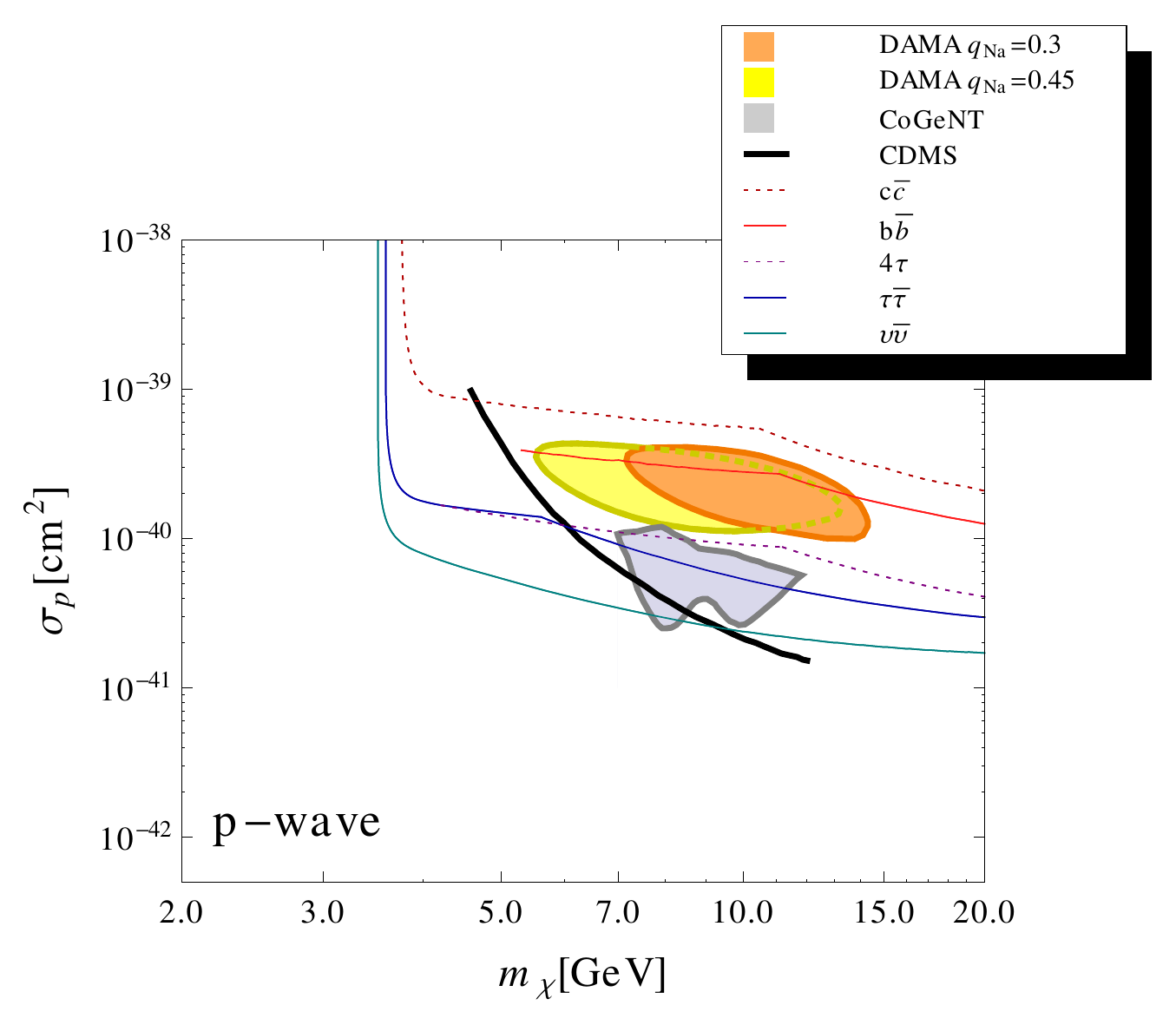}
\caption{Limits on the spin-dependent (top) and spin-independent (bottom) WIMP-nucleon cross
sections obtained from the Super-K data. The left (right) panels assume that the WIMP 
relic density is set by a pure s-wave (p-wave) annihilation cross section. Regions favored
by some direct detection experiments appear as shaded areas. These plots have been 
taken from Ref.~\cite{2011NuPhB.850..505K}.}
\label{fig:neut_direct_compl}
\end{figure*}

As already emphasized above, the neutrino signal induced by DM annihilation in celestial
bodies should exhibit a correlation with direct detection signals. For the Sun, the neutrino
signal is mostly proportional to the spin-dependent elastic scattering cross section, while
the scalar one may also, in some case, become important. Therefore, limits on the solar neutrino
flux can easily be translated in terms of elastic scattering cross sections and compared
with the sensitivities of direct detection experiments --- this is obviously 
DM model dependent. The bottom panels of \citefig{fig:neut_limits_sun}
nicely illustrate that, where the muon flux predictions of the top panel are replotted
in terms of WIMP-nucleon spin-dependent scattering cross section (bottom left) and of
scalar scattering cross section (bottom right). We clearly see that, for the former case, current 
neutrino telescopes are already very competitive with respect to existing direct detection 
experiments like KIM \cite{2007PhRvL..99i1301L} or COUPP \cite{2011PhRvL.106b1303B}. On the
other hand, the scalar cross section is more poorly constrained.

The complementarity between direct DM detection and indirect detection is further 
illustrated in \citefig{fig:neut_direct_compl}, taken from Ref. \cite{2011NuPhB.850..505K}, where 
the authors used the Super-K data to extract limits on generic WIMP models. The used
data set slightly differs from the previous example in that only fully contained events, \ie~upgoing
muons stopping within the detector, were considered to reduce the background and enhance 
sensitivity \cite{2009JCAP...01..032F}. The top (bottom) panels show the limits obtained on the
spin-dependent (scalar) cross section, while the left (right) panels concern models for which 
the relic abundance is set by a fully s-wave (p-wave) annihilation cross section. Note that
a dominating p-wave annihilation cross section makes indirect DM detection with other
messengers irrelevant since annihilation in galactic halos is then velocity suppressed; neutrinos
are therefore the only indirect detection possibility in this case. In the plots, the authors
have assumed different pure annihilation final states, which are made explicit. Finally, they
have reported as shaded zones the regions corresponding to the excess events found in two direct
detection experiments, DAMA \cite{2008EPJC..tmp..167B} and CoGeNT \cite{2011PhRvL.106m1301A}, 
together with the limits obtained by the PICASSO \cite{2009PhLB..682..185A}, COUPP 
\cite{2011PhRvL.106b1303B}, and CDMS \cite{2011PhRvD..84a1102A} collaborations.

% The Acknowledgements are also a un-numbered section
\section*{Acknowledgements}
% Acknowledgements text here
We would like to thank C. B{\oe}hm and P. Brax, the editors of this special issue, for having
invited us to express our point of view on the topic of WIMP indirect detection. PS expresses his 
gratitude to the Institut universitaire de France (IUF) for academically and financially supporting 
his research.
%
%%%%%%%%%%%%%%%%%%%%%%%%%%%%%%%%%%%%%%%%%%%%%%%%%%%%%%%%%%%%%%%%%%%%%%%%%%%%%%%%%%%%%%%%%%%%%%%%%%%
%%%%%%%%%%%%%%%%%%%%%%%%%%%%%%%%%%%%%%%%%%%%%%%%%%%%%%%%%%%%%%%%%%%%%%%%%%%%%%%%%%%%%%%%%%%%%%%%%%%
%
\bibliography{mybib}

\begin{thebibliography}{100}
\expandafter\ifx\csname url\endcsname\relax
  \def\url#1{\texttt{#1}}\fi
\expandafter\ifx\csname urlprefix\endcsname\relax\def\urlprefix{URL }\fi
\expandafter\ifx\csname href\endcsname\relax
  \def\href#1#2{#2} \def\path#1{#1}\fi


\bibitem{silk}
J.~Silk,
{Dark Matter: the astrophysical case},
this issue of Comptes Rendus de l'Acad\'emie des Sciences.

\bibitem{hoopertait}
D.~Hooper, T.~Tait,
{Theories of particle dark matter},
this issue of Comptes Rendus de l'Acad\'emie des Sciences.

\bibitem{armengaud}
E.~Armengaud,
{Direct detection of WIMPs},
this issue of Comptes Rendus de l'Acad\'emie des Sciences.


\bibitem{2002PhRvD..65b3002C}
F.~{Casse}, M.~{Lemoine}, G.~{Pelletier}, {Transport of cosmic rays in chaotic
  magnetic fields}, \prd 65~(2) (2002) 023002.
\newblock \href {http://arxiv.org/abs/astro-ph/0109223}
  {\path{arXiv:astro-ph/0109223}}, \href
  {http://dx.doi.org/10.1103/PhysRevD.65.023002}
  {\path{doi:10.1103/PhysRevD.65.023002}}.

\bibitem{2002astro.ph.12111M}
D.~{Maurin}, et~al., {Galactic Cosmic Ray Nuclei as a Tool for Astroparticle
  Physics}, Research Signposts, Recent Research Developments in Astronomy and
  Astrophysics 2 (2004) 193.
\newblock \href {http://arxiv.org/abs/astro-ph/0212111}
  {\path{arXiv:astro-ph/0212111}}.

\bibitem{1977A&A....54..973E}
R.~D. {Ekers}, R.~{Sancisi}, {The radio continuum halo in NGC 4631}, \aap 54
  (1977) 973.

\bibitem{1998ApJ...509..212S}
A.~W. {Strong}, I.~V. {Moskalenko}, {Propagation of Cosmic-Ray Nucleons in the
  Galaxy}, \apj 509 (1998) 212--228.
\newblock \href {http://arxiv.org/abs/astro-ph/9807150}
  {\path{arXiv:astro-ph/9807150}}, \href {http://dx.doi.org/10.1086/306470}
  {\path{doi:10.1086/306470}}.

\bibitem{2008JCAP...10..018E}
C.~{Evoli}, et~al., {Cosmic ray nuclei, antiprotons and gamma rays in the
  galaxy: a new diffusion model}, Journal of Cosmology and Astro-Particle
  Physics 10 (2008) 18--+.
\newblock \href {http://arxiv.org/abs/0807.4730} {\path{arXiv:0807.4730}},
  \href {http://dx.doi.org/10.1088/1475-7516/2008/10/018}
  {\path{doi:10.1088/1475-7516/2008/10/018}}.

\bibitem{2001ApJ...555..585M}
D.~{Maurin}, et~al., {Cosmic Rays below Z=30 in a Diffusion Model: New
  Constraints on Propagation Parameters}, \apj 555 (2001) 585--596.
\newblock \href {http://arxiv.org/abs/astro-ph/0101231}
  {\path{arXiv:astro-ph/0101231}}, \href {http://dx.doi.org/10.1086/321496}
  {\path{doi:10.1086/321496}}.

\bibitem{2004PhRvD..69f3501D}
F.~{Donato}, et~al.,
  {Antiprotons in cosmic rays from neutralino annihilation}, \prd 69~(6) (2004)
  063501--+.
\newblock \href {http://arxiv.org/abs/astro-ph/0306207}
  {\path{arXiv:astro-ph/0306207}}, \href
  {http://dx.doi.org/10.1103/PhysRevD.69.063501}
  {\path{doi:10.1103/PhysRevD.69.063501}}.

\bibitem{2010A&A...516A..66P}
A.~{Putze}, L.~{Derome}, D.~{Maurin}, {A Markov Chain Monte Carlo technique to
  sample transport and source parameters of Galactic cosmic rays. II. Results
  for the diffusion model combining B/C and radioactive nuclei}, \aap 516
  (2010) A66+.
\newblock \href {http://arxiv.org/abs/1001.0551} {\path{arXiv:1001.0551}},
  \href {http://dx.doi.org/10.1051/0004-6361/201014010}
  {\path{doi:10.1051/0004-6361/201014010}}.

\bibitem{1983JPhG....9.1289T}
L.~C. {Tan}, L.~K. {Ng}, {Parametrisation of hadron inclusive cross sections in
  p-p collisions extended to very low energies}, Journal of Physics G Nuclear
  Physics 9 (1983) 1289--1308.
\newblock \href {http://dx.doi.org/10.1088/0305-4616/9/10/015}
  {\path{doi:10.1088/0305-4616/9/10/015}}.

\bibitem{2007PhRvD..75h3006B}
T.~{Bringmann}, P.~{Salati}, {Galactic antiproton spectrum at high energies:
  Background expectation versus exotic contributions}, \prd 75~(8) (2007)
  083006.
\newblock \href {http://arxiv.org/abs/astro-ph/0612514}
  {\path{arXiv:astro-ph/0612514}}, \href
  {http://dx.doi.org/10.1103/PhysRevD.75.083006}
  {\path{doi:10.1103/PhysRevD.75.083006}}.

\bibitem{1982PhRvD..26.1179T}
L.~C. {Tan}, L.~K. {Ng}, {Parameterization of ? invariant cross section in p-p
  collisions using a new scaling variable}, \prd 26 (1982) 1179--1182.
\newblock \href {http://dx.doi.org/10.1103/PhysRevD.26.1179}
  {\path{doi:10.1103/PhysRevD.26.1179}}.

\bibitem{1983JPhG....9..227T}
L.~C. {Tan}, L.~K. {Ng}, {Calculation of the equilibrium antiproton spectrum},
  Journal of Physics G Nuclear Physics 9 (1983) 227--242.
\newblock \href {http://dx.doi.org/10.1088/0305-4616/9/2/015}
  {\path{doi:10.1088/0305-4616/9/2/015}}.

\bibitem{1999ApJ...526..215B}
L.~{Bergstr{\"o}m}, J.~{Edsj{\"o}}, P.~{Ullio}, {Cosmic Antiprotons as a Probe
  for Supersymmetric Dark Matter?}, \apj 526 (1999) 215--235.
\newblock \href {http://arxiv.org/abs/astro-ph/9902012}
  {\path{arXiv:astro-ph/9902012}}, \href {http://dx.doi.org/10.1086/307975}
  {\path{doi:10.1086/307975}}.

\bibitem{1967PhRvL..19..198A}
E.~W. {Anderson}, et~al., {Proton and Pion Spectra from Proton-Proton
  Interactions at 10, 20, and 30 BeV/c}, Physical Review Letters 19 (1967)
  198--201.
\newblock \href {http://dx.doi.org/10.1103/PhysRevLett.19.198}
  {\path{doi:10.1103/PhysRevLett.19.198}}.

\bibitem{2005PhRvD..71h3013D}
R.~{Duperray}, et~al., {Flux of light antimatter nuclei near Earth, induced by
  cosmic rays in the Galaxy and in the atmosphere}, \prd 71~(8) (2005) 083013.
\newblock \href {http://arxiv.org/abs/astro-ph/0503544}
  {\path{arXiv:astro-ph/0503544}}, \href
  {http://dx.doi.org/10.1103/PhysRevD.71.083013}
  {\path{doi:10.1103/PhysRevD.71.083013}}.

\bibitem{2008PhRvD..78d3506D}
F.~{Donato}, N.~{Fornengo}, D.~{Maurin}, {Antideuteron fluxes from dark matter
  annihilation in diffusion models}, \prd 78~(4) (2008) 043506--+.
\newblock \href {http://arxiv.org/abs/0803.2640} {\path{arXiv:0803.2640}},
  \href {http://dx.doi.org/10.1103/PhysRevD.78.043506}
  {\path{doi:10.1103/PhysRevD.78.043506}}.

\bibitem{2001ApJ...563..172D}
F.~{Donato}, D.~{Maurin}, P.~{Salati}, A.~{Barrau}, G.~{Boudoul}, R.~{Taillet},
  {Antiprotons from Spallations of Cosmic Rays on Interstellar Matter}, \apj
  563 (2001) 172--184.

\bibitem{2004PhRvL..93w1805A}
K.~{Agashe}, G.~{Servant}, {Warped Unification, Proton Stability, and Dark
  Matter}, Physical Review Letters 93~(23) (2004) 231805--+.
\newblock \href {http://arxiv.org/abs/hep-ph/0403143}
  {\path{arXiv:hep-ph/0403143}}, \href
  {http://dx.doi.org/10.1103/PhysRevLett.93.231805}
  {\path{doi:10.1103/PhysRevLett.93.231805}}.

\bibitem{2005JCAP...02..002A}
K.~{Agashe}, G.~{Servant}, {Baryon number in warped grand unified theories:
  model building and (dark matter related) phenomenology}, Journal of Cosmology
  and Astro-Particle Physics 2 (2005) 2--+.
\newblock \href {http://arxiv.org/abs/hep-ph/0411254}
  {\path{arXiv:hep-ph/0411254}}, \href
  {http://dx.doi.org/10.1088/1475-7516/2005/02/002}
  {\path{doi:10.1088/1475-7516/2005/02/002}}.

\bibitem{2005PhRvD..72f3507B}
A.~{Barrau}, et~al., {Kaluza-Klein dark matter and galactic antiprotons}, \prd
  72~(6) (2005) 063507.
\newblock \href {http://dx.doi.org/10.1103/PhysRevD.72.063507}
  {\path{doi:10.1103/PhysRevD.72.063507}}.

\bibitem{2010PhRvL.105l1101A}
O.~{Adriani}, et~al., {PAMELA Results on the Cosmic-Ray Antiproton Flux from 60
  MeV to 180 GeV in Kinetic Energy}, Physical Review Letters 105~(12) (2010)
  121101--+.
\newblock \href {http://arxiv.org/abs/1007.0821} {\path{arXiv:1007.0821}},
  \href {http://dx.doi.org/10.1103/PhysRevLett.105.121101}
  {\path{doi:10.1103/PhysRevLett.105.121101}}.

\bibitem{1997ApJ...490..493N}
J.~F. {Navarro}, C.~S. {Frenk}, S.~D.~M. {White}, {A Universal Density Profile
  from Hierarchical Clustering}, \apj 490 (1997) 493--+.
\newblock \href {http://arxiv.org/abs/astro-ph/9611107}
  {\path{arXiv:astro-ph/9611107}}, \href {http://dx.doi.org/10.1086/304888}
  {\path{doi:10.1086/304888}}.

\bibitem{1999ApJ...524L..19M}
B.~{Moore}, et~al., {Dark Matter Substructure within Galactic Halos}, \apjl 524
  (1999) L19--L22.
\newblock \href {http://arxiv.org/abs/astro-ph/9907411}
  {\path{arXiv:astro-ph/9907411}}, \href {http://dx.doi.org/10.1086/312287}
  {\path{doi:10.1086/312287}}.

\bibitem{2004MNRAS.349.1039N}
J.~F. {Navarro}, et~al., {The inner
  structure of {$\Lambda$}CDM haloes - III. Universality and asymptotic
  slopes}, MNRAS 349 (2004) 1039--1051.
\newblock \href {http://arxiv.org/abs/astro-ph/0311231}
  {\path{arXiv:astro-ph/0311231}}, \href
  {http://dx.doi.org/10.1111/j.1365-2966.2004.07586.x}
  {\path{doi:10.1111/j.1365-2966.2004.07586.x}}.

\bibitem{2004MNRAS.353..624D}
J.~{Diemand}, B.~{Moore}, J.~{Stadel}, {Convergence and scatter of cluster
  density profiles}, \mnras 353 (2004) 624--632.
\newblock \href {http://arxiv.org/abs/astro-ph/0402267}
  {\path{arXiv:astro-ph/0402267}}, \href
  {http://dx.doi.org/10.1111/j.1365-2966.2004.08094.x}
  {\path{doi:10.1111/j.1365-2966.2004.08094.x}}.

\bibitem{2004MNRAS.351..903G}
G.~{Gentile}, et~al., {The cored
  distribution of dark matter in spiral galaxies}, \mnras 351 (2004) 903--922.
\newblock \href {http://arxiv.org/abs/astro-ph/0403154}
  {\path{arXiv:astro-ph/0403154}}, \href
  {http://dx.doi.org/10.1111/j.1365-2966.2004.07836.x}
  {\path{doi:10.1111/j.1365-2966.2004.07836.x}}.

\bibitem{2004MNRAS.353L..17D}
F.~{Donato}, G.~{Gentile}, P.~{Salucci}, {Cores of dark matter haloes correlate
  with stellar scalelengths}, \mnras 353 (2004) L17--L22.
\newblock \href {http://arxiv.org/abs/astro-ph/0403206}
  {\path{arXiv:astro-ph/0403206}}, \href
  {http://dx.doi.org/10.1111/j.1365-2966.2004.08220.x}
  {\path{doi:10.1111/j.1365-2966.2004.08220.x}}.

\bibitem{2010JCAP...08..004C}
R.~{Catena}, P.~{Ullio}, {A novel determination of the local dark matter
  density}, \jcap 8 (2010) 4--+.
\newblock \href {http://arxiv.org/abs/0907.0018} {\path{arXiv:0907.0018}},
  \href {http://dx.doi.org/10.1088/1475-7516/2010/08/004}
  {\path{doi:10.1088/1475-7516/2010/08/004}}.

\bibitem{2010A&A...523A..83S}
P.~{Salucci}, et~al., {The dark matter
  density at the Sun's location}, \aap 523 (2010) A83+.
\newblock \href {http://arxiv.org/abs/1003.3101} {\path{arXiv:1003.3101}},
  \href {http://dx.doi.org/10.1051/0004-6361/201014385}
  {\path{doi:10.1051/0004-6361/201014385}}.

\bibitem{2011MNRAS.414.2446M}
P.~J. {McMillan}, {Mass models of the Milky Way}, \mnras 414 (2011) 2446--2457.
\newblock \href {http://arxiv.org/abs/1102.4340} {\path{arXiv:1102.4340}},
  \href {http://dx.doi.org/10.1111/j.1365-2966.2011.18564.x}
  {\path{doi:10.1111/j.1365-2966.2011.18564.x}}.

\bibitem{2012ApJ...756...89B}
J.~{Bovy}, S.~{Tremaine}, {On the Local Dark Matter Density}, \apj 756 (2012)
  89.
\newblock \href {http://arxiv.org/abs/1205.4033} {\path{arXiv:1205.4033}},
  \href {http://dx.doi.org/10.1088/0004-637X/756/1/89}
  {\path{doi:10.1088/0004-637X/756/1/89}}.

\bibitem{1980ApJS...44...73B}
J.~N. {Bahcall}, R.~M. {Soneira}, {The universe at faint magnitudes. I - Models
  for the galaxy and the predicted star counts}, \apjs 44 (1980) 73--110.
\newblock \href {http://dx.doi.org/10.1086/190685} {\path{doi:10.1086/190685}}.

\bibitem{2010Sci...327.1619C}
Z.~{Ahmed}, et~al., {Dark Matter Search Results from the CDMS II Experiment},
  Science 327 (2010) 1619--.
\newblock \href {http://arxiv.org/abs/0912.3592} {\path{arXiv:0912.3592}},
  \href {http://dx.doi.org/10.1126/science.1186112}
  {\path{doi:10.1126/science.1186112}}.

\bibitem{2011PhRvL.106m1301A}
C.~E. {Aalseth}, et~al., {Results from a Search for Light-Mass Dark Matter with
  a p-Type Point Contact Germanium Detector}, Physical Review Letters 106~(13)
  (2011) 131301--+.
\newblock \href {http://arxiv.org/abs/1002.4703} {\path{arXiv:1002.4703}},
  \href {http://dx.doi.org/10.1103/PhysRevLett.106.131301}
  {\path{doi:10.1103/PhysRevLett.106.131301}}.

\bibitem{2010PhRvD..82h1302L}
J.~{Lavalle}, {10 GeV dark matter candidates and cosmic-ray antiprotons}, \prd
  82~(8) (2010) 081302--+.
\newblock \href {http://arxiv.org/abs/1007.5253} {\path{arXiv:1007.5253}},
  \href {http://dx.doi.org/10.1103/PhysRevD.82.081302}
  {\path{doi:10.1103/PhysRevD.82.081302}}.

\bibitem{2012NuPhB.854..738C}
D.~G. {Cerde{\~n}o}, T.~{Delahaye}, J.~{Lavalle}, {Cosmic-ray antiproton
  constraints on light singlino-like dark matter candidates}, Nuclear Physics B
  854 (2012) 738--779.
\newblock \href {http://arxiv.org/abs/1108.1128} {\path{arXiv:1108.1128}},
  \href {http://dx.doi.org/10.1016/j.nuclphysb.2011.09.020}
  {\path{doi:10.1016/j.nuclphysb.2011.09.020}}.

\bibitem{2001JHEP...06..053U}
P.~{Ullio}, {Indirect detection of neutralino dark matter candidates in
  anomaly-mediated supersymmetry breaking scenarios}, Journal of High Energy
  Physics 6 (2001) 53.
\newblock \href {http://arxiv.org/abs/hep-ph/0105052}
  {\path{arXiv:hep-ph/0105052}}, \href
  {http://dx.doi.org/10.1088/1126-6708/2001/06/053}
  {\path{doi:10.1088/1126-6708/2001/06/053}}.

\bibitem{2004PhRvD..70i5004P}
S.~{Profumo}, C.~E. {Yaguna}, {Statistical analysis of supersymmetric dark
  matter in the minimal supersymmetric standard model after WMAP}, \prd 70~(9)
  (2004) 095004.
\newblock \href {http://arxiv.org/abs/hep-ph/0407036}
  {\path{arXiv:hep-ph/0407036}}, \href
  {http://dx.doi.org/10.1103/PhysRevD.70.095004}
  {\path{doi:10.1103/PhysRevD.70.095004}}.

\bibitem{2004PhRvL..92c1303H}
J.~{Hisano}, S.~{Matsumoto}, M.~M. {Nojiri}, {Explosive Dark Matter
  Annihilation}, Physical Review Letters 92~(3) (2004) 031303.
\newblock \href {http://arxiv.org/abs/hep-ph/0307216}
  {\path{arXiv:hep-ph/0307216}}, \href
  {http://dx.doi.org/10.1103/PhysRevLett.92.031303}
  {\path{doi:10.1103/PhysRevLett.92.031303}}.

\bibitem{2005PhRvD..71f3528H}
J.~{Hisano}, et~al., {Nonperturbative
  effect on dark matter annihilation and gamma ray signature from the galactic
  center}, \prd 71~(6) (2005) 063528--+.
\newblock \href {http://arxiv.org/abs/hep-ph/0412403}
  {\path{arXiv:hep-ph/0412403}}, \href
  {http://dx.doi.org/10.1103/PhysRevD.71.063528}
  {\path{doi:10.1103/PhysRevD.71.063528}}.

\bibitem{2009Natur.458..607A}
O.~{Adriani}, et~al., {An anomalous positron abundance in cosmic rays with
  energies 1.5-100GeV}, \nat 458 (2009) 607--609.
\newblock \href {http://arxiv.org/abs/0810.4995} {\path{arXiv:0810.4995}},
  \href {http://dx.doi.org/10.1038/nature07942}
  {\path{doi:10.1038/nature07942}}.

\bibitem{1998ApJ...493..694M}
I.~V. {Moskalenko}, A.~W. {Strong}, {Production and Propagation of Cosmic-Ray
  Positrons and Electrons}, \apj 493 (1998) 694--+.
\newblock \href {http://arxiv.org/abs/astro-ph/9710124}
  {\path{arXiv:astro-ph/9710124}}, \href {http://dx.doi.org/10.1086/305152}
  {\path{doi:10.1086/305152}}.

\bibitem{1999PhRvD..59b3511B}
E.~A. {Baltz}, J.~{Edsj{\"o}}, {Positron propagation and fluxes from neutralino
  annihilation in the halo}, \prd 59~(2) (1999) 023511.
\newblock \href {http://arxiv.org/abs/astro-ph/9808243}
  {\path{arXiv:astro-ph/9808243}}, \href
  {http://dx.doi.org/10.1103/PhysRevD.59.023511}
  {\path{doi:10.1103/PhysRevD.59.023511}}.

\bibitem{2008PhRvD..77f3527D}
T.~{Delahaye}, et~al., {Positrons from dark matter annihilation in the galactic
  halo: Theoretical uncertainties}, \prd 77~(6) (2008) 063527--+.
\newblock \href {http://arxiv.org/abs/0712.2312} {\path{arXiv:0712.2312}},
  \href {http://dx.doi.org/10.1103/PhysRevD.77.063527}
  {\path{doi:10.1103/PhysRevD.77.063527}}.

\bibitem{1977PhRvD..15..820B}
G.~D. {Badhwar}, R.~L. {Golden}, S.~A. {Stephens}, {Analytic representation of
  the proton-proton and proton-nucleus cross-sections and its application to
  the sea-level spectrum and charge ratio of muons}, \prd 15 (1977) 820--831.
\newblock \href {http://dx.doi.org/10.1103/PhysRevD.15.820}
  {\path{doi:10.1103/PhysRevD.15.820}}.

\bibitem{2006ApJ...647..692K}
T.~{Kamae}, et~al., {Parameterization
  of {$\gamma$}, $e^{\pm}$, and Neutrino Spectra Produced by p-p Interaction in
  Astronomical Environments}, \apj 647 (2006) 692--708.
\newblock \href {http://arxiv.org/abs/astro-ph/0605581}
  {\path{arXiv:astro-ph/0605581}}, \href {http://dx.doi.org/10.1086/505189}
  {\path{doi:10.1086/505189}}.

\bibitem{2011Sci...332...69A}
O.~{Adriani}, et~al., {PAMELA Measurements of Cosmic-Ray Proton and Helium
  Spectra}, Science 332 (2011) 69--.
\newblock \href {http://arxiv.org/abs/1103.4055} {\path{arXiv:1103.4055}},
  \href {http://dx.doi.org/10.1126/science.1199172}
  {\path{doi:10.1126/science.1199172}}.

\bibitem{2011arXiv1109.0834D}
T.~{Delahaye}, et~al., {Systematic effects in the estimate of the local
  gamma-ray emissivity}, ArXiv e-prints\href {http://arxiv.org/abs/1109.0834}
  {\path{arXiv:1109.0834}}.

\bibitem{2002PhR...366..331A}
{AMS Collaboration}, {The Alpha Magnetic Spectrometer (AMS) on the
  International Space Station: Part I - results from the test flight on the
  space shuttle}, \physrep 366 (2002) 331--405.
\newblock \href {http://dx.doi.org/10.1016/S0370-1573(02)00013-3}
  {\path{doi:10.1016/S0370-1573(02)00013-3}}.

\bibitem{2009PhRvL.102r1101A}
A.~A. {Abdo}, et~al., {Measurement of the Cosmic Ray $e^{+}+e^{-}$ Spectrum
  from 20GeV to 1TeV with the Fermi Large Area Telescope}, Physical Review
  Letters 102~(18) (2009) 181101--+.
\newblock \href {http://arxiv.org/abs/0905.0025} {\path{arXiv:0905.0025}},
  \href {http://dx.doi.org/10.1103/PhysRevLett.102.181101}
  {\path{doi:10.1103/PhysRevLett.102.181101}}.

\bibitem{2011MNRAS.414..985L}
J.~{Lavalle}, {Impact of the spectral hardening of TeV cosmic rays on the
  prediction of the secondary positron flux}, \mnras 414 (2011) 985--991.
\newblock \href {http://arxiv.org/abs/1011.3063} {\path{arXiv:1011.3063}},
  \href {http://dx.doi.org/10.1111/j.1365-2966.2011.18294.x}
  {\path{doi:10.1111/j.1365-2966.2011.18294.x}}.

\bibitem{2009A&A...501..821D}
T.~{Delahaye}, et~al., {Galactic secondary positron flux at the Earth}, \aap
  501 (2009) 821--833.
\newblock \href {http://arxiv.org/abs/0809.5268} {\path{arXiv:0809.5268}},
  \href {http://dx.doi.org/10.1051/0004-6361/200811130}
  {\path{doi:10.1051/0004-6361/200811130}}.

\bibitem{2008Natur.456..362C}
J.~{Chang}, et~al., {An excess of cosmic ray electrons at energies of
  300-800GeV}, \nat 456 (2008) 362--365.
\newblock \href {http://dx.doi.org/10.1038/nature07477}
  {\path{doi:10.1038/nature07477}}.

\bibitem{2008PhRvL.101z1104A}
F.~{Aharonian}, et~al., {Energy Spectrum of Cosmic-Ray Electrons at TeV
  Energies}, Physical Review Letters 101~(26) (2008) 261104--+.
\newblock \href {http://arxiv.org/abs/0811.3894} {\path{arXiv:0811.3894}},
  \href {http://dx.doi.org/10.1103/PhysRevLett.101.261104}
  {\path{doi:10.1103/PhysRevLett.101.261104}}.

\bibitem{2009A&A...508..561A}
F.~{Aharonian}, et~al., {Probing the ATIC peak in the cosmic-ray electron
  spectrum with H.E.S.S.}, \aap 508 (2009) 561--564.
\newblock \href {http://arxiv.org/abs/0905.0105} {\path{arXiv:0905.0105}},
  \href {http://dx.doi.org/10.1051/0004-6361/200913323}
  {\path{doi:10.1051/0004-6361/200913323}}.

\bibitem{2009APh....32..140G}
D.~{Grasso}, et~al., {On possible interpretations of the high energy
  electron-positron spectrum measured by the Fermi Large Area Telescope},
  Astroparticle Physics 32 (2009) 140--151.
\newblock \href {http://arxiv.org/abs/0905.0636} {\path{arXiv:0905.0636}},
  \href {http://dx.doi.org/10.1016/j.astropartphys.2009.07.003}
  {\path{doi:10.1016/j.astropartphys.2009.07.003}}.

\bibitem{2010A&A...524A..51D}
T.~{Delahaye}, et~al., {Galactic electrons and positrons at the Earth: new
  estimate of the primary and secondary fluxes}, \aap 524 (2010) A51+.
\newblock \href {http://arxiv.org/abs/1002.1910} {\path{arXiv:1002.1910}},
  \href {http://dx.doi.org/10.1051/0004-6361/201014225}
  {\path{doi:10.1051/0004-6361/201014225}}.

\bibitem{2004ApJ...613..962S}
A.~W. {Strong}, I.~V. {Moskalenko}, O.~{Reimer}, {Diffuse Galactic Continuum
  Gamma Rays: A Model Compatible with EGRET Data and Cosmic-Ray Measurements},
  \apj 613 (2004) 962--976.
\newblock \href {http://arxiv.org/abs/astro-ph/0406254}
  {\path{arXiv:astro-ph/0406254}}, \href {http://dx.doi.org/10.1086/423193}
  {\path{doi:10.1086/423193}}.

\bibitem{2009PhRvL.102g1301D}
F.~{Donato}, et~al., {Constraints
  on WIMP Dark Matter from the High Energy PAMELA {\= p}/p Data}, Physical
  Review Letters 102~(7) (2009) 071301--+.
\newblock \href {http://arxiv.org/abs/0810.5292} {\path{arXiv:0810.5292}},
  \href {http://dx.doi.org/10.1103/PhysRevLett.102.071301}
  {\path{doi:10.1103/PhysRevLett.102.071301}}.

\bibitem{2011ApJS..192...18K}
E.~{Komatsu}, et~al., {Seven-year Wilkinson
  Microwave Anisotropy Probe (WMAP) Observations: Cosmological Interpretation},
  \apjs 192 (2011) 18.
\newblock \href {http://arxiv.org/abs/1001.4538} {\path{arXiv:1001.4538}},
  \href {http://dx.doi.org/10.1088/0067-0049/192/2/18}
  {\path{doi:10.1088/0067-0049/192/2/18}}.

\bibitem{2003PhLB..571..121S}
P.~{Salati}, {Quintessence and the relic density of neutralinos}, Physics
  Letters B 571 (2003) 121--131.
\newblock \href {http://arxiv.org/abs/astro-ph/0207396}
  {\path{arXiv:astro-ph/0207396}}, \href
  {http://dx.doi.org/10.1016/j.physletb.2003.07.073}
  {\path{doi:10.1016/j.physletb.2003.07.073}}.

\bibitem{2003JCAP...11..006P}
S.~{Profumo}, P.~{Ullio}, {SUSY dark matter and quintessence}, \jcap 11 (2003)
  6.
\newblock \href {http://arxiv.org/abs/hep-ph/0309220}
  {\path{arXiv:hep-ph/0309220}}, \href
  {http://dx.doi.org/10.1088/1475-7516/2003/11/006}
  {\path{doi:10.1088/1475-7516/2003/11/006}}.

\bibitem{2009PhLB..681..151K}
G.~{Kane}, R.~{Lu}, S.~{Watson}, {PAMELA satellite data as a signal of
  non-thermal wino LSP dark matter}, Physics Letters B 681 (2009) 151--160.
\newblock \href {http://arxiv.org/abs/0906.4765} {\path{arXiv:0906.4765}},
  \href {http://dx.doi.org/10.1016/j.physletb.2009.09.053}
  {\path{doi:10.1016/j.physletb.2009.09.053}}.

\bibitem{2008NuPhB.800..204C}
M.~{Cirelli}, R.~{Franceschini}, A.~{Strumia}, {Minimal Dark Matter predictions
  for galactic positrons, anti-protons, photons}, Nuclear Physics B 800 (2008)
  204--220.
\newblock \href {http://arxiv.org/abs/0802.3378} {\path{arXiv:0802.3378}},
  \href {http://dx.doi.org/10.1016/j.nuclphysb.2008.03.013}
  {\path{doi:10.1016/j.nuclphysb.2008.03.013}}.

\bibitem{2009PhRvD..79a5014A}
N.~{Arkani-Hamed}, et~al., {A theory
  of dark matter}, \prd 79~(1) (2009) 015014--+.
\newblock \href {http://arxiv.org/abs/0810.0713} {\path{arXiv:0810.0713}},
  \href {http://dx.doi.org/10.1103/PhysRevD.79.015014}
  {\path{doi:10.1103/PhysRevD.79.015014}}.

\bibitem{2009PhLB..671..391P}
M.~{Pospelov}, A.~{Ritz}, {Astrophysical signatures of secluded dark matter},
  Physics Letters B 671 (2009) 391--397.
\newblock \href {http://arxiv.org/abs/0810.1502} {\path{arXiv:0810.1502}},
  \href {http://dx.doi.org/10.1016/j.physletb.2008.12.012}
  {\path{doi:10.1016/j.physletb.2008.12.012}}.

\bibitem{2008PhLB..662...53P}
M.~{Pospelov}, A.~{Ritz}, M.~{Voloshin}, {Secluded WIMP dark matter}, Physics
  Letters B 662 (2008) 53--61.
\newblock \href {http://arxiv.org/abs/0711.4866} {\path{arXiv:0711.4866}},
  \href {http://dx.doi.org/10.1016/j.physletb.2008.02.052}
  {\path{doi:10.1016/j.physletb.2008.02.052}}.

\bibitem{2009PhRvD..79h3523L}
M.~{Lattanzi}, J.~{Silk}, {Can the WIMP annihilation boost factor be boosted by
  the Sommerfeld enhancement?}, \prd 79~(8) (2009) 083523--+.
\newblock \href {http://arxiv.org/abs/0812.0360} {\path{arXiv:0812.0360}},
  \href {http://dx.doi.org/10.1103/PhysRevD.79.083523}
  {\path{doi:10.1103/PhysRevD.79.083523}}.

\bibitem{2009NuPhB.813....1C}
M.~{Cirelli}, et~al., {Model-independent
  implications of the $e^{+}$, $\bar{p}$ cosmic ray spectra on properties of
  Dark Matter}, Nuclear Physics B 813 (2009) 1--21.
\newblock \href {http://arxiv.org/abs/0809.2409} {\path{arXiv:0809.2409}},
  \href {http://dx.doi.org/10.1016/j.nuclphysb.2008.11.031}
  {\path{doi:10.1016/j.nuclphysb.2008.11.031}}.

\bibitem{2009PhRvL.102e1101A}
O.~{Adriani}, et~al., {New Measurement of the Antiproton-to-Proton Flux Ratio
  up to 100 GeV in the Cosmic Radiation}, Physical Review Letters 102~(5)
  (2009) 051101.
\newblock \href {http://arxiv.org/abs/0810.4994} {\path{arXiv:0810.4994}},
  \href {http://dx.doi.org/10.1103/PhysRevLett.102.051101}
  {\path{doi:10.1103/PhysRevLett.102.051101}}.

\bibitem{2009ApJ...699L..59B}
E.~{Borriello}, A.~{Cuoco}, G.~{Miele}, {Secondary Radiation from the
  Pamela/ATIC Excess and Relevance for Fermi}, \apjl 699 (2009) L59--L63.
\newblock \href {http://arxiv.org/abs/0903.1852} {\path{arXiv:0903.1852}},
  \href {http://dx.doi.org/10.1088/0004-637X/699/2/L59}
  {\path{doi:10.1088/0004-637X/699/2/L59}}.

\bibitem{2009JCAP...03..009B}
G.~{Bertone}, et~al., {Gamma-ray and radio
  tests of the $e^{\pm}$ excess from DM annihilations}, Journal of Cosmology
  and Astro-Particle Physics 3 (2009) 9--+.
\newblock \href {http://arxiv.org/abs/0811.3744} {\path{arXiv:0811.3744}},
  \href {http://dx.doi.org/10.1088/1475-7516/2009/03/009}
  {\path{doi:10.1088/1475-7516/2009/03/009}}.

\bibitem{2009PhRvD..79h1303B}
L.~{Bergstr{\"o}m}, et~al.,
  {Gamma-ray and radio constraints of high positron rate dark matter models
  annihilating into new light particles}, \prd 79~(8) (2009) 081303--+(R).
\newblock \href {http://arxiv.org/abs/0812.3895} {\path{arXiv:0812.3895}},
  \href {http://dx.doi.org/10.1103/PhysRevD.79.081303}
  {\path{doi:10.1103/PhysRevD.79.081303}}.

\bibitem{2010NuPhB.831..178M}
P.~{Meade}, et~al., {Dark Matter
  interpretations of the e$^{±}$ excesses after FERMI}, Nuclear Physics B 831
  (2010) 178--203.
\newblock \href {http://arxiv.org/abs/0905.0480} {\path{arXiv:0905.0480}},
  \href {http://dx.doi.org/10.1016/j.nuclphysb.2010.01.012}
  {\path{doi:10.1016/j.nuclphysb.2010.01.012}}.

\bibitem{2010NuPhB.840..284C}
M.~{Cirelli}, P.~{Panci}, P.~D. {Serpico}, {Diffuse gamma ray constraints on
  annihilating or decaying Dark Matter after Fermi}, Nuclear Physics B 840
  (2010) 284--303.
\newblock \href {http://arxiv.org/abs/0912.0663} {\path{arXiv:0912.0663}},
  \href {http://dx.doi.org/10.1016/j.nuclphysb.2010.07.010}
  {\path{doi:10.1016/j.nuclphysb.2010.07.010}}.

\bibitem{2011PhRvD..84b7302G}
S.~{Galli}, et~al., {Updated CMB constraints
  on dark matter annihilation cross sections}, \prd 84~(2) (2011) 027302.
\newblock \href {http://arxiv.org/abs/1106.1528} {\path{arXiv:1106.1528}},
  \href {http://dx.doi.org/10.1103/PhysRevD.84.027302}
  {\path{doi:10.1103/PhysRevD.84.027302}}.

\bibitem{2009PhRvD..80d3526S}
T.~R. {Slatyer}, N.~{Padmanabhan}, D.~P. {Finkbeiner}, {CMB constraints on WIMP
  annihilation: Energy absorption during the recombination epoch}, \prd 80~(4)
  (2009) 043526.
\newblock \href {http://arxiv.org/abs/0906.1197} {\path{arXiv:0906.1197}},
  \href {http://dx.doi.org/10.1103/PhysRevD.80.043526}
  {\path{doi:10.1103/PhysRevD.80.043526}}.

\bibitem{2007A&A...462..827L}
J.~{Lavalle}, et~al., {Clumpiness of dark matter and the positron annihilation
  signal}, \aap 462 (2007) 827--840.
\newblock \href {http://arxiv.org/abs/astro-ph/0603796}
  {\path{arXiv:astro-ph/0603796}}, \href
  {http://dx.doi.org/10.1051/0004-6361:20065312}
  {\path{doi:10.1051/0004-6361:20065312}}.

\bibitem{2008A&A...479..427L}
J.~{Lavalle}, et~al., {Full calculation of clumpiness boost factors for
  antimatter cosmic rays in the light of {$\Lambda$}CDM N-body simulation
  results. Abandoning hope in clumpiness enhancement?}, \aap 479 (2008)
  427--452.
\newblock \href {http://arxiv.org/abs/0709.3634} {\path{arXiv:0709.3634}},
  \href {http://dx.doi.org/10.1051/0004-6361:20078723}
  {\path{doi:10.1051/0004-6361:20078723}}.

\bibitem{2007MNRAS.374..455C}
D.~{Cumberbatch}, J.~{Silk}, {Local dark matter clumps and the positron
  excess}, \mnras 374 (2007) 455--465.
\newblock \href {http://arxiv.org/abs/astro-ph/0602320}
  {\path{arXiv:astro-ph/0602320}}, \href
  {http://dx.doi.org/10.1111/j.1365-2966.2006.11123.x}
  {\path{doi:10.1111/j.1365-2966.2006.11123.x}}.

\bibitem{2009PhRvD..79j3513H}
D.~{Hooper}, A.~{Stebbins}, K.~M. {Zurek}, {Excesses in cosmic ray positron and
  electron spectra from a nearby clump of neutralino dark matter}, \prd 79~(10)
  (2009) 103513.
\newblock \href {http://arxiv.org/abs/0812.3202} {\path{arXiv:0812.3202}},
  \href {http://dx.doi.org/10.1103/PhysRevD.79.103513}
  {\path{doi:10.1103/PhysRevD.79.103513}}.

\bibitem{2009PhRvD..80c5023B}
P.~{Brun}, et~al., {Cosmic ray
  lepton puzzle in the light of cosmological N-body simulations}, \prd 80~(3)
  (2009) 035023.
\newblock \href {http://arxiv.org/abs/0904.0812} {\path{arXiv:0904.0812}},
  \href {http://dx.doi.org/10.1103/PhysRevD.80.035023}
  {\path{doi:10.1103/PhysRevD.80.035023}}.

\bibitem{2008Natur.454..735D}
J.~{Diemand}, et~al., {Clumps and streams in the local dark matter distribution}, \nat
  454 (2008) 735--738.
\newblock \href {http://arxiv.org/abs/0805.1244} {\path{arXiv:0805.1244}},
  \href {http://dx.doi.org/10.1038/nature07153}
  {\path{doi:10.1038/nature07153}}.

\bibitem{2009JCAP...02..021I}
A.~{Ibarra}, D.~{Tran}, {Decaying dark matter and the PAMELA anomaly}, \jcap 2
  (2009) 21.
\newblock \href {http://arxiv.org/abs/0811.1555} {\path{arXiv:0811.1555}},
  \href {http://dx.doi.org/10.1088/1475-7516/2009/02/021}
  {\path{doi:10.1088/1475-7516/2009/02/021}}.

\bibitem{2009JCAP...08..017I}
A.~{Ibarra}, et~al., {Cosmic rays from
  leptophilic dark matter decay via kinetic mixing}, \jcap 8 (2009) 17.
\newblock \href {http://arxiv.org/abs/0903.3625} {\path{arXiv:0903.3625}},
  \href {http://dx.doi.org/10.1088/1475-7516/2009/08/017}
  {\path{doi:10.1088/1475-7516/2009/08/017}}.

\bibitem{2011JCAP...01..032G}
M.~{Garny}, et~al., {Gamma-ray lines from
  radiative dark matter decay}, \jcap 1 (2011) 32.
\newblock \href {http://arxiv.org/abs/1011.3786} {\path{arXiv:1011.3786}},
  \href {http://dx.doi.org/10.1088/1475-7516/2011/01/032}
  {\path{doi:10.1088/1475-7516/2011/01/032}}.

\bibitem{2009PThPh.122..553C}
C.~{Chen}, et~al., {Decaying Hidden
  Gauge Boson and the PAMELA and ATIC/PPB-BETS Anomalies}, Progress of
  Theoretical Physics 122 (2009) 553--559.
\newblock \href {http://arxiv.org/abs/0811.3357} {\path{arXiv:0811.3357}}.

\bibitem{2012CEJPh..10....1P}
S.~{Profumo}, {Dissecting cosmic-ray electron-positron data with Occam's razor:
  the role of known pulsars}, Central European Journal of Physics 10 (2012)
  1--31.
\newblock \href {http://arxiv.org/abs/0812.4457} {\path{arXiv:0812.4457}},
  \href {http://dx.doi.org/10.2478/s11534-011-0099-z}
  {\path{doi:10.2478/s11534-011-0099-z}}.

\bibitem{1997PhLB..409..313C}
P.~{Chardonnet}, J.~{Orloff}, P.~{Salati}, {Production of anti-matter in our
  galaxy}, Physics Letters B 409 (1997) 313--320.
\newblock \href {http://arxiv.org/abs/astro-ph/9705110}
  {\path{arXiv:astro-ph/9705110}}.

\bibitem{2000PhRvD..62d3003D}
F.~{Donato}, N.~{Fornengo}, P.~{Salati}, {Antideuterons as a signature of
  supersymmetric dark matter}, \prd 62~(4) (2000) 043003.
\newblock \href {http://arxiv.org/abs/hep-ph/9904481}
  {\path{arXiv:hep-ph/9904481}}, \href
  {http://dx.doi.org/10.1103/PhysRevD.62.043003}
  {\path{doi:10.1103/PhysRevD.62.043003}}.

\bibitem{Braun82}
M.~{Braun}, V.~{Vechernin}, {Fragmentation deuterons from nucleon pairing},
  Sov. J. Nucl. Phys. 36:3 (1982) 357--362.

\bibitem{2008ICRC....4..765C}
V.~{Choutko}, F.~{Giovacchini}, {Cosmic Rays Antideuteron Sensitivity for
  AMS-02 Experiment}, in: International Cosmic Ray Conference, Vol.~4 of
  International Cosmic Ray Conference, 2008, pp. 765--768.

\bibitem{2004NIMPB.214..122H}
C.~J. {Hailey}, et~al., {Development of the gaseous
  antiparticle spectrometer for space-based antimatter detection}, Nuclear
  Instruments and Methods in Physics Research B 214 (2004) 122--125.
\newblock \href {http://arxiv.org/abs/astro-ph/0306589}
  {\path{arXiv:astro-ph/0306589}}, \href
  {http://dx.doi.org/10.1016/j.nimb.2003.08.004}
  {\path{doi:10.1016/j.nimb.2003.08.004}}.

\bibitem{2006JCAP...01..007H}
C.~J. {Hailey}, et~al., {Accelerator
  testing of the general antiparticle spectrometer; a novel approach to
  indirect dark matter detection}, \jcap 1 (2006) 7.
\newblock \href {http://arxiv.org/abs/astro-ph/0509587}
  {\path{arXiv:astro-ph/0509587}}, \href
  {http://dx.doi.org/10.1088/1475-7516/2006/01/007}
  {\path{doi:10.1088/1475-7516/2006/01/007}}.

\bibitem{2008JPhCS.120d2011K}
J.~E. {Koglin}, et~al., {Antideuterons as an indirect dark matter signature:
  design and preparation for a balloon-born GAPS experiment}, Journal of
  Physics Conference Series 120~(4) (2008) 042011.
\newblock \href {http://dx.doi.org/10.1088/1742-6596/120/4/042011}
  {\path{doi:10.1088/1742-6596/120/4/042011}}.

\bibitem{2010AdSpR..46.1349A}
T.~{Aramaki}, et~al.,
  {Antideuterons as an indirect dark matter signature: Si(Li) detector
  development and a GAPS balloon mission}, Advances in Space Research 46 (2010)
  1349--1353.
\newblock \href {http://dx.doi.org/10.1016/j.asr.2010.06.036}
  {\path{doi:10.1016/j.asr.2010.06.036}}.

\bibitem{2010PhLB..683..248K}
M.~{Kadastik}, M.~{Raidal}, A.~{Strumia}, {Enhanced anti-deuteron Dark Matter
  signal and the implications of PAMELA}, Physics Letters B 683 (2010)
  248--254.
\newblock \href {http://arxiv.org/abs/0908.1578} {\path{arXiv:0908.1578}},
  \href {http://dx.doi.org/10.1016/j.physletb.2009.12.005}
  {\path{doi:10.1016/j.physletb.2009.12.005}}.

\bibitem{2010JHEP...11..017C}
Y.~{Cui}, J.~D. {Mason}, L.~{Randall}, {General analysis of antideuteron
  searches for dark matter}, Journal of High Energy Physics 11 (2010) 17.
\newblock \href {http://arxiv.org/abs/1006.0983} {\path{arXiv:1006.0983}},
  \href {http://dx.doi.org/10.1007/JHEP11(2010)017}
  {\path{doi:10.1007/JHEP11(2010)017}}.

\bibitem{2004PhRvD..70j3529F}
N.~{Fornengo}, L.~{Pieri}, S.~{Scopel}, {Neutralino annihilation into
  {$\gamma$} rays in the Milky Way and in external galaxies}, \prd 70~(10)
  (2004) 103529--+.
\newblock \href {http://arxiv.org/abs/hep-ph/0407342}
  {\path{arXiv:hep-ph/0407342}}, \href
  {http://dx.doi.org/10.1103/PhysRevD.70.103529}
  {\path{doi:10.1103/PhysRevD.70.103529}}.

\bibitem{1988PhRvD..37.3737B}
L.~{Bergstr{\"o}m}, H.~{Snellman}, {Observable monochromatic photons from
  cosmic photino annihilation}, \prd 37 (1988) 3737--3741.
\newblock \href {http://dx.doi.org/10.1103/PhysRevD.37.3737}
  {\path{doi:10.1103/PhysRevD.37.3737}}.

\bibitem{2003PhRvD..67g5014H}
J.~{Hisano}, S.~{Matsumoto}, M.~M. {Nojiri}, {Unitarity and higher-order
  corrections in neutralino dark matter annihilation into two photons}, \prd
  67~(7) (2003) 075014.
\newblock \href {http://arxiv.org/abs/hep-ph/0212022}
  {\path{arXiv:hep-ph/0212022}}, \href
  {http://dx.doi.org/10.1103/PhysRevD.67.075014}
  {\path{doi:10.1103/PhysRevD.67.075014}}.

\bibitem{1989PhLB..225..372B}
L.~{Bergstr{\"o}m}, {Radiative processes in dark matter photino annihilation},
  Physics Letters B 225 (1989) 372--380.
\newblock \href {http://dx.doi.org/10.1016/0370-2693(89)90585-6}
  {\path{doi:10.1016/0370-2693(89)90585-6}}.

\bibitem{2005PhRvL..94q1301B}
J.~F. {Beacom}, N.~F. {Bell}, G.~{Bertone}, {Gamma-Ray Constraint on Galactic
  Positron Production by MeV Dark Matter}, Physical Review Letters 94~(17)
  (2005) 171301--+.
\newblock \href {http://arxiv.org/abs/astro-ph/0409403}
  {\path{arXiv:astro-ph/0409403}}, \href
  {http://dx.doi.org/10.1103/PhysRevLett.94.171301}
  {\path{doi:10.1103/PhysRevLett.94.171301}}.

\bibitem{2005PhRvL..94m1301B}
L.~{Bergstr{\"o}m}, et~al., {Gamma Rays
  from Kaluza-Klein Dark Matter}, Physical Review Letters 94~(13) (2005)
  131301.
\newblock \href {http://arxiv.org/abs/astro-ph/0410359}
  {\path{arXiv:astro-ph/0410359}}, \href
  {http://dx.doi.org/10.1103/PhysRevLett.94.131301}
  {\path{doi:10.1103/PhysRevLett.94.131301}}.

\bibitem{2005hep.ph....7194B}
A.~{Birkedal}, et~al., {Robust Gamma Ray
  Signature of WIMP Dark Matter}, ArXiv High Energy Physics - Phenomenology
  e-prints\href {http://arxiv.org/abs/hep-ph/0507194}
  {\path{arXiv:hep-ph/0507194}}.

\bibitem{2005PhRvL..95x1301B}
L.~{Bergstr{\"o}m}, et~al., {Gamma Rays
  from Heavy Neutralino Dark Matter}, Physical Review Letters 95~(24) (2005)
  241301--+.
\newblock \href {http://arxiv.org/abs/hep-ph/0507229}
  {\path{arXiv:hep-ph/0507229}}, \href
  {http://dx.doi.org/10.1103/PhysRevLett.95.241301}
  {\path{doi:10.1103/PhysRevLett.95.241301}}.

\bibitem{1970RvMP...42..237B}
G.~R. {Blumenthal}, R.~J. {Gould}, {Bremsstrahlung, Synchrotron Radiation, and
  Compton Scattering of High-Energy Electrons Traversing Dilute Gases}, Reviews
  of Modern Physics 42 (1970) 237--271.

\bibitem{2009PhLB..671...10R}
L.~{Roszkowski}, et~al., {On prospects for
  dark matter indirect detection in the Constrained MSSM}, Physics Letters B
  671 (2009) 10--14.
\newblock \href {http://arxiv.org/abs/0707.0622} {\path{arXiv:0707.0622}},
  \href {http://dx.doi.org/10.1016/j.physletb.2008.11.061}
  {\path{doi:10.1016/j.physletb.2008.11.061}}.

\bibitem{1972PhRvL..28..985R}
M.~J. {Ryan}, J.~F. {Ormes}, V.~K. {Balasubrahmanyan}, {Cosmic-Ray Proton and
  Helium Spectra above 50 GeV}, Physical Review Letters 28 (1972) 985--988.
\newblock \href {http://dx.doi.org/10.1103/PhysRevLett.28.985}
  {\path{doi:10.1103/PhysRevLett.28.985}}.

\bibitem{2011A&A...531A..37D}
T.~{Delahaye}, et~al., {The GeV-TeV Galactic gamma-ray diffuse emission. I.
  Uncertainties in the predictions of the hadronic component}, \aap 531 (2011)
  A37.
\newblock \href {http://arxiv.org/abs/1102.0744} {\path{arXiv:1102.0744}},
  \href {http://dx.doi.org/10.1051/0004-6361/201116647}
  {\path{doi:10.1051/0004-6361/201116647}}.

\bibitem{1998ApJ...494..523S}
P.~{Sreekumar}, et~al., {EGRET Observations of the Extragalactic Gamma-Ray
  Emission}, \apj 494 (1998) 523.
\newblock \href {http://arxiv.org/abs/astro-ph/9709257}
  {\path{arXiv:astro-ph/9709257}}, \href {http://dx.doi.org/10.1086/305222}
  {\path{doi:10.1086/305222}}.

\bibitem{2004A&A...425L..13A}
F.~{Aharonian}, et~al., {Very high energy gamma rays from the direction of
  Sagittarius A$^{*}$}, \aap 425 (2004) L13--L17.
\newblock \href {http://arxiv.org/abs/astro-ph/0406658}
  {\path{arXiv:astro-ph/0406658}}, \href
  {http://dx.doi.org/10.1051/0004-6361:200400055}
  {\path{doi:10.1051/0004-6361:200400055}}.

\bibitem{2006Natur.439..695A}
F.~{Aharonian}, et~al., {Discovery of very-high-energy {$\gamma$}-rays from the
  Galactic Centre ridge}, \nat 439 (2006) 695--698.
\newblock \href {http://arxiv.org/abs/astro-ph/0603021}
  {\path{arXiv:astro-ph/0603021}}, \href
  {http://dx.doi.org/10.1038/nature04467} {\path{doi:10.1038/nature04467}}.

\bibitem{2011PhRvD..83d5024B}
L.~{Bergstr{\"o}m}, T.~{Bringmann}, J.~{Edsj{\"o}}, {Complementarity of direct
  dark matter detection and indirect detection through gamma rays}, \prd 83~(4)
  (2011) 045024.
\newblock \href {http://arxiv.org/abs/1011.4514} {\path{arXiv:1011.4514}},
  \href {http://dx.doi.org/10.1103/PhysRevD.83.045024}
  {\path{doi:10.1103/PhysRevD.83.045024}}.

\bibitem{1998A&A...335..161M}
H.~A. {Mayer-Hasselwander}, et~al., {High-energy gamma-ray emission from the
  Galactic Center}, \aap 335 (1998) 161--172.

\bibitem{2004APh....21..267C}
A.~{Cesarini}, et~al., {The
  Galactic center as a dark matter gamma-ray source}, Astroparticle Physics 21
  (2004) 267--285.
\newblock \href {http://arxiv.org/abs/astro-ph/0305075}
  {\path{arXiv:astro-ph/0305075}}, \href
  {http://dx.doi.org/10.1016/S0927-6505(04)00033-7}
  {\path{doi:10.1016/S0927-6505(04)00033-7}}.

\bibitem{1986ApJ...301...27B}
G.~R. {Blumenthal}, et~al., {Contraction
  of dark matter galactic halos due to baryonic infall}, \apj 301 (1986)
  27--34.
\newblock \href {http://dx.doi.org/10.1086/163867} {\path{doi:10.1086/163867}}.

\bibitem{1999PhRvL..83.1719G}
P.~{Gondolo}, J.~{Silk}, {Dark Matter Annihilation at the Galactic Center},
  Physical Review Letters 83 (1999) 1719--1722.
\newblock \href {http://arxiv.org/abs/astro-ph/9906391}
  {\path{arXiv:astro-ph/9906391}}, \href
  {http://dx.doi.org/10.1103/PhysRevLett.83.1719}
  {\path{doi:10.1103/PhysRevLett.83.1719}}.

\bibitem{2004ApJ...616...16G}
O.~Y. {Gnedin}, A.~V. {Kravtsov}, A.~A. {Klypin}, D.~{Nagai}, {Response of Dark
  Matter Halos to Condensation of Baryons: Cosmological Simulations and
  Improved Adiabatic Contraction Model}, \apj 616 (2004) 16--26.
\newblock \href {http://arxiv.org/abs/astro-ph/0406247}
  {\path{arXiv:astro-ph/0406247}}, \href {http://dx.doi.org/10.1086/424914}
  {\path{doi:10.1086/424914}}.

\bibitem{2010Natur.463..203G}
F.~{Governato}, C.~{Brook}, L.~{Mayer}, A.~{Brooks}, G.~{Rhee}, J.~{Wadsley},
  P.~{Jonsson}, B.~{Willman}, G.~{Stinson}, T.~{Quinn}, P.~{Madau}, {Bulgeless
  dwarf galaxies and dark matter cores from supernova-driven outflows}, \nat
  463 (2010) 203--206.
\newblock \href {http://dx.doi.org/10.1038/nature08640}
  {\path{doi:10.1038/nature08640}}.

\bibitem{2010PhRvL.104o1301F}
J.~L. {Feng}, M.~{Kaplinghat}, H.-B. {Yu}, {Halo-Shape and Relic-Density
  Exclusions of Sommerfeld-Enhanced Dark Matter Explanations of Cosmic Ray
  Excesses}, Physical Review Letters 104~(15) (2010) 151301.
\newblock \href {http://arxiv.org/abs/0911.0422} {\path{arXiv:0911.0422}},
  \href {http://dx.doi.org/10.1103/PhysRevLett.104.151301}
  {\path{doi:10.1103/PhysRevLett.104.151301}}.

\bibitem{2005ApJ...619..306A}
F.~{Aharonian}, A.~{Neronov}, {High-Energy Gamma Rays from the Massive Black
  Hole in the Galactic Center}, \apj 619 (2005) 306--313.
\newblock \href {http://arxiv.org/abs/astro-ph/0408303}
  {\path{arXiv:astro-ph/0408303}}, \href {http://dx.doi.org/10.1086/426426}
  {\path{doi:10.1086/426426}}.

\bibitem{2011ApJ...726...60C}
M.~{Chernyakova}, et~al., {The High-energy, Arcminute-scale Galactic Center Gamma-ray Source},
  \apj 726 (2011) 60.
\newblock \href {http://arxiv.org/abs/1009.2630} {\path{arXiv:1009.2630}},
  \href {http://dx.doi.org/10.1088/0004-637X/726/2/60}
  {\path{doi:10.1088/0004-637X/726/2/60}}.

\bibitem{2010JCAP...11..041A}
K.~N. {Abazajian}, et~al., {Conservative
  constraints on dark matter from the Fermi-LAT isotropic diffuse gamma-ray
  background spectrum}, \jcap 11 (2010) 41.
\newblock \href {http://arxiv.org/abs/1002.3820} {\path{arXiv:1002.3820}},
  \href {http://dx.doi.org/10.1088/1475-7516/2010/11/041}
  {\path{doi:10.1088/1475-7516/2010/11/041}}.

\bibitem{2011PhLB..697..412H}
D.~{Hooper}, L.~{Goodenough}, {Dark matter annihilation in the Galactic Center
  as seen by the Fermi Gamma Ray Space Telescope}, Physics Letters B 697 (2011)
  412--428.
\newblock \href {http://arxiv.org/abs/1010.2752} {\path{arXiv:1010.2752}},
  \href {http://dx.doi.org/10.1016/j.physletb.2011.02.029}
  {\path{doi:10.1016/j.physletb.2011.02.029}}.

\bibitem{2011PhRvD..84l3005H}
D.~{Hooper}, T.~{Linden}, {Origin of the gamma rays from the Galactic Center},
  \prd 84~(12) (2011) 123005.
\newblock \href {http://arxiv.org/abs/1110.0006} {\path{arXiv:1110.0006}},
  \href {http://dx.doi.org/10.1103/PhysRevD.84.123005}
  {\path{doi:10.1103/PhysRevD.84.123005}}.

\bibitem{2011PhLB..705..165B}
A.~{Boyarsky}, D.~{Malyshev}, O.~{Ruchayskiy}, {A comment on the emission from
  the Galactic Center as seen by the Fermi telescope}, Physics Letters B 705
  (2011) 165--169.
\newblock \href {http://arxiv.org/abs/1012.5839} {\path{arXiv:1012.5839}},
  \href {http://dx.doi.org/10.1016/j.physletb.2011.10.014}
  {\path{doi:10.1016/j.physletb.2011.10.014}}.

\bibitem{2010ApJ...724.1044S}
M.~{Su}, T.~R. {Slatyer}, D.~P. {Finkbeiner}, {Giant Gamma-ray Bubbles from
  Fermi-LAT: Active Galactic Nucleus Activity or Bipolar Galactic Wind?}, \apj
  724 (2010) 1044--1082.
\newblock \href {http://arxiv.org/abs/1005.5480} {\path{arXiv:1005.5480}},
  \href {http://dx.doi.org/10.1088/0004-637X/724/2/1044}
  {\path{doi:10.1088/0004-637X/724/2/1044}}.

\bibitem{2011PhRvL.106j1102C}
R.~M. {Crocker}, F.~{Aharonian}, {Fermi Bubbles: Giant, Multibillion-Year-Old
  Reservoirs of Galactic Center Cosmic Rays}, Physical Review Letters 106~(10)
  (2011) 101102.
\newblock \href {http://arxiv.org/abs/1008.2658} {\path{arXiv:1008.2658}},
  \href {http://dx.doi.org/10.1103/PhysRevLett.106.101102}
  {\path{doi:10.1103/PhysRevLett.106.101102}}.

\bibitem{2011PhRvL.107i1101M}
P.~{Mertsch}, S.~{Sarkar}, {Fermi Gamma-Ray ``Bubbles'' from Stochastic
  Acceleration of Electrons}, Physical Review Letters 107~(9) (2011) 091101.
\newblock \href {http://arxiv.org/abs/1104.3585} {\path{arXiv:1104.3585}},
  \href {http://dx.doi.org/10.1103/PhysRevLett.107.091101}
  {\path{doi:10.1103/PhysRevLett.107.091101}}.

\bibitem{2011PhRvD..83b3518P}
L.~{Pieri}, et~al., {Implications of
  high-resolution simulations on indirect dark matter searches}, \prd 83~(2)
  (2011) 023518--+.
\newblock \href {http://arxiv.org/abs/0908.0195} {\path{arXiv:0908.0195}},
  \href {http://dx.doi.org/10.1103/PhysRevD.83.023518}
  {\path{doi:10.1103/PhysRevD.83.023518}}.

\bibitem{2005A&A...441..513K}
J.~{Kn{\"o}dlseder}, et~al., {The all-sky distribution of 511 keV
  electron-positron annihilation emission}, \aap 441 (2005) 513--532.
\newblock \href {http://arxiv.org/abs/astro-ph/0506026}
  {\path{arXiv:astro-ph/0506026}}, \href
  {http://dx.doi.org/10.1051/0004-6361:20042063}
  {\path{doi:10.1051/0004-6361:20042063}}.

\bibitem{2004PhRvL..92j1301B}
C.~{Boehm}, et~al., {MeV Dark Matter:
  Has It Been Detected?}, Physical Review Letters 92~(10) (2004) 101301--+.
\newblock \href {http://arxiv.org/abs/astro-ph/0309686}
  {\path{arXiv:astro-ph/0309686}}, \href
  {http://dx.doi.org/10.1103/PhysRevLett.92.101301}
  {\path{doi:10.1103/PhysRevLett.92.101301}}.

\bibitem{2011RvMP...83.1001P}
N.~{Prantzos}, et~al., {The 511 keV emission from positron annihilation in the
  Galaxy}, Reviews of Modern Physics 83 (2011) 1001--1056.
\newblock \href {http://arxiv.org/abs/1009.4620} {\path{arXiv:1009.4620}},
  \href {http://dx.doi.org/10.1103/RevModPhys.83.1001}
  {\path{doi:10.1103/RevModPhys.83.1001}}.

\bibitem{2009PhRvL.103y1101A}
A.~A. {Abdo}, et~al., {Fermi Large Area Telescope Measurements of the Diffuse
  Gamma-Ray Emission at Intermediate Galactic Latitudes}, Physical Review
  Letters 103~(25) (2009) 251101--+.
\newblock \href {http://dx.doi.org/10.1103/PhysRevLett.103.251101}
  {\path{doi:10.1103/PhysRevLett.103.251101}}.

\bibitem{2012ApJ...750....3A}
M.~{Ackermann}, et~al., {Fermi-LAT Observations of the Diffuse {$\gamma$}-Ray
  Emission: Implications for Cosmic Rays and the Interstellar Medium}, \apj 750
  (2012) 3.
\newblock \href {http://arxiv.org/abs/1202.4039} {\path{arXiv:1202.4039}},
  \href {http://dx.doi.org/10.1088/0004-637X/750/1/3}
  {\path{doi:10.1088/0004-637X/750/1/3}}.

\bibitem{2011PhRvD..84j3007A}
K.~N. {Abazajian}, S.~{Blanchet}, J.~P. {Harding}, {Contribution of blazars to
  the extragalactic diffuse gamma-ray background and their future spatial
  resolution}, \prd 84~(10) (2011) 103007.
\newblock \href {http://arxiv.org/abs/1012.1247} {\path{arXiv:1012.1247}},
  \href {http://dx.doi.org/10.1103/PhysRevD.84.103007}
  {\path{doi:10.1103/PhysRevD.84.103007}}.

\bibitem{1989PhRvD..40.3168B}
A.~{Bouquet}, P.~{Salati}, J.~{Silk}, {gamma -ray lines as a probe for a
  cold-dark-matter halo}, \prd 40 (1989) 3168--3186.
\newblock \href {http://dx.doi.org/10.1103/PhysRevD.40.3168}
  {\path{doi:10.1103/PhysRevD.40.3168}}.

\bibitem{1994APh.....2..261B}
L.~{Bergstr{\"o}m}, J.~{Kaplan}, {Gamma ray lines from TeV dark matter},
  Astroparticle Physics 2 (1994) 261--268.
\newblock \href {http://arxiv.org/abs/hep-ph/9403239}
  {\path{arXiv:hep-ph/9403239}}, \href
  {http://dx.doi.org/10.1016/0927-6505(94)90005-1}
  {\path{doi:10.1016/0927-6505(94)90005-1}}.

\bibitem{2008JHEP...01..049B}
T.~{Bringmann}, L.~{Bergstr{\"o}m}, J.~{Edsj{\"o}}, {New gamma-ray
  contributions to supersymmetric dark matter annihilation}, Journal of High
  Energy Physics 1 (2008) 49--+.
\newblock \href {http://arxiv.org/abs/0710.3169} {\path{arXiv:0710.3169}},
  \href {http://dx.doi.org/10.1088/1126-6708/2008/01/049}
  {\path{doi:10.1088/1126-6708/2008/01/049}}.

\bibitem{2010PhRvL.104i1302A}
A.~A. {Abdo}, et~al., {Fermi Large Area Telescope Search for Photon Lines from
  30 to 200 GeV and Dark Matter Implications}, Physical Review Letters 104~(9)
  (2010) 091302.
\newblock \href {http://arxiv.org/abs/1001.4836} {\path{arXiv:1001.4836}},
  \href {http://dx.doi.org/10.1103/PhysRevLett.104.091302}
  {\path{doi:10.1103/PhysRevLett.104.091302}}.

\bibitem{2011JCAP...05..027V}
G.~{Vertongen}, C.~{Weniger}, {Hunting dark matter gamma-ray lines with the
  Fermi LAT}, \jcap 5 (2011) 27.
\newblock \href {http://arxiv.org/abs/1101.2610} {\path{arXiv:1101.2610}},
  \href {http://dx.doi.org/10.1088/1475-7516/2011/05/027}
  {\path{doi:10.1088/1475-7516/2011/05/027}}.

\bibitem{2011NuPhB.844...55G}
J.~{Goodman}, et~al., {Gamma ray line constraints on effective theories of dark matter},
  Nuclear Physics B 844 (2011) 55--68.
\newblock \href {http://arxiv.org/abs/1009.0008} {\path{arXiv:1009.0008}},
  \href {http://dx.doi.org/10.1016/j.nuclphysb.2010.10.022}
  {\path{doi:10.1016/j.nuclphysb.2010.10.022}}.

\bibitem{2012JCAP...07..054B}
T.~{Bringmann}, et~al., {Fermi LAT
  search for internal bremsstrahlung signatures from dark matter annihilation},
  \jcap 7 (2012) 54.
\newblock \href {http://arxiv.org/abs/1203.1312} {\path{arXiv:1203.1312}},
  \href {http://dx.doi.org/10.1088/1475-7516/2012/07/054}
  {\path{doi:10.1088/1475-7516/2012/07/054}}.

\bibitem{2011PhRvL.107x1303G}
A.~{Geringer-Sameth}, S.~M. {Koushiappas}, {Exclusion of Canonical Weakly
  Interacting Massive Particles by Joint Analysis of Milky Way Dwarf Galaxies
  with Data from the Fermi Gamma-Ray Space Telescope}, Physical Review Letters
  107~(24) (2011) 241303.
\newblock \href {http://arxiv.org/abs/1108.2914} {\path{arXiv:1108.2914}},
  \href {http://dx.doi.org/10.1103/PhysRevLett.107.241303}
  {\path{doi:10.1103/PhysRevLett.107.241303}}.

\bibitem{2009ApJ...703.1249A}
A.~A. {Abdo}, et~al., {Fermi LAT Observation of Diffuse Gamma Rays Produced
  Through Interactions Between Local Interstellar Matter and High-energy Cosmic
  Rays}, \apj 703 (2009) 1249--1256.
\newblock \href {http://arxiv.org/abs/0908.1171} {\path{arXiv:0908.1171}},
  \href {http://dx.doi.org/10.1088/0004-637X/703/2/1249}
  {\path{doi:10.1088/0004-637X/703/2/1249}}.

\bibitem{2010PhRvL.104j1101A}
A.~A. {Abdo}, et~al., {Spectrum of the Isotropic Diffuse Gamma-Ray Emission
  Derived from First-Year Fermi Large Area Telescope Data}, Physical Review
  Letters 104~(10) (2010) 101101.
\newblock \href {http://arxiv.org/abs/1002.3603} {\path{arXiv:1002.3603}},
  \href {http://dx.doi.org/10.1103/PhysRevLett.104.101101}
  {\path{doi:10.1103/PhysRevLett.104.101101}}.

\bibitem{2010JCAP...04..014A}
A.~A. {Abdo}, et~al., {Constraints on cosmological dark matter annihilation
  from the Fermi-LAT isotropic diffuse gamma-ray measurement}, \jcap 4 (2010)
  14.
\newblock \href {http://arxiv.org/abs/1002.4415} {\path{arXiv:1002.4415}},
  \href {http://dx.doi.org/10.1088/1475-7516/2010/04/014}
  {\path{doi:10.1088/1475-7516/2010/04/014}}.

\bibitem{2012PhRvD..85d3509A}
K.~N. {Abazajian}, S.~{Blanchet}, J.~P. {Harding}, {Current and future
  constraints on dark matter from prompt and inverse-Compton photon emission in
  the isotropic diffuse gamma-ray background}, \prd 85~(4) (2012) 043509.
\newblock \href {http://arxiv.org/abs/1011.5090} {\path{arXiv:1011.5090}},
  \href {http://dx.doi.org/10.1103/PhysRevD.85.043509}
  {\path{doi:10.1103/PhysRevD.85.043509}}.

\bibitem{2011JCAP...01..011A}
C.~{Arina}, M.~H.~G. {Tytgat}, {Constraints on light WIMP candidates from the
  isotropic diffuse gamma-ray emission}, \jcap 1 (2011) 11--+.
\newblock \href {http://arxiv.org/abs/1007.2765} {\path{arXiv:1007.2765}},
  \href {http://dx.doi.org/10.1088/1475-7516/2011/01/011}
  {\path{doi:10.1088/1475-7516/2011/01/011}}.

\bibitem{2012MNRAS.421L..87S}
P.~D. {Serpico}, et~al.,
  {Extragalactic gamma-ray signal from dark matter annihilation: a power
  spectrum based computation}, \mnras 421 (2012) L87--L91.
\newblock \href {http://arxiv.org/abs/1109.0095} {\path{arXiv:1109.0095}},
  \href {http://dx.doi.org/10.1111/j.1745-3933.2011.01212.x}
  {\path{doi:10.1111/j.1745-3933.2011.01212.x}}.

\bibitem{2010ApJ...712..147A}
A.~A. {Abdo}, et~al., {Observations of Milky Way Dwarf Spheroidal Galaxies with
  the Fermi-Large Area Telescope Detector and Constraints on Dark Matter
  Models}, \apj 712 (2010) 147--158.
\newblock \href {http://arxiv.org/abs/1001.4531} {\path{arXiv:1001.4531}},
  \href {http://dx.doi.org/10.1088/0004-637X/712/1/147}
  {\path{doi:10.1088/0004-637X/712/1/147}}.

\bibitem{2011PhRvL.107x1302A}
M.~{Ackermann}, et~al., {Constraining Dark Matter Models from a Combined
  Analysis of Milky Way Satellites with the Fermi Large Area Telescope},
  Physical Review Letters 107~(24) (2011) 241302.
\newblock \href {http://arxiv.org/abs/1108.3546} {\path{arXiv:1108.3546}},
  \href {http://dx.doi.org/10.1103/PhysRevLett.107.241302}
  {\path{doi:10.1103/PhysRevLett.107.241302}}.

\bibitem{2011PhRvL.106p1301A}
A.~{Abramowski}, et~al., {Search for a Dark Matter Annihilation Signal from the
  Galactic Center Halo with H.E.S.S.}, Physical Review Letters 106~(16) (2011)
  161301.
\newblock \href {http://arxiv.org/abs/1103.3266} {\path{arXiv:1103.3266}},
  \href {http://dx.doi.org/10.1103/PhysRevLett.106.161301}
  {\path{doi:10.1103/PhysRevLett.106.161301}}.

\bibitem{2012JCAP...01..041A}
K.~N. {Abazajian}, J.~P. {Harding}, {Constraints on WIMP and
  Sommerfeld-enhanced dark matter annihilation from HESS observations of the
  galactic center}, \jcap 1 (2012) 41.
\newblock \href {http://arxiv.org/abs/1110.6151} {\path{arXiv:1110.6151}},
  \href {http://dx.doi.org/10.1088/1475-7516/2012/01/041}
  {\path{doi:10.1088/1475-7516/2012/01/041}}.

\bibitem{2009JHEP...12..052M}
P.~{Meade}, M.~{Papucci}, T.~{Volansky}, {Dark matter sees the light}, Journal
  of High Energy Physics 12 (2009) 52.
\newblock \href {http://arxiv.org/abs/0901.2925} {\path{arXiv:0901.2925}},
  \href {http://dx.doi.org/10.1088/1126-6708/2009/12/052}
  {\path{doi:10.1088/1126-6708/2009/12/052}}.

\bibitem{2010JCAP...03..014P}
M.~{Papucci}, A.~{Strumia}, {Robust implications on dark matter from the first
  FERMI sky {$\gamma$} map}, \jcap 3 (2010) 14.
\newblock \href {http://arxiv.org/abs/0912.0742} {\path{arXiv:0912.0742}},
  \href {http://dx.doi.org/10.1088/1475-7516/2010/03/014}
  {\path{doi:10.1088/1475-7516/2010/03/014}}.

\bibitem{2011MNRAS.418.1526C}
A.~{Charbonnier}, et~al., {Dark matter profiles and annihilation in dwarf spheroidal
  galaxies: prospectives for present and future {$\gamma$}-ray observatories -
  I. The classical dwarf spheroidal galaxies}, \mnras 418 (2011) 1526--1556.
\newblock \href {http://arxiv.org/abs/1104.0412} {\path{arXiv:1104.0412}},
  \href {http://dx.doi.org/10.1111/j.1365-2966.2011.19387.x}
  {\path{doi:10.1111/j.1365-2966.2011.19387.x}}.

\bibitem{2011ApJ...733L..46W}
M.~G. {Walker}, et~al.,
  {Dark Matter in the Classical Dwarf Spheroidal Galaxies: A Robust Constraint
  on the Astrophysical Factor for {$\gamma$}-Ray Flux Calculations}, \apjl 733
  (2011) L46.
\newblock \href {http://arxiv.org/abs/1104.0411} {\path{arXiv:1104.0411}},
  \href {http://dx.doi.org/10.1088/2041-8205/733/2/L46}
  {\path{doi:10.1088/2041-8205/733/2/L46}}.

\bibitem{2008ApJ...679..428A}
J.~{Albert}, et~al., {Upper Limit for {$\gamma$}-Ray Emission above 140 GeV
  from the Dwarf Spheroidal Galaxy Draco}, \apj 679 (2008) 428--431.
\newblock \href {http://arxiv.org/abs/0711.2574} {\path{arXiv:0711.2574}},
  \href {http://dx.doi.org/10.1086/529135} {\path{doi:10.1086/529135}}.

\bibitem{2010ApJ...720.1174A}
V.~A. {Acciari}, et~al., {VERITAS Search for VHE Gamma-ray Emission from Dwarf
  Spheroidal Galaxies}, \apj 720 (2010) 1174--1180.
\newblock \href {http://arxiv.org/abs/1006.5955} {\path{arXiv:1006.5955}},
  \href {http://dx.doi.org/10.1088/0004-637X/720/2/1174}
  {\path{doi:10.1088/0004-637X/720/2/1174}}.

\bibitem{2011APh....34..608H}
A.~{Abramowski}, et~al., {H.E.S.S. constraints on dark matter annihilations
  towards the sculptor and carina dwarf galaxies}, Astroparticle Physics 34
  (2011) 608--616.
\newblock \href {http://arxiv.org/abs/1012.5602} {\path{arXiv:1012.5602}},
  \href {http://dx.doi.org/10.1016/j.astropartphys.2010.12.006}
  {\path{doi:10.1016/j.astropartphys.2010.12.006}}.

\bibitem{2012arXiv1203.2954C}
I.~{Cholis}, P.~{Salucci}, {Extracting limits on Dark Matter annihilation from
  dwarf Spheroidal galaxies at gamma-rays}, ArXiv e-prints\href
  {http://arxiv.org/abs/1203.2954} {\path{arXiv:1203.2954}}.

\bibitem{2000PhRvD..61b3514B}
E.~A. {Baltz}, et~al., {Detection of
  neutralino annihilation photons from external galaxies}, \prd 61~(2) (1999)
  023514--+.
\newblock \href {http://arxiv.org/abs/astro-ph/9909112}
  {\path{arXiv:astro-ph/9909112}}, \href
  {http://dx.doi.org/10.1103/PhysRevD.61.023514}
  {\path{doi:10.1103/PhysRevD.61.023514}}.

\bibitem{2004APh....20..467F}
A.~{Falvard}, et~al., {Supersymmetric dark matter in M31: can one see neutralino
  annihilation with CELESTE?}, Astroparticle Physics 20 (2004) 467--484.
\newblock \href {http://arxiv.org/abs/astro-ph/0210184}
  {\path{arXiv:astro-ph/0210184}}, \href
  {http://dx.doi.org/10.1016/S0927-6505(03)00215-9}
  {\path{doi:10.1016/S0927-6505(03)00215-9}}.

\bibitem{2003A&A...400..153A}
F.~A. {Aharonian}, et~al., {Search for TeV gamma ray emission from the
  Andromeda galaxy}, \aap 400 (2003) 153--159.
\newblock \href {http://arxiv.org/abs/astro-ph/0302347}
  {\path{arXiv:astro-ph/0302347}}, \href
  {http://dx.doi.org/10.1051/0004-6361:20021895}
  {\path{doi:10.1051/0004-6361:20021895}}.

\bibitem{2006A&A...450....1L}
J.~{Lavalle}, et~al., {Indirect search for dark matter in M 31 with the CELESTE
  experiment}, \aap 450 (2006) 1--8.
\newblock \href {http://arxiv.org/abs/astro-ph/0601298}
  {\path{arXiv:astro-ph/0601298}}, \href
  {http://dx.doi.org/10.1051/0004-6361:20054340}
  {\path{doi:10.1051/0004-6361:20054340}}.

\bibitem{2008ApJ...678..594W}
M.~{Wood}, et~al., {A Search for Dark Matter Annihilation with the Whipple 10 m
  Telescope}, \apj 678 (2008) 594--605.
\newblock \href {http://arxiv.org/abs/0801.1708} {\path{arXiv:0801.1708}},
  \href {http://dx.doi.org/10.1086/529421} {\path{doi:10.1086/529421}}.

\bibitem{2006Sci...314.1424A}
F.~{Aharonian}, et~al., {Fast Variability of Tera-Electron Volt {$\gamma$} Rays
  from the Radio Galaxy M87}, Science 314 (2006) 1424--1427.
\newblock \href {http://arxiv.org/abs/astro-ph/0612016}
  {\path{arXiv:astro-ph/0612016}}, \href
  {http://dx.doi.org/10.1126/science.1134408}
  {\path{doi:10.1126/science.1134408}}.

\bibitem{2012ApJ...746..141A}
E.~{Aliu}, et~al., {VERITAS Observations of Day-scale Flaring of M 87 in 2010
  April}, \apj 746 (2012) 141.
\newblock \href {http://arxiv.org/abs/1112.4518} {\path{arXiv:1112.4518}},
  \href {http://dx.doi.org/10.1088/0004-637X/746/2/141}
  {\path{doi:10.1088/0004-637X/746/2/141}}.

\bibitem{2012A&A...539L...2A}
J.~{Aleksi{\'c}}, et~al., {Detection of very-high energy {$\gamma$}-ray
  emission from <ASTROBJ>NGC 1275</ASTROBJ> by the MAGIC telescopes}, \aap 539
  (2012) L2.
\newblock \href {http://arxiv.org/abs/1112.3917} {\path{arXiv:1112.3917}},
  \href {http://dx.doi.org/10.1051/0004-6361/201118668}
  {\path{doi:10.1051/0004-6361/201118668}}.

\bibitem{2010A&A...523L...2A}
A.~A. {Abdo}, et~al., {Fermi Large Area Telescope observations of Local Group
  galaxies: detection of M 31 and search for M 33}, \aap 523 (2010) L2.
\newblock \href {http://arxiv.org/abs/1012.1952} {\path{arXiv:1012.1952}},
  \href {http://dx.doi.org/10.1051/0004-6361/201015759}
  {\path{doi:10.1051/0004-6361/201015759}}.

\bibitem{2010A&A...523A..46A}
A.~A. {Abdo}, et~al., {Detection of the Small Magellanic Cloud in gamma-rays
  with Fermi/LAT}, \aap 523 (2010) A46.
\newblock \href {http://dx.doi.org/10.1051/0004-6361/201014855}
  {\path{doi:10.1051/0004-6361/201014855}}.

\bibitem{2010A&A...512A...7A}
A.~A. {Abdo}, et~al., {Observations of the Large Magellanic Cloud with Fermi},
  \aap 512 (2010) A7.
\newblock \href {http://dx.doi.org/10.1051/0004-6361/200913474}
  {\path{doi:10.1051/0004-6361/200913474}}.

\bibitem{2006A&A...455...21C}
S.~{Colafrancesco}, S.~{Profumo}, P.~{Ullio}, {Multi-frequency analysis of
  neutralino dark matter annihilations in the Coma cluster}, \aap 455 (2006)
  21--43.
\newblock \href {http://arxiv.org/abs/astro-ph/0507575}
  {\path{arXiv:astro-ph/0507575}}, \href
  {http://dx.doi.org/10.1051/0004-6361:20053887}
  {\path{doi:10.1051/0004-6361:20053887}}.

\bibitem{2011PhRvD..84l3509P}
A.~{Pinzke}, C.~{Pfrommer}, L.~{Bergstr{\"o}m}, {Prospects of detecting
  gamma-ray emission from galaxy clusters: Cosmic rays and dark matter
  annihilations}, \prd 84~(12) (2011) 123509.
\newblock \href {http://arxiv.org/abs/1105.3240} {\path{arXiv:1105.3240}},
  \href {http://dx.doi.org/10.1103/PhysRevD.84.123509}
  {\path{doi:10.1103/PhysRevD.84.123509}}.

\bibitem{2011JCAP...12..011S}
M.~A. {S{\'a}nchez-Conde}, et~al., {Dark matter searches with Cherenkov telescopes: nearby dwarf
  galaxies or local galaxy clusters?}, \jcap 12 (2011) 11.
\newblock \href {http://arxiv.org/abs/1104.3530} {\path{arXiv:1104.3530}},
  \href {http://dx.doi.org/10.1088/1475-7516/2011/12/011}
  {\path{doi:10.1088/1475-7516/2011/12/011}}.

\bibitem{2012JCAP...01..042H}
X.~{Huang}, G.~{Vertongen}, C.~{Weniger}, {Probing dark matter decay and
  annihilation with Fermi LAT observations of nearby galaxy clusters}, \jcap 1
  (2012) 42.
\newblock \href {http://arxiv.org/abs/1110.1529} {\path{arXiv:1110.1529}},
  \href {http://dx.doi.org/10.1088/1475-7516/2012/01/042}
  {\path{doi:10.1088/1475-7516/2012/01/042}}.

\bibitem{2012PhRvD..85f3517C}
C.~{Combet}, D.~{Maurin}, E.~{Nezri}, E.~{Pointecouteau}, J.~A. {Hinton},
  R.~{White}, {Decaying dark matter: Stacking analysis of galaxy clusters to
  improve on current limits}, \prd 85~(6) (2012) 063517.
\newblock \href {http://arxiv.org/abs/1203.1164} {\path{arXiv:1203.1164}},
  \href {http://dx.doi.org/10.1103/PhysRevD.85.063517}
  {\path{doi:10.1103/PhysRevD.85.063517}}.

\bibitem{2012arXiv1203.1165N}
E.~{Nezri}, R.~{White}, C.~{Combet}, J.~A. {Hinton}, D.~{Maurin},
  E.~{Pointecouteau}, {gamma-rays from annihilating dark matter in galaxy
  clusters: stacking vs single source analysis}, ArXiv e-prints\href
  {http://arxiv.org/abs/1203.1165} {\path{arXiv:1203.1165}}.

\bibitem{2012arXiv1203.1166M}
D.~{Maurin}, C.~{Combet}, E.~{Nezri}, E.~{Pointecouteau}, {Disentangling
  cosmic-ray and dark matter induced gamma-rays in galaxy clusters?}, ArXiv
  e-prints\href {http://arxiv.org/abs/1203.1166} {\path{arXiv:1203.1166}}.

\bibitem{2010JCAP...05..025A}
M.~{Ackermann}, et~al., {Constraints on dark matter annihilation in clusters of
  galaxies with the Fermi large area telescope}, \jcap 5 (2010) 25.
\newblock \href {http://arxiv.org/abs/1002.2239} {\path{arXiv:1002.2239}},
  \href {http://dx.doi.org/10.1088/1475-7516/2010/05/025}
  {\path{doi:10.1088/1475-7516/2010/05/025}}.

\bibitem{2011arXiv1110.6863Z}
S.~{Zimmer}, J.~{Conrad}, {for the Fermi-LAT Collaboration}, A.~{Pinzke}, {A
  Combined Analysis of Clusters of Galaxies - Gamma Ray Emission from Cosmic
  Rays and Dark Matter}, ArXiv e-prints\href {http://arxiv.org/abs/1110.6863}
  {\path{arXiv:1110.6863}}.

\bibitem{2012arXiv1201.1003H}
J.~{Han}, et~al., {Evidence for
  extended gamma-ray emission from galaxy clusters}, ArXiv e-prints\href
  {http://arxiv.org/abs/1201.1003} {\path{arXiv:1201.1003}}.

\bibitem{2010MNRAS.409..449P}
A.~{Pinzke}, C.~{Pfrommer}, {Simulating the {$\gamma$}-ray emission from galaxy
  clusters: a universal cosmic ray spectrum and spatial distribution}, \mnras
  409 (2010) 449--480.
\newblock \href {http://arxiv.org/abs/1001.5023} {\path{arXiv:1001.5023}},
  \href {http://dx.doi.org/10.1111/j.1365-2966.2010.17328.x}
  {\path{doi:10.1111/j.1365-2966.2010.17328.x}}.

\bibitem{1985ApJ...296..679P}
W.~H. {Press}, D.~N. {Spergel}, {Capture by the sun of a galactic population of
  weakly interacting, massive particles}, \apj 296 (1985) 679--684.
\newblock \href {http://dx.doi.org/10.1086/163485} {\path{doi:10.1086/163485}}.

\bibitem{1985PhRvL..55..257S}
J.~{Silk}, K.~{Olive}, M.~{Srednicki}, {The photino, the sun, and high-energy
  neutrinos}, Physical Review Letters 55 (1985) 257--259.
\newblock \href {http://dx.doi.org/10.1103/PhysRevLett.55.257}
  {\path{doi:10.1103/PhysRevLett.55.257}}.

\bibitem{1986PhLB..167..295F}
K.~{Freese}, {Can scalar neutrinos or massive Dirac neutrinos be the missing
  mass?}, Physics Letters B 167 (1986) 295--300.
\newblock \href {http://dx.doi.org/10.1016/0370-2693(86)90349-7}
  {\path{doi:10.1016/0370-2693(86)90349-7}}.

\bibitem{1986PhRvD..33.2079K}
L.~M. {Krauss}, M.~{Srednicki}, F.~{Wilczek}, {Solar System constraints and
  signatures for dark-matter candidates}, \prd 33 (1986) 2079--2083.
\newblock \href {http://dx.doi.org/10.1103/PhysRevD.33.2079}
  {\path{doi:10.1103/PhysRevD.33.2079}}.

\bibitem{1986PhRvD..34.2206G}
T.~K. {Gaisser}, G.~{Steigman}, S.~{Tilav}, {Limits on cold-dark-matter
  candidates from deep underground detectors}, \prd 34 (1986) 2206--2222.
\newblock \href {http://dx.doi.org/10.1103/PhysRevD.34.2206}
  {\path{doi:10.1103/PhysRevD.34.2206}}.

\bibitem{2004PhRvD..70b3006B}
G.~D. {Barr}, T.~K. {Gaisser}, P.~{Lipari}, S.~{Robbins}, T.~{Stanev},
  {Three-dimensional calculation of atmospheric neutrinos}, \prd 70~(2) (2004)
  023006.
\newblock \href {http://arxiv.org/abs/astro-ph/0403630}
  {\path{arXiv:astro-ph/0403630}}, \href
  {http://dx.doi.org/10.1103/PhysRevD.70.023006}
  {\path{doi:10.1103/PhysRevD.70.023006}}.

\bibitem{2011PhRvD..83l3001H}
M.~{Honda}, T.~{Kajita}, K.~{Kasahara}, S.~{Midorikawa}, {Improvement of low
  energy atmospheric neutrino flux calculation using the JAM nuclear
  interaction model}, \prd 83~(12) (2011) 123001.
\newblock \href {http://arxiv.org/abs/1102.2688} {\path{arXiv:1102.2688}},
  \href {http://dx.doi.org/10.1103/PhysRevD.83.123001}
  {\path{doi:10.1103/PhysRevD.83.123001}}.

\bibitem{2008JPhCS.136d2063C}
J.~{Carr}, et~al.,
  {the KM3NeT Constortium}, {Detector design studies for a cubic kilometre Deep
  Sea neutrino telescope KM3NeT}, Journal of Physics Conference Series 136~(4)
  (2008) 042063.
\newblock \href {http://arxiv.org/abs/0711.2145} {\path{arXiv:0711.2145}},
  \href {http://dx.doi.org/10.1088/1742-6596/136/4/042063}
  {\path{doi:10.1088/1742-6596/136/4/042063}}.

\bibitem{2010RScI...81h1101H}
F.~{Halzen}, S.~R. {Klein}, {Invited Review Article: IceCube: An instrument for
  neutrino astronomy}, Review of Scientific Instruments 81~(8) (2010) 081101.
\newblock \href {http://arxiv.org/abs/1007.1247} {\path{arXiv:1007.1247}},
  \href {http://dx.doi.org/10.1063/1.3480478} {\path{doi:10.1063/1.3480478}}.

\bibitem{1968JETP...26..984P}
B.~{Pontecorvo}, {Neutrino Experiments and the Problem of Conservation of
  Leptonic Charge}, Soviet Journal of Experimental and Theoretical Physics 26
  (1968) 984.

\bibitem{2006hep.ph....6054S}
A.~{Strumia}, F.~{Vissani}, {Neutrino masses and mixings and...}, ArXiv High
  Energy Physics - Phenomenology e-prints\href
  {http://arxiv.org/abs/hep-ph/0606054} {\path{arXiv:hep-ph/0606054}}.

\bibitem{2008PhR...460....1G}
M.~C. {Gonzalez-Garcia}, M.~{Maltoni}, {Phenomenology with massive neutrinos},
  \physrep 460 (2008) 1--129.
\newblock \href {http://arxiv.org/abs/0704.1800} {\path{arXiv:0704.1800}},
  \href {http://dx.doi.org/10.1016/j.physrep.2007.12.004}
  {\path{doi:10.1016/j.physrep.2007.12.004}}.

\bibitem{2008PhR...458..173B}
J.~K. {Becker}, {High-energy neutrinos in the context of multimessenger
  astrophysics}, \physrep 458 (2008) 173--246.
\newblock \href {http://arxiv.org/abs/0710.1557} {\path{arXiv:0710.1557}},
  \href {http://dx.doi.org/10.1016/j.physrep.2007.10.006}
  {\path{doi:10.1016/j.physrep.2007.10.006}}.

\bibitem{2003PhRvD..67g3024B}
G.~{Barenboim}, C.~{Quigg}, {Neutrino observatories can characterize cosmic
  sources and neutrino properties}, \prd 67~(7) (2003) 073024.
\newblock \href {http://arxiv.org/abs/hep-ph/0301220}
  {\path{arXiv:hep-ph/0301220}}, \href
  {http://dx.doi.org/10.1103/PhysRevD.67.073024}
  {\path{doi:10.1103/PhysRevD.67.073024}}.

\bibitem{1978PhRvD..17.2369W}
L.~{Wolfenstein}, {Neutrino oscillations in matter}, \prd 17 (1978) 2369--2374.
\newblock \href {http://dx.doi.org/10.1103/PhysRevD.17.2369}
  {\path{doi:10.1103/PhysRevD.17.2369}}.

\bibitem{1986NCimC...9...17M}
S.~P. {Mikheev}, A.~I. {Smirnov}, {Resonant amplification of neutrino
  oscillations in matter and solar-neutrino spectroscopy}, Nuovo Cimento C
  Geophysics Space Physics C 9 (1986) 17--26.
\newblock \href {http://dx.doi.org/10.1007/BF02508049}
  {\path{doi:10.1007/BF02508049}}.

\bibitem{2000PhLB..474..153O}
T.~{Ohlsson}, H.~{Snellman}, {Neutrino oscillations with three flavors in
  matter: Applications to neutrinos traversing the Earth}, Physics Letters B
  474 (2000) 153--162.
\newblock \href {http://arxiv.org/abs/hep-ph/9912295}
  {\path{arXiv:hep-ph/9912295}}, \href
  {http://dx.doi.org/10.1016/S0370-2693(00)00008-3}
  {\path{doi:10.1016/S0370-2693(00)00008-3}}.

\bibitem{2008JCAP...01..021B}
M.~{Blennow}, J.~{Edsj{\"o}}, T.~{Ohlsson}, {Neutrinos from WIMP annihilations
  obtained using a full three-flavor Monte Carlo approach}, \jcap 1 (2008) 21.
\newblock \href {http://arxiv.org/abs/0709.3898} {\path{arXiv:0709.3898}},
  \href {http://dx.doi.org/10.1088/1475-7516/2008/01/021}
  {\path{doi:10.1088/1475-7516/2008/01/021}}.

\bibitem{2005NuPhB.727...99C}
M.~{Cirelli}, et~al., {Spectra of neutrinos from dark matter annihilations}, Nuclear
  Physics B 727 (2005) 99--138.
\newblock \href {http://arxiv.org/abs/hep-ph/0506298}
  {\path{arXiv:hep-ph/0506298}}, \href
  {http://dx.doi.org/10.1016/j.nuclphysb.2005.08.017}
  {\path{doi:10.1016/j.nuclphysb.2005.08.017}}.

\bibitem{2007PhRvD..76i5008B}
V.~{Barger}, et~al., {High energy
  neutrinos from neutralino annihilations in the Sun}, \prd 76~(9) (2007)
  095008.
\newblock \href {http://arxiv.org/abs/0708.1325} {\path{arXiv:0708.1325}},
  \href {http://dx.doi.org/10.1103/PhysRevD.76.095008}
  {\path{doi:10.1103/PhysRevD.76.095008}}.

\bibitem{2004PhRvD..70f3503B}
G.~{Bertone}, et~al., {Neutrinos from dark matter
  annihilations at the galactic center}, \prd 70~(6) (2004) 063503.
\newblock \href {http://arxiv.org/abs/astro-ph/0403322}
  {\path{arXiv:astro-ph/0403322}}, \href
  {http://dx.doi.org/10.1103/PhysRevD.70.063503}
  {\path{doi:10.1103/PhysRevD.70.063503}}.

\bibitem{2010PhRvD..81i6007E}
A.~E. {Erkoca}, et~al., {Muon fluxes and
  showers from dark matter annihilation in the Galactic center}, \prd 81~(9)
  (2010) 096007.
\newblock \href {http://arxiv.org/abs/1002.2220} {\path{arXiv:1002.2220}},
  \href {http://dx.doi.org/10.1103/PhysRevD.81.096007}
  {\path{doi:10.1103/PhysRevD.81.096007}}.

\bibitem{2011PhRvD..84b2004A}
R.~{Abbasi}, et~al., {Search for dark matter from the Galactic halo with the
  IceCube Neutrino Telescope}, \prd 84~(2) (2011) 022004.
\newblock \href {http://arxiv.org/abs/1101.3349} {\path{arXiv:1101.3349}},
  \href {http://dx.doi.org/10.1103/PhysRevD.84.022004}
  {\path{doi:10.1103/PhysRevD.84.022004}}.

\bibitem{1990PhRvL..64..615G}
K.~{Griest}, M.~{Kamionkowski}, {Unitarity limits on the mass and radius of
  dark-matter particles}, Physical Review Letters 64 (1990) 615--618.
\newblock \href {http://dx.doi.org/10.1103/PhysRevLett.64.615}
  {\path{doi:10.1103/PhysRevLett.64.615}}.

\bibitem{2010PhRvD..81h3506S}
P.~{Sandick}, et~al.,
  {Sensitivity of the IceCube neutrino detector to dark matter annihilating in
  dwarf galaxies}, \prd 81~(8) (2010) 083506.
\newblock \href {http://arxiv.org/abs/0912.0513} {\path{arXiv:0912.0513}},
  \href {http://dx.doi.org/10.1103/PhysRevD.81.083506}
  {\path{doi:10.1103/PhysRevD.81.083506}}.

\bibitem{1987NuPhB.283..681G}
K.~{Griest}, D.~{Seckel}, {Cosmic asymmetry, neutrinos and the sun}, Nuclear
  Physics B 283 (1987) 681--705.
\newblock \href {http://dx.doi.org/10.1016/0550-3213(87)90293-8}
  {\path{doi:10.1016/0550-3213(87)90293-8}}.

\bibitem{1987ApJ...321..560G}
A.~{Gould}, {Weakly interacting massive particle distribution in and
  evaporation from the sun}, \apj 321 (1987) 560--570.
\newblock \href {http://dx.doi.org/10.1086/165652} {\path{doi:10.1086/165652}}.

\bibitem{1995PhLB..357..595E}
J.~{Edsj{\"o}}, P.~{Gondolo}, {WIMP mass determination with neutrino
  telescopes}, Physics Letters B 357 (1995) 595--601.
\newblock \href {http://arxiv.org/abs/hep-ph/9504283}
  {\path{arXiv:hep-ph/9504283}}, \href
  {http://dx.doi.org/10.1016/0370-2693(95)00930-J}
  {\path{doi:10.1016/0370-2693(95)00930-J}}.

\bibitem{1995APh.....3...65B}
A.~{Bottino}, et~al., {Signals of neutralino
  dark matter from Earth and Sun}, Astroparticle Physics 3 (1995) 65--75.
\newblock \href {http://arxiv.org/abs/hep-ph/9408391}
  {\path{arXiv:hep-ph/9408391}}, \href
  {http://dx.doi.org/10.1016/0927-6505(94)00028-2}
  {\path{doi:10.1016/0927-6505(94)00028-2}}.

\bibitem{1996APh.....5..333B}
V.~{Berezinsky}, et~al., {Searching for relic neutralinos using neutrino telescopes},
  Astroparticle Physics 5 (1996) 333--352.
\newblock \href {http://arxiv.org/abs/hep-ph/9603342}
  {\path{arXiv:hep-ph/9603342}}, \href
  {http://dx.doi.org/10.1016/0927-6505(96)00035-7}
  {\path{doi:10.1016/0927-6505(96)00035-7}}.

\bibitem{1997PhDT.........5E}
J.~{Edsj{\"o}}, {Aspects of Neutrino Detection of Neutralino Dark Matter},
  Ph.D. thesis, , Uppsala Univ.~(preprint hep-ph/9704384), (1997) (Oct. 1997).
\newblock \href {http://arxiv.org/abs/hep-ph/9704384}
  {\path{arXiv:hep-ph/9704384}}.

\bibitem{1998PhRvD..58j3519B}
L.~{Bergstr{\"o}m}, J.~{Edsj{\"o}}, P.~{Gondolo}, {Indirect detection of dark
  matter in km-size neutrino telescopes}, \prd 58~(10) (1998) 103519.
\newblock \href {http://arxiv.org/abs/hep-ph/9806293}
  {\path{arXiv:hep-ph/9806293}}, \href
  {http://dx.doi.org/10.1103/PhysRevD.58.103519}
  {\path{doi:10.1103/PhysRevD.58.103519}}.

\bibitem{2004JCAP...07..008G}
P.~{Gondolo}, et~al., {DarkSUSY: computing supersymmetric dark matter properties
  numerically}, \jcap 7 (2004) 8--+.
\newblock \href {http://arxiv.org/abs/astro-ph/0406204}
  {\path{arXiv:astro-ph/0406204}}, \href
  {http://dx.doi.org/10.1088/1475-7516/2004/07/008}
  {\path{doi:10.1088/1475-7516/2004/07/008}}.

\bibitem{2006PhRvD..73l3507H}
F.~{Halzen}, D.~{Hooper}, {Prospects for detecting dark matter with neutrino
  telescopes in light of recent results from direct detection experiments},
  \prd 73~(12) (2006) 123507.
\newblock \href {http://arxiv.org/abs/hep-ph/0510048}
  {\path{arXiv:hep-ph/0510048}}, \href
  {http://dx.doi.org/10.1103/PhysRevD.73.123507}
  {\path{doi:10.1103/PhysRevD.73.123507}}.

\bibitem{2009PhRvD..80i5019N}
V.~{Niro}, et~al., {Investigating light neutralinos at neutrino telescopes},
  \prd 80~(9) (2009) 095019.
\newblock \href {http://arxiv.org/abs/0909.2348} {\path{arXiv:0909.2348}},
  \href {http://dx.doi.org/10.1103/PhysRevD.80.095019}
  {\path{doi:10.1103/PhysRevD.80.095019}}.

\bibitem{2009JCAP...01..032F}
J.~L. {Feng}, et~al., {Testing the Dark
  Matter interpretation of the DAMA/LIBRA result with Super-Kamiokande}, \jcap
  1 (2009) 32.
\newblock \href {http://arxiv.org/abs/0808.4151} {\path{arXiv:0808.4151}},
  \href {http://dx.doi.org/10.1088/1475-7516/2009/01/032}
  {\path{doi:10.1088/1475-7516/2009/01/032}}.

\bibitem{2011NuPhB.850..505K}
R.~{Kappl}, M.~W. {Winkler}, {New limits on dark matter from Super-Kamiokande},
  Nuclear Physics B 850 (2011) 505--521.
\newblock \href {http://arxiv.org/abs/1104.0679} {\path{arXiv:1104.0679}},
  \href {http://dx.doi.org/10.1016/j.nuclphysb.2011.05.006}
  {\path{doi:10.1016/j.nuclphysb.2011.05.006}}.

\bibitem{2011JHEP...11..133F}
N.~{Fornengo}, V.~{Niro}, {Downward-going tau neutrinos as a new prospect of
  detecting dark matter}, Journal of High Energy Physics 11 (2011) 133.
\newblock \href {http://arxiv.org/abs/1108.2630} {\path{arXiv:1108.2630}},
  \href {http://dx.doi.org/10.1007/JHEP11(2011)133}
  {\path{doi:10.1007/JHEP11(2011)133}}.

\bibitem{1996PhR...267..195J}
G.~{Jungman}, M.~{Kamionkowski}, K.~{Griest}, {Supersymmetric dark matter},
  \physrep 267 (1996) 195--373.
\newblock \href {http://arxiv.org/abs/hep-ph/9506380}
  {\path{arXiv:hep-ph/9506380}}, \href
  {http://dx.doi.org/10.1016/0370-1573(95)00058-5}
  {\path{doi:10.1016/0370-1573(95)00058-5}}.

\bibitem{1987ApJ...321..571G}
A.~{Gould}, {Resonant enhancements in weakly interacting massive particle
  capture by the earth}, \apj 321 (1987) 571--585.
\newblock \href {http://dx.doi.org/10.1086/165653} {\path{doi:10.1086/165653}}.

\bibitem{1996APh.....6...87L}
J.~D. {Lewin}, P.~F. {Smith}, {Review of mathematics, numerical factors, and
  corrections for dark matter experiments based on elastic nuclear recoil},
  Astroparticle Physics 6 (1996) 87--112.
\newblock \href {http://dx.doi.org/10.1016/S0927-6505(96)00047-3}
  {\path{doi:10.1016/S0927-6505(96)00047-3}}.

\bibitem{1997PhRvC..56..535R}
M.~T. {Ressell}, D.~J. {Dean}, {Spin-dependent neutralino-nucleus scattering
  for A\~{}127 nuclei}, \prc 56 (1997) 535--546.
\newblock \href {http://arxiv.org/abs/hep-ph/9702290}
  {\path{arXiv:hep-ph/9702290}}, \href
  {http://dx.doi.org/10.1103/PhysRevC.56.535}
  {\path{doi:10.1103/PhysRevC.56.535}}.

\bibitem{1988ApJ...328..919G}
A.~{Gould}, {Direct and indirect capture of weakly interacting massive
  particles by the earth}, \apj 328 (1988) 919--939.
\newblock \href {http://dx.doi.org/10.1086/166347} {\path{doi:10.1086/166347}}.

\bibitem{2010PhRvD..82b3534L}
F.~{Ling}, {Is the dark disc contribution to dark matter signals important?},
  \prd 82~(2) (2010) 023534--+.
\newblock \href {http://arxiv.org/abs/0911.2321} {\path{arXiv:0911.2321}},
  \href {http://dx.doi.org/10.1103/PhysRevD.82.023534}
  {\path{doi:10.1103/PhysRevD.82.023534}}.

\bibitem{2008PhRvD..77f5026E}
J.~{Ellis}, K.~A. {Olive}, C.~{Savage}, {Hadronic uncertainties in the elastic
  scattering of supersymmetric dark matter}, \prd 77~(6) (2008) 065026--+.
\newblock \href {http://arxiv.org/abs/0801.3656} {\path{arXiv:0801.3656}},
  \href {http://dx.doi.org/10.1103/PhysRevD.77.065026}
  {\path{doi:10.1103/PhysRevD.77.065026}}.

\bibitem{2001PhRvD..63d5024F}
J.~L. {Feng}, K.~T. {Matchev}, F.~{Wilczek}, {Prospects for indirect detection
  of neutralino dark matter}, \prd 63~(4) (2001) 045024.
\newblock \href {http://arxiv.org/abs/astro-ph/0008115}
  {\path{arXiv:astro-ph/0008115}}, \href
  {http://dx.doi.org/10.1103/PhysRevD.63.045024}
  {\path{doi:10.1103/PhysRevD.63.045024}}.

\bibitem{2002EPJC...26..111B}
V.~{Bertin}, E.~{Nezri}, J.~{Orloff}, {Neutrino indirect detection of
  neutralino dark matter in the CMSSM}, European Physical Journal C 26 (2002)
  111--124.
\newblock \href {http://arxiv.org/abs/hep-ph/0204135}
  {\path{arXiv:hep-ph/0204135}}, \href
  {http://dx.doi.org/10.1140/epjc/s2002-01043-0}
  {\path{doi:10.1140/epjc/s2002-01043-0}}.

\bibitem{2009JCAP...08..034T}
R.~{Trotta}, R.~{Ruiz de Austri}, C.~{P{\'e}rez de los Heros}, {Prospects for
  dark matter detection with IceCube in the context of the CMSSM}, \jcap 8
  (2009) 34.
\newblock \href {http://arxiv.org/abs/0906.0366} {\path{arXiv:0906.0366}},
  \href {http://dx.doi.org/10.1088/1475-7516/2009/08/034}
  {\path{doi:10.1088/1475-7516/2009/08/034}}.

\bibitem{2003PhRvD..67e5003H}
D.~{Hooper}, G.~D. {Kribs}, {Probing Kaluza-Klein dark matter with neutrino
  telescopes}, \prd 67~(5) (2003) 055003.
\newblock \href {http://arxiv.org/abs/hep-ph/0208261}
  {\path{arXiv:hep-ph/0208261}}, \href
  {http://dx.doi.org/10.1103/PhysRevD.67.055003}
  {\path{doi:10.1103/PhysRevD.67.055003}}.

\bibitem{2010JCAP...01..018B}
M.~{Blennow}, H.~{Melb{\'e}us}, T.~{Ohlsson}, {Neutrinos from Kaluza-Klein dark
  matter in the Sun}, \jcap 1 (2010) 18.
\newblock \href {http://arxiv.org/abs/0910.1588} {\path{arXiv:0910.1588}},
  \href {http://dx.doi.org/10.1088/1475-7516/2010/01/018}
  {\path{doi:10.1088/1475-7516/2010/01/018}}.

\bibitem{1991ApJ...382..652S}
D.~{Seckel}, T.~{Stanev}, T.~K. {Gaisser}, {Signatures of cosmic-ray
  interactions on the solar surface}, \apj 382 (1991) 652--666.
\newblock \href {http://dx.doi.org/10.1086/170753} {\path{doi:10.1086/170753}}.

\bibitem{1996PhRvD..54.4385I}
G.~{Ingelman}, M.~{Thunman}, {High energy neutrino production by cosmic ray
  interactions in the Sun}, \prd 54 (1996) 4385--4392.
\newblock \href {http://arxiv.org/abs/hep-ph/9604288}
  {\path{arXiv:hep-ph/9604288}}, \href
  {http://dx.doi.org/10.1103/PhysRevD.54.4385}
  {\path{doi:10.1103/PhysRevD.54.4385}}.

\bibitem{2011ApJ...742...78T}
T.~{Tanaka}, et~al., {An Indirect Search for Weakly Interacting Massive
  Particles in the Sun Using 3109.6 Days of Upward-going Muons in
  Super-Kamiokande}, \apj 742 (2011) 78.
\newblock \href {http://arxiv.org/abs/1108.3384} {\path{arXiv:1108.3384}},
  \href {http://dx.doi.org/10.1088/0004-637X/742/2/78}
  {\path{doi:10.1088/0004-637X/742/2/78}}.

\bibitem{2012PhRvD..85d2002A}
R.~{Abbasi}, et~al., {Multiyear search for dark matter annihilations in the Sun
  with the AMANDA-II and IceCube detectors}, \prd 85~(4) (2012) 042002.
\newblock \href {http://arxiv.org/abs/1112.1840} {\path{arXiv:1112.1840}},
  \href {http://dx.doi.org/10.1103/PhysRevD.85.042002}
  {\path{doi:10.1103/PhysRevD.85.042002}}.

\bibitem{2011PhRvD..84a1102A}
Z.~{Ahmed}, et~al., {Combined limits on WIMPs from the CDMS and EDELWEISS
  experiments}, \prd 84~(1) (2011) 011102.
\newblock \href {http://arxiv.org/abs/1105.3377} {\path{arXiv:1105.3377}},
  \href {http://dx.doi.org/10.1103/PhysRevD.84.011102}
  {\path{doi:10.1103/PhysRevD.84.011102}}.

\bibitem{2011PhRvL.107m1302A}
E.~{Aprile}, et~al., {Dark Matter Results from 100 Live Days of XENON100 Data},
  Physical Review Letters 107~(13) (2011) 131302.
\newblock \href {http://arxiv.org/abs/1104.2549} {\path{arXiv:1104.2549}},
  \href {http://dx.doi.org/10.1103/PhysRevLett.107.131302}
  {\path{doi:10.1103/PhysRevLett.107.131302}}.

\bibitem{2007PhRvL..99i1301L}
H.~S. {Lee}, et~al., {Limits on Interactions between Weakly Interacting Massive
  Particles and Nucleons Obtained with CsI(Tl) Crystal Detectors}, Physical
  Review Letters 99~(9) (2007) 091301.
\newblock \href {http://arxiv.org/abs/0704.0423} {\path{arXiv:0704.0423}},
  \href {http://dx.doi.org/10.1103/PhysRevLett.99.091301}
  {\path{doi:10.1103/PhysRevLett.99.091301}}.

\bibitem{2011PhRvL.106b1303B}
E.~{Behnke}, others., {Improved Limits on Spin-Dependent WIMP-Proton
  Interactions from a Two Liter CF$_{3}$I Bubble Chamber}, Physical Review
  Letters 106~(2) (2011) 021303.
\newblock \href {http://arxiv.org/abs/1008.3518} {\path{arXiv:1008.3518}},
  \href {http://dx.doi.org/10.1103/PhysRevLett.106.021303}
  {\path{doi:10.1103/PhysRevLett.106.021303}}.

\bibitem{2008EPJC..tmp..167B}
R.~{Bernabei}, et~al., {First results from DAMA/LIBRA and the combined results
  with DAMA/NaI}, European Physical Journal C (2008) 167--+\href
  {http://arxiv.org/abs/0804.2741} {\path{arXiv:0804.2741}}, \href
  {http://dx.doi.org/10.1140/epjc/s10052-008-0662-y}
  {\path{doi:10.1140/epjc/s10052-008-0662-y}}.

\bibitem{2009PhLB..682..185A}
S.~{Archambault}, et~al., {Dark matter spin-dependent limits for WIMP
  interactions on $^{19}$F by PICASSO}, Physics Letters B 682 (2009) 185--192.
\newblock \href {http://arxiv.org/abs/0907.0307} {\path{arXiv:0907.0307}},
  \href {http://dx.doi.org/10.1016/j.physletb.2009.11.019}
  {\path{doi:10.1016/j.physletb.2009.11.019}}.

\end{thebibliography}

\end{document}
%
%%%%%%%%%%%%%%%%%%%%%%%%%%%%%%%%%%%%%%%%%%%%%%%%%%%%%%%%%%%%%%%%%%%%%%%%%%%%%%%%%%%%%%%%%%%%%%%%%%%
%%%%%%%%%%%%%%%%%%%%%%%%%%%%%%%%%%%%%%%%%%%%%%%%%%%%%%%%%%%%%%%%%%%%%%%%%%%%%%%%%%%%%%%%%%%%%%%%%%%
%%%%%%%%%%%%%%%%%%%%%%%%%%%%%%%%%%%%%%%%%%%%%%%%%%%%%%%%%%%%%%%%%%%%%%%%%%%%%%%%%%%%%%%%%%%%%%%%%%%
%%%%%%%%%%%%%%%%%%%%%%%%%%%%%%%%%%%%%%%%%%%%%%%%%%%%%%%%%%%%%%%%%%%%%%%%%%%%%%%%%%%%%%%%%%%%%%%%%%%